\RequirePackage{tikz}
\documentclass[sn-mathphys]{sn-jnl}
\pdfoutput=1
\usepackage{hyperref}
\usepackage{amsmath,amssymb,amsthm,amscd}
\usepackage[all,cmtip]{xy}
\usepackage{xcolor}
\usepackage{blkarray}
\usepackage{float}
\usepackage{lscape}
 
\newcommand{\ts}{{\tilde S} }
\newcommand{\be}{\begin{equation}}
\newcommand{\ee}{\end{equation}}
\newcommand{\ut}{\underline t}
\newcommand{\uz}{\underline z}
\newcommand{\uPi}{\underline \Pi}
\newcommand{\mF}{\check Y_{\underline z}}
\newcommand{\mM}{\check Y}

\usepackage{tikz,textcomp}
\usetikzlibrary{decorations.pathreplacing,calc,fadings,fit,shapes,arrows,arrows.meta,positioning,chains,matrix}

\newtheorem{conj}{Conjecture}
\newtheorem{claim}{Claim}

\numberwithin{equation}{section}

\DeclareFontFamily{U}{wncy}{}
\DeclareFontShape{U}{wncy}{m}{n}{<->wncyr10}{}
\DeclareSymbolFont{mcy}{U}{wncy}{m}{n}
\DeclareMathSymbol{\Sh}{\mathord}{mcy}{"58} 

\numberwithin{equation}{section}
\renewcommand{\theequation}{\thesection.\arabic{equation}}

\title{Topological Strings on Non-Commutative Resolutions}

\author[1]{\fnm{Sheldon} \sur{Katz}}\email{katzs@illinois.edu}
\author[2,3]{\fnm{Albrecht} \sur{Klemm}}\email{aklemm@th.physik.uni-bonn.de}
\author[4]{\fnm{Thorsten} \sur{Schimannek}}\email{schimannek@lpthe.jussieu.fr}
\author[5]{\fnm{Eric} \sur{Sharpe}}\email{ersharpe@vt.edu}
\affil[1]{\small\orgdiv{Department of Mathematics}, \orgname{University of Illinois Urbana-Champaign}, \orgaddress{\city{Urbana}, \state{IL}, \postcode{61801}, \country{USA}}}
\affil[2]{\small\orgdiv{Bethe Center for Theoretical Physics}, \orgname{Universit\"at Bonn},\\ \orgaddress{\city{Bonn}, \postcode{D-53115}, \country{Germany}}}
\affil[3]{\small\orgdiv{Hausdorff Center for Mathematics}, \orgname{Universit\"at Bonn},\\ \orgaddress{\city{Bonn}, \postcode{D-53115}, \country{Germany}}}
\affil[4]{\small\orgdiv{Laboratoire de Physique Th\'eorique et Hautes Energies (LPTHE), UMR 7589}, \orgname{CNRS-Sorbonne Universit\'e}, \orgaddress{\street{Campus Pierre et Marie Curie, 4 place Jussieu}, \postcode{F-75005}, \city{Paris}, \country{France}}}
\affil[5]{\small\orgdiv{Department of Physics (MC 0435)}, \orgname{Virginia Tech}, \orgaddress{\street{850 West Campus Drive}, \city{Blacksburg}, \state{VA}, \postcode{24061}, \country{USA}}}

\abstract{
In this paper we propose a definition of torsion refined Gopakumar-Vafa (GV) invariants for Calabi-Yau threefolds with terminal nodal singularities that do not admit K\"ahler crepant resolutions. Physically, the refinement takes into account the charge of five-dimensional BPS states under a discrete gauge symmetry in M-theory.
We propose a mathematical definition of the invariants in terms of the geometry of all non-K\"ahler crepant resolutions taken together.
The invariants are encoded in the A-model topological string partition functions associated to non-commutative (nc) resolutions of the Calabi-Yau.
Our main example will be a singular degeneration of the generic Calabi-Yau double cover of $\mathbb{P}^3$ and leads to an enumerative interpretation of the topological string partition function of a hybrid Landau-Ginzburg model.
Our results generalize a recent physical proposal made in the context of torus fibered Calabi-Yau manifolds by one of the authors and clarify the associated enumerative geometry.

}

\jyear{2022}

\begin{document}
\maketitle

\flushbottom

\tableofcontents

\section{Introduction}
Topological string theory describes BPS protected subsectors of Type II string compactifications, as well as of M-theory, F-theory and other string compactifications that are related via a rich network of dualities~\cite{hori2003mirror,Neitzke:2004ni}.
The theory is usually studied on generically smooth families of Calabi-Yau 3-folds with no torsion in their homology and trivial Neveu-Schwarz B-field background in their large volume limit.
Already here the complexification of the classical K\"ahler parameter by the B-field in string theory leads to the notion of string geometry and string theory behaves very differently on singular geometries than a point particle.

Early  examples appear in~\cite{Aspinwall:1993nu,Witten:1993yc}, where it was argued  solely on the genus zero results that in the complexified K\"ahler moduli space one can pass through classically singular 
geometries, while the instanton corrected correlation functions determining the effective action can remain perfectly smooth.
In the $(2,2)$ supersymmetric gauged linear $\sigma$ model (GLSM) description of the world-sheet theory, this was explained by a second order phase transition in which a non-vanishing $B$ field (a Fayet-Iliopoulos term from the GLSM perspective) protects the correlation function from acquiring singularities~\cite{Witten:1993yc}.  
However, in these examples the B-field is a continuous parameter and in the large volume limit all the singularities of the geometry are resolved.

In this paper we study topological strings on singular Calabi-Yau 3-folds $X$ that do not have a K\"ahler parameter associated to the resolution of the singularities. Instead the choice of B-field becomes quantized and different values correspond to distinct large volume limits.
From a physical perspective, following~\cite{Schimannek:2021pau}, we will use mirror symmetry to study the topological string A-model on backgrounds 
$(\widehat{X},[k])$, where $\widehat{X}$ is a small~\footnote{We will use the terms small- and crepant resolution interchangeably, to refer to a birational resolution $\pi:\widehat{X}\rightarrow X$ of a singular space $X$ that does not affect the canonical class, i.e. $K_{\widehat{X}}=\pi^* K_X$. Similarly, a resolution is large if the K\"ahler class changes.} non-K\"ahler resolution of a nodal K\"ahler Calabi-Yau 3-fold and 
\begin{align}
	[k]\in \text{Tors}\,H^3(\widehat{X},\mathbb{Z})\simeq\mathbb{Z}_N\,,
\end{align}
is the cohomology class of a flat but topologically non-trivial B-field on $\widehat{X}$.
In the following such a B-field will be referred to as \textit{fractional}.
As we will discuss in detail, this background admits an interpretation in terms of two distinct but closely related notions of non-commutative 
resolutions from mathematics, Kuznetsov's \textit{crepant categorical resolutions} (CCR)~\cite{kuznetsov2008lefschetz} 
on the one hand (see e.g.~\cite{thomas-rev} for a review) and van den Bergh's \textit{non-commutative crepant resolutions} (NCCR)~\cite{bondalorlov,vdb02,vdB:nc-rev} on the other.

Recently an enumerative interpretation of the A-model topological string partition function on non-commutative resolutions of certain compact torus fibered Calabi-Yau threefolds with nodal singularities has been proposed by one of the authors~\cite{Schimannek:2021pau}.
We are going to sharpen this proposal, apply it to non-torus fibered Calabi-Yau threefolds and conjecture a precise mathematical definition of the associated invariants in terms of twisted derived categories of non-K\"ahler Calabi-Yau threefolds.

In open string theory, non-commutative geometry generally\footnote{
We should also mention the work \cite{Roggenkamp:2003qp,Roggenkamp:2008jm}
describing spectral triples in conformal field theory,
realizing Connes's ideas \cite{Connes:1994yd} in physics.  Furthermore, beyond D-branes, there is also a question of the interpretation of closed-string sectors.} arises in the presence of a non-vanishing B-field along the worldvolume of a D-brane~\cite{Seiberg:1999vs}.
If the B-field is large, the worldvolume theory of a stack of D-branes is governed by non-commutative Yang-Mills theory~\cite{Connes:1987ue,Connes:1997cr,Douglas:1997fm}. 
On the other hand, in the simplest case of a constant B-field on flat space, one can show that the commutator of the coordinates $x_i(\tau)$ of the open string boundary takes the form~\cite{Schomerus:1999ug,Seiberg:1999vs}
\begin{align}
	[x_i(\tau),x_j(\tau)]=iB_{ij}\,.
	\label{eq:ncdef}
\end{align}
The algebra~\eqref{eq:ncdef} can be interpreted as defining a deformation of $\mathbb{R}^n$ on which the multiplication in the algebra of functions is replaced by the non-commutative Moyal star-product~\cite{Connes:1994yd}.
D-branes on the deformed space are represented by complexes of modules over this non-commutative algebra~\cite{Kapustin:1999di}.

The non-commutative description is valid when the B-field is large relative to the metric~\cite{Seiberg:1999vs}.
In the extreme case, where the volume of a subspace shrinks to zero but the B-field is non-vanishing, this leads to the concept of non-commutative resolutions of singularities~\cite{Berenstein:2000ux,Berenstein:2000jh,Berenstein:2001jr}.
A classical example are compactifications on orbifolds $M/\Gamma$ with discrete torsion~\cite{Vafa:1994rv}, classified by
\begin{align}
	\gamma\in H^2\left(\Gamma,U(1)\right)\,.
\end{align}
The presence of discrete torsion obstructs the deformation of some of the orbifold singularities while at the same time it can not be fully resolved via blow-ups.
Nevertheless, the closed string worldsheet theory is regular while the D-branes see a non-commutative deformation of the singularity~\cite{Douglas:1998xa,Berenstein:2000ux}.
One can in this sense say that discrete torsion ``stabilizes'' some of the singularities.
Moreover, on the smooth locus of $M/\Gamma$, a non-zero choice of discrete torsion maps to the cohomology class of a flat topologically non-trivial B-field.
This suggests the interpretation of discrete torsion as a fractional B-field on the entire orbifold $M/\Gamma$~\cite{Sharpe:2000ki}.

It has already been observed in~\cite{Vafa:1994rv} that orbifolds with discrete torsion can admit complex structure deformations to Calabi-Yau manifolds $X$ with conifold singularities that are similarly stabilized due to the presence of a B-field.
The corresponding worldsheet theories can, at least in principle, be interpreted and studied as infrared fixed points of non-linear sigma models on a large resolution $\tilde{X}$~\cite{Aspinwall:1995rb}, replacing each node by $\mathbb{P}^1\times\mathbb{P}^1$.
The large resolution $\tilde{X}$ is not Calabi-Yau but, in the example studied in~\cite{Aspinwall:1995rb}, exhibits torsional 3-cohomology and therefore supports a fractional B-field.

There also exist analytic small resolutions $\widehat{X}$ that are Calabi-Yau, in the sense that their first Chern class vanishes, but are not K\"ahler.
The small resolutions have the same torsion in cohomology as $\tilde{X}$, thus equally support a fractional B-field, and it has been suggested that they flow to the same string sigma model~\cite{caldararuThesis}.
This expectation was motivated by the observation that for certain examples the category of topological B-branes on $\widehat{X}$ in the presence of a fractional B-field $[k]\in H^3(\widehat{X},\mathbb{Z})$
is derived equivalent to the category of topological B-branes on some different smooth Calabi-Yau 3-fold $Y$ without a B-field.
Mathematically speaking, this amounts to a so-called twisted derived equivalence
\begin{align}
	D^b(\widehat{X},[k])\simeq D^b(Y)\,,
	\label{eqn:twistedeq}
\end{align}
and, at least in all examples that we are aware of, $\widehat{X}$ admits an interpretation as a coarse moduli space of sheaves on $Y$~\cite{caldararu2001derived,addington2009derived}.

The sigma model on $(\widehat{X},[k])$ is expected to flow to a sigma model on the singular Calabi-Yau $X$ itself, that is stabilized by a corresponding B-field background.
We will argue that this infrared fixed point can naturally be interpreted as a non-commutative resolution $X_{\text{n.c.}k}$ of $X$.
Since the non-K\"ahler volume of the exceptional curves, but not the B-field holonomy, decouples from the topological A-model, we will usually not distinguish between $(\widehat{X},[k])$ and $X_{\text{n.c.}k}$ and call $(\widehat{X},[k])$ itself a non-commutative resolution of $X$.
 In the absence of a B-field, the decoupling implies that $X_{\text{n.c.}0}=X$.
This provides a unified perspective on several mathematically rather different notions of non-commutative resolutions and in particular suggests a local interpretation in terms of the non-commutative conifold studied e.g. in~\cite{szendrHoi2008non}.

Let us point out that strictly speaking, actual non-commutativity only arises in the open string sector, although the closed string sector is also non-singular in the presence of the B-field.
We will use the term non-commutative resolution to refer to a singular Calabi-Yau background with a fractional B-field, independent of the type of strings that are being considered.

A first class of examples arises from the IR fixed points of Landau-Ginzburg $\mathbb{Z}_2$-orbifolds with quadratic superpotentials that are fibered over Fano threefolds~\cite{Caldararu:2010ljp}.
These models naturally correspond to non-commutative resolutions of singular double covers in terms of sheaves of Clifford algebras on the base which encode the endomorphisms of 0-branes~\cite{buchweitz-eisenbud-herzog,Kapustin:2002bi,dyckerhoff-cg,teleman-mf}, \cite[chapter 14]{yoshino}.
In mathematics, such non-commutative spaces appeared in the context of homological projective duality~\cite{kuz0,kuznetsov2008derived} and are expected to fit into the framework of crepant categorical resolutions~\cite{kuzso}.
Nevertheless, we will argue that these backgrounds can also be interpreted in terms of small analytic resolutions $\widehat{X}$ with a fractional B-field $[1]\in H^3(\widehat{X},\mathbb{Z})_{\text{tors.}}=\mathbb{Z}_2$ such that $(\widehat{X},[1])\simeq X_{\text{n.c.}1}$. 
A combinatorial construction of these non-commutative resolutions was developed in~\cite{borisovzhan} and leads to a rich class of geometries with duals $Y$ that are complete intersections in toric ambient spaces.
More general examples have been obtained from GLSMs with a non-Abelian gauge group and are twisted derived equivalent to Pfaffian Calabi-Yaus~\cite{Hori:2013gga}.
We refer to the non-commutative resolutions that arise in this context as being of \textit{Clifford type}.

A second class of examples arises in the context of torus fibrations.
The elliptic Jacobian fibration $X=J(Y)$ that is associated to a smooth genus one fibered Calabi-Yau 3-fold $Y$ without a section is generically singular.
However, a small analytic resolution $\widehat{X}$ of $X$ again exhibits torsion 3-cocycles and is twisted derived equivalent to $Y$~\cite{caldararu2001derived}.
It was recently discovered that the corresponding non-commutative resolutions $X_{\text{n.c.}k}$ are also realized as large volume limits in the stringy K\"ahler moduli spaces of $Y$ and of other torus fibered Calabi-Yau 3-fold that share the same Jacobian~\cite{Schimannek:2021pau}.
The torsion that can arise in these examples is more general and includes cases with $\mathbb{Z}_n$ for $n=2,\ldots,5$.
Moreover, the presence of the singularities as well as of a fractional B-field along the exceptional curves in $\widehat{X}$ can be explicitly demonstrated using extremal transitions. On the other hand, finding an explicit realization of the non-commutative resolution in terms of a sheaf of non-commutative algebras is mostly an open problem and we refer to those resolutions as being of \textit{general type}.
The only exceptions in this class are given by certain singular genus one fibrations with a 2-section that, as we discuss in Section~\ref{sec:ellfibrations}, are also singular double covers of uniruled  threefolds (i.e. $\mathbb{P}^1$ bundles) and admit a Clifford type non-commutative resolution.

In all of the previously mentioned examples, there is a twisted derived equivalence between the non-commutative resolution $X_{\text{n.c.}k}$ and some dual smooth Calabi-Yau 3-fold $Y$, that is physically realized by brane transport along paths between inequivalent large volume limits in the stringy K\"ahler moduli space.
This implies that both $X_{\text{n.c.}k}$ and $Y$ share the same mirror, which itself is in general an ordinary smooth Calabi-Yau.
The usual machinery of mirror symmetry can then be applied to $Y$ and, via analytic continuation, also be used to study topological strings on $X_{\text{n.c.}k}$~\cite{Sharpe:2012ji,Schimannek:2021pau}.
In particular, standard techniques (see e.g.~\cite{CANDELAS199121,Hosono:1993qy,Hosono:1994ax,Huang:2006hq}) can in principle be used to calculate the A-model topological string free energies $F_g(\underline{t})$ associated to $X_{\text{n.c.}k}$ as a function of the complexified K\"ahler parameters $\underline{t}$.
However, the question arises if there is an enumerative interpretation for the coefficients in the instanton expansion.

On an ordinary smooth Calabi-Yau threefold, the GV-GW correspondence relates the topological string amplitudes, which can mathematically defined as generating functions of rational Gromov-Witten (GW) invariants, to integer Gopakumar-Vafa (GV) invariants $n^\beta_g\in\mathbb{Z}$.
From a physical perspective, integrality follows from the definition of GV-invariants as traces of multiplicities of BPS states in a five dimensional M-theory compactification~\cite{Gopakumar:1998ii,Gopakumar:1998jq}.
The invariants depend on the class $\beta\in H_2(X)$ of a curve in the Calabi-Yau, physically corresponding to the electric $U(1)^{b_2(X)}$ charge of the BPS particles, and the genus $g\in\mathbb{N}$, that determines the representation under the M-theory $SU(2)_L\subset SO(4)$ little group and is related to the genus of the corresponding topological string string amplitude.
When defined via this correspondence in terms of GW invariants, it is by now proven that the GV-invariants are indeed integers~\cite{gvintegral} and further satisfy the so-called Castelnuovo vanishing condition, which posits the existence of a function $g_{\text{max}}(\beta)$ such that $n^\beta_g=0$ for $g>g_{\text{max}}(\beta)$~\cite{gvfinite}.  A mathematical formula for the GV-invariants for genera that are very close to this Castelnuovo bound was given in~\cite{Katz:1999xq} and can be used to explicitly calculate such invariants in many cases. 
Finding a general mathematical definition of GV-invariants has proven to be a difficult problem but significant progress has been made in recent years.
In particular, a definition in terms of the cohomology of a certain perverse sheaf on the moduli space of stable sheaves supported on a curve was proposed in~\cite{MT}.\footnote{The consistency of the more general mathematical definition with the formulae of~\cite{Katz:1999xq} was proven in \cite{Zhao} for local $\mathbb{P}^2$, but the argument applies more generally.}

Using duality with F-theory, it was argued in~\cite{Schimannek:2021pau} that for a large class of nodal torus fibered Calabi-Yau threefolds $X$, the topological string partition function
\begin{align}
\begin{split}
	Z_{\text{top.},k}(\underline{t},\lambda)=\exp\left(\sum\limits_{g=0}^\infty\lambda^{2g-2}F^{{\rm top}.,k}_g(\underline{t})\right)\,,
	\end{split}
	\label{eq:ZTSdef} 
\end{align}
on $X_{\text{n.c.}k}$, where $\lambda$ is the topological string coupling, encodes torsion refinements $n_g^{\beta,\ell}$ of the usual Gopakumar-Vafa invariants from~\cite{Gopakumar:1998ii,Gopakumar:1998jq}. These integer invariants are labelled not only by the genus $g$ and a $U(1)^{b_2(X)}$ charge $\beta\in H_2(X,\mathbb{Z})$, but also by a discrete charge $\ell\in\{0,\ldots,N-1\}$ under a $\text{Tors}\,H^3(\widehat{X},\mathbb{Z})\simeq\mathbb{Z}_N$ gauge symmetry which M-theory develops on the singular Calabi-Yau $X$.
More precisely, it was proposed that the topological string partition function takes the form
\begin{align}
\begin{split}
	&\log\left[Z_{\text{top.},k}\left(\underline{t},\lambda\right)\right]\\=&\sum\limits_{g=0}^\infty\sum\limits_{\beta\in H_2(X)}\sum\limits_{l=0}^{N-1}\sum\limits_{m=1}^\infty n_{g}^{\beta,l}\cdot\frac{1}{m}\left(2\sin\frac{m\lambda}{2}\right)^{2g-2}e^{2\pi i mlk/N}q^{m\beta}\,,
\end{split}
\label{eqn:torsiongv}
\end{align}
where $q^\beta=e^{2\pi i \underline{t}\cdot\beta}$.
Knowledge of $Z_{\text{top.},k}$ for all $k=0,\ldots N-1$ then enables one to extract the individual integer invariants $n_g^{\beta,l}$.  By the universal coefficient theorem $\text{Tors}\,H^3(\widehat{X},\mathbb{Z})\simeq\text{Tors}\,H_2(\widehat{X},\mathbb{Z})$, and the full M-theory charge lattice can be identified with $H_2(\widehat{X},\mathbb{Z})$.
In the ordinary smooth case, $k=0$ and the phase is trivial such that~\eqref{eqn:torsiongv} reduces to the usual GV-GW correspondence.

Note that the appearance of the fractional B-field as a phase in the Type IIA instanton action was already anticipated in~\cite{Aspinwall:1995rb}, before the discovery of Gopakumar-Vafa invariants, as well as for the Gopakumar-Vafa expansion on smooth Calabi-Yau 3-folds with torsion homology~\cite{Dedushenko:2014nya}, but had only been applied in one example to genus zero Gromov-Witten invariants in the presence of torsion classes on a smooth Calabi-Yau 3-fold~\cite{Braun:2007tp,Braun:2007xh,Braun:2007vy}.
On the other hand, torsion-refined Gopakumar-Vafa invariants for many singular torus-fibered examples have been calculated in~\cite{Schimannek:2021pau}.
Particularly interesting is the case $N=5$, where $X_{\text{n.c.}k}$ with $k=1,\ldots 4$ lead to free energies with irrational coefficients that take values in $\mathbb{Q}[\sqrt{5}]$.
 Only after the contributions from all the large volume limits associated to $X_{\text{n.c.}k},\,k=0,\ldots,4$ are correctly combined one finds integer invariants $n^{\beta,l}_g$.
From an M-theory perspective, the phase in~\eqref{eqn:torsiongv} has recently been interpreted in terms of a discrete holonomy along the M-theory circle~\cite{Dierigl:2022zll}.

In this paper, we will propose a mathematical definition of the refined Gopakumar-Vafa invariants in terms of the derived categories $D^b(\widehat{X})$ of any small non-K\"ahler resolution $\widehat{X}$.
A striking implication will be that the invariants do not capture the geometry of one particular small non-K\"ahler resolution but, in a sense that will be made precise, see all of the possible resolutions simultaneously.
We will also derive closed expressions for the constant map contributions to the topological string partition function on $X_{\text{n.c.}k}$, using the non-commutative conifold as a local model around the singularities and specializing the corresponding Donaldson-Thomas partition function from~\cite{szendrHoi2008non} to fractional values of the B-field.

Our main example will be the \textit{singular octic double solid} $X$, a singular Calabi-Yau double coverof $\mathbb{P}^3$ that is ramified over the vanishing locus of the determinant of an $8\times 8$ matrix $A_{8\times 8}(\vec{x})$ with entries that are linear in the homogeneous coordinates.
The geometry has $84$ conifold singularities that do not admit a crepant K\"ahler resolution.
The dual Calabi-Yau $Y=X^{(1,65)}_{2222}(1^8)$ is a complete intersection of four quadrics in $\mathbb{P}^7$~\cite{addington2009derived} while the smooth deformation $X_{\text{def.}}=X^{(1,149)}_8(11114)$ of $X$ is a generic octic hypersurface in $\mathbb{P}(1,1,1,1,4)$.
A non-commutative resolution $X_{\text{n.c.}1}$ of $X$ in terms of a sheaf of Clifford algebras on $\mathbb{P}^3$ was constructed explicitly in~\cite{kuz0} and the GLSM description as well as D0-brane probes were studied in~\cite{Caldararu:2010ljp,Addington:2012zv,Sharpe:2012ji}.
We will argue that $X_{\text{n.c.}1}$ can be equivalently interpreted in terms of a non-K\"ahler resolution $\widehat{X}$ of $X$ with a fractional B-field
\begin{align}
[1]\in H^3(\widehat{X},\mathbb{Z})_{\text{tors.}}\simeq\mathbb{Z}_2\,.
\end{align}

We use mirror symmetry and analytic continuation from the large complex structure limit that is mirror to the dual Calabi-Yau $Y$ to study the topological string A-model on $X_{\text{n.c.}1}$.
Directly integrating the holomorphic anomaly equations~\cite{Bershadsky:1993cx} we then calculate the free energies on $X_{\text{n.c.}1}$.
We then apply the proposal from~\cite{Schimannek:2021pau}, taking into account the free energies associated to the smooth deformation $X_{\text{n.c.}0}=X_{\text{def.}}$ from~\cite{Huang:2006hq}, to obtain the $\mathbb{Z}_2$ refined Gopakumar-Vafa invariants associated to the singular Calabi-Yau $X$.
Following the analysis in~\cite{Huang:2006hq}, the usual boundary behaviour of the free energies and the Castelnuovo vanishing for $X^{(1,65)}_{2222}(1^8)$ can be used to fix the holomorphic ambiguity up to genus $17$.
However, we find that the $\mathbb{Z}_2$-refined GV-invariants also exhibit Castelnuovo like vanishing and this allows us to explicitly carry out the direct integration and fix the holomorphic ambiguity, that arises at each genus in the direct integration method, up to genus $32$.
In many cases we can calculate the prediction from our proposed enumerative definition of the invariants and find exact agreement with the results from mirror symmetry.

The outline of the paper is as follows.
In Section~\ref{sec:ncconi} we review the mathematical notions of non-commutative crepant resolutions and crepant categorical resolutions as well as the non-commutative conifold.
We relate this to our notion of non-commutative resolution in terms of a fractional B-field and provide an overview of the examples that have appeared in the literature and that we study in this paper.
Section~\ref{sect:prop} is the heart of the paper where we discuss the physical notion of torsion refined GV-invariants and propose a mathematical definition. We also review their relation to the A-model topological string partition functions on non-commutative resolutions and derive closed expressions for the constant map contributions.
In Section~\ref{sec:octic} we discuss in detail the geometry of our main example, the singular octic double solid, a nodal Calabi-Yau double cover of $\mathbb{P}^3$ that is ramified over a determinantal octic.
We also give a brief introduction to the Brauer group, twisted sheaves and Azumaya algebras and review how a sheaf of non-commutative algebras arises from a physical perspective, by considering D0-branes in a hybrid Landau-Ginzburg models with quadratic superpotential.
In Section~\ref{sect:octicdoub:b-model} we use mirror symmetry and integrate the holomorphic anomaly equations to calculate $\mathbb{Z}_2$-refined GV-invariants for the singular octic double.
We also discuss the behaviour of the topological B-branes under monodromies and homological projective duality.
Section~\ref{sec:ellfibrations} briefly reviews a class of torus fibered examples and also elaborates on the relation of the fractional B-field to discrete torsion.
In Section~\ref{sec:geometric} we apply our proposed mathematical definition of the torsion refined GV-invariants to various examples and find perfect agreement with the predictions from mirror symmetry.
Section~\ref{sec:outlook} outlines open question that we would like to address in future work.

Some additional aspects are relegated to appendices.
Appendix~\ref{app:brauer} contains the derivation of the Brauer group for a small analytic resolution of the singular octic double solid.
Appendix~\ref{app:algebraicKahlergauge} summarizes the direct integration procedure and the transformation of the propagator ambiguities under K\"ahler transformations.
Appendix~\ref{app:otherexamples} summarizes four additional examples that arise as duals of Pfaffian Calabi-Yau threefolds.
In Appendix~\ref{app:ctgv} we list some of the $\mathbb{Z}_2$-refined GV-invariants for a torus fibered example.

The torsion refined Gopakumar-Vafa invariants for the singular octic double solid, as well as the corresponding topological string free energies for the non-commutative resolution, the generic degree eight hypersurface and the dual complete intersection of four quadrics are provided online~\cite{datalink}.
\section{Non-commutative resolutions and the conifold}
\label{sec:ncconi}
This section has two goals.  The first goal is to introduce the various notions of non-commutative resolution that will appear in this paper.
From the mathematical side, the two relevant notrions are non-commutative crepant resolutions (NCCRs) as discussed in 
e.g.~\cite{vdb02,vdb02-3d,vdB:nc-rev},
and crepant categorical resolutions (CCRs) as discussed in 
e.g.~\cite{kuznetsov2008derived,kuz0,kuzgrass,kuzso,perry1,kuzp1,kuzp2,kuzp3,kuz4,kuznetsov2008lefschetz}.
In Section~\ref{sec:geooutline} we will then clarify our use of the term non-commutative resolution as referring to a singular Calabi-Yau with a fractional B-field and also provide an overview of the different examples that have appeared in the literature and that are covered by our proposal, some of which we are then studying in more detail in the rest of the paper.

The second goal is to review how these different notions can be applied to the conifold singularity
\begin{align}
	\{xy-zw=0\}\subset \mathbb{C}^4\,,
\end{align}
and naturally arise from physics.
In fact, they are related by a duality and we therefore collectively refer to them as the non-commutative conifold.
However, they are different from the two ordinary small resolutions $A_+,A_-$ that are in turn both isomorphic to
\begin{align}
    \mathcal{O}(-1)\oplus\mathcal{O}(-1)\rightarrow\mathbb{P}^1\,,
\end{align}
and related by a flop.
As we will argue in Section~\ref{sect:fbnc}, the non-commutative conifold appears to describe the local geometry around  non-commutatively resolved nodes in compact Calabi-Yaus.

\subsection{Non-commutative crepant resolutions}  \label{subsec:nccr}
The first type of resolution that we discuss arises in physics from worldvolume theories of D-branes on singularities that admit a description in terms of a quiver gauge theory~\cite{Douglas:1996sw}.
The nodes of the quiver $Q$ correspond to factors of the gauge group while arrows encode bi-fundamental matter.
The theory is also equipped with a superpotential $W$.
Vacua of the theory, and therefore stable brane configurations, correspond to (stable) representations of a certain non-commutative algebra, the Jacobi algebra associated to $(Q,W)$.

Mathematically, this is a special case of the more general notion of a non-commutative crepant resolution (NCCR). The mathematical definition in \cite{vdb02} is local and defined for an affine variety $X=\operatorname{Spec}(R)$, where $R$ is a normal Gorenstein domain (which simply means that $X$ is reduced and irreducible, and has a well-defined canonical bundle $K_X$):\vspace{.2cm}
\begin{quotation}
A \emph{non-commutative crepant resolution} of $R$ is a 
homologically homogenous $R$-algebra of the form $A=\operatorname{End}_R(M)$ where $M$ is a
reflexive $R$-module.
\end{quotation}\vspace{.2cm}
Recall that an $R$-algebra $A$ is homologically homogeneous if all simple $A$-modules have projective dimension $n=\operatorname{dim}(R)$
and an $R$-module $M$ is reflexive if the natural map $M\to\operatorname{Hom}_R(\operatorname{Hom}_R(M,R),R)$ is an isomorphism of $R$-modules.
The first part of the definition is motivated by the fact that an irreducible algebraic variety is non-singular exactly if the coordinate ring is homologically homogeneous.
Similarly, reflexivity should be thought of as generalizing the notion of crepancy in this context.
For introductions we refer to~\cite{vdb02,Leuschke+2012+293+364,wemyss,vdB:nc-rev}.

In the case of the conifold
\begin{align}
        \{xy-zw=0\}\subset \mathbb{C}^4\,,
        \label{eqn:conifold}
\end{align}
$R=\mathbb{C}[x,y,w,z]/(xy-wz)$ and the NCCR of the conifold is defined
by the ring \cite{vdb02,vdb02-3d}
\begin{equation}
A \: = \: {\rm End}_R(R \oplus I) \: = \:
\left[ \begin{array}{ll} R & I \\ I^{-1} & R \end{array} \right].
\end{equation}
This algebra coincides with the Jacobi algebra of a two-node quiver,
with (Klebanov-Witten) superpotential
\begin{align}
        W \: = \: \frac{1}{2} \epsilon^{ij} \epsilon^{k\ell}
A_i B_k A_j B_{\ell},   \label{eq:conifold:W}
\end{align}
as in \cite[equ'n (17)]{Klebanov:1998hh}, and it is in this form that this algebra arises in worldvolumes of D-branes at conifolds \cite{Klebanov:1998hh}. 

It was noted by Szendr{\H{o}}i in~\cite{szendrHoi2008non} that the NCCR of the conifold is also realized in the stringy K\"ahler moduli space of the resolved conifold and that there is a well defined notion of Donaldson-Thomas invariants as well as a corresponding topological string partition function.
The stringy K\"ahler moduli space is a three-punctured sphere
\begin{align}
	\mathcal{M}_{\text{K}}=\mathbb{P}^1\backslash\{0,1,\infty\}\,,
	\label{eqn:conikaehler}
\end{align}
that is parametrized by $z=e^{2\pi i t}$, with $t=b+iv$ being the complexified volume of the exceptional 2-cycle.
The two small projective resolutions $A_\pm$ respectively correspond to the limits $t\rightarrow \pm i\infty$.
At the conifold point $t=0$ the exceptional curve has zero volume and the conifold singularity is restored.
However, turning on a non-trivial B-field $t\in\mathbb{R}_{\ne 0}$ regularizes the string worldsheet theory without affecting the classical volume. 

The only non-vanishing Gopakumar-Vafa invariants on either resolution $A_\pm$ are $n_{g=0}^{d=1}=1$, corresponding to the $\mathbb{P}^1$ base, and the constant map contribution $n_{g=0}^{d=0}=-2$, given by minus the Euler characteristic of the resolved conifold~\cite{Gopakumar:1998ki}.
The topological string partition function in the large volume limits $\text{Im}(t)\rightarrow\pm i\infty$ then takes the form
\begin{align}
	Z_{A_\pm}(t,\lambda)=\prod\limits_{d=0,1}\prod\limits_{k=1}^\infty\left(1-e^{-k\lambda}e^{\pm 2\pi i d t}\right)^{k n_{0}^d}\,,
\end{align}
and the Gopakumar-Vafa invariants can be interpreted in terms of Donaldson-Thomas invariants on $A_\pm$.
However, on the equator $\text{Im}(t)=0$ the partition function becomes
\begin{align}
	Z_{A}(t,\lambda)=\prod\limits_{k=1}^\infty\left[\left(1-e^{-k\lambda}\right)^{-2}\left(1-e^{-k\lambda}e^{2\pi i t}\right)\left(1-e^{-k\lambda}e^{-2\pi i t}\right)\right]^k\,,
	\label{eqn:coniztop}
\end{align}
and encodes non-commutative Donaldson-Thomas invariants associated to the conifold quiver, essentially counting semistable representations of the non-commutative algebra $A$~\cite{szendrHoi2008non,MORRISON20122065}.
Formally, the non-commutative partition function $Z_A$ appears to count the compact curves of both $A_+$ and $A_-$ simultaneously.
Indeed one can check that there are representations of the algebra which correspond to branes that probe either of the associated exceptional curves.

There is a natural extension of the notion of a NCCR to compact projective varieties $X$.  If $X$ admits a projective small resolution, then an NCCR of $X$ exists globally \cite{vdb02-3d}.
However, it is not clear whether NCCRs always exist for projective varieties with terminal singularities.
Nevertheless, we will argue in Section~\ref{sect:fbnc} that the non-commutative conifold is at least a good local model for non-commutative resolutions of certain nodal compact Calabi-Yau 3-folds.

The relationship of NCCRs to orbifolds with discrete torsion has been discussed in~\cite{Berenstein:2001jr}, while a systematic construction of the quiver with potential from matrix factorizations of the singularity was described in~\cite{Aspinwall:2010mw}.
For a more detailed discussion of the non-commutative Donaldson-Thomas invariants of the conifold and other local Calabi-Yau singularities we refer e.g. to~\cite{Mozgovoy:2020has} and the references therein.

\subsection{Crepant categorical resolutions}
\label{subsec:CCR}   \label{sect:probes}

The second relevant class of ``non-commutative'' resolutions are the categorical resolutions, which we outline next.
Examples naturally arise in physics as infrared fixed points of hybrid gauged linear sigma models, in which a Landau-Ginzburg $\mathbb{Z}_2$-orbifold with quadratic superpotential is fibered over a Fano threefold base~\cite{Aspinwall:1995rb,Caldararu:2010ljp}.

The mathematical definition is slightly more involved. Let $X$ be any singular algebraic variety, and let $D^b(X)$ be its derived category.   We also let $D^b(X)^{\mathrm{perf}}\subset D^b(X)$ denote the subcategory of complexes of sheaves which can be represented by a finite complex of vector bundles.\footnote{On a smooth variety, $D^b(X)^{\mathrm{perf}}= D^b(X)$ but this is not true for general $X$.}   Following \cite{kuznetsov2008lefschetz}, a \emph{categorical resolution} of $D^b(X)$ is a smooth triangulated category $\tilde{\mathcal{D}}$ and a pair of adjoint functors
\begin{equation}
\pi_*:\tilde{\mathcal{D}} \to D^b(X),\qquad \pi^*:D^b(X)^{\mathrm{perf}}\to \tilde{\mathcal{D}}, 
\end{equation}
$\pi^*$ being the left adjoint of $\pi_*$,
such that the natural adjunction morphism $\mathrm{id}_{D^b(X)}^{\mathrm{perf}} \to \pi_*\pi^*$ is an isomorphism of functors.
A categorical resolution is \emph{crepant} if $\pi^*$ is also the right adjoint of $\pi_*$.

The justification for the terminology is that if $\pi:\tilde{X}\to X$ is an ordinary resolution of singularities, then $\mathcal{D}=D^b(\tilde{X})$ together with the usual maps $\pi_*,\pi^*$ is a categorical resolution of $D^b(X)$.   If $\pi$ is crepant, then $D^b(\tilde{X})$ is a crepant categorical resolution.
A common situation for us is when $X$ has a sheaf $\mathcal{C}$ of non-commutative $\mathcal{O}_X$-algebras such that the derived category $D^b(X,\mathcal{C})$ of coherent sheaves of $\mathcal{C}$-modules is a crepant categorical resolution of $D^b(X)$.

While there are competing notions of the smoothness of $\mathcal{D}$ in the literature, we ignore these issues here and simply assert that the smoothness condition can be checked whenever we state that we have a crepant categorical resolution.

It is often difficult to prove that a triangulated category is a crepant categorical resolution, and many of the expected examples are still conjectural.  Following~\cite{kuzso}, in which it is stated that $(X,\mathcal{C})$ can be thought of as a crepant categorical resolution, we will refer to this example as a crepant categorical resolution or nc-resolution of $X$, even though this has not been proven to our knowledge.

The derived category of an NCCR is conjectured to also be a crepant categorical resolution.  The conjecture is proven if $\text{dim}\, X\le 3$ \cite{vdB:nc-rev}, and therefore holds for the non-commutative conifold discussed above.
We emphasize however, that the crucial feature for us in a crepant categorical resolution is the category of branes. 
In principle we could have the same category of branes corresponding to different nc-resolutions.

Crepant categorical resolutions naturally arise as hybrid phases in GLSMs, see
e.g.~\cite{Caldararu:2010ljp,Sharpe:2012ji,Addington:2012zv,Hori:2011pd,Halverson:2013eua,Sharpe:2013bwa,Sharpe:2010iv,Sharpe:2010zz,Ballard:2013fxa,Hori:2013gga,Hori:2016txh,Wong:2017cqs,Kapustka:2017jyt,Parsian:2018fhm,Chen:2018qww,Chen:2020iyo,Guo:2021aqj}, essentially
as a prediction of homological projective duality 
\cite{kuznetsov2008derived,kuz0,kuzgrass,kuzso,perry1,kuzp1,kuzp2,kuzp3,kuz4,kuznetsov2008lefschetz}.
The basic pattern \cite{Caldararu:2010ljp} 
is that in some GLSMs, one has phases given by
hybrid Landau-Ginzburg models in which, generically on the base $B$,
there is a ${\mathbb Z}_2$ one-form symmetry.  The presence of such a symmetry
implies a decomposition \cite{Hellerman:2006zs} into two universes, 
which in this context
means that the hybrid Landau-Ginzburg model RG flows to a double cover,
branched over the locus where the ${\mathbb Z}_2$ one-form symmetry
is broken.

In such phases, one often finds apparent singularities in the target space while the GLSM is perfectly regular, as can be verified by the absence of a non-compact Coulomb-branch.
This implies that the CFT in the IR limit is seeing some sort of
resolution.
A detailed analysis of the branes in the theory, which can be described as families of equivariant matrix factorizations over $B$, demonstrates that they are modules over a sheaf of Clifford-algebras
${\cal B}$ on $B$ (of derived automorphisms of D0-branes)~\cite{buchweitz-eisenbud-herzog,Kapustin:2002bi,dyckerhoff-cg,Caldararu:2010ljp,teleman-mf}, \cite[chapter 14]{yoshino}.
The pair $(B,\mathcal{B})$ then conjecturally defines the categorical resolution.
We will discuss this in greater detail in Section~\ref{sect:octicdoub:nc-res:glsm}.

\subsection{Non-commutative conifolds from GLSMs}
\label{sec:ncconifromglsm}
Both a crepant categorical resolution as well as the NCCR of the conifold naturally arise in GLSMs and their equivalence can be seen as a special case of a Seiberg-like duality~\cite{Hori:2011pd}.
While the former description can be used to explicitly construct matrix factorizations associated to 0-branes that probe the exceptional curves from both small resolutions, the relation to the latter demonstrates the presence of the fractional B-field.

A GLSM that realizes a categorical resolution of the conifold in a hybrid phase has been described in~\cite[section 2.4]{Hori:2011pd}.
Briefly, this is a 
$U(1)$ GLSM
with six fields of charges
\begin{center}
\begin{tabular}{cccccc}
$p$ & $x$ & $y$ & $a$ & $b$ & $c$ \\ \hline
-2 & 1 & 1 & 0 & 0 & 0
\end{tabular}\,,
\end{center}
with superpotential
\begin{equation}  \label{eq:glsm:hori-conifold:W}
W \: = \: p\left( a x^2 + 2 c x y + b y^2 \right)\,.
\end{equation}
The phase with FI-parameter $r \gg 0$ describes an ordinary small resolution of the conifold;
the $r \ll 0$ phase, for the same reasons as in \cite{Caldararu:2010ljp},
describes a categorical resolution of a branched double cover of 
${\mathbb C}^3 = {\rm Spec}\,{\mathbb C}[a,b,c]$, branched over
the locus
\begin{equation}
\{ c^2 - ab = 0 \} \: \subset \: {\mathbb C}^3\,.
\end{equation}
The hybrid point exhibits a $\mathbb{Z}_2$-quantum symmetry that acts by exchanging the two sheets of the cover.
The matrix factorizations that correspond to 0-branes have been analyzed in~\cite[section 5.1]{Addington:2012zv}.
Away from the conifold singularity $a=b=c=0$, the branes indeed see the branched double cover.
However, at the singularity itself one can construct two $\mathbb{P}^1$ families of branes which can be interpreted as 0-branes that probe the exceptional curves in each of the two small resolutions.

On the other hand, the typical Atiyah flop can be described by a $U(1)$ GLSM with four fields of charges
\begin{center}
\begin{tabular}{cccccc}
$x_1$&$x_2$&$x_3$&$x_4$ \\ \hline
-1 &-1 & 1 & 1
\end{tabular}\,,
\end{center}
and with vanishing superpotential~\cite[section 5.5]{Witten:1993yc}.
The complexified K\"ahler parameter is not renormalized and can be directly expressed as
\begin{align}
    t=\theta+ir\,,
\end{align}
in terms of the FI-parameter $r$ and the theta angle $\theta$.
Following the discussion in Section~\ref{subsec:nccr}, the NCCR of the conifold is realized at the values $t\in\mathbb{R}_{\ne 0}$.
It was argued in~\cite{Hori:2011pd}, that the hybrid phase of the first GLSM corresponds in fact to the value $t=1/2$.
This can be seen utilizing  a Seiberg-like duality~\cite[section 4.2]{Hori:2011pd}.
It also follows by observing that $t=1/2$ is the only smooth point with the
correct symmetries, namely a ${\mathbb Z}_2$ action $t\rightarrow -t$ which is identified with the quantum ${\mathbb Z}_2$ of the LG orbifold
at the $r \ll 0$ limit~\cite[section 2.4]{Hori:2011pd}.

\subsection{Non-commutative resolutions of compact Calabi-Yau}
\label{sec:geooutline}

\begin{figure}[t!]
\centering
\begin{tikzpicture}[remember picture,scale=.6, every node/.style={scale=0.6},node distance=4mm, >=latex', ]
\begin{scope}[shift={(0,0)},scale=1.2]
	\node[align=center] at (7.2,4) {$X_{\text{def.}}$\\[.3em]$G_{5d}=U(1)^{r}$};
	\node[align=center] at (12,4) {$X$\\[.3em]$G_{5d}=U(1)^{r}\times\mathbb{Z}_N$};
	\node[align=center] at (12,1) {$X_{\text{n.c.}k}\simeq D^b(\widehat{X},k\alpha)$};
	\node[align=center] at (15.8,1) {$Y$};
 {degeneration} ++ (-2.2,0) ;
	\draw[<-,dashed,black] (8.3,4.2) to (10.5,4.2);
	\node[align=center] at (9.4,4.73) {cplx. structure\\deformation};
	\node[align=left] at (14.8,2.5) {$\frac{k}{N}$ B-field along exceptional curves};
	\node[align=left] at (18.1,1) {Smooth CY 3-fold,\\derived equivalent to $X_{\text{n.c.}k}$};
	\node[align=left] at (16.8,4) {CY 3-fold with \textbf{terminal nodal singularities}\\that can not be crepantly resolved \textit{algebraically}.\\\textbf{M-theory develops $\mathbb{Z}_N$ gauge symmetry}};
	\node[align=right] at (7.0,1) {\textbf{Singularities can be stabilized by B-field}\\$\rightarrow$ \textit{``non-commutative resolution''}};
	\node[align=right] at (4.9,4.0) {Smooth CY 3-fold\\{\small$r=h^{1,1}(X_{\text{def.}})$}};
	\draw[->,black] (12,3.3) to (12,1.5);
	\draw[<->,black] (13.6,1) to (15.3,1);
	\node[align=center] at (14.5,0.3) {cplx. K\"ahler\\deformation};
\end{scope}
\end{tikzpicture}
\caption{The generic relationships between the various geometries that are associated to a Calabi-Yau 3-fold $X$ with terminal nodal singularities that are resolved by torsion curves in a non-K\"ahler small resolution $\widehat{X}$.}
\label{fig:geooutline}
\end{figure}
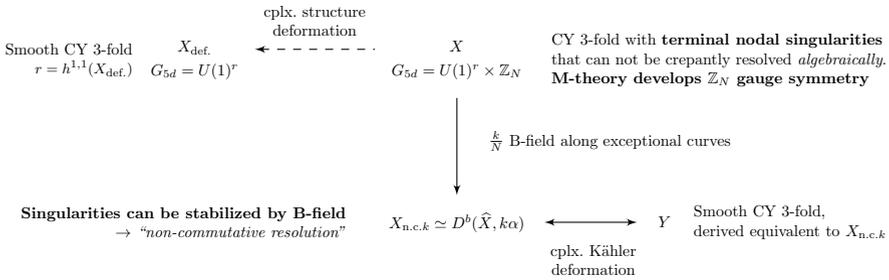

In most of the paper we will use the term ``non-commutative resolution'', as opposed to non-commutative crepant resolution or crepant categorical resolution, to refer to a Calabi-Yau $X$ with terminal nodal singularities that is equipped with a fractional B-field, as well as to a small analytic resolution $\widehat{X}$ of $X$ that carries a lift $\alpha\in\text{Br}(\widehat{X})$ of that B-field.
We expect that a sheaf of non-commutative algebras on $X$ can be constructed from a pair $(\widehat{X},\alpha)$ and that the twisted derived categories $D^b(\widehat{X},\alpha)$ are examples of crepant categorical resolutions.
Moreover, we will argue that locally around the nodes the geometry is described by a non-commutative crepant resolution of the conifold.
We will discuss several examples and provide both mathematical and physical evidence for these claims.
However, we will not attempt to make these statements fully precise or even to prove them in general.

We now give an overview of the different types of geometries and their non-commutative resolutions that have been studied in the literature and that are covered by our proposal.
In each case, there is associated to a Calabi-Yau threefold $X$ with terminal nodal singularities a set of closely related geometries $(X_{\text{def.}},X_{\text{n.c.}k},Y)$.
Their relationship is outlined in Figure~\ref{fig:geooutline}.
We distinguish between two classes:
\paragraph{Clifford type (CT) non-commutative resolutions} The unresolved Calabi-Yau $X$ is a singular double cover over a Fano base $B$.
	The non-commutative resolution can be described in terms of a sheaf $\mathcal{S}$ of Clifford algebras on $B$.
In addition, $\mathcal{S}$ is the pushforward to $B$ of a sheaf $\tilde{\mathcal{S}}$ of non-commutative algebras on $X$, so that $(X,\tilde{\mathcal{S}})$ is a non-commutative variety resolving $X$. It is expected that they are examples of \textit{crepant categorical resolutions}~\cite{kuzso}.
	Moreover, we expect that the corresponding torsion in a small analytic resolution is always
	\begin{align}
	    H_2(\widehat{X},\mathbb{Z})=H_2(X,\mathbb{Z})\oplus\mathbb{Z}_2\,,
	\end{align}
	and there exists a twisted derived equivalence
	\begin{align}
	    D^b(\widehat{X},\alpha)\simeq D^b(Y)\,,
	    \label{eqn:twdeq1}
	\end{align}
	with some smooth Calabi-Yau 3-fold $Y$.
	In many cases the derived equivalence is known to be an example of homological projective duality~\cite{kuz0}.
	Physically, the large volume limits associated to $(\widehat{X},\alpha)$ can be realized as infrared fixed points of hybrid gauged linear sigma models, where a $\mathbb{Z}_2$ orbifold of a Landau-Ginzburg model with quadratic superpotential is fibered over a non-linear sigma model on $B$~\cite{Caldararu:2010ljp}.\\[.5em]
	\phantom{ss}\textbf{Examples}
	\begin{itemize}
	    \item Our main example is a singular octic double solid $X$, i.e. a degenerate double cover of $\mathbb{P}^3$~\cite{Clemens}, and the Clifford nc-resolution was constructed in~\cite{kuznetsov2008derived}.
	    This will be studied in detail in Sections~\ref{sec:octic} and~\ref{sect:octicdoub:b-model}.
	    The GLSM realization has been constructed in~\cite{Caldararu:2010ljp} and the derived equivalence~\eqref{eqn:twdeq1}, where $Y$ is a complete intersection of four quadrics in $\mathbb{P}^7$, was proven in~\cite{addington2009derived}.
	    \item An nc-resolution of a singular double cover of $\mathbb{P}^1\times\mathbb{P}^2$ was constructed in~\cite{Calabrese2016DerivedEC}.
	    As we discuss in Section~\ref{sec:ellfibrations}, this can also be interpreted as a singular genus one fibration with a $2$-section and an analogous example, where the base is a $\mathbb{P}^1$ fibration over $\mathbb{P}^2$, has already been studied in detail in~\cite{Schimannek:2021pau}.
	    In each case the dual $Y$ is a smooth genus one fibration over $\mathbb{P}^2$ that is not elliptic but exhibits a $4$-section.
	    Another example of this type corresponds to the nc-resolution of a singular double cover of $\mathbb{P}^1\times\mathbb{P}^1\times\mathbb{P}^1$ and arose in~\cite{Vafa:1994rv} from a deformation of a $T^6/\mathbb{Z}_2\times\mathbb{Z}_2$ orbifold with discrete torsion.
	     This example has also been studied using a large resolution in~\cite{Aspinwall:1995rb} and using the GLSM realization in~\cite{Caldararu:2010ljp}.
	    \item All of the previous examples are instances of Clifford double mirrors~\cite{borisov2018clifford}.
	    This is a combinatorial construction of Clifford nc-resolutions with duals $Y$ that are complete intersections in toric ambient spaces and provides a rich class of examples.
	    \item We also consider four hybrid GLSMs with non-Abelian gauge group constructed in~\cite{Hori:2013gga}. These are expected to realize nc-resolutions of singular double covers where the dual $Y$ is not a complete intersection in a toric ambient space, but a Pfaffian Calabi-Yau 3-fold in a (weighted) projective space.
	    They will be discussed in Appendix~\ref{app:otherexamples}.
	    
	\end{itemize}
	\paragraph{General type (GT) non-commutative resolutions.} 
	Most of the singular torus fibered examples studied in~\cite{Schimannek:2021pau} can not be described as degenerate double covers over some base.
	In particular, an explicit non-commutative resolution is not yet known and can not take the form of a sheaf of Clifford algebras on some base.
	Nevertheless, in many cases the existence of a twisted derived equivalence
	\begin{align}
	    D^b(\widehat{X},\alpha)\simeq D^b(Y)\,,
	    \label{eqn:twdeq2}
	\end{align}
	where the dual $Y$ is a smooth genus one fibered Calabi-Yau 3-fold, has been proven in~\cite{caldararu2001derived}.
	A common example arises when $Y$ does not have a section but only an $n$-section and the Jacobian fibration $X=J(Y)$ has terminal nodal singularities that are resolved by $n$-torsion curves in a small non-K\"ahler resolution $\widehat{X}$.
	More generally, the singular Calabi-Yau $X$ can be a torus fibration that shares the same Jacobian fibration as $Y$.
	The large volume limits associated to $(\widehat{X},\alpha)$ and $Y$ exist in the same stringy K\"ahler moduli space.
	However, from a GLSM perspective the limit associated to $(\widehat{X},\alpha)$ lies on a singular phase boundary.
	Nevertheless, topological strings on $(\widehat{X},\alpha)$ can be studied, for example, using modularity, mirror symmetry and analytic continuation.
	Examples with
		\begin{align}
	    H_2(\widehat{X},\mathbb{Z})=H_2(X,\mathbb{Z})\oplus\mathbb{Z}_N\,,
	\end{align}
	for $N=2,\ldots, 5$ have been studied in~\cite{Schimannek:2021pau}.

\section{Topological strings on non-commutative resolutions}
\label{sect:prop}
We now turn to the study of general properties of the topological string A-model on non-commutative resolutions of K\"ahler Calabi-Yau threefolds with terminal nodal singularities.
More precisely, we focus on resolutions that can be obtained, roughly speaking, by turning on a fractional B-field along torsional curves in a small analytic resolution.  

\subsection{Discrete symmetries in M-theory on nodal Calabi-Yau threefolds}
\label{sec:nodalcy3}
Let us first specify our geometric setup.
We consider a K\"ahler Calabi-Yau threefold $X$ with terminal nodal singularities and denote the number of nodes by $m_s$.
Here \textit{terminal} means that the singularities do not admit a K\"ahler resolution that preserves the Calabi-Yau property.
Note that given any nodal K\"ahler Calabi-Yau threefold we can perform partial small K\"ahler resolutions such that the remaining nodes, if they exist, are terminal~\cite{Kawamata}.
Then it has been shown that $X$ always admits a smooth deformation that we will denote by $X_{\text{def.}}$~\cite{Namikawa1995}.
Some of the properties of terminal singularities and the related physical aspects have for example also been discussed in the context of F-theory in~\cite{Arras:2016evy,ARRAS201871}.

Using an analytic change of coordinates, each nodal singularity can be locally identified with the conifold~\eqref{eqn:conifold}.
In particular, one can perform analytic blow-ups and resolve each of the singularities with an exceptional $\mathbb{P}^1$ which, taking into account the local flops, leads to $2^{m_s}$ different \textit{small analytic resolutions}.
We choose any one of those and denote it by $\widehat{X}$.
Although $\widehat{X}$ has a vanishing first Chern class it can not, by assumption, be K\"ahler.
It can be easily shown using the analysis of Clemens~\cite{Clemens} or Werner~\cite{Werner}, that each exceptional curve in $\widehat{X}$ is either trivial or torsion,\footnote{(\ref{eqn:h2rel}) does not hold as written in complete generality, as $H_2(X,\mathbb{Z})$ itself can have torsion, in which case $H_2(X,\mathbb{Z})$ need not be a summand.  However, in all of our examples, $H_2(X,\mathbb{Z})$ is torsion free and (\ref{eqn:h2rel}) is correct.}
\begin{align}
	H_2(\widehat{X},\mathbb{Z})=H_2(X,\mathbb{Z})\oplus\mathbb{Z}_{N_1}\oplus\ldots\oplus\mathbb{Z}_{N_l}\,,\quad \mathbb{Z}_N=\mathbb{Z}/N\mathbb{Z}\,,
	\label{eqn:h2rel}
\end{align}
with the torsion subgroup being generated by the exceptional curves.
We will focus on the case
\begin{align}
	H_2(X,\mathbb{Z})=\mathbb{Z}^{b_2(X)}\,,\quad H_2(\widehat{X},\mathbb{Z})=\mathbb{Z}^{b_2(X)}\oplus\mathbb{Z}_{N}\,,
	\label{eqn:h2ZN}
\end{align}
but the generalization to multiple discrete subgroups is straightforward.  The claim about the exceptional curves will be explained in an example in Section~\ref{sect:octicdoub:small-analytic}.  The same analysis can immediately be adapted for the general case.

A typical example where this can arise is the Jacobian fibration $X=J(Y)$ of a smooth genus one fibered Calabi-Yau threefold $Y$ that does not admit a section but only an $N$-section that intersects the generic fiber $N$ times~\cite{DG,caldararu2001derived}.
Based on physical arguments from F-theory it is then expected that M-theory on $X$ develops a
\begin{align}
	G_{5d}\simeq U(1)^{b_2(X)}\times \mathbb{Z}_N\,,
	\label{eqn:u1zn}
\end{align}
gauge symmetry, see e.g.~\cite{Braun:2014oya}.
Moreover, this is conjecturally related to the presence of $\mathbb{Z}_N$ torsion curves in the small analytic resolution $\widehat{X}$~\cite{Weigand} and motivated the definition of torsion refined Gopakumar-Vafa invariants in~\cite{Schimannek:2021pau}.
This is consistent with the mathematical results on the Brauer group of $\widehat{X}$ from~\cite{caldararu2001derived,DG}.
We will review the physical definition in Section~\ref{sec:gvtorsion} and present a proposal for a geometric definition of the torsion refined invariants in Section~\ref{subsec:gv}.

As we discuss in Section~\ref{sec:octic}, another source of examples are Calabi-Yau threefolds that are double covers of Fano bases with a ramification locus that itself develops isolated nodal singularities.
Our successful calculation of torsion refined Gopakumar-Vafa invariants in all of these examples leads us to the following claim:
\begin{claim}\label{conj:mthg}
	The gauge group of M-Theory on a nodal K\"ahler Calabi-Yau threefold $X$ at a generic point of the K\"ahler cone is equal to
	\begin{align}
		\normalfont G_{5d}=\text{Hom}\left(H_2(\widehat{X},\mathbb{Z}),U(1)\right)\,,
		\label{eqn:conjMG}
	\end{align}
	for any small analytic resolution $\widehat{X}$. Moreover, the electric charge lattice is populated by M2-branes wrapping curves in $H_2(\widehat{X},\mathbb{Z})$ with either of the two possible orientations, leading to pairs of states with opposite charges.
 In particular, given $X$ such that~\eqref{eqn:h2rel} holds the gauge symmetry will be~\eqref{eqn:u1zn}.
\end{claim}

Let us stress that this is not only a highly non-trivial statement about physics but also suggests new techniques to calculate Brauer groups of nodal threefolds by using the dictionary between Higgs transitions in physics and extremal transitions in geometry.

Assuming for simplicity again that the torsion in $H_2(\widehat{X},\mathbb{Z})$ is $\mathbb{Z}_N$, note that choosing a basis of the gauge group~\eqref{eqn:conjMG}
\begin{align}
	G_{5d}\simeq U(1)^{b_2(X)}\times \mathbb{Z}_N\,,
	\label{eqn:gbasis}
\end{align}
induces a corresponding basis of the charge lattice $H_2(\widehat{X},\mathbb{Z})\simeq\mathbb{Z}^{b_2(X)}\times\mathbb{Z}_N$.
Taking into account the orientation, this assigns a pair of $\mathbb{Z}_N$ charges $[\pm \ell]$ for $\ell=0,\ldots, \lfloor N/2\rfloor$ to each of the $m_s$ exceptional curves on $\widehat{X}$ and we denote the multiplicity of exceptional curves with charge $[\pm \ell]$ by $m^{(\pm \ell)}$.
In the following we will often implicitly assume that such a basis has been chosen.

\subsection{B-fields in the presence of torsion}\label{sec:Btorsion}
The non-canonical choice of basis for the 5-dimensional gauge group~\eqref{eqn:gbasis} is closely related to a similar choice that is necessary to define the Wilson lines associated to a flat B-field for Type II strings in the presence of torsion curves.

Let us first consider flat B-fields on an arbitrary manifold $X$.\footnote{In this subsection only, $X$ will denote a general manifold rather than a nodal Calabi-Yau threefold as defined in Section~\ref{sec:nodalcy3}.}
As was pointed out in~\cite{Aspinwall:1995rb}, and is from a more modern perspective formalized in the language of differential cohomology~\cite{Freed:2006yc}, flat B-fields correspond to elements of $H^2(X,U(1))$.  From the exponential short exact sequence
\begin{equation}
0 \rightarrow \mathbb{Z} \rightarrow \mathbb{R}\rightarrow U(1) \rightarrow 0\,,
\end{equation}
we get the exact sequence
\begin{equation}  \label{eq:bfield:shortexact}
H^2(X,\mathbb{Z}) \rightarrow H^2(X,\mathbb{R}) \rightarrow H^2(X,U(1)) \stackrel{c}{\rightarrow} 
H^3(X,\mathbb{Z}) \rightarrow H^3(X,\mathbb{R})\,,
\end{equation}
so that the image of $H^2(X,U(1))$ in $H^3(X,\mathbb{Z})$ is precisely the torsion subgroup $H^3(X,\mathbb{Z})_{\mathrm{tors}}$.
Thus $H^2(X,U(1))$ decomposes into cosets of its subgroup $H^2(X,\mathbb{R})/H^2(X,\mathbb{Z})$, and these cosets are indexed by $H^3(X,\mathbb{Z})_{\mathrm{tors}}$.   The trivial coset $H^2(X,\mathbb{R})/H^2(X,\mathbb{Z})$ is identified with the space of topologically trivial B-fields, but there are additional cosets in the presence of torsion.
Let us point out here that for simply connected spaces the universal coefficient theorem implies that
\begin{align}\label{eq:uct}
H_2(X,\mathbb{Z})_{\mathrm{tors}}\simeq H^3(X,\mathbb{Z})_{\mathrm{tors}}\,.
\end{align}

Now suppose that $X$ is Calabi-Yau.  For each $\rho\in H^3(X,\mathbb{Z})_{\mathrm{tors}}$  we get a distinct large radius limit as follows.  First, we recall the description of the ordinary large radius limit.  The set
\begin{equation}
\mathcal{K}=\left\{B+iJ\mid B\in H^2(X,\mathbb{R})/H^2(X,\mathbb{Z}),\ J>>0\right\}\,,
\end{equation}
is topologically a punctured polydisc $(\Delta^*)^r$, where $r=b_2(X)$.  The large radius limit is the origin of $\Delta^r$.

The situation is almost the same for any $\rho\in H^3(X,\mathbb{Z})_{\mathrm{tors}}$.  The set
\begin{equation}\label{eq:krho}
\mathcal{K}_\rho=\left\{B+iJ\mid B\in H^2(X,U(1)), c(B)=\rho,\ J>>0\right\}\,,
\end{equation}
is a principal homogeneous space for $\mathcal{K}$ with its natural group structure, so it is isomorphic to $\mathcal{K}$.  For $\rho\ne0$, the isomorphism  cannot be canonical, simply because if $c(B)=\rho$, then $B$ cannot be zero and there is no natural origin.
However, $\mathcal{K}_\gamma$ is still topologically a punctured polydisc $(\Delta^*)^r$, and therefore the corresponding large radius limit is unambiguously identified with the origin of $\Delta^r$.

To construct an isomorphism between $\mathcal{K}$ and $\mathcal{K}_\rho$ we proceed as follows.
For simplicity, we assume that $H^3(X,\mathbb{Z})_{\mathrm{tors}}=\mathbb{Z}_N$ and pick any B-field with $c(B)=\rho_0$ of order $N$ such that $\rho=k\cdot \rho_0$ for some $k=0,\ldots,N-1$.  Since $c(N\cdot B)=0$, the B-field $N\cdot B$ is a topologically trivial B-field $N\cdot B\in H^2(X,\mathbb{R})/H^2(X,\mathbb{Z})\simeq U(1)^r$.  Since $U(1)^r$ is divisible, we can choose $B'\in H^2(X,\mathbb{R})/H^2(X,\mathbb{Z})$ with $N\cdot B'=N\cdot B$, and there are $N^r$ possible choices for such a $B'$.  Then $B_0:=B-B'$ satisifies $N\cdot B_0=0$ as well as $c(B_0)=\rho_0$ and one obtains an isomorphism
\begin{align}
	\mathcal{K}\rightarrow\mathcal{K}_\rho=\mathcal{K}_{k\cdot \rho_0}\,,\quad B+iJ\mapsto k\cdot B_0+B+iJ\,.
	\label{eqn:Kiso}
\end{align}
We will often implicitly assume that a choice of $B_0$ has been fixed and denote the topological class of a B-field background by a coset $[k]\in\mathbb{Z}_N$.

The choice of reference B-field $B_0$ is actually equivalent to a splitting of the M-theory charge lattice $H_2(X,\mathbb{Z})$ and correspondingly to a choice of isomorphism~\eqref{eqn:gbasis} for the gauge symmetry.
We have the natural cap product pairing
\begin{align}
H^2(X,U(1)) \times H_2(X,\mathbb{Z}) \to U(1)\,,
	\label{eqn:cup}
\end{align}
which computes the holonomy of a B-field around a 2-cycle.  We will write this pairing as $\langle B,\beta\rangle$.
Since $N\cdot B_0=0$ we have for any curve class $\beta\in H_2(X,\mathbb{Z})$ that
\begin{align}
\langle B_0, \beta\rangle^N=1\,,
\end{align}
which implies that $\langle B_0, \beta\rangle$ is a $k$th root of unity.
If we have
\begin{equation}\label{eq:charge}
\langle B_0,\beta\rangle =e^{2\pi i \ell/N}\,,
\end{equation}
we say that $\beta$ lies in $\mathbb{Z}_N$-charge sector $\ell$.
This induces an isomorphism of the charge lattice
\begin{align}
	H_2(X,\mathbb{Z})=\mathbb{Z}^r\oplus\mathbb{Z}_N\,.
\end{align}
In the following we will always assume that a consistent choice has been made and allows us to freely talk about the charges associated to given curves as well as their Wilson lines in a flat B-field background.
Note that the above discussion and (\ref{eq:uct}) nicely ties together the roles of torsion in $H^3(X,\mathbb{Z})$ and $H_2(X,\mathbb{Z})$.

We caution the reader that $\langle B,\beta\rangle$ is usually written in the physics literature as $\mathrm{exp}(2\pi i\int_\beta B)$.
Consistent with this notation, we will refer to $B_0$ as a fractional B-field, as $\int_\beta B_0$ is fractional for all $\beta$, well-defined up to integral shifts.
We trust that no confusion will arise.

The cup product~\eqref{eqn:cup} leads to a natural generalization of the usual instanton action that serves as an expansion variable for the A-model topological string free energies which are generating functions of Gromov-Witten invariants.
Given a choice of K\"ahler class $J\in H^{1,1}(X,\mathbb{C})$ this takes the form
\begin{align}
q^\beta = \mathrm{exp}\left( 2\pi i\int_\beta B+iJ
\right), \qquad \beta\in H_2(X,\mathbb{Z})\,,
\end{align}
Thus on any $\mathcal{K}_\rho$ we have instrinsic expansion variables
\begin{align}
q_\rho^\beta = \langle B,\beta\rangle\mathrm{exp}\left(- 2\pi \int_\beta J
\right)\,.
\end{align}
If we can make sense of Gromov-Witten invariants $N_g^\beta$ for $\beta\in H_2(X,\mathbb{Z})$, then it is natural to  define the partition function
\begin{align}\label{eq:rhofe}
F_\rho = \sum N_g^\beta q_\rho^\beta\lambda^{2g-2}\,,
\end{align}
on $\mathcal{K}_\rho$.
Using $B_0$ with $\rho=k\cdot B_0$ to identify $\mathcal{K}_\rho$ and $\mathcal{K}$ via~\eqref{eqn:Kiso} leads to the expansion
\begin{align}\label{eqrhofeb0}
F_\rho = \sum N_g^\beta \lambda^{2g-2} \langle B_0,\beta\rangle^k q^\beta\,.
\end{align}
This is essentially the proposal in \cite{Aspinwall:1995rb}.  Our improvement is that $F_\rho$ has been intrinsically defined in (\ref{eq:rhofe}) as an expansion on $\mathcal{K}_\rho$; (\ref{eqrhofeb0}) simply expresses this in a choice of coordinates.

An important point is that the charge assignments depend on the choice of $B_0$.  For example, suppose that $r=1$ and $k=2$, so that $H_2(X,\mathbb{Z})=\mathbb{Z}\oplus\mathbb{Z}_2$ and there are two possible $B_0$ with $c(B_0)=1$ and $2{B}_0=0$.  For one choice, $(a,b)\in \mathbb{Z}\oplus\mathbb{Z}_2$ has $\mathbb{Z}_2$-charge $b$.  For the other choice, the $\mathbb{Z}_2$-charge is $a+b$ mod 2.  Neither choice is canonical.

\subsection{Fractional B-fields on non-K\"ahler Calabi-Yau and nc-resolutions}
\label{sect:fbnc}
Let us now come back to a Calabi-Yau threefold $X$ with terminal nodal singularities as considered in Section~\ref{sec:nodalcy3}.
The discussion of the B-field on manifolds with torsion from the previous section applies with a simple modification also to non-K\"ahler small resolutions $\widehat{X}$ of $X$.
Terminality implies that the pullback map $H^2(X,\mathbb{C})\to H^2(\widehat{X},\mathbb{C})$ is an isomorphism.
In (\ref{eq:krho}) we then simply replace the condition that $J$ is a sufficiently large K\"ahler class on $\widehat{X}$ with the condition that $J$ is the pullback of a sufficiently large K\"ahler class on ${X}$.

It was suggested that the small analytic resolution $\widehat{X}$ equipped with a B-field $B\in H^2(\widehat{X},\mathbb{R})/H^2(\widehat{X},\mathbb{Z})$ with non-trivial fractional part $c(B)\in H^3(\widehat{X},\mathbb{Z})_{\mathrm{tors}}$ should be interpreted as a non-commutative crepant resolution of $X$~\cite{Aspinwall:1995rb,caldararu2001derived,Caldararu:2010ljp,Schimannek:2021pau}.
We will now argue that this can be understood by interpreting the non-commutative conifold discussed in Section~\ref{subsec:nccr} as a local model. 

Recall that the conifold describes the local geometry around each of the nodal singularities in a compact Calabi-Yau threefold.
If such a node can be resolved by performing a small projective resolution of the entire Calabi-Yau, this opens up a conifold transition~\cite{Candelas:1987kf,Strominger:1995cz}.
The complexified K\"ahler parameter of the local conifold then lifts to a corresponding K\"ahler parameter 
\begin{align}
    t=b+iv\,\in\,\mathbb{C}/(t\sim t+1)\,,
    \label{eqn:conikaehler2}
\end{align}
of the resolved Calabi-Yau, where $v$ is the volume of the exceptional curve $\beta$ and $b$ measures the Wilson line $\langle B,\beta\rangle=e^{2\pi i b}$ of the B-field along it.

However, such a projective resolution does not exist for $X$ and, as discussed in Section~\ref{sec:nodalcy3}, the exceptional curve classes in a small analytic resolution $\widehat{X}$ will be either homologically trivial or $N$-torsion for some $N>1$.
Let us assume again for simplicity that $H_2(\widehat{X},\mathbb{Z})_{\text{tors}}\simeq\mathbb{Z}_N$.
The choice of a fractional B-field corresponds to an element $[k]\in\mathbb{Z}_N$, and according to (\ref{eqn:Kiso}) the holonomy of an exceptional curve $\beta$ is given by
\begin{align}
	\langle kB_0,\beta\rangle =e^{2\pi i k \ell/N}\,,
\end{align}
where we have chosen a generator $B_0$ such that $\langle B_0,\beta\rangle =e^{2\pi i  \ell/N}$.
Recall that the holonomy being restricted to an $N$-th root of unity is a direct consequence of the curve being $N$-torsion.
One can in this way say that the complexified ``K\"ahler parameters''~\eqref{eqn:conikaehler2} of the corresponding local conifolds are not absent but just restricted to take values 
\begin{align}
        t\overset{!}{=}k/N\,,\quad k\in\{0,\ldots,N-1\}\,.
        \label{eqn:localkahler}
\end{align}
In particular, they lie on the non-commutative equator of the moduli space~\eqref{eqn:conikaehler2} of the local conifold.
This picture was indeed verified in~\cite{Schimannek:2021pau} at the level of the topological string partition function by studying extremal transitions between torus fibered Calabi-Yau threefold.

On the other hand, a given non-commutative  crepant resolution $X_{\text{n.c.}}$ of $X$ locally induces an nc-resolution of each of the conifold singularities and therefore a corresponding value~\eqref{eqn:localkahler} of the K\"ahler parameter.
This suggests a general correspondence between non-commutative crepant resolutions of $X$ and flat B-fields on $\widehat{X}$.

We will provide further evidence for this idea in the derivation of the constant map contributions to the topological string partition function in Section~\ref{sec:constantmap} and in the examples studied in later sections. 

\subsection{Torsion refined Gopakumar-Vafa invariants}\label{sec:gvtorsion}
In this section we develop the general physical notion of torsion refined Gopakumar-Vafa invariants associated to a K\"ahler Calabi-Yau threefold $X$ with terminal nodal singularities that has first been introduced in~\cite{Schimannek:2021pau}.
To streamline the exposition we will again focus on the case~\eqref{eqn:h2ZN}, such that $H_2(X,\mathbb{Z})=\mathbb{Z}^r$ and a given small analytic resolution $\widehat{X}$ has
\begin{align}
	H_2(\widehat{X},\mathbb{Z})=\mathbb{Z}^r\oplus \mathbb{Z}_N\,,
	\label{eqn:gvcspl}
\end{align}
for some $N\in\mathbb{N}$ with $N\ge 2$.
Assuming Claim~\ref{conj:mthg}, the M-theory gauge group then takes the form
\begin{align}
	G_{5d}=U(1)^r\times \mathbb{Z}_N\,.
	\label{eq:5dgauge}
\end{align}

As pointed out in~\cite{Schimannek:2021pau}, the presence of the discrete gauge symmetry does not affect the usual construction of Gopakumar-Vafa invariants~\cite{Gopakumar:1998ii,Gopakumar:1998jq}.
The five-dimensional massive little group decomposes as
\begin{align}\label{eq:littlegroup}
	SO(4)=SU(2)_L\times SU(2)_R\,,
\end{align}
and we can denote the multiplicity of BPS particles in the representation
\begin{align}
	\left[\left(\frac12,0\right)\oplus 2(0,0)\right]\otimes (j_L,j_R)\,,
\end{align}
that carry a charge $\beta \in H_2(\widehat{X},\mathbb{Z})$ by $N^\beta_{j_L,j_R}$.
The \textit{torsion refined Gopakumar-Vafa invariants} $n_{g}^\beta$ are then obtained by tracing over the right spins, as usual.
More precisely, using $I_g=([\frac12]+2[0])^g$, they are defined by the relation
\begin{align}
	\sum\limits_{g=0}^\infty n_{g}^\beta I_g=\sum\limits_{j_L,j_R}(-1)^{2j_R}(2j_R+1)N^\beta_{j_L,j_R}\cdot [j_L]\,.
	\label{eq:spingenus} 
\end{align}
Using the splitting of the charge lattice~\eqref{eqn:gvcspl}, we also write $n_{g}^{\beta,l}=n_{g}^{(\beta',l)}$ where $\beta\in H_2(X,\mathbb{Z})$ is the image of $(\beta',l)$ under the  projection $\widehat{X}\rightarrow X$.
As will expanded upon in Section~\ref{sect:octicdoub:small-analytic}, 
the invariants satisfy the reflection identity
\begin{align}
	n_g^{\beta,l}=n_g^{\beta,-l}\,.
	\label{eqn:reflection}
\end{align}

As discussed in the previous sections, $\text{Tors}\,H^3(\widehat{X},\mathbb{Z})=\mathbb{Z}_N$ and therefore the topological type of a flat B-field background on the small analytic resolution corresponds to a choice $[k]\in\mathbb{Z}_N$ with $k=0,\ldots N-1$.
We denote the corresponding string background by $X_{\text{n.c.}k}$, referring to the interpretation in terms of a non-commutative resolution, and the topological string A-model partition function by $Z_{\text{top}.,k}$.
Following~\cite{Schimannek:2021pau}, the torsion refined GV-invariants are encoded in
\begin{align}
\begin{split}
        &\log\left[Z_{\text{top.},k}\left(t,\lambda\right)\right]\\=&\sum\limits_{g=0}^\infty\sum\limits_{\beta\in H_2(X)}\sum\limits_{l=0}^{N-1}\sum\limits_{m=1}^\infty n_{g}^{\beta,l}\cdot\frac{1}{m}\left(2\sin\frac{m\lambda}{2}\right)^{2g-2}e^{2\pi i mlk/N}q^{m\beta}\,.
        \label{eq:Znc}
    \end{split}
\end{align}
The relation~\eqref{eqn:reflection} implies that $Z_{\text{top.},k}=Z_{\text{top.},N-k}$.
To extract the invariants, one therefore needs to know the partition functions $Z_{\text{top.},k}$ for $k=0,\ldots,\lfloor N/2\rfloor$.

Given a choice of B-field $[k]\in \mathbb{Z}_N$, there will in general be exceptional curves on $\widehat{X}$ that are $k$-torsion.
The Wilson line along those curves is trivial and the local conifold picture suggests that   $X_{\text{n.c.}k}$ only provides a partial resolution.
However, one can then perform a complex structure deformation $\overline{X}_{\text{n.c.}k}$ of $X_{\text{n.c.}k}$ that removes all of the corresponding nodes and is smooth.
Since the topological string A-model is not sensitive to complex structure deformations, we use the topological string partition function on $\overline{X}_{\text{n.c.}k}$ to define $Z_{\text{top.},k}$.
In particular, we use the smooth deformation $X_{\text{def.}}$ of $X$ to define $Z_{\text{top.},N}=Z_{\text{top.},0}$.
A less trivial example of this behaviour can be found in the context of genus one fibrations with $4$-sections discussed in~\cite{Schimannek:2021pau}, see also Section~\ref{sec:ellfibrations}.
We always assume that such a deformation exists.

\subsection{Gopakumar-Vafa invariants from non-K\"ahler small resolutions}\label{subsec:gv}
We now propose a mathematical definition of the torsion refined Gopakumar-Vafa invariants. 
Let $X$ be a Calabi-Yau threefold with only terminal conifold singularities and no K\"ahler small resolution.  
Let $\mathcal{S}$ be the set of all small resolutions $\widehat{X}$ of $X$.   
Then $\mathrm{Br}(\widehat{X})$ is independent of $\widehat{X}\in\mathcal{S}$ (a stronger statement is proven in the projective case in \cite[Corollary 5.2.6]{C-TS}).  Let $\alpha\in \mathrm{Br}(\widehat{X})$, and let $\mathcal{K}_\alpha$ be the corresponding component of the K\"ahler moduli space.  

If $\widehat{X},\widehat{X}'\in\mathcal{S}$,  we let
$\Gamma=\widehat{X}\times_X\times\widehat{X}'\subset \widehat{X}\times\widehat{X}'$ be the closure of the graph of the natural
 birational isomorphism between them, giving rise to a derived equivalence
$\psi:D^b(\widehat{X})\to D^b(\widehat{X}')$ defined by the Fourier-Mukai transform with kernel $\mathcal{O}_\Gamma$.  The same form of the Fourier-Mukai transform gives a derived equivalence
$\psi:D^b(\widehat{X},\alpha)\to D^b(\widehat{X}',\alpha')$, where $\alpha'\in\operatorname{Br}(\widehat{X}')$ corresponds to $\alpha\in\operatorname{Br}(\widehat{X})$ under the isomorphism $\operatorname{Br}(\widehat{X})\simeq \operatorname{Br}(\widehat{X}')$.  We will illustrate the use of such a Fourier-Mukai transform in Section~\ref{sec:geometric}.

Let $\mathrm{Coh}_{\le1}(\widehat{X})$ be the category of coherent sheaves on $\widehat{X}$ whose support has dimension at most 1.   Since the Brauer group of a curve is trivial, $\mathrm{Coh}_{\le1}(\widehat{X})$ can be viewed as a subcategory of  $D^b(\widehat{X},\alpha)$. Furthermore, using the above derived equivalences, we can identify $\mathrm{Coh}_{\le1}(\widehat{X}')$ with a subcategory of $D^b(\widehat{X},\alpha)$, for any $\widehat{X}'\in\mathcal{S}$.  Thus the union $\cup_{\widehat{X}\in\mathcal{S}}\mathrm{Coh}_{\le1}(\widehat{X})$ is a well-defined subcategory of $D^b(\widehat{X},\alpha)$, as well as of $D^b(\widehat{X})$ itself.  We emphasize that $D^b(\widehat{X},\alpha)$ is independent of $\widehat{X}\in\mathcal{S}$ up to equivalence, and so the category $\cup_{\widehat{X}\in\mathcal{S}}\mathrm{Coh}_{\le1}(\widehat{X})$ is well-defined up to equivalence.

\begin{conj}
\begin{enumerate}
\item $D^b(\widehat{X},\alpha)$ supports Bridgeland stability conditions that are parametrized by $\mathcal{K}_\alpha$.
\item Among these Bridgeland stability conditions are some whose hearts contain
$$\bigcup_{\widehat{X}\in\mathcal{S}}\mathrm{Coh}_{\le1}(\widehat{X})\,.$$
\item On each $\mathrm{Coh}_{\le1}(\widehat{X})$, the Bridgeland stability condition coincides with Gieseker stability.
\end{enumerate}
\end{conj}

Fixing any $\widehat{X}\in\mathcal{S}$, we have a support map 
\begin{equation}
\bigcup_{\widehat{X}\in\mathcal{S}}\mathrm{Coh}_{\le1}(\widehat{X})\to H_2(\widehat{X},\mathbb{Z})\,.
\end{equation}
Let $\Lambda$ be the image of this map, consisting of the union of all effective curve classes of all small resolutions.\footnote{The definition of $\Lambda$ is independent of the choice of $\widehat{X}\in\mathcal{S}$ up to isomorphism.}  For $\beta\in\Lambda$, we want to define the Gopakumar-Vafa invariants $n_g^\beta(X)$.  Before defining these invariants, we recall some of the ingredients of the mathematical definition of the Gopakumar-Vafa invariants in the smooth projective case, following \cite{MT}.

Note that here we associated the torsion refined Gopakumar-Vafa invariants directly to the singular Calabi-Yau $X$.
This is natural from a physical perspective, since, as we discussed above, M-theory on $X$ itself is expected to develop a discrete gauge symmetry and in particular doesn't see the non-commutative resolutions that are related to the presence of a B-field in Type IIA string theory.
However, the reader who is more comfortable with defining the invariants on a smooth space can equivalently replace $X$ with the set of all its small analytic resolutions $\widehat{X}$ or even with a non-commutative resolution $(\widehat{X},\alpha)$, in which case the invariants will not depend on the choice of $\alpha$.

For any smooth projective Calabi-Yau threefold $Y$, we let $\beta\in H_2(Y,\mathbb{Z})$ and let $M_\beta(Y)$ be the moduli space of stable sheaves $F$ on $Y$ of dimension 1 with support class $\beta$ and $\chi(F)=1$.  We let $\mathrm{Chow}(\beta)$ be the Chow variety of 1-dimensional cycles (formal $\mathbb{Z}$-linear combinations of irreducible curves) on $Y$.  There is a natural support map
\begin{equation}
    \pi_\beta:M_\beta(Y)\to \mathrm{Chow}(\beta)\,.
\end{equation}

In \cite{Gopakumar:1998jq}, $M_\beta(Y)$ is the space of D2-D0 branes and $\mathrm{Chow}(\beta)$ is the space of curves.  The little group $SU(2)_L\times SU(2)_R$ of M-theory compactified on $Y$ is identified with the Lefschetz $SU(2)_L$ action on the cohomology of the fibers of $\pi_\beta$ and the Lefschetz $SU(2)_R$-action on the cohomology of $\mathrm{Chow}(\beta)$.  While it may be tempting to package both $SU(2)$ actions in terms of the sheaf $R(\pi_\beta)_*\mathbb{C}$ on $\mathrm{Chow}(\beta)$, whose generic stalks are the cohomologies of the fibers, this approach does not work.  Instead, we have to use perverse sheaves and perverse cohomology.  See \cite{MT} for the mathematical details, or \cite{Huang:2020dbh} for a summary exposition written for physicists. 

For this reason, we rewrite the GV invariant of $Y$ as
\begin{equation}
    n_g^\beta(Y)=n_g(M_\beta(Y))\,,
    \label{eq:gvnotation}
\end{equation}
in recognition of the central role of $M_\beta(Y)$ in the definition of $n_g^\beta$.  It should be noted that two important ingredients are suppressed from the notation: the Chow variety $\mathrm{Chow}(\beta)$, and a certain perverse sheaf of vanishing cycles on $M_\beta(Y)$.  Modulo important technical details and caveats associated with the map $\pi_\beta$ and the Chow variety, we think of $n_g^\beta$ being an invariant of $M_\beta(Y)$.

We can now define 
\begin{equation}\label{eq:geometrictorsioninvariants}
n_g^\beta(X)=n_g\left(\bigcup_{\widehat{X}'\in\mathcal{S}}(M_\beta(\widehat{X})\right),
\end{equation}
In other words, we replace $M_\beta(Y)$ in (\ref{eq:gvnotation}) with the union $\bigcup_{\widehat{X}'\in\mathcal{S}}(M_\beta(\widehat{X}))$ of the moduli spaces of sheaves on small resolutions $\widehat{X}$.  Via the canonical derived equivalences between each of the small resolutions, we can view the union as a moduli space of objects in the derived category of any one small resolution.  The meaning of the right-hand side of  (\ref{eq:geometrictorsioninvariants}) is the application of the method of \cite{MT} to the morphism from this union to a corresponding union of Chow varieties.

A definition of the form (\ref{eq:geometrictorsioninvariants}) does not make sense when $X$ has a projective (K\"ahler)
small resolution, because in that situation the D2-brane charge of an exceptional curve changes sign under a flop and so we can't have a single Bridgeland stability condition containing the structure sheaf of a curve and its flop.  However, from results 
\cite[cor. B.1]{Li:1998hba} for the behavior of Gopakumar-Vafa invariants
under conifold transitions, if we add all the Gopakumar-Vafa invariants
in each of the charge sectors, one recovers the Gopakumar-Vafa invariants
of the generic smoothing $X_{\text{def.}}$. 

In the case of a torsion exceptional curve of charge $k$,
then the flopped curve has charge $-k$.  While there is still a sign change, a torsion class cannot affect the central charge and there is no contradiction.

In our examples, for $\widehat{X},\widehat{X}'\in\mathcal{S}$, the corresponding moduli spaces $M_\beta(\widehat{X})$ and $M_\beta(\widehat{X}')$ are either identified by the derived equivalence or are disjoint.  It follows that $n_g^\beta(X)$ can be computed as a sum of $n_g^\beta(\widehat{X})$ for a set of representative small resolutions $\widehat{X}$.  We will see many example calculations in Section~\ref{sec:geometric}, where we will find perfect match with the B-model.

Returning for simplicity to the situation where $H_2(\widehat{X},\mathbb{Z})\simeq \mathbb{Z}^r\oplus \mathbb{Z}_N$ and expressing classes in $H_2(\widehat{X},\mathbb{Z})$ as a pair $(\beta,\ell)\in \mathbb{Z}^r\oplus \mathbb{Z}_N$, the GV invariants $n_g^{\beta,\ell}$ defined as above simultaneously generate the $N$ partition functions $Z_{\text{top.},k}(t,\lambda)$ as expressed in (\ref{eq:Znc}).

We anticipate that there is an analogue of Donaldson-Thomas invariants of ideal sheaves in this context.  The natural definition would involve D6-D2-D0-branes on nc-resolutions with 1 unit of D6-brane charge. Then for each $X_{\text{n.c.,k}}$, we would expect an MNOP-type relationship between these DT invariants and the GV invariants.  However, given an analytic subspace $Z\subset\widehat{X}$ of dimension at most one, the ideal sheaf $\mathcal{I}_{Z,\widehat{X}}$ is not an $\alpha$-twisted sheaf, so it is not clear at present how to describe these  DT invariants in terms of the $D^b(\widehat{X},\alpha)$.  This complication is the reason why we focus on GV invariants in this paper, since the Brauer class does not complicate matters for sheaves of dimension at most 1.

\subsection{Constant map contributions on nc-resolutions}
\label{sec:constantmap}
We will now discuss the contribution of constant maps to the free energies of the topological A-model on $X_{\text{n.c.}k}$.
On a smooth Calabi-Yau threefold $Y$, these are the contributions to the free energies that arise from maps of the worldsheet to a generic point.
In the context of the non-commutative resolutions that we consider, the contribution from maps to the singularities of $X$ will be modified.

Following our discussion in Section~\ref{sect:fbnc}, we expect that the effect can be understood by replacing the nodal singularities with non-commutative conifolds.
At the level of Gopakumar-Vafa/Donaldson-Thomas invariants we therefore propose that the shift from each such replacement is captured by Szendrői's partition function~\eqref{eqn:coniztop}~\cite{szendrHoi2008non}, evaluated at rational values of the complexified K\"ahler parameter.
This leads to closed expressions for the constant map contributions that we verify later in concrete examples.

Let us start by considering the genus zero free energy on a smooth Calabi-Yau threefold $Y$ with complex K\"ahler parameters $t^i$.
As will be further discussed in Section~\ref{sect:octicdoub:b-model}, special  
geometry applies by mirror symmetry and the prepotential or genus zero
free energy is 
\begin{align}
\begin{split}
        \mathcal{F}_0=&-\frac{\kappa_{ijk}}{6}t^it^jt^k -
         \frac{\sigma_{ij}}{2} t^i t^j +\gamma_jt^j\\
         +&\frac{\zeta(3)}{(2\pi i)^3}\frac{\chi}{2}-\frac{1}{(2\pi i)^3}\!\!\!\! \sum\limits_{\beta\in H_2(Y)}\!\!\!\! n^{0}_\beta\text{Li}_3\left(q^\beta\right)\,.
         \end{split}
        \label{eqn:genuszeroconst}
\end{align}
Here $\kappa_{ijk}=D_i\cdot D_j \cdot D_k$ is the triple intersection 
on $Y$, $\gamma_j= c_2(TY)\cdot D_j/24$ and up to an integral  
symplectic transformation one can chose $\sigma_{ij}/2=\kappa_{iij}\ {\rm mod}\ 2$.
Since $\zeta(3)=\text{Li}_3(1)$, one can interpret the constant part in~\eqref{eqn:genuszeroconst} as the contribution from constant maps of the worldsheet into the Calabi-Yau~\cite{Bershadsky:1993cx}.
In general, the Gopakumar-Vafa invariant $n_g^C$ associated to an irreducible rational curve $C$ is given by
\begin{align}
        n_g^C=(-1)^d\chi({\mathcal{M}})\,,
\end{align}
where $\mathcal{M}$ is the moduli space of $C$ and $d=\text{dim}(\mathcal{M})$~\cite{Katz:1999xq}.
The moduli space of such maps is the Calabi-Yau $Y$ itself and therefore the Euler characteristic of the moduli space is $\chi=\chi(Y)$.

However, this raises the question why the contribution to~\eqref{eqn:genuszeroconst} is $\chi/2$ and not $-\chi$.
To avoid this subletly, one can instead consider the Donaldson-Thomas partition function~\cite{maulik2006gromov}.
Up to a change of variables, this only differs from the topological string partition function in that the constant map contributions are weighted with $-\chi$ instead of $\chi/2$.

The constant term in the higher genus free energies takes the form
\begin{align}
        \mathcal{F}_{(g\ge 2)}^{\text{const.}}=(-1)^{g-1}\frac{B_{2g}B_{2g-2}}{2g(2g-2)\left[(2g-2)!\right]}\frac{\chi}{2}\,.
        \label{eqn:cmapsgg2}
\end{align}
The coefficient of $-\chi/2$ can be interpreted as the limit $q\rightarrow 1$ of
\begin{align}
\begin{split}
        &\sum\limits_{m=1}^\infty\frac{1}{m}\frac{q^m}{\left(2\sin\left(\frac{m\lambda}{2}\right)\right)^2}\\=&\lambda^{-2}\cdot\text{Li}_3(q)+\sum\limits_{g=1}^\infty\lambda^{2g-2}\frac{(-1)^{g-1}B_{2g}}{2g\left[(2g-2)!\right]}\text{Li}_{3-2g}(q)\,,
        \label{eqn:mcpl}
        \end{split}
\end{align}
by performing $\zeta$-function regularization using $\text{Li}_s(1)=\zeta(s)$, valid only for $\text{Re}(s)>1$, and analytically continuing to
\begin{align}
        \zeta(3-2g)=-\frac{B_{2g-2}}{2g-2}\,,\quad g\ge 2\,.
\end{align}
Again we see that formally identifying the genus $0$ degree $0$ Gopakumar-Vafa invariant with $-\chi/2$ reproduces the correct result~\eqref{eqn:cmapsgg2}.

How does this change in the presence of the singularities that are non-commutatively resolved?
The correction to the constant map contributions on $X$ comes from the exceptional curves on $\widehat{X}$.
As discussed in Section~\ref{sect:fbnc}, $X_{\text{n.c.}k}$ corresponds to a choice of B-field on $\widehat{X}$ such that the action of a worldsheet instanton wrapped on an exceptional curve $\beta$ with charge $\ell$ is given by~\eqref{eq:charge}
\begin{align}
	\langle k\cdot B_0,\beta\rangle =e^{2\pi i k \ell/N}\,.
\end{align}
We denote the number of exceptional curve with $\mathbb{Z}_N$ charge $\ell$ again by $m^{(\pm \ell)}$.
Keeping in mind the factor $1/2$ between the constant map contributions to $Z_{\text{top.}}$ and $Z_{\text{DT}}$, we expect that
\begin{align*}
	\log(Z_{\text{top.},k})\big\vert_{\text{const.}}=\log(Z_{\text{top.},0})\big\vert_{\text{const.}}-\frac12\sum\limits_{l=1}^{\lfloor N/2\rfloor}m^{(\pm \ell)}\log\left[Z_{\text{DT},A}(k\ell/N,\lambda)\right]\,,
\end{align*}
where $Z_{\text{top.},0}$ is the topological string partition function associated to the smooth deformation $X_{\text{def.}}$ of $X$ and $Z_{\text{DT},A}(t,\lambda)$ is the Donaldson-Thomas partition function~\eqref{eqn:coniztop} of the non-commutative conifold.

Let us simplify this expression for the case $N=2$ that is the main interest in this paper.
Smoothness of $X_{\text{n.c.},1}$ requires that $m^{(0)}=0$, i.e. there are no exceptional curves on $\widehat{X}$ that are homologically trivial.
Then all exceptional curves are $2$-torsion and the total number of nodes is $m_s=m^{(\pm1)}$.
One can rewrite the Donaldson-Thomas partition function~\eqref{eqn:coniztop} of the non-commutative conifold as
\begin{align}
        \log\left[Z_{\text{DT},A}(b,\lambda)\right]=\sum\limits_{m=1}^\infty\frac{1}{m}\frac{e^{2\pi imb}-2+e^{-2\pi i mb}}{\left(2\sin\left(\frac{m\lambda}{2}\right)\right)^2}\,.
\end{align}
Together with~\eqref{eqn:mcpl} and the relations
\begin{align}
        \text{Li}_3(-1)=-\frac{3}{4}\zeta(3)\,,\quad \text{Li}_{3-2g}(-1)=\frac{1-2^{2g-2}}{2g-2}B_{2g-2}\,,
        \label{eqn:plm1}
\end{align}
as well as $\chi(\widehat{X})=\chi(X_{\text{def.}})+2 m_s$, one obtains the expressions
\begin{align}
\begin{split}
        &\mathcal{F}_{(g)}^{\text{const.}}(X_{\text{n.c.},1})\\=&\left\{\begin{array}{cl}
                \frac{\zeta(3)}{(2\pi i)^3}\left(\frac{\chi(\widehat{X})}{2}+\frac34m_s\right)&\text{ for }g=0\,,\\
                (-1)^{g-1}\frac{B_{2g}B_{2g-2}}{2g(2g-2)\left[(2g-2)!\right]}\left(\frac{\chi(\widehat{X})}{2}+(1-2^{2g-2})m_s\right)&\text{ for }g\ge 2\,.
        \end{array}\right.
        \label{eqn:cmapcontrz2}
\end{split}
\end{align}
Note that the identities~\eqref{eqn:plm1} hold without any regularization.
As will be discussed in Section~\ref{sect:octicdoub:b-model}, we obtain the free energies for the nc-resolution of the singular octic double cover by integrating the holomorphic anomaly equations up to genus $g=32$ and verify that~\eqref{eqn:cmapcontrz2} indeed holds.
\section{The octic double solid}\label{sec:octic}

Double covers of $\mathbb{P}^3$ branched over degree 8 hypersurfaces were studied in~\cite{Clemens}, where they were called octic double solids.
In this section, we apply the ideas discussed earlier in the paper to a particular octic double solid $X$ with terminal conifold singularities.

In Section~\ref{sec:singdouble} we discuss the geometry of
the singular Calabi-Yau $X$, which has no K\"ahler small resolutions.
In Section~\ref{sect:octicdoub:small-analytic} we discuss
non-K\"ahler small resolutions of $X$, which we denote by $\widehat{X}$.
In Sections~\ref{subsec:spinor} and~\ref{subsec:brauer} we give some background on spinor sheaves and on the Brauer group, both of which are needed to understand the derived equivalence  between $D^b(\widehat{X},\alpha)$ and $D^b(Y)$, where $\alpha\in\text{Br}(\widehat{X})$ and the smooth dual Calabi-Yau threefold $Y$ is a complete intersection of four quadrics in $\mathbb{P}^7$, in Section~\ref{subsec:derived}.
In Section~\ref{sect:octicdoub:nc-res:glsm} we describe an
nc resolution of $X$, denoted $X_{\rm n.c.}$, and its physical realization via
a GLSM.  The GLSM implicitly provides a derived equivalence between
$D^b(X_{\rm n.c.})$ and $D^b(Y)$ and suggests the relation of $X_{\text{n.c.}}$ and $(\widehat{X},\alpha)$ under renormalization group flow.
Later in the paper we will discuss curve-counting in this model, first using mirror symmetry in Section~\ref{sect:octicdoub:b-model},
and then, in Section~\ref{sec:geometric}, we verify some of our results directly in the A-model geometry.

\subsection{The singular double cover $X$ of $\mathbb{P}^3$}
\label{sec:singdouble}
A generic degree eight hypersurface in the weighted projective space $\mathbb{P}_{11114}$ is a smooth Calabi-Yau threefold $X_{\text{def.}}$.
Denoting the homogeneous coordinates by $[x_1:x_2:x_3:x_4:w]$, the defining equation can always be brought into the form
\begin{align}\label{eq:doublecover}
        w^2=p_8(x_1,x_2,x_3,x_4)\,,
\end{align}
where $p_8(x_1,\ldots,x_4)$ is a homogeneous polynomial of degree eight.
This makes it obvious that $X_{\text{def.}}$ is a double cover of $\mathbb{P}^3$, which is ramified over the degree eight hypersurface
\begin{align}
        S_{\text{ram.}}=\{p_8(x_1,\ldots,x_4)=0\}\subset\mathbb{P}^3\,.
\end{align}
For general $p_8$, $X_{\text{def.}}$ is a smooth octic double solid.

All deformations of $X_{\text{def.}}$ arise as deformations of the weight 8 hypersurface \cite{Batyrev:1993oya}, so we can complete the square to put the equation in the form (\ref{eq:doublecover}) and count parameters.  Substracting the 16 parameters of a general linear transformation on the homogeneous coordinates on $\mathbb{P}^3$ from the 165 monomials of degree eight in four variables leaves $h^{2,1}=149$ independent complex structure deformations.
The only divisor on $X_{\text{def.}}$ is induced from the ambient space such that $h^{1,1}=1$ and therefore $\chi=-296$.

We are interested in the sublocus of the complex structure moduli space where
\begin{align}
        p_8(x_1,\ldots,x_4)=\det\,A_{8\times 8}(x_1,\ldots,x_4)\,,
        \label{eqn:detram}
\end{align}
and $A_{8\times 8}$ is a symmetric $8\times 8$ matrix with entries linear in $x_1,\ldots,x_4$.
The ramification locus then has ordinary double points (ODPs), that is nodal singularities, where the corank of the matrix is greater than one.  We denote the corresponding singular Calabi-Yau by $X$.

In local analytic coordinates $(x,y,z)$ on $\mathbb{P}^3$ near one of the $A_1$ singularities, we can write $p_8$ as $x^2+y^2+z^2$, so (\ref{eq:doublecover}) takes the form
\begin{align}
w^2=x^2+y^2+z^2\,,
\end{align}
and we see that $X$ has conifold points over the $A_1$ singularities and is smooth elsewhere.  Since $X$ is obtained from $X_{\text{def.}}$ up to homotopy  by contracting $S^3$'s to the conifolds, we have \cite{Clemens}
\begin{align}
H_2(X,\mathbb{Z}) = H_2(X_{\text{def.}},\mathbb{Z})=\mathbb{Z}\,.
\end{align}

In general, if $A$ is a symmetric $n\times n$ matrix with entries that are sections of a line bundle $L$ on some compact oriented manifold, the class of the locus where the corank is at least $r$ is given by~\cite{HARRIS198471}
\begin{align}
        \left[\det\,A\right]=\left(\prod\limits_{k=0}^{r-1}\left(\begin{array}{c}n+k\\r-k\end{array}\right)\big\slash\left(\begin{array}{c}2k+1\\k\end{array}\right)\right)c_1(L)^{\scriptsize\left(\begin{array}{c}r+1\\2\end{array}\right)}\,.
\end{align}
Applied to $A_{8\times 8}$, this implies the ramification locus has $n_s=84$ nodes.
It is also easy to determine the number of complex structure deformations that preserve the singularities.
To this end one notes that one can use a generic matrix $S\in \text{SL}(8,\mathbb{C})$ and transform
\begin{align}
A_{8\times 8}\rightarrow S^\intercal A_{8\times 8} S\,,
\end{align}
without affecting $p_8(x_1,\ldots,x_4)$.
Taking into account the toric automorphisms, this leaves
\begin{align}\label{eq:cxmoduli}
        n_{\text{cplx.}}=144-63-16=65\,,
\end{align}
deformations in the subspace of the complex structure moduli space where the ramification locus is of the form~\eqref{eqn:detram}.

We digress for a moment to consider generalities.  Let $\mathcal{M}$ be any moduli space of hypersurfaces of any dimension, and let $\mathcal{M}_n\subset\mathcal{M}$ be the closure of the set of hypersurfaces with $n$ nodes and no other singularities.  Then the codimension of $\mathcal{M}_n$ in $\mathcal{M}$ is at most $n$, as the condition on the coefficients for the hypersurface to have a node is given by the vanishing of a single resultant.
Returning to our example, since $h^{2,1}(X_{\text{def.}})=149$ while the singular $X$ has 65 complex moduli, we see that the 84 nodes collectively impose the maximum number 84 of independent conditions.

\subsection{Small analytic resolutions}
\label{sect:octicdoub:small-analytic}

A bit more generally, let $X$ be any octic double solid with a set $S$ of conifold points and otherwise smooth.  Let $\mathbb{Z}^{S}$ be the free abelian group generated by $S$.   Let $\widehat{X}$ be an analytic small resolution of $X$, with the exceptional curve $C_p\simeq\mathbb{P}^1\subset \widehat{X}$ corresponding to the conifold point $p$. Such resolutions always exist~\cite{Clemens}, and have trivial $K_{\widehat{X}}$.

Since $X$ can be obtained from $\widehat{X}$ up to homotopy by gluing the boundary of a 3-disk to $C_p$, we have an exact sequence~\cite{Clemens}
\begin{equation}
0 \rightarrow H_3(\widehat{X},\mathbb{Z}) \rightarrow  H_3(X,\mathbb{Z}) \stackrel{k}{\rightarrow}  \mathbb{Z}^{S} \stackrel{\alpha}{\rightarrow} H_2(\widehat{X},\mathbb{Z}) \rightarrow  H_2(X,\mathbb{Z}) \rightarrow 0\,,
\end{equation}
where $\alpha(p)=[C_p]$.   We also have the isomorphism
\begin{equation}
H_4(\widehat{X},\mathbb{Z}) \simeq  H_4(X,\mathbb{Z})\,,
\end{equation}
or equivalently by, Poincar\'e duality,
\begin{equation}
H^2(\widehat{X},\mathbb{Z}) \simeq  H^2(X,\mathbb{Z})\,.
\end{equation}
Since $H^2(X,\mathbb{Z})$ is known to be generated by the pullback of the hyperplane class of $\mathbb{P}^3$, it follows that $H^2(\widehat{X},\mathbb{Z})$ is generated by the pullback of the hyperplane class of $\mathbb{P}^3$ as well.

Let $n=\vert S\vert$ be the number of conifold points and let $\delta=n-\mathrm{rank}(k)$ be the corank of $k$ (called the \emph{defect} in \cite{Clemens}).  We have the relation between the homology of $\widehat{X}$ and $X_{\text{def.}}$ (see also \cite{Werner}):
\begin{equation}\label{eq:conifoldtrans}
b_3(\widehat{X}) = b_3(X_{\text{def.}}) -2n+2\delta\,.
\end{equation}

In our situation of the determinantal octic, it was computed that $\delta=0$ \cite{Addington:2009thesis}.  Therefore
$\mathbb{Z}^{S}/\mathrm{Im}(k)$ is a finite abelian group and we
 have a short exact sequence
\begin{equation}
0 \rightarrow \mathbb{Z}^{S}/\mathrm{Im}(k)\rightarrow H_2(\widehat{X},\mathbb{Z}) \rightarrow  H_2(X,\mathbb{Z})=\mathbb{Z} \rightarrow 0\,.
\end{equation}
This sequence identifies $\mathbb{Z}^{S}/\mathrm{Im}(k)$ with the torsion subgroup of  $H_2(\widehat{X},\mathbb{Z})$.
Since each $[C_p]$ is in this subgroup, it is a torsion class (possibly trivial a priori).
In particular $\widehat{X}$ is not K\"ahler, since $\int_{C_p}\omega$ would be strictly positive for a K\"ahler form $\omega$, but vanishes since $[C_p]$ is torsion.

We also have $\mathbb{Z}^{S}/\mathrm{Im}(k)\simeq\mathbb{Z}_2$, as will be shown in Appendix~\ref{app:brauer}, using an argument outlined for us by N. Addington.  Thus
\begin{align}\label{eq:torsionh2}
H_2(\widehat{X},\mathbb{Z})\simeq \mathbb{Z}\oplus\mathbb{Z}_2\,.
\end{align}
It will be important to note that while the $\mathbb{Z}_2$ torsion subgroup of $H_2(\widehat{X},\mathbb{Z})$ is canonical, the splitting in (\ref{eq:torsionh2}) is not canonical.

We now claim that each exceptional curve $C_p$ represents the nontrivial torsion class.  Since the $\{[C_p]\}$ generate the $\mathbb{Z}_2$ torsion subgroup, at least one of the $[C_p]$ is the nontrivial class.  We then see that all $[C_p]$ are nontrivial by a monodromy argument.  We consider the space of pairs
\begin{align}
I=\left\{(p,A) \mid \operatorname{rank}(A(p))=6
\right\}\,,
\end{align}
where $A$ is an $8\times 8$ symmetric matrix of linear forms on $\mathbb{P}^3$.  Fixing $A$, If $(p,A)\in I$ then $p$ can only be one of the 84 nodes, so $I$ is an 84-sheeted cover of the space of symmetric $8\times 8$ matrices $A$.
However, if we project $I$ to $\mathbb{P}^3$ by sending $(p,A)$ to $p$, then all fibers are isomorphic to the determinantal variety in the space of symmetric $8\times 8$ matrices defined by the vanishing of the $7\times 7$ minors.  This space is well-known to be irreducible (for example the proof in \cite[Chapter II.2]{ACGH} is readily adapted from the context of general matrices to symmetric matrices).  It follows that $I$ is irreducible.  Thus each of the 84 exceptional curves $C_p$ are related by monodromy, so none of the $[C_p]$ can vanish.

We can now explain the reflection identity (\ref{eqn:reflection}).  We consider the moduli space $\mathcal{M}$ of determinantal octic double solids and its $2^{84}$-sheeted cover $\widetilde{M}$ of small resolutions $\widehat{X}$. In the absence of any obvious way to choose a distinguished proper subset of these small resolutions over the entire moduli space $\mathcal{M}$, we anticipate that $\mathcal{M}$ is connected.  As a consequence, any exceptional curve $C_p$ would be connected by monodromy to its flopped exceptional curve $C_p'$.  Continuing to identify the cohomologies of the small resolutions as we have been doing, we have $[C_p']=-[C_p]$, so its $\mathbb{Z}_2$ charge changes sign.  The degree $\beta$ of a curve is independent of the choice of small resolution, so is a monodromy invariant.  The reflection identity~\eqref{eqn:reflection} follows immediately from the deformation invariance of GV invariants.

While this argument has been provided specifically for the determinantal octic double solid, it readily adapts for all examples with terminal nodal singularities, by arguing that any of these small resolutions are deformation equivalent to one of their flops.

\subsection{Spinor sheaves} \label{subsec:spinor}
Our matrix $A_{8\times 8}(x_1,x_2,x_3,x_4)$ can be viewed as a family of symmetric $8\times 8$ matrices (up to scalar multiplication) parametrized by $(x_1,\ldots,x_4)\in \mathbb{P}^3$.  Equivalently, this data can be interpreted as a linear system $Q$ of quadric hypersurfaces in $\mathbb{P}^7$, or quadratic forms on $\mathbb{C}^8$ (up to scalar multiplication), parametrized by $\mathbb{P}^3$.  The base locus of this linear system is a complete intersection $Y$ of four quadric hypersurfaces in $\mathbb{P}^7$, a familiar example of a Calabi-Yau threefold with $h^{1,1}=1$.   We also naturally get a sheaf of Clifford algebras over $\mathbb{P}^3$.

Fix $p\in \mathbb{P}^3$ and let $Q_p\subset\mathbb{P}^7$ be the associated quadric hypersurface and $\mathrm{Cl}_p$ the associated Clifford algebra with 8 generators.  Let $w_1,w_2,w_3,w_4$ be a basis for a subspace $W$ which is isotropic for the quadratic form, or, equivalently, $\mathbb{P}^3\simeq\mathbb{P}(W)\subset Q_p\subset\mathbb{P}^7$.  Consider the left ideal in $\mathrm{Cl}_p$,
\begin{align}
I = \mathrm{Cl}_p\cdot w_1w_2w_3w_4\,,
\end{align}
which is seen to only depend on $W=\mathrm{span}(w_i)$ since $W$ is isotropic.  The $\mathbb{Z}_2$-grading $\mathrm{Cl}_p=\mathrm{Cl}_p^0\oplus \mathrm{Cl}_p^1$ determines a $\mathbb{Z}_2$ grading $I=I_{\mathrm{even}}\oplus I_{\mathrm{odd}}$.  Both $I_{\mathrm{even}}$ and $I_{\mathrm{odd}}$ are $\mathrm{Cl}_p^0$-modules.

Identifying $\mathbb{P}^7$ with the Grassmannian $G(1,8)$, the bundle $\mathcal{O}_{\mathbb{P}^7}(-1)$ is identified with the universal subbundle of the trivial bundle $\mathbb{P}^7\times\mathbb{C}^8$ on $\mathbb{P}^7$.  Therefore, left multiplication by elements of $\mathbb{C}^8$ acting on $I$ can be interpreted as a mapping of vector bundles on
$\mathbb{P}^7$,
\begin{align}
\phi:\mathcal{O}_{\mathbb{P}^7}(-1) \otimes I_{\mathrm{odd}}\to \mathcal{O}_{\mathbb{P}^7}(-1) \otimes I_{\mathrm{even}}\,.
\end{align}
The spinor sheaf $S$ associated to $W$ is the cokernel of $\phi$,
\begin{align}
S=\mathrm{coker}(\phi)\,,
\end{align}
and is supported on $Q_p$.

If the quadratic form has rank 8, so that $Q_p$ is smooth, then there are two connected families of $\mathbb{P}^3$'s contained in $Q_p$.  The isomorphism class of the spinor sheaf associated to the $\mathbb{P}^3$ only depends on which of the two families the $\mathbb{P}^3$ is contained in \cite{addington2009derived}.  Thus there are two spinor sheaves supported on $Q_p$.

If the quadratic form has rank 7, then the family of $\mathbb{P}^3$'s contained in $Q_p$ is connected, and there is only one spinor sheaf supported on $Q_p$ \cite{addington2009derived}.

The situation is more interesting if the quadratic form has rank~6, so that $\mathrm{Sing}(Q_p)=L\simeq\mathbb{P}^1$.  In this case, $Q_p$ can be described as the cone over a smooth $Q_p'\subset\mathbb{P}^5$ with vertex $L$.

Now pick a point $p\in L$ and a $\mathbb{P}^2\simeq P\subset Q_p'$.
Then $p$ and $P$ span a 3-plane in $Q_p$ and therefore determine a spinor sheaf.
It is shown in \cite{addington2009derived} that these spinor sheaves only depend on the point $p$ and the family of $\mathbb{P}^2$'s on $Q_p'$ that $P$ belongs to (there are two such families) \cite{addington2009derived}.
Thus the set of spinor sheaves supported on $Q_p$ is a disjoint union of two $\mathbb{P}^1$'s.

In summary, the set of spinor sheaves is set-theoretically the singular octic double solid $X$, with each conifold replaced by two $\mathbb{P}^1$'s.  \footnote{This set is not an algebraic variety.  To obtain a good moduli space, we need a notion of stability.  If we choose Gieseker stability, it can be seen that the spinor sheaves corresponding to either $\mathbb{P}^1$ are all S-equivalent and therefore get blown back down to the conifold point.  So the moduli space of Gieseker stable sheaves up to S-equvalence is the singular octic double solid $X$, and the moduli space is not a resolution of singularities.}

The spinor sheaves on the $Q_p$ can all be restricted to the base locus $Y$.  If we make a choice $\widehat{X}$ of small resolution (so we are choosing one of the two $\mathbb{P}^1$s at each conifold), then these spinor sheaves on $Y$ fit into a reasonably nice family of sheaves parametrized by $\widehat{X}$.  If there were a universal sheaf on $\widehat{X}\times Y$, it could be used to construct a Fourier-Mukai transform from $D^b(X)$ to $D^b(Y)$.  While there is no universal sheaf, there is a universal twisted sheaf $\mathcal{E}$, with twisting determined by a Brauer class $\alpha\in \mathrm{Br}(\widehat{X})$.

The construction of the universal twisted sheaf in \cite{addington2009derived} relies on the construction of families of 4-dimensional isotropic subspaces, locally defined over $\widehat{X}$.  The space of such isotropic subspaces forms an $\mathrm{OGr}^+(4,8)$-bundle over $\widehat{X}$.\footnote{This is clear away from the exceptional curves.  We will explore a finer analysis over the exceptional curves in Appendix~\ref{app:brauer}, where we give an equivalent formulation of the Brauer class.}  The Brauer class $\alpha$ is the obstruction to finding a rational section of this $\mathrm{OGr}^+(4,8)$-bundle.   The Brauer class $\alpha$ is nontrivial, as the discussion in \cite{addington2009derived} makes it clear that there is no rational section.

We digress to quickly review Brauer groups and twisted sheaves before completing the description of the derived equivalence of \cite{addington2009derived}.

\subsection{The Brauer group and twisted sheaves}\label{subsec:brauer}

The Brauer group $\text{Br}(X)$ of a complex manifold $X$ is generated by projective bundles or Azumaya algebras.  Here we outline the description in terms of projective bundles.  The description in terms of Azumaya algebras is similar.

Let $P\to X$ be a $\mathbb{P}^{r-1}$ bundle.  We choose an open cover $\{U_\alpha\}$ over which $P$ restricts to a trivial bundle:
\begin{align}
\phi_\alpha:P\vert_{U_\alpha}\simeq U_\alpha\times\mathbb{P}^{r-1}\,.
\end{align}
Since $\mathrm{Aut}(\mathbb{P}^{r-1})\simeq\operatorname{PGL}(r)$, the automorphism $\phi_\beta\circ\phi_\alpha^{-1}$ of $(U_\alpha\cap U_\beta)\times \mathbb{P}^{r-1}$ determines a map
\begin{align}
\psi_{\alpha\beta}:U_\alpha\cap U_\beta\to \operatorname{PGL}(r)\,.
\end{align}
Since $\psi_{\alpha\gamma}=\psi_{\alpha\beta}\circ\psi_{\beta\gamma}$, the $\psi_{\alpha\beta}$ determine a cohomology class $\psi\in H^1(X,\operatorname{PGL}(r))$, where with some abuse of notation $\operatorname{PGL}(r)$ denotes the sheaf of holomorphic  $\operatorname{PGL}(r)$-valued functions.

Letting $\mu_r$ denote the sheaf of $r$th roots of unity, we infer from the exact sequence
\[
0\to \mu_r\to \operatorname{GL}(r)\to \operatorname{PGL}(r)\to 0\,,
\]
a coboundary map
\[
H^1(X,\operatorname{PGL}(r)) \to H^2(X,\mu_r)\,,
\]
which can be followed by the map $H^2(X,\mu_r)\to H^2(X,\mathcal{O}_X^*)$ induced from the inclusion $\mu_r\hookrightarrow \mathcal{O}_X^*$ to yield a map
\begin{align}\label{eq:cobound+}
H^1(X,\operatorname{PGL}(r))\to H^2(X,\mathcal{O}_X^*)\,.
\end{align}
Since the group $H^2(X,\mu_r)$ is $r$-torsion, the image of $\psi$ under (\ref{eq:cobound+}) is also $r$-torsion.

Grothendieck asked if every torsion class in $H^2(X,\mathcal{O}_X^*)$ is in the image of (\ref{eq:cobound+}) for some $r$.  This is proven for algebraic varieties in \cite{GdJ}, but we have to say a bit more in our sitution since a small resolution $\widehat{X}$ of the determinantal octic double is not even K\"ahler.  

We return to the non-algebraic situation momentarily after providing some clarification.  We note that the group $H^2(X,\mathcal{O}_X^*)_{\mathrm{tors}}$ is called the cohomological Brauer group, which is clearly represented by projective bundles.  We do not need the formal definition of $\text{Br}(X)$, but the result of \cite{GdJ} as stated there is that a natural map
\begin{align}\label{eq:cohbrauer}
\operatorname{Br}(X) \to H^2(X,\mathcal{O}_X^*)_{\mathrm{tors}}\,,
\end{align}
is an isomorphism.  If an element of $\operatorname{Br}(X)$ is represented by a projective bundle, then its image via (\ref{eq:cohbrauer}) is identified with the image of the projective bundle via (\ref{eq:cobound+}).

In the general context of complex manifolds $X$, the natural map (\ref{eq:cohbrauer}) is at least an inclusion \cite{schroer}.
From the exact sequence
\begin{align}
0\to \mathbb{Z}\to \mathcal{O}_X\to \mathcal{O}_X^*\to 0\,,
\end{align}
we get maps
\begin{align}
H^2(X,\mathcal{O}_X)\to H^2(X,\mathcal{O}_X^*)\to H^3(X,\mathbb{Z})\to H^3(X,\mathcal{O}_X)\,.
\end{align}
If $H^2(X,\mathcal{O}_X)=0$, which holds for a Calabi-Yau threefold, we deduce that
\begin{align}
H^2(X,\mathcal{O}_X^*)_{\mathrm{tors}}\simeq H^3(X,\mathbb{Z})_{\mathrm{tors}}\,.
\end{align}
This isomorphism holds for a small resolution of the octic double solid as well by \cite[Proposition 1.1]{schroer} since $H^2(\widehat{X},\mathbb{Z})\simeq\mathrm{Pic}(\widehat{X})$, as $H^2(\widehat{X},\mathbb{Z})$ is generated by the pullback of $\mathcal{O}_{\mathbb{P}^3}(1)$.
In Appendix \ref{app:brauer}, we show that $\mathbb{Z}_2\subset\mathrm{Br}(\widehat{X})$ and $H^3(X,\mathbb{Z})_{\mathrm{tors}}\simeq\mathbb{Z}_2$.  It follows that also for the non-K\"ahler small resolution $\widehat{X}$
\begin{align}\label{eq:brauereqs}
\operatorname{Br}(\widehat{X}) \simeq H^2(\widehat{X},\mathcal{O}_{\widehat{X}}^*)_{\mathrm{tors}}\simeq H^3(X,\mathbb{Z})_{\mathrm{tors}}\,.
\end{align}

\smallskip
We say a few words here about Azumaya algebras.  Briefly, an Azumaya algebra is a bundle of $r\times r$ matrix algebras $M_{r\times r}$.  Since $\operatorname{Aut}(M_{r\times r})\simeq \operatorname{PGL}(r)$ by conjugation, the subsequent analysis coincides with what was written above starting with projective bundles.  Said differently, there is a 1-1 correspondence between projective bundles and Azumaya algebras up to isomorphism.

 We now describe twisted sheaves, which are determined by a class $c\in H^2(X,\mathcal{O}_X^*)$.  Choose a \v{C}ech cocycle $c_{\eta\theta\iota}\in Z^2(\{U_\eta\},\mathcal{O}_X^*)$ representing $c$.

A $c$-twisted sheaf $\mathcal{F}$ is a collection of sheaves
 $\mathcal{F}_\eta$ on each $U_\eta$ together with isomorphisms
\[
\phi_{\eta\theta}:\mathcal{F}_\eta\vert_{U_\eta\cap U_\theta} \simeq \mathcal{F}_\theta\vert_{U_\eta\cap U_\theta}
\]
such that the isomorphism $\phi_{\eta\iota}^{-1}\circ \phi_{\theta\iota}\circ \phi_{\eta\theta}$ is multiplication by $c_{\eta\theta\iota}$.  The categories of $c$-twisted sheaves determined by different representative cocycles $c_{\eta\theta\iota}$ and $c_{\eta\theta\iota}'$ are readily checked to be equivalent.

A twisted sheaf $\mathcal{F}$ is called a twisted bundle if each $\mathcal{F}_\eta$ is a vector bundle.  It is straightforward to check that nonzero $c$-twisted bundles exist only if $c\in H^2(X,\mathcal{O}_X^*)$ is torsion.  For this reason, we only consider twisted sheaves associated to Brauer classes $\alpha\in \text{Br}(X)$, where the twisting class associated to $\alpha$ is determined by (\ref{eq:cohbrauer}).  In practice, $c$ can arise physically from a flat topologically nontrivial B-field, as illustrated in equation~(\ref{eq:bfield:shortexact}), and D-branes in such a background are well-known to be described by sheaves of modules over Azumaya algebras, see e.g.~\cite{Kapustin:1999di}.

Associated to the category of twisted sheaves we obtain the derived category of $\alpha$-twisted sheaves $D^b(\widehat{X},\alpha)$, which is the category of topological B-branes in the presence of the B-field.

\subsection{The derived equivalence}\label{subsec:derived}

We can finally describe the derived equivalence $D^b(\widehat{X},\alpha) \cong
D^b(Y)$ of  \cite{addington2009derived}.  
Recall that $\widehat{X}$ parametrizes certain spinor sheaves on $Y$.  It was shown in \cite{addington2009derived} that $\widehat{X}$ can be covered by open subsets $U_\eta$ with the property that the spinor sheaves parametrized by $U_\eta\subset\widehat{X}$ have a universal sheaf $\mathcal{E}_\eta$ on $U_\eta\times Y$.    These locally defined universal sheaves cannot be glued to a sheaf on $\widehat{X}\times Y$, but they can be glued to form an $\alpha$-twisted sheaf on $\widehat{X}\times Y$, with twisting associated to a Brauer class $\alpha\in\operatorname{Br}(\widehat{X})$.

Then it is shown that the Fourier-Mukai transform with kernel $\mathcal{E}$ is a derived equivalence
\begin{align}\label{eq:derivedequiv}
D^b(\widehat{X},\alpha) = D^b(Y)\,.
\end{align}
Furthermore, it is shown in~\cite{kuznetsov2008derived} that we have a derived equivalence
\begin{equation}
D^b(\mathbb{P}^3,\mathcal{B})= D^b(Y)\,,
\label{eqn:depby}
\end{equation}
where $\mathcal{B}$ is a certain sheaf of non-commutative algebras on $\mathbb{P}^3$.
This is an example of homological projective duality.

Together,~\eqref{eq:derivedequiv} and~\eqref{eqn:depby} imply that $D^b(\widehat{X},\alpha)=D^b(\mathbb{P}^3,\mathcal{B})$.
The derived category $D^b(\mathbb{P}^3,\mathcal{B})$ of modules over $\mathcal{B}$ is expected to be an example of a categorical resolution of $X$~\cite{kuzso}.
As we will discuss in the next section, both $(\mathbb{P}^3,\mathcal{B})$ and $(\widehat{X},\alpha)$ naturally arise from the same hybrid phase in a gauged linear sigma model.
This motivates our interpretation of the pair $(\widehat{X},\alpha)$ as a non-commutative resolution $X_{\mathrm{n.c.}}$ of $X$.

Before that, let us briefly turn to the torsion refined GV invariants of $X$, which we understand geometrically as 1-dimensional sheaves of $\mathcal{B}$-modules on $\mathbb{P}^3$, or equivalently, as 1-dimensional sheaves of $\mathcal{C}$-modules on $X$.   The crucial point is that spinor sheaves are described in terms of modules over the even parts of Clifford algebras, which is precisely the same thing as a 0-dimensional $\mathcal{B}$-module.  

Since the Brauer group of a curve $C$ is trivial (as $H^2(C,\mathcal{O}_C^*)=H^3(C,\mathbb{Z})=0$) there is no essential distinction between considering twisted sheaves or sheaves on curves.  So if we let $\mathrm{Coh}_{\le 1}(\widehat{X})\subset \mathrm{Coh}(\widehat{X})$ denote the subcategory of sheaves on $\widehat{X}$ supported in dimension at most 1, we have
\begin{align}\label{eq:dim1untwisted}
\mathrm{Coh}_{\le 1}(\widehat{X})\subset D^b(\widehat{X},\alpha)\,.
\end{align}
Let $\mathcal{S}$ be the set of all $2^{84}$ small resolutions of $X$.  Then~\eqref{eq:derivedequiv} and~\eqref{eq:dim1untwisted} hold for any $\widehat{X}\in\mathcal{S}$.

Roughly speaking~\footnote{After all, on an ordinary variety, a sheaf on a curve has more information than a 1-parameter family of points.  For instance, the sheaf could depend on a choice of line bundle supported on the curve.}, the analysis of section~\ref{subsec:spinor} shows that 1-dimensional sheaves on $\widehat{X}$ correspond to 1-parameter families of spinor sheaves on $Y$.
However, the discussion in Section~\ref{subsec:spinor} also makes clear that, since our objective is to compute the torsion refined GV invariants of $X$, we have to use all spinor sheaves, hence all  small resolutions of ${X}$.
This discussion provides justification for the proposal of
section~\ref{subsec:gv} in the case of the singular octic double solid.

\subsection{The nc-resolution and its physical realization}
\label{sect:octicdoub:nc-res:glsm}
As mentioned above, a categorical resolution of the singular octic double solid $X$ was constructed by Kuznetsov in the context of homological projective duality~\cite{kuznetsov2008derived,kuznetsov2008lefschetz} and is given by a sheaf of Clifford algebras $\mathcal{B}$ on $\mathbb{P}^3$.
Away from the nodes, this is equivalent to a sheaf of Azumaya algebras on the singular double cover itself~\cite{kuznetsov2008derived}.
Sheaves of Azumaya algebras on the other hand exactly correspond to the non-commutative deformations of the structure sheaf in the presence of a non-trivial flat B-field in string theory~\cite{Berenstein:2001jr}.

The nc resolution of the octic double solid also emerges as a phase of a gauged linear sigma model, as discussed in
\cite{Caldararu:2010ljp,Addington:2012zv,Sharpe:2012ji}, 
and which we review in this section.

Consider a GLSM describing the complete intersection $Y$.
This GLSM has gauge group $G=U(1)$ and four chiral fields $p_i,\,i=1,\ldots,4,$ of charge $-2$ as well as eight chiral fields $x_j,\,j=1,\ldots,8$ of charge $1$.
The Fayet-Iliopoulos parameter $r$ can be combined with the theta angle $\theta$ into the complex parameter
\begin{align}
        t=\frac{\theta}{2\pi}+ir\,.
\end{align}
A generic superpotential takes the form
\begin{align} \label{eq:mass-matrix}
        W=p_i Q^i(x_1,\ldots,x_8)=\vec{x}^{\,\intercal} A_{8\times 8}(p)\vec{x}\,,
\end{align}
where $Q^{i,\,i=1,\ldots,4}$ are sufficiently generic quadratic homogeneous polynomials in $x_{j=1,\ldots,8}$ and one can choose respective R-charges $-2$ and $0$ for the fields $p_i$ and $x_j$ such that $W$ carries the required R-charge $-2$.
The F-term equations then read
\begin{align}
        \partial_{p_i}W=0\,,\quad i=1,\ldots,4\,,\quad \partial_{x_j}W=0\,,\quad j=1,\ldots,8\,,
        \label{eqn:fterm}
\end{align}
while the D-term equation is given by
\begin{align}
        \sum\limits_{j=1}^8\vert x_j\vert^2-\sum\limits_{i=1}^4\vert p_i\vert^2=r\,.
        \label{eqn:dterm}
\end{align}

In the phase $r\gg0$, the D-term equation~\eqref{eqn:dterm} requires that at least one of the $x_j$ is non-vanishing.
If the superpotential is sufficiently generic, the F-term equations can only be solved if all $p$-fields vanish simultaneously.
Identifying vacua that are related by the $U(1)$ gauge transformation, one finds that the theory flows to a non-linear sigma model on the intersection of four quadrics
\begin{align}
     Y \: = \:   \left\{Q_1=\ldots=Q_4=0\right\}\subset\mathbb{P}^7\,.
\end{align}

On the other hand, solving the F- and D-term equations simultaneously in the phase $r\ll 0$ forces us to set $x_j=0,\,j=1,\ldots,8$ while the $p$-fields become homogeneous coordinates on a $\mathbb{P}^3$.
The $x_j$ become sections of the fractional bundle $\mathcal{O}_{\mathbb{P}^3}\left(-\frac12\right)$ and the GLSM flows to a $\mathbb{Z}_2$ quotient of a Landau-Ginzburg theory with quadratic superpotential that is fibered over a $\mathbb{P}^3$.
Phases corresponding to fibrations of Landau-Ginzburg models over some non-trivial base space are usually referred to as hybrid phases.

We can analyze this hybrid model locally over the space of $\vec{p}$'s in
a Born-Oppenheimer approximation.  At generic points, the
superpotential~(\ref{eq:mass-matrix}) is determined by the a mass matrix $A_{8\times 8}(p)$ for the fields
$\vec{x}$.  At low energies, one then generically has a $U(1)$
gauge theory with only fields of charge $-2$, and so there is a
trivially-acting ${\mathbb Z}_2$ subgroup of the gauged $U(1)$, meaning
a global $B {\mathbb Z}_2$ one-form symmetry.
This is the setup for decomposition~\cite{Hellerman:2006zs}, hence such a
theory is equivalent (locally in Born-Oppenheimer) to a disjoint union of
$\vert \hat{\mathbb Z}_2 \vert = 2$ theories, each a copy of the space of
$\vec{p}$'s, which gives us a double cover structure.  
Over the locus where the mass matrix develops a zero eigenvalue,
the one-form symmetry disappears, and the two sheets of the cover collide.
Furthermore, one can show \cite[section 2.3]{Caldararu:2010ljp} that due
to Berry phases, monodromies around the branch loci result in interchanging
the sheets. 

The resulting structure is, for the most part, a branched double cover
of ${\mathbb P}^3$ (the space of $\vec{p}$'s), branched over the locus
\begin{equation}
\{ \det A_{8\times 8}(p) = 0 \}\subset {\mathbb P}^3\,,
\end{equation}
where the mass matrix has zero eigenvalues.
However, there is a slight complication in this description.
The branch locus consists generically of points describing
quadrics in ${\mathbb P}^7$ (defined by the
superpotential) with a corank 1 degeneration.
Within the branch loci are nongeneric points where the quadrics defined by 
the superpotential have a corank 2 degeneration.
Geometrically, these points represent singularities of the
branched double cover.  

Physically, however, the GLSM remains regular even at the points where the branched double cover
is singular.
A physical GLSM singularity would manifest itself as an unbounded direction
in the solutions to the F-terms, which we can write as
\begin{equation}
\sum_j A^{ij}_{8\times 8}(p) x_j = 0\,,
\: \: \:
\sum_{ij} x_i \frac{\partial A^{ij}_{8\times 8}}{\partial p_k} x_j = 0\,.
\end{equation}
Solving the first equation requires that $\vec{x}$ is a zero eigenvector
of the mass matrix $A$, consistent with the geometrical observation that
singularities of the branched double cover can only arise along the branch locus.
However, solving the second equation requires that the same eigenvectors
are also eigenvectors of the matrix $(\partial A^{ij} / \partial p_k)$
for every $k$, which has no solutions for generic quadrics.

This mismatch between singularities in geometry and physics is interpreted in terms
of a non-commutative resolution.  We can see that structure by examining the
D-branes in this theory.
The GLSM is a hybrid model,
fibered over
${\mathbb P}^3$, in which the fibers are (orbifolds of)
Landau-Ginzburg models with
quadratic superpotentials.
As explained in \cite[section 2.6.2]{Caldararu:2010ljp}, see also~\cite{Addington:2012zv}, taking into account the orbifold, the topological B-branes are elements of the derived category of modules over a sheaf ${\cal B}$ on ${\mathbb P}^3$ of even parts of Clifford algebras, defined by the quadratic form implicit in the superpotential. 

Let us walk through this in more detail.
Consider a hybrid Landau-Ginzburg model on the total space of an even-rank\footnote{
As noted in \cite{Caldararu:2010ljp,Hori:2011pd},
one only gets a branched double cover in the case
that the rank is even.
} vector bundle
$E \rightarrow {\cal G}$, where ${\cal G}$ is a
${\mathbb Z}_2$ gerbe over a space $B$,
and $E$ is not a pullback from a bundle on $B$,
and work locally in a Born-Oppenheimer approximation.
Suppose each fiber of the hybrid model is
a Landau-Ginzburg
model on\footnote{
In \cite{Addington:2012zv}, the convention is to work with Landau-Ginzburg
models on $[{\mathbb C}^{n+1}/{\mathbb Z}_2]$, so $n$ shifts by one.
} $[ V / {\mathbb Z}_2]$,
where $V = {\mathbb C}^n$ and
${\mathbb Z}_2$ acts by sign flips, with superpotential
\begin{equation} \label{eq:sup:hybrid}
W \: = \: \sum_{ij} A^{ij}(p) \phi_i \phi_j,
\end{equation}
as in \cite[section 4.2.1]{Addington:2012zv}, where the
$\phi_i$ span ${\mathbb C}^{n}$, and $p$ is a local coordinate
on the base space $B$.
In general, the B-branes 
are modules over the sheaf of
automorphisms of 0-branes.
From 
\cite{buchweitz-eisenbud-herzog,Kapustin:2002bi,dyckerhoff-cg,Caldararu:2010ljp,teleman-mf}, \cite[chapter 14]{yoshino},
as this is a quadratic superpotential over any
point on the base $B$, the sheaf of automorphisms ${\cal B}$ of 0-branes is expected to be a sheaf of even parts of Clifford algebras, 
which we will identify with the sheaf defining the categorical resolution. We can therefore obtain categorical resolutions from this class of hybrid Landau-Ginzburg models.

The 0-branes themselves in these hybrid Landau-Ginzburg models
 are quasi-isomorphism classes of `point-like' matrix
factorizations, as described in
\cite[section 3.3]{Addington:2012zv}.
Globally, from the structure of the superpotential~(\ref{eq:sup:hybrid}),
we can describe the 0-branes by particularly degenerate matrix
factorizations, namely skyscraper sheaves supported at the origin,
as observed in \cite{Addington:2012zv}.  Let ${\cal S}$ denote the
torsion sheaf $\iota_* {\cal O}_{\cal G}$, supported\footnote{
In principle, one should specify a ${\mathbb Z}_2$-equivariant structure
on the sheaf to uniquely identify it; however, that structure will cancel
out when we compute automorphisms momentarily.
} along the zero section
of the bundle $E \rightarrow {\cal G}$.
The local\footnote{
We can understand the global structure as follows.
Let $V$ be our 8-dimensional vector space and $Q \in H^0({\mathbb P}^3,S^2(V)^* \otimes {\cal O}_{{\mathbb P}^3}(1))$ the family of quadrics on ${\mathbb P}^3$.
The claim is, that a sheaf of modules over Kuznetsov's sheaf of algebras is equivalent to
ordinary sheaves $S_0$ and $S_1$ on ${\mathbb P}^3$, together with maps of sheaves
\begin{displaymath}
V \otimes S_0 \to S_1,
\: \: \:
V \otimes S_1 \to S_0 \otimes {\cal O}_{{\mathbb P}^3}(1).
\end{displaymath}
Interpreting these maps as operations $v \cdot$ from $S_0$ to $S_1$ and from $S_1$ to $S_0\otimes {\cal O}_{{\mathbb P}^3}(1)$, these are required to have the property that 
$(v \cdot)^2: S_i \to S_i \otimes {\cal O}(1)$ is just multiplication by $Q(v)$ for $i=0,1$ and any $v \in V$.
If the $S_i$ are skyscraper sheaves at a point $p$, then over the fiber, this is just the same as an ordinary matrix factorization for $Q(p)$.
In this language, there is a natural auto-equivalence $T : (S_0,S_1) \mapsto (S_1,S_0 \otimes {\cal O}(1))$
with a natural definition of $v\cdot$ on $S_1$ and $S_0 \otimes {\cal O}(1)$ induced from the given actions on $S_0, S_1$.  Then $T^2$ is visibly just tensoring $S_0$ and $S_1$ with ${\cal O}(1)$. 
}
automorphisms of ${\cal S}$, whose graded components are the sheaves
$\underline{\rm Ext}^k({\cal S},{\cal S})$, have the form\footnote{
The reader may find it helpful to compare the structure in a single
fiber.  There, the D0-brane is a skyscraper sheaf at the origin of
a vector space $V = {\mathbb C}^{n}$,
for which
\begin{equation*}
{\rm Ext}^k \left({\cal O}_p, {\cal O}_p \right) \: \cong \:
\wedge^k V.
\end{equation*}
}
(see e.g.~\cite[appendix A]{Katz:2002gh})
\begin{equation}
\underline{\rm Ext}^k({\cal S},{\cal S}) \: \cong \: 
\wedge^k N_{{\cal G}/E} \: \cong \: \wedge^k E\vert_{\cal G},
\end{equation}
and so we see that the sheaf of automorphisms of the sheaf ${\cal S}$ of
D0-branes has associated graded algebra\footnote{
From \cite[prop. 1.2]{lawson-michelsohn}, the exterior algebra is
an associated graded algebra of the Clifford algebra.
}
\begin{equation}
\oplus_k \underline{\rm Ext}^k({\cal S},{\cal S}) \: \cong \:
\oplus_k  \wedge^k E\vert_{\cal G}.
\end{equation}
Only the even degree components are relevant, and they descend to a sheaf
on $B$:  $\wedge^{2k} E \: = \: p^* \wedge^{2k} \tilde{E}$,
for $p: {\cal G} \rightarrow B$ the projection and $\tilde{E} \rightarrow B$
a vector bundle on $B$.  This gives the associated graded ${\cal B}_{\rm ag}$ of the sheaf ${\cal B}$ over $B$
which is the analogue of the structure sheaf for the
categorical crepant resolution, explicitly
\begin{equation}\label{eq:ncsheaf}
{\cal B}_{\rm ag} \: = \: \oplus_k \wedge^{2k} \tilde{E},
\end{equation}
which we have now explicitly identified with the associated graded of the sheaf of automorphisms of the sheaf of 0-branes (in passing, since the associated graded loses information about the multiplication, we emphasize that (\ref{eq:ncsheaf}) only determines ${\cal B}$ up to an isomorphism of sheaves of $\mathcal{ O}_X$-modules, not up to an isomorphism of $\mathcal{O}_X$-algebras).

Returning to the example of the GLSM for the octic double solid,
the $r \ll 0$ limit is a hybrid Landau-Ginzburg model, defined by a 
superpotential over the total space of the vector bundle
$E = {\cal O}(-1/2)^{\oplus 8} \rightarrow {\mathbb P}^3_{[2,2,2,2]}$, where ${\mathbb P}^3_{[2,2,2,2]}$ is a weighted projective stack (weighted projective space with non-minimal weights).
Here, $\wedge^{2k} E = p^* \wedge^{2k} V \otimes {\cal O}(-k)$ for
$V = {\mathbb C}^8$, where $p: {\mathbb P}^3_{[2,2,2,2]} \rightarrow
{\mathbb P}^3$ is the projection.
Thus, following the analysis above for this case,
the sheaf of automorphisms of 0-branes is 
\begin{equation}
{\cal B} \: = \: \oplus_k \wedge^{2k} V \otimes {\cal O}(-k) 
\: \longrightarrow \: {\mathbb P}^3\,,
\end{equation}
which has the same $\mathcal{O}_{\mathbb{P}^3}$-module structure  as the sheaf of Clifford algebras defining the nc resolution,
as given in \cite{kuznetsov2008derived}.

This structure $({\mathbb P}^3, {\cal B})$ was used in
\cite{kuznetsov2008derived} to define a non-commutative resolution of the branched double
cover, related to the complete intersection of four quadrics in
${\mathbb P}^7$ by homological projective duality \cite{kuz0},
which in physics is now believed to relate phases of gauged linear sigma
models.  It was also observed in \cite{kuznetsov2008derived} that the center of $\mathcal{B}$ is
\begin{equation}
Z(\mathcal{B}) = \mathcal{O} \oplus \left(\wedge^8V\otimes \mathcal{O}(-4)\right)\simeq \mathcal{O} \oplus  \mathcal{O}(-4),
\end{equation}
where the non-obvious multiplication $\mathcal{O}(-4)\otimes \mathcal{O}(-4)\to \mathcal{O}$ is given as multiplication by $\det(A)$.  Since
$X=\operatorname{Spec}_{\mathbb{P}^3}( \mathcal{O} \oplus  \mathcal{O}(-4))$, we see that we have an nc sheaf $\mathcal{C}$ on $X$ such that
\begin{equation}
\mathcal{B} = \pi_*\mathcal{C}\,,
\end{equation}
where $\pi:X\to\mathbb{P}^3$ is the double cover.  The sheaf $\mathcal{C}$ is a sheaf of Azumaya algebras on the complement of the conifolds in $X$.
This is consistent in physics
with the presence of a flat topologically-nontrivial $B$ field,
corresponding to an element of the Brauer group, as we discussed in section~\ref{subsec:brauer}.
The reader should also note that the GLSM for the nc resolution is in a different component of the moduli space than the GLSM for the singular octic double solid, by virtue of nontriviality of the Brauer group.

From the GLSM perspective, the interpretation in terms of a fractional B-field on $X$ also arises explicitly after applying a Seiberg-like duality, as has been shown in~\cite{Hori:2011pd}.
The discussion is similar to the one that we reviewed in the context of the conifold in Section~\ref{sec:ncconifromglsm} and we refer to~\cite{Hori:2011pd} for details on the compact case.

\section{The predictions from the topological B-model}
\label{sect:octicdoub:b-model}
In Section~\ref{sect:prop} we have discussed the torsion refined Gopakumar-Vafa invariants and general properties of the topological string A-model on non-commutative resolutions of compact Calabi-Yau threefolds.
We will now use mirror symmetry and the holomorphic anomaly equations to explicitly calculate the free energies for the non-commutative resolution $X_{\text{n.c.}}$ of the singular octic double solid $X$ introduced in Section~\ref{sec:octic} and, combining this with the information from the corresponding free energies of the generic smooth octic double solid $X_{\text{def.}}=X^{(1,149)}_8(11114)$, extract the torsion refined Gopakumar-Vafa invariants.
We will see that, in addition to the constant map contributions derived in Section~\ref{sec:constantmap}, the torsion refined Gopakumar-Vafa invariants exhibit a Castelnuovo-like vanishing that allows us to perform the direct integration of the holomorphic anomaly equations and fix the holomorphic ambiguities for the free energies of $X_{\text{n.c.}}$ and the dual Calabi-Yau $Y=X^{(1,65)}_{2222}(1^8)$ up to genus $g=32$.

\subsection{Mirror symmetry and special geometry}
\label{sec:specialgeometry} 

We first review the situation for a smooth compact Calabi-Yau threefold $Y$ with mirror $\mM$.
The coefficients in the defining equations of $\mM$ give rise to $h^{2,1}(\mM)$ {\sl algebraic coordinates ${\underline z}$} on the complex structure moduli space.
We can then calculate
\begin{equation} 
Z(\uz,\lambda) =\exp\left(\sum_{g=0} \lambda^{2g-2} F_{g}(\uz)\right)\,,
\end{equation} 
perturbatively with  $B$-model techniques, at least up to some maximal genus, and use mirror symmetry to obtain the A-model free energies that, in the so-called holomorphic limit, are generating functions of the enumerative invariants associated to $Y$. 

Due to special geometry, the {\sl genus zero} generating 
function $F_0({\underline z})$ is expressible in terms of the periods $\Pi_\Gamma(\uz)=\int_\Gamma \Omega(\uz)$ 
of the holomorphic $(3,0)$-form $\Omega$ on $\check{Y}$ over an integral symplectic basis of three cycles 
$\{\Gamma\}=\{A^I,B_J\}\in H_3(\mF,\mathbb{Z})$, with $I,J=0,\ldots, h^{2,1}(\mF)$.
The latter are covariantly constant with respect to the flat Gauss-Manin connection, or, equivalently, in the kernel of 
the Picard-Fuchs differential ideal $\{{\cal D}_{\uz} \}$ of the family in an {\sl integral monodromy 
basis} (of course  up to ${\rm Sp}(b_3(\mF),\mathbb{Z})$ transformations).
The locus where the periods develop singularities is given by the vanishing locus of the {\sl discriminant polynomial} $\Delta(\underline{z})$.

The corresponding period vector takes the form
\begin{align}
    {\underline \Pi}(z)^T=(\mathfrak{F}_J,\mathfrak{X}^I)=\left(\int_{B_J} \Omega(\uz),\int_{A^I} \Omega(\uz)\right)\,,
\end{align}
and by the local Torelli theorem the $\mathfrak{X}^I(\uz)$ can be used as {\sl homogenous 
coordinates} on ${\cal M}(\mM)$.
The moduli space is K\"ahler and the K\"ahler potential can be expressed as  
\begin{align}
    e^{-K}=i \int_{\mF} \Omega\wedge \bar\Omega=\uPi^\dagger\Sigma \uPi =
i (\mathfrak{X}^I {\bar {\mathfrak{F}}}_I-{\bar {\mathfrak{X}}}^I \mathfrak{F}_I)\,,
\end{align}
with the corresponding Weil-Petersson metric $G_{i\bar \jmath}=\partial_i\bar \partial_{\bar\jmath} K(\uz)$.

We define $C_{\underline{i}}=\int_{\mF}\Omega\partial_{\underline{i}}\Omega$, so that the 3-point functions are $C_{ijk}(\uz) = \uPi^T \Sigma \partial_i\partial_j\partial_k \uPi$, where $\Sigma$ is the symplectic intersection matrix and $\partial_i=\partial/\partial {z^i}$ with $i=1,\ldots h^{2,1}(\mM)$. 
The $C_{ijk}(\uz)$ are rational functions in the algebraic coordinates $\uz$. 
On the other hand, $C_i$ and $C_{ij}$ vanish
due to Griffiths transversality, which together with the existence of the unique $(3,0)$-form  is the origin of special geometry~\cite{MR717607}, and this implies that there is a so-called prepotential 
\be 
F_0(\uz)=\frac{1}{2}(\mathfrak{X}^I\mathfrak{F}_I)\,,
\label{eq:F0}
\ee
homogeneous of degree two in $\underline{\mathfrak{X}}$, such that $\mathfrak{F}_I=\partial_I F_0$ with $\partial_I=\partial/\partial \mathfrak{X}_I$.

Note that the definition of $K$ implies 
that $\Omega$ and therefore $\mathfrak{X}^I$ as well as $\mathfrak{F}_I$ are sections of the K\"ahler line bundle 
${\cal L}^{-1}$, while $F_0$ is a section~\footnote{One says a section of ${\cal L}^m$ has 
K\"ahler weight $m$.}  of ${\cal L}^{-2}$, and from the definition of the 
$3$-point functions it follows that they are holomorphic sections of ${\cal L}^{-2} \otimes {\rm Sym}^3(T^*{\cal M}(\mM))$. Similarly $C_{\bar\jmath}^{ik}\equiv e^{2K} {\bar C}_{\bar i\bar k \bar \jmath} G^{\bar \imath i} G^{\bar k k}$ are  sections of ${\cal L}^{2} \otimes {\rm Sym}^2(T{\cal M}(\mM))\otimes {\overline { T^*{\cal M}}}(\mM))$.

Let $V_{j\bar \jmath}$ be a section of $T^*{\cal M}(\mM)\otimes {\overline { T^*{\cal M}(\mM)}} \otimes {\cal L}^m  \otimes 
{\bar {\cal L}}^n$.
Then the covariant derivatives with respect to the Christoffel connection $\Gamma^l_{ij}$ associated to the Weil-Petersson  metric and the connection $\partial_i K$ on the K\"ahler line bundle, are 
$D_{i} V_{j\bar \jmath}=\partial_i V_{j\bar \jmath}-\Gamma_{ij}^l V_{l\bar \jmath}-m K_i V_{j\bar \jmath}$ 
    and 
$D_{\bar \imath} V_{j\bar \jmath}= \bar \partial_{\bar \imath} V_{j\bar \jmath}-\Gamma_{\bar i\bar j}^{\bar l} V_{j\bar l}-n K_{\bar \imath} V_{j\bar \jmath}$  with 
$K_{\bar \imath}=\bar \partial_{\bar \imath} K$ and  $K_{\bar \imath}=\bar \partial_{\bar \imath} K$. 
Griffiths tranversality, the Tian-Todorov theorem and the definition of $e^{-K}$ imply 
that $\chi_i\equiv D_i\Omega$ span $H^{2,1}(\mM)$. Hence $V=(\Omega, \chi_i,{\bar \chi}_{\bar i},\bar \Omega)$ 
spans $H^3(\mM)$, respecting its Hodge decomposition $\oplus_{i=1}^3 H^{3-i,i}(\mM)$.
Then with the above and the definition of $C_{ijk}$ it follows that
\be 
D_i\left(\begin{array}{c} V_0\\ V_j\\V_{\bar \jmath} \\\bar V_0\end{array}\right)=
\left(\begin{array}{cccc} 
0&{\delta}_{ik}&{ 0} &0\\ 
0&{0}&-i e^K C_{ijk}G^{k\bar k} &0\\
0&{0}&{0}& G_{i\bar \jmath}\\ 
0&{0}&{0} &0\end{array} 
\right)\left(\begin{array}{c} V_0\\ V_k\\V_{\bar k} \\\bar V_0\end{array}\right)\,,
\ee
and therefore
\be 
[D_i,D_{\bar \jmath}]^{\ k}_l=-R_{i\bar \jmath \ l}^{\ \ k}=\partial_{\bar \jmath} \Gamma^{k}_{il}= \delta_l^k G_{i\bar \jmath}+\delta_i^k G_{l\bar \jmath}-C_{\bar \jmath}^{km}C_{ilm}\, .
\label{eq:specialKaehler}  
\ee

To make contact  with the topological string A-model on $Y$,
and extract the enumerative information from the $F_g(\uz)$, we also need the {\sl inhomogenous coordinates} $t^a(\uz)=\mathfrak{X}^a/\mathfrak{X}^0$ .
Again by the local Torelli theorem, for a suitable choice of the {\sl symplectic frame} 
of A- and B-cycles, the map from the algebraic to the inhomogenous coordinates is locally 
invertible,  up to multi coverings choices, and referred to as the {\sl local mirror map}. 
By picking a K\"ahler gauge one introduces the K\"ahler weight zero generating function 
${\cal F}(\ut)= F_0/(\mathfrak{X}^0)^2[\uz(\ut)]$ in the inhomogeneous coordinates.
One then calculates ${\underline \Pi}^T=\mathfrak{X}^0 (2 {\cal F}-t^a\partial_a {\cal F},\partial_a {\cal F},1,t^a)$ and in this gauge $C_{abc}=-\partial_a\partial_b\partial_c {\cal F}$.
Knowing the solution in this gauge, one can form the period vector ${\cal V}=
(2 {\cal F}-t^c\partial_c {\cal F}, \partial_j (2 {\cal F}-t^c\partial_c {\cal F}),t^j,1)^T$ 
and with ${\cal V}^j={\cal V}_{b_3(\mM)/2+j}$, ${\cal V}^0={\cal V}_{b_3(\mM)}$ one concludes  
\be 
\partial_i \left(\begin{array}{c} {\cal V}_0\\ {\cal V}_j\\  {\cal V}^j\\{\cal V}^0\end{array}\right) =
\left(\begin{array}{cccc} 
0&\delta_{ik}&{ 0} &0\\ 
0&{0}&C_{ijk} &0\\
0&{0}&{0}& \delta_{i}^j\\ 
0&{0}&{0} &0\end{array} 
\right)\left(\begin{array}{c}  {\cal V}_0\\ {\cal V}_k\\ {\cal V}^k \\{\cal V}^0\end{array}\right)\ .
\ee
The discussion so far has been independent of the point that one considers in the moduli space and we will sometimes use $\underline{t}_p$ to denote the local flat coordinates around $p$.
\subsection{Special geometry at higher genus}
\label{sec:highergenus} 
The {\sl genus one} generating function $F_1(\uz)$, which is manifestly of K\"ahler weight zero, obeys the holomorphic anomaly equation 
\be 
\partial_i\bar \partial_{\bar \jmath}F_1(\uz)=\frac{1}{2}C_{ikl}C^{kl}_{\bar \jmath}-\left(\frac{\chi(\mM)}{24}-1\right)G_{i\bar \jmath}\,,
\label{eq:holF1} 
\ee 
for the Ray Singer Torsion~\cite{Bershadsky:1993ta}.
Using~\eqref{eq:specialKaehler} and standard formulas for the K\"ahler curvature this can be integrated to
\be 
F_1=-\frac{1}{2}\log \det (G_{i\bar\jmath})+\left(\frac{b_3(\mM)}{2}-\frac{\chi(\mM)}{24}+1\right)K +\log\|f_1(\uz)\|^2\,,
\label{eq:F1}
\ee
with a holomorphic ambiguity given by the holomorphic function $f_1(\uz)$. 

The {\sl higher genus}  generating functions $F_{g>1}(\uz)$, which are of K\"ahler weight $2g-2$, are obtained recursively 
from the generating functions at genus $g=0,1$ by solving the holomorphic anomaly equations~\cite{Bershadsky:1993cx} 
\be 
{\bar \partial}_{\bar\imath} F_g(\uz) = \frac{1}{2}C^{jk}_{\bar \imath}\left(D_j D_k F_{g-1}(\uz) +\sum_{h=1}^{g-1} D_j F_h(\uz) D_k F_{g-h}(\uz)\right)\,,
\label{eq:holan}
\ee
using the {\sl direct integration method}. This method uses a fact that also follows  from special geometry, namely that the $F_g(\{G\},\uz)$ are weighted homogeneous polynomials in a finite number of independent non-holomorphic modular generators $\{G\}$~\cite{Yamaguchi:2004bt,Grimm:2007tm,Alim:2007qj}. 
The non-holomorphic derivatives of the generators w.r.t $\bar z_{\bar \imath}$ 
are proportional to $C^{jk}_{\bar \imath}$ and  they close under 
the derivatives\footnote{Covariant under the  Weil-Petersson and the  K\"ahler connection.} $D_i$ in~\eqref{eq:holan}. Therefore the left hand side is converted into derivatives of $F_g(\{G\},\uz)$ 
w.r.t. the $\{G\}$, while the expression in the bracket on the r.h.s. is converted 
into a polynomial in $\{G\}$ and the $C^{jk}_{\bar \imath}$ cancel. 
As a consequence,~\eqref{eq:holan} is solved by direct polynomial integration 
w.r.t. the $\{G\}$, up to holomorphic ambiguities $f_g(\uz)$ that arise at each genus.

As described in the next section, the ambiguities are fixed at low genus by boundary conditions on local expansions of the holomorphic limit $F^{\text{top.}}_g(t)$, that will be defined below, at points with critical fibres in ${\cal M}(\mM)$.
One novel type of them that we consider corresponds to non-commutative resolutions in the $A$-model and yields new boundary conditions for the direct integration.
As a consequence of the $\mathbb{Z}_2$ torsion group appears an algebraic change in the K\"ahler gauge over ${\cal M}(\mM)$ with a $\mathbb{Z}_2$ branching  behaviour, which has also not been encountered before.
We relegate the description of the behaviour of the generators and their ambiguities under this K\"ahler gauge choice and how it affects the general procedure of the direct integration to Appendix~\ref{app:algebraicKahlergauge}.

\subsection{Boundary conditions for the holomorphic ambiguity}
\label{sec:boundaryconditions} 
Modular invariance requires the $f_{g>1}(\uz)$ to be  rational functions of the algebraic complex structure parameters ${\uz}$.
The $f_g(\uz)$ are restricted by the behaviour of the $F_g$ around the position of the critical fibres of $\mF$ in ${\cal M}(\mM)$. 
Ignoring eventually occurring apparent singularities, the critical fibres arise over the vanishing loci of the discriminant $\Delta$ of $\{{\cal D}_{\uz} \}$ in ${\cal M}(\mM)$.
In a suitable model of ${\cal M}(\mM)$ this is a union of normal crossing divisors. 
To fix the $f_g(\uz)$, the  behaviour of the $F_g$ has to be analysed in terms of {\sl local flat inhomogeneous coordinates}
$t^i_*={\mathfrak{X}}^i_*/{\mathfrak{X}}^0_*$ and in the weight zero K\"ahler gauge.
In addition, one has to take the holomorphic limit $\bar t^i_* \rightarrow t^i_{*,0}$  which needs to be defined only for the
generators $\{G\}$ and\footnote{Note that $\partial_i\log \det (G_{i\bar \jmath})=C_{ijk}S^{jk}$ can also be expressed in terms of the generators $S^{ij}$ chosen below.} $\log \det (G_{i\bar \jmath})$. 
We refer to the resulting quantities as the {\sl genus $g$ topological string free energies}
\be 
F^{\text{top.}}_g(\ut_*)= [\mathfrak{X}_*^0(\uz(\ut_*))]^{2g-2}\lim_{\bar t_*^i\rightarrow t^i_{*,0}}  F_g(\uz(\ut_*))\, .
\ee
Note that $F_0(\underline{t})$ is already holomorphic and $F_g^{\text{top.}}(\underline{t})=\mathcal{F}(\underline{t})$.

Important examples for boundary conditions are the conifold gap and the expansion 
at points of maximal unipotent monodromy.
The conifold fibre ${\mM}_{z_c}$ has a nodal 
singularity with a shrunken 3-cycle $\nu$ with the topology of a $S^3$. As a result, the period $\Pi_\nu$ vanishes at $z=z_c$ and if 
one period $\Pi_{\Gamma_0}$ with $\Gamma_0\cap \nu=0$ stays constant it has been argued in~\cite{Huang:2006hq}, based on the assumption of the contribution of single 
massless hypermultiplet to the Schwinger Loop contributing 
to the $R_+^2 (F_+)^{2g-2}$ amplitude, that in the local flat inhomogeneous 
coordinates $t_c=\Pi_\nu/\Pi_{\Gamma_0}$ the  local expansion of $F^{\text{top.}}_g$ has to have the gap 
\be 
F^{\text{top.}}_{g>1}(t_c)=\frac{(-1)^{g-1} B_{2g}}{2 g (2g-2) (t_c/n)^{2g-2}} + O(t_c^0)\, ,
\ee
which gives $2g-2$ boundary conditions for the ambiguities. We call this boundary the {\sl conifold  
gap condition}. 
The leading coefficient has already been conjectured by~\cite{Ghoshal:1995wm}.
On the other hand, at genus $g=1$ a computation of the gravitational index gives the leading behaviour~\cite{Vafa:1995ta}
\be 
F^{\text{top.}}_1(t_c)=\frac{1}{12} \log(t_c) + O(t_c^0)\,.
\label{eq:F1cn} 
\ee

The large volume limit of $Y$ is dual to a singular boundary point of $\mathcal{M}(\mM)$ around which the periods experience maximally unipotent monodromy, a so-called MUM point.
The local flat coordinates are ratios $t^i=X^i/X_0$, where $X^0(\uz)$
is the unique homolomorphic period at the MUM point and $X^i(\uz)=\frac{1}{2 \pi i} X^0 \log(z_i)+{\cal O}(\uz)$ are $h^{2,1}$ ``single logarithmic'' ones.
They are identified under mirror symmetry with the complexified K\"ahler parameters in the A-model.
This predicts the  prepotential  by mirror symmetry, using the integral monodromy basis as given by~\eqref{eqn:genuszeroconst} \cite{CANDELAS199121,Hosono:1993qy,Hosono:1994ax}. The leading terms are also directly calculable using the $\hat \Gamma$-class in the $A$-model,
see \cite{MR3965409} for a review. The genus one boundary conditions at the MUM 
point were determined in \cite{Bershadsky:1993ta} to be
\be 
F_1^{\text{top.}}(\ut)=- t^i\frac{c_2 \cdot D_i}{24}+ {\cal O}(q) \,. 
\label{eq:F1mum} 
\ee
By combining~\eqref{eq:F1cn} and~\eqref{eq:F1mum} it is  typically possible to 
fix the coefficients $e_i$ in the ansatz $f_1\sim \prod_i \Delta_i^{e_i}$, where  
the product runs over all components of the critical locus.  

As discussed in Section~\ref{sec:constantmap}, for $g\ge 2$ the leading behaviour of $F^{\text{top.}}_g$ can be deduced from the Gopakumar-Vafa expansion of the topological string partition function.
For smooth Calabi-Yau threefolds, the corresponding specialization of~\eqref{eq:Znc} with $N=1,k=0$ of~\eqref{eq:Znc} has been obtained by Gopakumar and Vafa~\cite{Gopakumar:1998jq} from a Schwinger loop calculation for the  $R_+^2 (F_+)^{2g-2}$ terms in the effective action that, following~\cite{Antoniadis:1993ze}, encode all of $Z_{\text{top}.}(\ut,\lambda)$. 
The formula for the constant map contributions~\eqref{eqn:cmapsgg2} follows using the $\zeta$-function regularisation and assuming $n_g(0)=\chi/2$. 
The latter fact can be argued also from  Gromov-Witten calculations for the map from the worldsheet $\Sigma_g$ to $\mM$~\cite{MR1728879}. Like in the conifold case, the light states running in a loop determine the loop integral. 
At the MUM-point, the light states are identified with bound states of a single D6 brane with D2 branes in 
the class $\beta\in H_2(X,\mathbb{Z})$ as well as $k\in \mathbb{N}$ D0 branes.
Including them 
explicitly as 5d particles running in the Schwinger loop yields \eqref{eq:Znc} with $N=1$ and $k=0$~\cite{Gopakumar:1998jq}. As explained  already in \eqref{eq:5dgauge} the comparison relates
the charge of the particle under the 5d gauge group to the curve class $\beta\in H_2(X,\mathbb{Z})$
and an index in the spin representations to the genus \eqref{eq:spingenus}.

The resulting index
$n_{g}^\beta$ of the physical states has been geometrically understood~\cite{Gopakumar:1998jq}
in terms of a Lefschetz decomposition in the cohomology of the moduli space of 
M2 branes wrapping a curve in the class $\beta$~\cite{Gopakumar:1998jq,Katz:1999xq}.
In particular if there is no curve of genus $g$ in the class $\beta$, which 
by the adjunction formula will be if $\beta^2\lesssim\sqrt{g}$, then $n_g^\beta=0$, which
gives $\sim \sqrt{g}$ boundary conditions. Indeed the vanishing of the $n_g^d$ can be
analysed geometrically by the Castelnuovo criterium. Moreover the information 
of $n_{g}^d=(-1)^{{\rm dim}(\chi({\cal M}_C))}\chi({\cal M}_C)$ for smooth genus 
curves $C$ (for which ${\cal M}_C$ is also smooth projective and easily 
calculable)~\cite{Katz:1999xq} may also serve as a well established bound. 
We refer to the combined boundary conditions including the constant map contributions  
at points of maximal unipotent monodromy as the {\sl Castelnuovo bounds}.

At regular points in ${\cal M}(\mM)$ we 
expect the $F_g^{\text{top.}}(t_r)$ not to develop singularities.
From these considerations one can conclude in the one modulus case, assuming furthermore that there is a single conifold point, that the holomorphic ambiguity is of the form $f_g=p_{3g-3}(z)/\Delta_c(z)^{2g-2}$, with $3g-2$ unkowns in the
coefficients of the polynomial $p_{3g-3}$. After incorporating the gap condition, one needs still to determine $g$ variables.
The linear growth is eased in the presence of loci with finite branch 
coverings in the canonical Batyrev variables $z$ of ${\cal M}(\mM)$ and therefore
a finite monodromy, the {\sl  orbifold points}, where the propagators can be singular 
and the regularity restriction is stronger.
We call this boundary condition {\sl orbifold regularity}.
In all cases there is a maximal genus after which the boundary conditions are insufficient to fix the $f_g$, see~\cite{Huang:2006hq} and also~\cite{BHTS} for more details on the bounds in the case of hyper geometric one parameter families.  

\subsection{New boundary conditions from nc-resolutions}
\label{sec:mirrorsymmetry} 
In the  presence of other types of degenerations of $\mM$, the situation with the boundary 
conditions changes. As in the examples above the properties of the massless spectrum  
will be  determined by the local degeneration of the mixed Hodge structure, which has 
been studied mathematically with very strong results, see~\cite{MR4012553} for recent work. 
This analysis has been used to support the so called landscape distance conjectures, claiming that if the distance to the singular locus is infinite in the Weil-Petersson metric an infinite tower of light particles appears~\cite{Grimm:2018ohb,Joshi:2019nzi}.
For one parameter Calabi-Yau families the types of degenerations have been classified~\cite{MR3822913} 
and come in three types: the MUM points, the conifold points and the K-points.
The type is determined by the four local exponents $\mathfrak{l}$ associated to the Picard-Fuchs operator, which are of the form 
$\mathfrak{l}=(a,a,a,a)$, $\mathfrak{l}=(a,b,b,c)$ and $\mathfrak{l}=(a,a,b,b)$ respectively, where $a,b\in \mathbb{Q}$.

Let us recall that for the fourteen hypergeometric one parameter families the forth order Picard-Fuchs operator is given by 
\be 
D \; = \;\theta^4- \mu^{-1} z \prod_{k\; = \;1}^4(\theta+a_k)\,,
\label{eq:diffgeneral} 
\ee  
with $\theta=z \frac{d}{dz}$, where $\mu,a_i\in \mathbb{Q}$ and  
for the fourteen cases are summarized e.g. in Table 1 adapted 
from~\cite{Bonisch:2022mgw}. 
$D$ generates the Picard-Fuchs ideal and has the Riemann symbol 
\be
{\cal  P}\left\{\begin{array}{ccc}
0& \mu& \infty\\ \hline
0& 0 & a_1\\
0& 1 & a_2\\
0& 1 & a_3\\
0& 2 & a_4 
\end{array}\right\}\, .
\label{riemannsymbolgeneral}
\ee 
The latter summarizes that (\ref{eq:diffgeneral}) has three regular 
singular points at $z\in \{0,\mu,\infty \}$, so that the parameter space of $z$ 
is ${\cal M}(\mM)\; = \;\mathbb{P}^1\setminus \{0,\mu,\infty\}$, while the local 
exponents $\mathfrak{l}^T$ appear as columns under the singular point.
For $z=0$ they are $(0,0,0,0)$ which identifies this as a MUM-point.
From this it follows that the so-called Frobenius basis of solutions has one regular solution $X^0\sim 1+{\cal O}(z)$, in fact 
a Hypergeometric function of type $X^0=_4F_3(a_1,a_2,a_3,a_4,1,1,1, \mu z)$,  and a logarithmic solution $X^1\sim X^0 \log(z)+{\cal O}(z)$, as well as a double logarithmic solution $F_1\sim X^0\log^2(z)+\ldots$ and a triple logarithmic solution $F_0\sim X^0 \log^3(z)+\ldots$. 
This implies that the monodromy has one maximal $4\times 4$ Jordan block and is maximally unipotent.
It then also follows that the point $z=0$ is at infinite distance with respect to the Weil-Petersson metric.   

Similarly at $z=\mu$, the conifold point, one has a constant 
holomorphic solution $X^0_c\sim 1+{\cal O}(\delta^3)$  a vanishing solution 
$X^1_c\sim \delta+{\cal O}(\delta^2)$, a logarithmic solution 
$X^1_c\sim X^0_c\log(\delta)+{\cal O}(z^0)$ and a further power 
series solution. The irreducible Jordan block of the monodromy 
is $2\times 2$ and conifolds are at finite distance.

The bi-cubic in $\mathbb{P}^5$ has at $w=1/z=0$ a  
K-point with $\mathfrak{l}=(\frac{1}{3},\frac{1}{3},\frac{2}{3},\frac{2}{3})$, hence we 
have two fractional power series solution $X^1_k=w^\frac{1}{3}+{\cal O}(w^\frac{4}{3})$, 
$X^2_k=w^\frac{2}{3}+{\cal O}(w^\frac{5}{3})$ and the two corresponding log solutions
$F_1=X^1_k\log(z)+ {\cal O}(w^\frac{1}{3})$ and  $F_2=X^2_k\log(z)+ {\cal O}(w^\frac{1}{3})$.
The monodromy at $k$-points has two irreducible $2\times 2$ Jordan blocks and K-points 
are at infinite distance. The 14 cases have seven orbifold points 
at $1/z=0$, three conifolds, three K-points and one MUM-point, all with additional finite branch order on top of the logarithmic structure. 

For us the most interesting case is the one parameter mirror family to  the complete intersection of 
four quadrics in $\mathbb{P}^7$.
The complete intersection of these quadrics is abbreviated by $X^{(1,65)}_{2222}(1^8)$, see Table~\ref{tab:cystab}. This family has a non-standard MUM-point at $w=0$ with $\mathfrak{l}=(\frac{1}{2},\frac{1}{2},\frac{1}{2},\frac{1}{2})$ and $\mu=2^{-8}$. The 
geometry of such MUM-points with a finite branch cover\footnote{In~\cite{Joshi:2019nzi} 
it has been noticed that the model has a symmetry by which one can associate to both
MUM points the same BPS numbers, yielding the infinite states for the distance 
conjecture. While the conclusion regarding the infinite states in accordance with  
the distance conjecture is correct, it misses decisively the subtle geometrical interpretation for the second MUM point.} had not been understood before even though  they are quite frequent in the list of Calabi-Yau 3 fold one parameter operators described in~\cite{MR3822913}. 

As reviewed in Section~\ref{sect:octicdoub:nc-res:glsm}, it is known that its mirror is identified with the non-commutative 
resolution of the singular Calabi-Yau double cover of $\mathbb{P}^3$, branched over a singular determinantal octic, with 84 terminal nodal singularities.
As explained in Section~\ref{sec:singdouble}, the latter enjoys a complex 
structure deformation to the generic octic $X^{(1,149)}_8(11114)$, 
whose mirror is another hypergeometric one parameter family. With the conifold gap, the 
Castelnuovo bounds, as well as the orbifold regularity the latter model can be solved to genus $g=60$.
However for the mirror of $X^{(1,65)}_{2222}(1^8)$, without understanding the information at the non-standard MUM point at $1/z=0$, the constant map contributions and the Castelnuovo vanishing from the MUM-point at $z=0$ together with the conifold gap only allow to solve the model to genus $g=17$.

Our new insight is to provide an enumerative interpretation for the corresponding topological string free energies in terms of the torsion refined Gopakumar-Vafa invariants, using the proposal first made in~\cite{Schimannek:2021pau}.
The correct interpretation of the expansion at the non-standard MUM-point as non-commutative resolution, allows to use new Castelnuovo bounds at $1/z=0$ for the torsion refined BPS invariants, the traditional Castelnuovo bounds $z=0$ and the conifold gap to fix the holomorphic ambiguity to solve the topological string for this mirror pair to genus $32$.
Let us stress that to split the two torsion sectors and extract the integer $\mathbb{Z}_2$-refined Gopakumar-Vafa invariants, and therefore to use the corresponding Castelnuovo vanishing, we also need the information of the Gopakumar-Vafa invariants of $X^{(1,149)}_8(11114)$. 

Let us end this discussion with a few summarizing remarks: We learn that we need to distinguish 
two types of MUM-points in mirror symmetry: {\sl commutative MUM points} and 
{\sl non-commutative MUM-points}, depending on 
whether the mirror has a geometrical interpretation in terms of a commutative or a non-commutative Calabi-Yau category.
To extract all local information in a complex structure moduli space ${\cal M}(\hat X)$  that contains both types of MUM points, i.e. by a local expansion at the {\sl commutative MUM-points} in inhomogeneous variables $t_{cM}$ as well as at  
the {\sl non-commutative MUM-points} in variables $t_{ncM}$, one has to consider in general, on top of the K\"ahler weight $0$ gauge choice, different algebraic K\"ahler gauge choices at these points. 
Changing the latter acts in a nontrivial way in particular on the $f_g(\uz)$, the propagators and their ambiguities as explained in Appendix \ref{app:algebraicKahlergauge}. One needs to understand this gauge change to reconstruct from the  local expansion the global ambiguities $f_g(\uz)$ 
in ${\cal M}(\hat X)$. 

If one encounters a {\sl non-commutative MUM-point} in the complex structure moduli space of  a $B$-model 
description, its first hallmark is that in the local BPS expansion no naive integrality of the BPS indices exists~\footnote{Although the naive genus $0$ GV-invariants at such a point can be integral~\cite{Sharpe:2012ji}, this breaks down at genera $g\ge 1$.}.
In fact there are examples in~\cite{Schimannek:2021pau} of non-commutative MUM-points, where the local expansion is defined over an algebraic extension of the type $\mathbb{Q}[\sqrt{5}]$.
Only if the local expansions, in these cases at two inequivalent non-commutative MUM-points, are combined with the expansion at a {\sl commutative 
MUM-point} in a deformed geometry $X_{\text{def.}}$, which exists in the moduli space of a different mirror manifold, one can absorb theses roots precisely in fifths unit root associated with the expansion~\ref{eq:Znc} of  $\mathbb{Z}_5$-torsion  
refined integer BPS indices $n_g^{\beta,l}\in \mathbb{Z}$.
These invariants then allow the natural interpretation in terms of  degeneracies of five-dimensional BPS 
invariants with $U(1)^r\times \mathbb{Z}_5$ gauge symmetry. This implies 
that non-integer BPS expansions at MUM-points are great indicators of 
a subtle but important gobal arithmetic symmetry structure with direct  
physical implications in the effective theory.

\subsection{Brane charges and monodromies}
\label{sec:gamma}
As a consequence of the derived equivalence of the non-commutative resolution $X_{\text{n.c.}}$ of the singular octic double solid $X$ and the complete intersection $Y=X^{(1,65)}_{2222}(1^8)$ of four quadrics in $\mathbb{P}^7$, both share the same mirror $\check{Y}$.
Before proceeding with the direct integration for $X_{\text{n.c.}}$, it is necessary to spend some time studying the periods and the complex structure moduli space of $\check{Y}$.
We will first discuss the integral basis of periods and the choice of prepotential around both MUM-points.
Using analytic continuation, this will allow us to fix the K\"ahler transformation that connects the gauge choices at both point.

Although not necessary for the direct integration, we will also use the opportunity to discuss the central charges of an integral basis of topological B-branes on $X_{\text{n.c.}}$, their monodromies and their transformation under homological projective duality.
For a review on topological branes we refer to~\cite{Aspinwall:2004jr} and a brief introduction can be found in~\cite{Cota:2019cjx}.
The topological invariants, as well as the indicials at infinity, for $Y$, $X_{\text{n.c.}}$ and the smooth deformation $X_{\text{def.}}$ of $X$ are listed in Table~\ref{sec:gamma}.
\begin{table}[ht]
\label{tab:cystab}
\centering
	\begin{tabular}{|c|c|c|c|c|c|c|}
		\hline
	       $a_1,a_2,a_3,a_4$& $1/\mu$	& MUM Mirror	& at $z$ &$\kappa$	& $c_2 \cdot D$	& $\chi(M)$\\\hline
    $\frac{1}{2},\frac{1}{2},\frac{1}{2},\frac{1}{2}$ & $2^8$		& $Y=X^{(1,65)}_{2,2,2,2}(1^8)$& $0$ & $16$	& $64$ & $-128$\\[2mm]
    $\frac{1}{2},\frac{1}{2},\frac{1}{2},\frac{1}{2}$ &  $2^8$       & nc-res. $X_{\text{n.c.}}$ of $X$&        $\infty$     & $2$       & $44$   & $-128$\\[2mm]\hline
    $\frac{1}{8},\frac{3}{8},\frac{5}{8},\frac{7}{8}$			& $2^{16}$	& $X_{\text{def.}}=X^{(1,149)}_{8}(1^4 4^1)$& $0$ 		& $2$		& $44$		& $-296$\\[2mm]
		\hline
	 \end{tabular}\vspace{.2cm}
\caption{
Data of the three relevant MUM-points in two hypergeometric one--parameter Calabi--Yau families. 
 }
\end{table}

To find an integral basis of periods around the MUM-point associated to $X_{\text{n.c.}}$, we first consider the homologically projective dual $Y$ and then use analytic continuation.
The fundamental period of the mirror of the complete intersection of two generic quadrics in $\mathbb{P}^7$ takes the form
\begin{align}
        \varpi_0=\sum\limits_{k=0}^\infty \frac{(2k!)^4}{(k!)^8}z^k={_4}F_3\left(\frac12,\frac12,\frac12,\frac12;1,1,1;2^8z\right)\,,
        \label{eqn:w0fundY}
\end{align}
and is annihilated by the Picard-Fuchs operator
\begin{align}
        \mathcal{D}_1=\theta^4-2^4z(2\theta+1)^4\,.
        \label{eqn:pfX2}
\end{align}
The moduli space contains one conifold point at $z=2^{-8}$ and the discriminant takes the form
\begin{align}
        \Delta=1-2^8z\,.
        \label{eqn:discY}
\end{align}
Note that $\varpi_0$ is the Hadamard product of the corresponding period ${_2}F_1\left(\frac12,\frac12;1;2^4z\right)$ for the degree $4$ normal curve $X_{2,2}(1^4)$ with itself.
The relevant topological invariants are
\begin{align}
        \chi_Y=-128\,,\quad \kappa_Y=\int_Y J\cdot J\cdot J=16\,,\quad  b_Y=\int_Y c_2(Y)\cdot J=64\,,
\end{align}
where $J$ is the generator of the K\"ahler cone and the Yuakawa coupling is
\begin{align}
        C_{zzz}=\frac{16}{z^3\Delta}\,.
\end{align}

To obtain an integral basis of periods around the MUM-point associated to $Y$, we first choose a basis of topological B-branes in the derived category $D^b(Y)$, that generate the charge lattice on $Y$, as follows.
As a 6-brane we choose the structure sheaf $\mathcal{E}_6=\mathcal{O}_Y$ while the 0-brane corresponds to the skyscraper sheaf $\mathcal{E}_0=\mathcal{O}_{\text{pt.}}$.
The 4-brane $\mathcal{E}_4=\mathcal{O}_J$ is homotopy equivalent to the complex
\begin{align}
        0\rightarrow\mathcal{O}_Y(-J)\rightarrow\mathcal{O}_Y\rightarrow 0\,,
\end{align}
while the 2-brane $\mathcal{E}_2=\iota_! \mathcal{O}_{C}(K_{C}^{1/2})$ is given by a certain K-theoretic push-forward of the structure sheaf on $C=J^2$, twisted by a spin-structure as in~\cite{Gerhardus:2016iot}.
Using the formula for the asymptotic central charge~\cite{IRITANI20091016,Halverson:2013qca}
\begin{align}
        Z_{\text{asy.}}(\mathcal{E})=\int_Y e^\omega\Gamma_{\mathbb{C}}(Y)\text{ch}(\mathcal{E})^\vee\,,
\end{align}
in terms of the Gamma-class that for a Calabi-Yau 3-fold takes the form
\begin{align}
    \Gamma_{\mathcal{C}}=1+\frac{1}{24}c_2(Y)+\frac{\zeta(3)}{(2\pi i)^3}c_3(Y)\,,
\end{align}
and with $\bullet^\vee$ acting on elements $\gamma\in H^{i,i}$ by multiplication with $(-1)^i$,
we find the leading terms $\Pi^{(i)}_1=Z_{\text{asy.}}(\mathcal{E}_i)$ of the central charges
\begin{align}
        \begin{split}
                \Pi^{(6)}_1=&\frac16\kappa_Yt^3+\frac{b_Y}{24}t+\frac{1}{(2\pi i)^3}\zeta(3)\chi_Y\,,\\
                \Pi^{(4)}_1=&-\frac12\kappa_Yt^2-\frac12\kappa_Yt-\frac{1}{24}(b_Y+4\kappa_Y)\,,\\
                \Pi^{(2)}_1=&t\,,\quad\Pi^{(0)}_1=-1\,.
        \end{split}
\end{align}
The open string index can be calculated using
\begin{align}
    \chi(\mathcal{E},\mathcal{F})=\int_Y\text{Td}(Y)\text{ch}(\mathcal{E})^\vee\text{ch}(\mathcal{F})\,,
\end{align}
and with respect to our basis $(\mathcal{E}_6,\mathcal{E}_4,\mathcal{E}_2,\mathcal{E}_0)$ we find the matrix
\begin{align}
        \Sigma_2=\left(\begin{array}{cccc}
                0&8&0&1\\-8&0&-1&0\\0&1&0&0\\-1&0&0&0
        \end{array}\right)\,.
\end{align}
Mirror symmetry relates the central charges to periods of the holomorphic $(3,0)$-form $\Omega$ on the mirror $\tilde{Y}$ of $Y$ over a basis of integral 3-cycles.
The monodromies around the large volume limit $z=0$, the conifold point $z=2^{-8}$ and the hybrid point at infinity act on the period vector respectively as $\vec{\Pi}_1\rightarrow M_\bullet\cdot\vec{\Pi}_1$, with
\begin{equation}
\begin{gathered}
        M_{\text{LV}}=\left(
\begin{array}{cccc}
 1 & -1 & 0 & 0 \\
 0 & 1 & -16 & 16 \\
 0 & 0 & 1 & -1 \\
 0 & 0 & 0 & 1 \\
\end{array}
\right)\,,\quad M_{\text{C}}=\left(
\begin{array}{cccc}
 1 & 0 & 0 & 0 \\
 8 & 1 & 0 & 0 \\
 0 & 0 & 1 & 0 \\
 1 & 0 & 0 & 1 \\
\end{array}
        \right)\,,\\ M_{\text{H}}=\left(
\begin{array}{cccc}
 -7 & 1 & 16 & 0 \\
 -8 & 1 & 16 & 0 \\
 -1 & 0 & 1 & 1 \\
 -1 & 0 & 0 & 1 \\
\end{array}
\right)\,,
\end{gathered}
\end{equation}
such that $M_H\cdot M_C\cdot M_{\text{LV}}=1$.

To obtain the central charges of an integral basis of branes on the non-commutative resolution $X_{\text{n.c.}}$ of the singular octic double solid, we change coordinates as
\begin{align}
        z=\frac{1}{2^{16}v^2}\,,
        \label{eqn:ztov}
\end{align}
such that $v=0$ corresponds to the hybrid point associated to $X_{\text{n.c.}}$.
The importance of using $v$ instead of $\sqrt{v}\propto 1/z$ as a local coordinate was pointed out in~\cite{Schimannek:2021pau} and is necessary to extract the $\mathbb{Z}_2$-refined GV-invariants.
As a result, this leads to the fundamental period
\begin{align}
        \varpi_0=\sum\limits_{k=0}^\infty\frac{(2k!)^4}{(k!)^8}v^{2k}={_4}F_3\left(\frac12,\frac12,\frac12,\frac12;1,1,1;2^8v^2\right)\,.
        \label{eqn:w0fundXnc}
\end{align}
This is annihilated by the Picard-Fuchs operator
\begin{align}
        \mathcal{D}_2=\theta^4-2^8v^2(\theta+1)^4\,,
\end{align}
with discriminant $\Delta=\Delta_1\cdot\Delta_2$, where
\begin{align}
        \Delta_1=1-2^4v\,,\quad \Delta_2=1+2^4v\,.
        \label{eqn:discXnc}
\end{align}

Following~\cite{Schimannek:2021pau}, we expect that the triple intersection number on $X_{\text{n.c.}}$ as well as the intersection with the second Chern class should be unaffected by the nodes are equal to those of the generic degree $8$ hypersurface $X_{\text{def.}}=X^{(1,149)}_8(11114)$ in $\mathbb{P}^4(1,1,1,1,4)$, such that
\begin{align}
        \kappa_{\text{n.c.}}=\int_{X_8}J\cdot J\cdot J=2\,,\quad b_{\text{n.c.}}=\int_{X_8}c_2(X_8)\cdot J=44\,.
\end{align}
The stringy Euler characteristic on the other hand is a global property and sensitive to the fractional B-field that prevent the deformation of the nodes.
Due to the derived equivalence with $Y$ it should be equal to $\chi_{\text{n.c.}}=\chi_Y=-128$\,.
Using~\eqref{eqn:genuszeroconst} and accounting for the constant map contribution~\eqref{eqn:cmapcontrz2},
with the number of nodes $m_s=84$, we can then fix the prepotential
\begin{align}
\begin{split}
    F_0=&-\frac16\kappa_{\text{n.c.}}t^3+\frac{b_{\text{n.c.}}}{24}t+\frac{\zeta(3)}{(2\pi i)^3}\left(\frac{\chi_{\text{n.c.}}}{2}+\frac{3}{4}m_s\right)+\mathcal{O}(e^{2\pi i t})\\
    =&-\frac13t^3-\frac{11}{6}t-\frac{i\zeta(3)}{8\pi^3}+\mathcal{O}(e^{2\pi i t})\,,
    \end{split}
    \label{eqn:xncf0}
\end{align}
such that a symplectic basis of periods is given by
\begin{align}
        \Pi_{\text{sp}}=(1,\,t,\,\partial_t F_0,\,2F_0-t\partial_t F_0)\,.
        \label{eqn:xncsymp}
\end{align}
The prepotential~\eqref{eqn:xncf0} is the correct choice to, combining it with the corresponding prepotential for $X_{\text{def.}}$, extract the $\mathbb{Z}_2$-refined Gopakumar-Vafa invariants at genus $0$.
However, it turns out that~\eqref{eqn:xncsymp} does not correspond to an integral basis of brane charges.

The fractional B-field implies that the 0-brane charge on $X_{\text{n.c.}}$ is not primitive anymore, but can decompose into that of two 2-branes wrapping $2$-torsion curves, each having a central charge $-\frac12$.
Charge quantization then suggests that the correct generator with 6-brane charge has a central charge that is twice as large as one would expect from the leading behaviour of the central charge associated to the structure sheaf of $X_{\text{def.}}$.
Indeed we find that the basis
\begin{align}
        \begin{split}
                 \Pi^{(6)}_2=&\frac{2}{3}t^3-\frac{1}{12}t-4\frac{1}{(2\pi i)^3}\zeta(3)\,,\\
                \Pi^{(4)}_2=&-t^2-\frac12t-\frac{1}{24}\,,\\
                \Pi^{(2)}_2=&t\,,\quad\Pi^{(0)}_2=-\frac12\,,
        \end{split}
\end{align}
for the asymptotic central charges of a basis of topological B-branes on $X_{\text{n.c.}}$ leads to an integral transfer matrix
\begin{align}
        \Pi_2=\frac{1}{128v}T\cdot\Pi_1\,,\quad T=\left(\begin{array}{cccc}
                0&0&-1&0\\
                1&0&-2&-1\\
                -1&0&4&0\\
                2&-1&-8&4
        \end{array}\right)\,,
\end{align}
where the overall factor of $1/(128v)$ corresponds to a K\"ahler transformation
\begin{align}
        e^{-K}\rightarrow f(v)e^{-K}\,,\quad f(v)=\frac{1}{2^7v}\,.
        \label{eqn:xncktrafo}
\end{align}
This determines also the Yukawa coupling
\begin{align}
        C_{vvv}=\frac{2}{v^3\Delta_1\Delta_2}=f(v)^2\left(\frac{\partial z}{\partial v}\right)^3C_{zzz}\,.
\end{align}
The determinant of the transfer matrix is $\text{det}\,T=1$, which confirms that $\vec{\Pi}_2$ corresponds to an integral basis of branes.
It would be very interesting to try to better understand this basis of branes and the corresponding central charges, for examples using the techniques from~\cite{Knapp:2020oba}.

Note that the analogous transfer matrix between the large volume limit associated to a smooth genus one fibration with a 2-section and a non-commutative resolution of the corresponding Jacobian fibration was calculated in~\cite[equ'n (6.49)]{Schimannek:2021pau} and is also rendered integral by changing the normalization of the elementary ``0-brane'' charge to $1/2$ while multiplying the naive 6-brane charge by 2.

Let us denote the monodromy around $v=0$ acting on $\Pi_2$ by $W_{\text{LV}}$ and define
\begin{align}
        W_{\text{H}}=T\cdot M_H\cdot T^{-1}=\left(\begin{array}{cccc}
 -1 & 1 & 0 & 0 \\
 0 & -1 & 1 & -1 \\
 0 & 0 & -1 & 1 \\
 0 & 0 & 0 & -1 \\
        \end{array}\right)\,.
\end{align}
which acts as $t\rightarrow t-\frac12$ and satisfies $W_{\text{H}}^2=W_{\text{LV}}$.
Up to a homological shift, this corresponds to the action of the $\mathbb{Z}_2$ quantum symmetry of the hybrid model.
Note that $t$ is the central charge of a degree one curve that has four tangencies with the degree $8$ ramification locus of the double cover of $\mathbb{P}^3$ and is supported on one of the two sheets.
On the other hand, as discussed above, the $\frac12$ should be interpreted as the central charge of a $2$-brane that wraps a $2$-torsion curve.
It was argued in~\cite{Hori:2011pd} that the $\mathbb{Z}_2$ quantum symmetry of the Landau-Ginzburg fiber acts by exchanging the two sheets of the cover.
This suggests, that the difference between a degree one curve and its image under the involution of the cover is a $2$-torsion curve.

We can also see the monodromy $W_H$ directly in the A-model as an autoequivalence of $D^b(\mathbb{P}^3,\mathcal{B})$ following \cite{kuznetsov2008derived}, where our sheaf $\mathcal{B}$ is denoted as $\mathcal{B}_0$. It is shown that there is a graded sheaf (denoted by $\mathcal{B}$ \cite{kuznetsov2008derived}) whose category of graded modules has a quotient category which is equivalent to the category of $\mathcal{B}_0$-modules.  Let $\mathcal{T}$ be the autoequivalence of the quotient category of graded $\mathcal{B}$-modules which shifts degrees by 1.  Then in our situation, \cite[Proposition~3.17]{kuznetsov2008derived} says that $\mathcal{T}^2$ is equal to tensoring with $\mathcal{O}_{\mathbb{P}^3}(1)$.  Since tensoring with $\mathcal{O}_{\mathbb{P}^3}(1)$ is the A-model version of $W_{\text{LV}}$, we conclude that $\mathcal{T}$ is the A-model version of $W_{\text{H}}$.  We have previously given a different description of $\mathcal{T}$  in Section~\ref{sect:octicdoub:nc-res:glsm}.

\subsection{Direct integration for $X_{\text{n.c.}}$}
\label{sec:directintegration} 
We can now carry out the direct integration of the holomorphic anomaly equations to obtain the higher genus topological string free energies for $X_{\text{n.c.}}$.
The $\mathbb{Z}_2$-refined Gopakumar-Vafa invariants associated to the singular octic double solid $X$ can then be extracted using~\eqref{eq:Znc}, by combining the information in the free energies on the non-commutative resolution $X_{\text{n.c.}}$ and those on the smooth deformation $X_{\text{def.}}$ of $X$.
The general method for the direct integration method for smooth Calabi-Yau threefolds in terms of a set of propagators $\tilde{S}^{zz},\tilde{S}^z,\tilde{S}$ is reviewed in Appendix~\ref{app:algebraicKahlergauge} and here we focus on the differences that arise for the non-commutative resolution $X_{\text{n.c.}}$.

The direct integration for $X_{\text{def.}}$ and $Y$ has been carried out for low genera in~\cite{Huang:2006hq}.
We can use the transformation of the propagator ambiguities that are discussed in Appendix~\ref{app:algebraicKahlergauge} under the K\"ahler transformation~\eqref{eqn:xncktrafo} and the change of coordinate~\eqref{eqn:ztov} to obtain the corresponding propagators around the MUM-point of $X_{\text{n.c.}}$.
This allows us to transform the entire free energies of $Y$ to the corresponding free energies of $X_{\text{n.c.}}$ .
Alternatively, we can also carry out the direct integration directly at the MUM-point associated to $X_{\text{n.c.}}$ and, taking into account the modified constant map contributions~\eqref{eqn:cmapcontrz2} as well as the correct values of the topological invariants and the corresponding normalization of the prepotential discussed above, the procedure is the same as for an ordinary smooth Calabi-Yau threefold.

To start the direct integration, we first need the free energies at genus $g=1$. Using the Ansatz~\eqref{eq:F1}, together with the boundary conditions~\eqref{eq:F1cn},~\eqref{eq:F1mum} 
 and the topological invariants that are listed in Table~\ref{tab:cystab}, we find that the genus one free energies for $Y$ and $X_{\text{n.c.}}$ are respectively given by
\begin{align}
\begin{split}
    F_1^Y(z)=& \frac{22}{3}K-\frac12\log\det(G_{z\bar{z}})-\frac{1}{12}\log(\Delta)-\frac{19}{6}\log(z)\,,\\
    F_1^{X_{\text{n.c.}}}(v)=& \frac{22}{3}K-\frac12\log\det(G_{v\bar{v}})-\frac{1}{12}\log(\Delta_1\Delta_2)-\frac{7}{3}\log(v)\,,
\end{split}
\end{align}
where the discriminants are respectively given in~\eqref{eqn:discY} and~\eqref{eqn:discXnc}.
 The holomorphic limit of the K\"ahler potential is given by $K\sim-\log(\varpi_0)$ in terms of the respective fundamental periods~\eqref{eqn:w0fundY} and~\eqref{eqn:w0fundXnc}.

A suitable choice for the propagator ambiguities, that are discussed in Appendix~\ref{app:algebraicKahlergauge}, around the MUM-point $z=0$ associated to $Y$ is given by
\begin{align}
        {\small s^z_{zz}=-\frac{1}{z}\,,\quad h^{zz}_z=-\frac{z}{2^5}\,,\quad h^z_z=0\,,\quad h_z=-2+\frac{1}{2^9z}\,,\quad h_{zz}=\frac{1}{4z^2}\,,}
        \label{eqn:X2amb1}
\end{align}
while the corresponding transformed ambiguities for $X_{\text{n.c.}}$ take the form
\begin{align}
        {\small s^v_{vv}=-\frac{3}{v}\,,\,\,\, h^{vv}_v=2^8v^3\,,\,\,\, h^v_v=-2^8v^2\,,\,\,\, h_v=\frac{1}{v}+2^6v\,,\,\,\, h_{vv}=\frac{2}{v^2}\,.}
\end{align}

The direct integration for $Y$ and $X_{\text{def.}}$ has been carried out in~\cite{Huang:2006hq}. As discussed above, the constant map contributions at the respective MUM-points, the Castelnuovo vanishing of the Gopakumar-Vafa invariants -- together with the explicit value of the invariants that saturate the bound -- the gap condition at the conifold and the branching at infinity are sufficient to, at least in principle, fix the holomorphic ambiguities $f^{(g)}$ that appear in the direct integration for $Y$ and $X_{\text{def.}}$ respectively up to genus $17$ and $60$.

Using the K\"ahler transformation and considering now the free energies for $X_{\text{n.c.}}$, which from the perspective of $Y$ lies at the MUM-point at infinity, we get new boundary conditions from the modified constant map contributions~\eqref{eqn:cmapcontrz2} but it also turns out that the $\mathbb{Z}_2$-refined GV-invariants exhibit a Castelnuovo-like vanishing.
Including these additional conditions, this allows us to fix the holomorphic ambiguities for $X_{\text{n.c.}}$ and $Y$ up to genus $32$.
The resulting invariants are listed in Tables~\ref{tab:nc0} and~\ref{tab:nc1}.
In Section~\ref{sec:geometric} we explicitly derive some of those numbers from the mathematical definition proposed in Section~\ref{subsec:gv}.
In all of those cases we find agreement with the predictions from mirror symmetry and the corresponding invariants are highlighted in the tables.

To illustrate the general structure and to make it easier for the reader to reproduce our calculation, we give the result for the genus $2$ free energy on $Y$
\begin{align}
\small\begin{split}
 F_2^{Y}(z)=&\frac{-11+11000 z-3472320 z^2+352763904 z^3}{45 \Delta ^2}+\frac{128 (1-264 z) }{9 z \Delta }\tilde{S}^z\\
 &+\frac{2 \left(19-10080 z+1348608 z^2\right)}{9 z^2 \Delta
   ^2}\tilde{S}^{zz}+\frac{304 }{9}\tilde{S}-\frac{128
   }{3 z^3 \Delta }\tilde{S}^{zz}\tilde{S}^z\\
   &-\frac{64 (1-288 z) }{3 z^4 \Delta ^2}\tilde{S}^{zz}\tilde{S}^{zz}+\frac{160}{3 z^6 \Delta ^2}\tilde{S}^{zz}\tilde{S}^{zz}\tilde{S}^{zz}\,,
   \end{split}
\end{align}
as well as the corresponding result on $X_{\text{n.c.}}$, that is obtained after the K\"ahler and coordinate transformation,
\begin{align}
    \small\begin{split}
 F_2^{X_{\text{n.c.}}}(v)=&-\frac{21531-13889280 v^2+2883584000 v^4-188978561024 v^6}{720 \Delta ^2}\\
 +&\frac{ \left(421-202752 v^2+25165824 v^4\right)}{72 v^2 \Delta
   ^2}\tilde{S}^{vv}+\frac{8\left(5-1536 v^2\right)}{9 v \Delta }\tilde{S}^v\\
   &+\frac{5}{6 v^6 \Delta ^2}\tilde{S}^{vv}\tilde{S}^{vv}\tilde{S}^{vv}+\frac{2}{3 v^4 \Delta ^2}\tilde{S}^{vv}\tilde{S}^{vv}-\frac{16}{3 v^3
   \Delta }\tilde{S}^{vv}\tilde{S}^v+\frac{304}{9}\tilde{S}\,.
   \end{split}
\end{align}

Let us stress again, that the refined invariants are only obtained after combining the information from the free energies associated to $X_{\text{n.c.}}$ and $X_{\text{def.}}$.
The ``naive'' unrefined GV-invariants that one could extract from the free energies of $X_{\text{n.c.}}$ do not exhibit a Castelnuovo-like vanishing.
We therefore observe a highly non-trivial interplay between the topological string partition functions at three different MUM-points that lie in the complex structure moduli spaces of two different mirror manifolds.

{\small \begin{table}[h!]
	\centering
        \begin{gather*}
\begin{array}{c|ccccc}
n^{\beta,0}_g&\beta=1&2&3&4&5\\\hline
g=0&\textbf{14752}&64415616&711860273440&11596528004344320&233938237312624658400\\
1&\textbf{0}&20160&10732175296&902646044328864&50712027457008177856\\
2&\textbf{0}&\textbf{504}&-8275872&6249833130944&2700746768622436448\\
3&\textbf{0}&\textbf{0}&\textbf{-88512}&-87429839184&10292236849965248\\
4&\textbf{0}&\textbf{0}&\textbf{0}&198065872&-337281112359424\\
5&\textbf{0}&\textbf{0}&\textbf{0}&157306&6031964134528\\
6&\textbf{0}&\textbf{0}&\textbf{0}&\textbf{1632}&-43153905216\\
7&\textbf{0}&\textbf{0}&\textbf{0}&\textbf{24}&18764544\\
8&\textbf{0}&\textbf{0}&\textbf{0}&\textbf{0}&\textbf{177024}\\
9&\textbf{0}&\textbf{0}&\textbf{0}&\textbf{0}&\textbf{0}
\end{array}\\
\begin{array}{c|ccc}
n^{\beta,0}_g&\beta=6&7&8\\\hline
g=11&238896&1365366811756288&-4736147442532505835596\\
12&\textbf{4536}&-17274516630240&241184169662887498000\\
13&\textbf{0}&97442213760&-9638193260013962094\\
14&\textbf{0}&-33988608&289481324332786944\\
15&\textbf{0}&\textbf{-324544}&-6124171374649000\\
16&\textbf{0}&\textbf{0}&81787112365664\\
17&\textbf{0}&\textbf{0}&-550148288563\\
18&\textbf{0}&\textbf{0}&845297088\\
19&\textbf{0}&\textbf{0}&441736\\
20&\textbf{0}&\textbf{0}&\textbf{3200}\\
21&\textbf{0}&\textbf{0}&\textbf{64}\\
22&\textbf{0}&\textbf{0}&\textbf{0}
\end{array}\\
\begin{array}{c|ccc}
n^{\beta,0}_g&\beta=9&10&11\\\hline
g=23&48799616&-31102670029152533520&2292168771748652654604291968\\
24&\textbf{531072}&773847175064950328&-158373976042091789355897472\\
25&\textbf{0}&-13810315238046256&9491926931091427533124736\\
26&\textbf{0}&158750107627640&-486491652797947678259968\\
27&\textbf{0}&-942809065216&20941105449951277658112\\
28&\textbf{0}&1362250176&-739377404194595085312\\
29&\textbf{0}&259056&20736415070762709120\\
30&\textbf{0}&\textbf{11592}&-441204207287440640\\
31&\textbf{0}&\textbf{0}&6632918677947904\\
32&\textbf{0}&\textbf{0}&-62184411813632
\end{array}
        \end{gather*}
	\caption{Torsion refined Gopakumar-Vafa invariants for $\mathbb{Z}_2$-charge $q=0$ the degenerate double cover of $\mathbb{P}^3$, ramified over a determinantal octic. See Table~\ref{tab:nc1} for charge $q=1$. The highlighted numbers will be confirmed by a direct enumerative computation in Section~\ref{subsec:veroctic}.}
	\label{tab:nc0}
\end{table}}

{\small \begin{table}[h!]
	\centering
        \begin{gather*}
\begin{array}{c|ccccc}
n^{\beta,1}_g&\beta=1&2&3&4&5\\\hline
g=0&\textbf{14752}&64419296&711860273440&11596528020448992&233938237312624658400\\
1&\textbf{0}&21152&10732175296&902646048376992&50712027457008177856\\
2&\textbf{0}&\textbf{360}&-8275872&6249834146800&2700746768622436448\\
3&\textbf{0}&\textbf{6}&\textbf{-88512}&-87429664640&10292236849965248\\
4&\textbf{0}&\textbf{0}&\textbf{0}&198149928&-337281112359424\\
5&\textbf{0}&\textbf{0}&\textbf{0}&144144&6031964134528\\
6&\textbf{0}&\textbf{0}&\textbf{0}&\textbf{2520}&-43153905216\\
7&\textbf{0}&\textbf{0}&\textbf{0}&\textbf{0}&18764544\\
8&\textbf{0}&\textbf{0}&\textbf{0}&\textbf{0}&\textbf{177024}\\
9&\textbf{0}&\textbf{0}&\textbf{0}&\textbf{0}&\textbf{0}
\end{array}\\
\begin{array}{c|ccc}
n^{\beta,1}_g&\beta=6&7&8\\\hline
g=11&284538&1365366811756288&-4736147442578975095808\\
12&\textbf{2496}&-17274516630240&241184169689646798464\\
13&\textbf{40}&97442213760&-9638193272233061776\\
14&\textbf{0}&-33988608&289481328620120504\\
15&\textbf{0}&\textbf{-324544}&-6124172499145888\\
16&\textbf{0}&\textbf{0}&81787327067664\\
17&\textbf{0}&\textbf{0}&-550176980192\\
18&\textbf{0}&\textbf{0}&847830320\\
19&\textbf{0}&\textbf{0}&305424\\
20&\textbf{0}&\textbf{0}&\textbf{7560}\\
21&\textbf{0}&\textbf{0}&\textbf{0}\\
22&\textbf{0}&\textbf{0}&\textbf{0}
\end{array}\\
\begin{array}{c|ccc}
n^{\beta,1}_g&\beta=9&10&11\\\hline
g=23&48799616&-31102669207528659588&2292168771748652654604291968\\
24&\textbf{531072}&773847033066507008&-158373976042091789355897472\\
25&\textbf{0}&-13810295800263040&9491926931091427533124736\\
26&\textbf{0}&158748050119808&-486491652797947678259968\\
27&\textbf{0}&-942646032272&20941105449951277658112\\
28&\textbf{0}&1352987680&-739377404194595085312\\
29&\textbf{0}&615592&20736415070762709120\\
30&\textbf{0}&\textbf{3072}&-441204207287440640\\
31&\textbf{0}&\textbf{96}&6632918677947904\\
32&\textbf{0}&\textbf{0}&-62184411813632
\end{array}
        \end{gather*}
	\caption{Torsion refined Gopakumar-Vafa invariants for $\mathbb{Z}_2$-charge $q=1$ the degenerate double cover of $\mathbb{P}^3$, ramified over a determinantal octic. See Table~\ref{tab:nc0} for charge $q=0$. The highlighted numbers will be confirmed directly by a direct enumerative computation in Section~\ref{subsec:veroctic}.}
        \label{tab:nc1}
\end{table}}

\section{Torus fibered examples}
\label{sec:ellfibrations}

In this section we discuss the relative situation that arises when a complete intersection of two quadrics in $\mathbb{P}^3$ is fibered over a Fano surface such that the total space is a Calabi-Yau threefold.
Note that this has already been studied at length in~\cite{Schimannek:2021pau}.
Our goal here is therefore only to highlight the connection to non-commutative resolutions in terms of sheaves of Clifford algebras as well as to the construction of Clifford double mirrors by Borisov, Li~\cite{borisovzhan} and an example of a twisted derived equivalence by Calabrese, Thomas~\cite{Calabrese2016DerivedEC}.

\subsection{Family of GLSMs and their analysis}
Let us first consider a smooth Calabi-Yau threefold $Y$ that is genus one fibered over $\mathbb{P}^2$ with a $4$-section and $h^{1,1}(X)=2$.
Each fiber can be realized as the intersection of two quadrics in $\mathbb{P}^3$ and for simplicity we assume that the corresponding $\mathbb{P}^3$ bundle over $\mathbb{P}^2$ is toric.

We can then construct a gauged linear sigma model with gauge group $G=U(1)_F\times U(1)_B$ and chiral fields $p_1,p_2,x_{i=1,\ldots,4},y_{i=1,\ldots,3}$ that carry the charges given in Table~\ref{tab:4secglsm}, where the Calabi-Yau condition amounts to the requirement that $3+\sum_i q_i=0$.
\begin{table}[h!]
        \centering
        \begin{align*}
                \begin{array}{c|cc|cccc|c}
                        &p_1&p_2&x_1&x_2&x_3&x_4&y_{i=1,2,3}\\\hline
                        U(1)_F&-2&-2&1&1&1&1&0\\
                        U(1)_B& q_1&q_2&q_3&q_4&q_5&0&1
                \end{array}
        \end{align*}
        \label{tab:4secglsm}
        \caption{The field content of a GLSM that is associated to a generic Calabi-Yau threefold which is genus one fibered over $\mathbb{P}^2$ and exhibits a $4$-section.}
\end{table}
We denote the FI-theta parameters associated to $U(1)_{F/B}$ respectively by $t_F$ and $t_B$ and for $t_F,t_B\gg 0$ the GLSM flows to a non-linear sigma model on $Y$.
On the other hand, following the arguments from~\cite{Caldararu:2010ljp}, choosing $t_F\ll 0$ while maintaining $t_B\gg 0$ produces a non-commutative resolution of a certain singular double cover $X$ of a $\mathbb{P}^1$ fibration over $\mathbb{P}^2$.
From a more mathematical perspective, the limits are related by a fiberwise application of homological projective duality~\cite{kuz0}.
Here the duality can easily be verified to be an instance of the Clifford double mirror construction that has subsequently been developed in~\cite{borisovzhan}.~\footnote{To this end one just notes that the relevant decompositions of the degree element of the Gorenstein cone on page 45 of~\cite{borisovzhan} are available for all GLSMs of the form given in Table~\ref{tab:4secglsm}.}

To construct the singular double cover $X$, let us write down the generic superpotential
\begin{align} \label{eq:ct:sup}
        W_Y=p_1 Q_1(\vec{x},\vec{y})+p_2 Q_2(\vec{x},\vec{y})=\vec{x}^{\,\intercal} \left[p_1A^1(\vec{y})+p_2A^2(\vec{y})\right] \vec{x}\,,
\end{align}
where $A_i$ are $4\times 4$ matrices with entries that are polynomials in $y_{i=1,\ldots,3}$ such that $W_Y$ is neutral.
More precisely, the degrees are given by
\begin{align}
        \text{deg}_y\,A^i=-q_i-\left(\begin{array}{cccc}
                2q_3&q_3+q_4&q_3+q_5&q_3\\
                q_3+q_4&2q_4&q_4+q_5&q_4\\
                q_3+q_5&q_4+q_5&2q_5&q_5\\
                q_3&q_4&q_5&0
        \end{array}\right)\,,
\end{align}
while the degrees of the determinant polynomials are
\begin{align}
        \tilde{q}_i=\text{deg}\,\det A^i=-4q_i-2\hat{q}\,,\quad i=1,2\,,
\end{align}
where we have introduced $\hat{q}=q_3+q_4+q_5$.
One can now consider $[p_1:p_2:y_1:y_2:y_3]$ as homogenous coordinates on a projective bundle
\begin{align}
        B=\mathbb{P}\left(\mathcal{O}(q_1)\oplus\mathcal{O}(q_2)\right)\rightarrow\mathbb{P}^2\,,
        \label{eqn:projbundle}
\end{align}
and construct the double cover $X$ over $B$ with ramification locus
\begin{align}
        \det\left(p_1A^1+p_2A^2\right)=0\,.
\end{align}
The unresolved double cover can again be realized by a GLSM, now with chiral fields $P,z,p_1,p_2,y_{i=1,\ldots,3}$ of $U(1)_F\times U(1)_B$ charges given in Table~\ref{tab:dcglsm}.
\begin{table}[h!]
        \centering
        \begin{align*}
                \begin{array}{c|c|c|cc|c}
                        &P&z&p_1&p_2&y_{i=1,\ldots,3}\\\hline
                        U(1)_F&-4& 2& 1& 1& 0\\
                        U(1)_B&2\hat{q}&-\hat{q}&q_1&q_2&1
                \end{array}
        \end{align*}
        \caption{The field content of a GLSM that is associated to a Calabi-Yau double cover of the projective bundle $B$ from~\eqref{eqn:projbundle}. Here we use again $\hat{q}=q_3+q_4+q_4$.}
        \label{tab:dcglsm}
\end{table}

The corresponding superpotential now takes the form
\begin{align}
        W_X=P\cdot\left[z^2+\det\left(p_1A^1+p_2A^2\right)\right]\,.
\end{align}
This superpotential is degenerate and the GLSM is singular.
However, it can be deformed to a smooth GLSM with superpotential
\begin{align}
        W_{X_\text{def.}}=P\cdot\left[z^2+\sum\limits_{k=0}^4p_1^{4-k}p_2^k\widetilde{Q}_{(4-k)q_1+kq_2+2\hat{q}}(y)\right]\,,
\end{align}
where $\widetilde{Q}_{-d}(y)$ are homogeneous polynomials of degree $d$ in $y_{i=1,\ldots,3}$.
In the geometric phase $t_F,t_B\gg 0$ the GLSM flows to a non-linear sigma model on a smooth deformation $X_{\text{def.}}$ of $X$.
One can interpret $p_1,p_2,z$ as homogeneous coordinates on the fibers of a fibration of weighted projective spaces $\mathbb{P}_{112}$ and $X_{\text{def}.}$ is a fibration of degree $4$ hypersurfaces over $\mathbb{P}^2$.
This leads to a torus fibration of $X_{\text{def}.}$ over $\mathbb{P}^2$, that is not elliptic but has a two-section corresponding e.g. to $p_1=0$.

As was argued in~\cite{Schimannek:2021pau}, when we degenerate $X_{\text{def.}}$ to $X$ an M-theory compactification will develop an additional $\mathbb{Z}_2$ gauge symmetry.
In particular, due to the absence of a corresponding continuous Coulomb branch, we expect the singularities to be terminal and an analytic small resolution $\widehat{X}$ of $X$ to exhibit
\begin{align}
H^3(\widehat{X},\mathbb{Z})\simeq\mathbb{Z}^2\oplus\mathbb{Z}_2\,,
\end{align}
such that the discussion from Section~\ref{sect:prop} applies.
The A-model topological string partition function on the non-commutative resolution $D^b(\widehat{X},\alpha)$ together with the partition function $X_{\text{def.}}$ can be used to extract the corresponding Gopakumar-Vafa invariants.

The example with charge vector
\begin{align}
        \vec{q}=(-2,\,0,-1,\,0,\,0)\,,
\end{align}  
including the calculation of the $\mathbb{Z}_2$-refined GV-invariants via mirror symmetry and using a modular bootstrap approach, has been discussed in detail in~\cite{Schimannek:2021pau}.
For those invariants for which we can evaluate our proposed mathematical definition from Section~\ref{subsec:gv} explicitly, the results are in perfect agreement.

\subsection{The Calabrese-Thomas example}\label{subsec:CalabreseThomas}

Another example of this form turns out to correspond to a non-commutative resolution and twisted derived equivalence that was already studied by Calabrese-Thomas in~\cite{Calabrese2016DerivedEC}.
The corresponding charge vector is
\begin{align}
        \vec{q}=(-1,-1,-1,\,0,\,0)\,,
        \label{eqn:ctcharges}
\end{align}
and from the GLSM (table~\eqref{tab:dcglsm}) one can see that we obtain a Calabi-Yau threefold $X$ that is a singular double cover of $\mathbb{P}^1\times\mathbb{P}^2$, ramified over a degree $(6,4)$ determinantal surface.
Any small resolution $\widehat{X}$ of $X$ is necessarily non-K\"ahler \cite[lemma 3.12]{Calabrese2016DerivedEC}, just as in the octic double solid example, and a resolution in terms of a sheaf of Clifford algebras on $\mathbb{P}^1\times\mathbb{P}^2$ has also been discussed in~\cite{Calabrese2016DerivedEC,borisov2018clifford}.

Moreover, it was shown in~\cite{Calabrese2016DerivedEC} that there is an order two element $\alpha \in H^3(\widehat{X},\mathbb{Z})_{\text{tors.}}$, and a twisted derived equivalence
\begin{align}
    D^b(\widehat{X},\alpha)\simeq D^b(Y)\,.
\end{align}
where the dual $Y$ is a smooth Calabi-Yau threefold.
Of course $Y$ is nothing but the genus one fibration over $\mathbb{P}^2$ with a 4-section that we realized in the $t_F,t_B\gg 0$ phase of the GLSM~\eqref{tab:4secglsm}.
For physical reasons, discussed in~\cite{Schimannek:2021pau}, we expect that $H^3(\widehat{X},\mathbb{Z})\simeq \mathbb{Z}^2\oplus\mathbb{Z}_2$ and that M-theory on $X$ develops a $\mathbb{Z}_2$ gauge symmetry.
The calculation of the $\mathbb{Z}_2$-refined GV-invariants associated to $X$ is analogous to the discussion in~\cite{Schimannek:2021pau} and we list some of the invariants in Appendix~\ref{app:ctgv}.

Let us briefly match the realization of $Y$ as a complete intersection of two quadrics in $\mathbb{P}^3$ that is fibered over $\mathbb{P}^2$ with the construction from~\cite{Calabrese2016DerivedEC}.
In their notation, we denote the vector spaces spanned by the vacuum expectation values of the scalar fields $x_{2,\ldots,4},\,y_{1,\ldots,3}$ in the $t_F,t_B\gg 0$ phase of GLSM~\eqref{tab:4secglsm} respectively by $V$ and $W$.
The toric ambient space $Z$ that is spanned by the gauge inequivalent vacuum configurations of $x_{1,\ldots,4},\,y_{1,\ldots,3}$ is a blowup of ${\mathbb P}^5$ along ${\mathbb P}^2$,
\begin{equation}
Z \: = \: {\rm Bl}_{{\mathbb P}V}\left(
{\mathbb P}(V \oplus W) = {\mathbb P}^5 \right),
\end{equation}
and the Calabi-Yau $Y$ itself can be seen as a blowup of a degree $(3,3)$ hypersurface in ${\mathbb P}^5$.

\subsection{Relation to discrete torsion}
The previous discussion can easily be generalized to genus one fibered Calabi-Yau threefolds with 4-sections over other bases than $\mathbb{P}^2$.
If we choose the base of the double cover to be $\mathbb{P}^1\times\mathbb{P}^1\times\mathbb{P}^1$, we obtain an example that has also been studied in~\cite{Vafa:1994rv,Aspinwall:1995rb} in the context of discrete torsion.

The starting point of the construction is an orbifold $X'=(E_1\times E_2\times E_3)/(\mathbb{Z}_2\times\mathbb{Z}_2)$, where $E_i=\mathbb{C}/\langle\tau_i,1\rangle$ for some $\tau_i\in\mathbb{H}$ and the action on the corresponding coordinates $(z_1,z_2,z_3)$
is generated by multiplications with $(-1,1,-1)$ and $(1,-1,-1)$.
The orbifold admits discrete torsion
\begin{align}
    H^2(\mathbb{Z}_2\times\mathbb{Z}_2,U(1))\simeq \mathbb{Z}_2\,.
\end{align}
A generic smoothing of $X'$ takes the form of a Calabi-Yau double cover $X_{\text{def.}}$ over $\mathbb{P}^1\times\mathbb{P}^1\times\mathbb{P}^1$ with $h^{2,1}=115$ complex structure deformations.
However, turning on non-trivial discrete torsion obstructs 64 of the corresponding moduli and the generic deformation is a singular Calabi-Yau $X$ with $64$ conifold singularities.

Again the nc-resolution of $X$ falls under the Clifford double-mirror construction from~\cite{borisovzhan}.
In particular, there is a twisted derived equivalence
\begin{align}
    D^b(\widehat{X},\alpha)\simeq D^b(Y)\,,
\end{align}
where $\widehat{X}$ is a small analytic resolution and $Y$ is a smooth genus one fibration over $\mathbb{P}^1\times\mathbb{P}^1$ with a $4$-section.
Following~\cite{Schimannek:2021pau}, we expect that M-theory on $X$ develops a $\mathbb{Z}_2$ gauge symmetry and therefore that $\widehat{X}$ exhibits
\begin{align}
    H^3(\widehat{X},\mathbb{Z})\simeq \mathbb{Z}^3\oplus\mathbb{Z}_2\,.
\end{align}
The fractional B-field on $X$ also stabilizes the $64$ nodes and reduces the number of complex structure deformations to $h^{2,1}(Y)=115-64=51$.
We therefore expect that $(\widehat{X},\alpha)$ is a complex structure deformation of $X'$ with discrete torsion.
In this example we thus find a concrete realization of the idea that the fractional B-field can be seen as a generalization of discrete torsion.
\section{Geometric verifications}\label{sec:geometric}

In this section, 
we check the proposal of Section~\ref{subsec:gv} for the description of the torsion refined Gopakumar-Vafa invariants of Calabi-Yau threefolds with terminal nodal singularities in terms of non-K\"ahler small resolutions.  

\subsection{The octic double solid}\label{subsec:veroctic}
We will make frequent use of the conifold transition from $\widehat{X}$ to $X_{\text{def.}}$.  The induced map $H_2(\widehat{X},\mathbb{Z})\to H_2(X_{\text{def.}},\mathbb{Z})$ has kernel $\mathbb{Z}_2$.  So the Gopakumar-Vafa invariants of $X_{\text{def.}}$ are obtained from the invariants of $\widehat{X}$ by summing over two $\mathbb{Z}_2$ sectors.  In other words, we can view the $\mathbb{Z}_2$ as a discrete K\"ahler class and this result follows from~\cite[Cor.\ B.1]{Li:1998hba}.

We use the methods of~\cite{Katz:1999xq} to compute the Gopakumar-Vafa invariants.  These invariants can be defined mathematically with certain hypotheses using perverse sheaves~\cite{MT}.  The methods of~\cite{Katz:1999xq} can be derived from the definitions in~\cite{MT} in many situations using the Decomposition Theorem~\cite{Zhao}.

We start by analyzing curves $C\subset X_{\text{def.}}$ of degree $\beta=1$.  Letting $\pi:X_{\text{def.}}\to \mathbb{P}^3$ be the double cover and noting that $H^2({X}_{\text def.},\mathbb{Z})$ is generated by the pullback of the hyperplane class of $\mathbb{P}^3$, we see that $\pi(C)$ also has degree 1, hence is a line $L$.  Since $C\subset \pi^{-1}(L)$ and $\pi^{-1}(L)$ has degree 2, we see that $\pi^{-1}(L)$ must be a union of two components, $C$ and another curve $C'$ of degree~1.  However, $\pi^{-1}(L)$ is reducible if and only if $L$ is tangent to $B$ at each of 4 points $q_i$, accounting for all of the 8 intersection points of $L$ with $B$ including multiplicity.

By classical enumerative geometry, for a generic degree 8 hypersurface $B\subset \mathbb{P}^3$ there are exactly 14752 lines $L$  which are tangent to $B$ at each of 4 points $q_i$~\cite{schubert:1879}.  It is convenient to change notation and write $\pi^{-1}(L)=L_1\cup L_2$, with $L_1$ and $L_2$ curves of degree 1.   Counting both $L_1$ and $L_2$ for each of the 14752 lines $L$ to get the count of degree 1 curves in $ X_{\text{def.}}$, we obtain $n^1_0=2\cdot 14752=29504$.

We now return to $\widehat{X}$.  In Section~\ref{sec:Btorsion}, we explained how a topologically nontrivial but flat B-field defines a projection map
\begin{equation}
H_2(\widehat{X},\mathbb{Z})\to \mathbb{Z}_2\,.
\end{equation}
This projection will be used to divide curve classes $[C]\in H_2(\widehat{X},\mathbb{Z})$ into two sectors.  In M-theory, the sectors are distinguished by the $\mathbb{Z}_2$ charge of M2-branes wrapping $C$.  The only thing we need to observe now is that the nontrivial torsion class in $H_2(\widehat{X},\mathbb{Z})$ maps to 1.

The total count of curves does not change, but we have to take the $\mathbb{Z}_2$ into consideration.  From Tables~\ref{tab:nc0} and~\ref{tab:nc1}, we see that $L_1$ and $L_2$ are in different charge sectors and so must differ by a torsion class.  So we learn that for each line $L$, the difference $[L_1]-[L_2]\in H_2(\widehat{X},\mathbb{Z})$ is the nontrivial torsion class.  It would be of interest to verify this directly by geometry.

It follows that $[L_1]-[L_2]=[L_2]-[L_1]$, so the roles of $L_1$ and $L_2$ are interchangeable.

The $\beta=1$ columns of Tables~\ref{tab:nc0} and \ref{tab:nc1} are verified immediately: for each of the 14752 4-tangent lines, exactly one of $L_1,L_2$ has charge 0 and exactly one has charge 1.

This precisely matches the indeterminacy in labeling the charge sectors which was noted in Section~\ref{sec:Btorsion}. We have $H_2(\widehat{X},\mathbb{Z})\simeq \mathbb{Z}\oplus \mathbb{Z}_2$, with $ H_2(\widehat{X},\mathbb{Z})_{\mathrm{tors}}$ identified with the $\mathbb{Z}_2$ summand.   The degree 1 curve classes $[L_1]$ and $[L_2]$  are identified with
$(1,0)$ and $(1,1)$ in some order.  Since we have two direct sum decompositions
\begin{equation}
H_2(\widehat{X},\mathbb{Z}) \simeq \mathbb{Z}\cdot [L_1] \oplus \mathbb{Z}_2\simeq \mathbb{Z}\cdot [L_2] \oplus \mathbb{Z}_2\,,
\end{equation}
neither curve $L_1$ or $L_2$ is distinguished.  For one choice of reference B-field $\tilde{B}$, the charge assignments of $L_1$ and $L_2$ are 0 and 1 respectively, while for the other choice of $\tilde{B}$, the charge assignments are respectively 1 and 0.

We can now verify many other Gopakumar-Vafa invariants geometrically.

For a curve $C$ of degree $\beta=2$, we have that $\pi(C)$ has degree 2 including multiplicity.  So either $C$ is a 2-1 cover of a line $L$ via $\pi$, or $C$ maps isomorphically to a degree 2 curve, in which case $C$ has genus~0.  Now note that a general line $L\subset \mathbb{P}^3$ meets $B$ in 8 distinct points, so pulls back to a double cover of $L$ branched at 8 points, which has genus 3.  Thus for degree $\beta=2$, the maximum genus is $g_{\rm max}=3$, which we refer to as the Castelnuovo bound.  Thus, the moduli space $\mathcal{M}$ of $\beta=2$ curves which contribute to $n_3^2$ is just the Grassmannian $G(2,4)$  of lines in $\mathbb{P}^3$.   By~\cite{Gopakumar:1998jq}, we have for any Calabi-Yau threefold and any smooth moduli space $\mathcal{M}$ of curves class $\beta$ and genus $g_{\rm max}$
\begin{equation}\label{eq:ng}
n_{g_{\rm max}}^\beta = \left(-1
\right)^{\dim\left(\mathcal{M}\right)}e\left(\mathcal{M}\right)\,.
\end{equation}
Applying this to $X_{\text{def.}}$, we get $n^2_3=6$.

Turning to $\widehat{X}$, it remains to indentify how the GV invariant distributes into the charge sectors as $6=n^2_3=n^{2,0}_3+n^{2,1}_3$.

Since the Grassmannian parametrizes a connected family of lines $L$, all of the curve classes $\pi^{-1}(L)$ are homologous, so we only have to identify the charge of $\pi^{-1}(L)$ for any $L$.  A convenient choice is one of the four-tangent lines.  In this case $\pi^{-1}(L)=L_1\cup L_2$.  Since one of the two curves $L_i$ has charge 1 and the other has charge 0, we conclude that $L_1\cup L_2$ has charge 1, and the entire GV invariant is concentrated in the charge 1 sector.   This checks against the numbers  $n_3^{2,0}=0$ in Table~\ref{tab:nc0} and $n_3^{2,1}=6$ in Table~\ref{tab:nc1}.

The above method generalizes to compute $n^{g_{\rm max}}_{2d}$ for any even degree.  In terms of $\mathbb{P}_{11114}$, the double cover of a line is a complete intersection of two weighted hypersurfaces of degree 1.  A similar argument to the above shows that the Castelnuovo bound is realized by complete intersections of weighted hypersurfaces of degrees 1 and $d$.  By the adjunction formula, these curves $C$ have canonical bundle $K_C=\mathcal{O}_C(d+1)$.  Since the degree of this bundle is $2d(d+1)$, the genus of our curves satisfy $2g-2=2d(d+1)$, so
\begin{equation}\label{eq:gmax2d}
g=g_{\rm max}=d^2+d+1\,.
\end{equation}
The moduli space of these Castelnuovo curves is fibered over the $\mathbb{P}^3$ of hyperplanes in $\mathbb{P}^3$.  The fiber can be identified with the space of weighted hypersurfaces of degree $d$ in $\mathbb{P}_{1114}$, up to the ambiguity of adding a multiple of the octic defining the Calabi-Yau.  The generating function of the dimensions of these vector spaces is then
\begin{equation}
    \frac{(1-t^8)}{(1-t)^3(1-t^4)}=\frac{(1+t^4)}{(1-t)^3}=\sum_d(d^2-d+4)t^d\,,
\end{equation}
and so the moduli space is a $\mathbb{P}^{d^2-d+3}$-bundle over $\mathbb{P}^3$, which is always even-dimensional.
So we get for $X_{\text{def.}}$
\begin{equation}\label{eq:hyperelliptic}
n_{g_{\rm max}}^{2d}=4\left(d^2-d+4\right)\,,\quad d\ge 2\,.
\end{equation}

For $\widehat{X}$ we have to identify the charge sector.  Again, these curves form a continuous family, so have the same homology classes, and we only have to find the charge sector for one curve, which we can choose to be the double cover of a union of $d$ lines, or equivalently, the union of $d$ double covers of lines.  Since the double covers of lines have charge 1, we conclude that the curves under consideration have charge $d$ mod 2.  So (\ref{eq:hyperelliptic}) gets refined to
\begin{equation}\label{eq:hyperelliptictorsion}
n_{g_{\rm max}}^{2d,d}=4\left(d^2-d+4\right)\,,\ n_{g_{\rm max}}^{2d,d+1}=0\,,
\end{equation}
where the $d$ in the second part of the superscript in $n_{g_{\rm max}}^{2d,d}$ is understood as $d$ mod 2, and similarly for the $d+1$ in $n_{g_{\rm max}}^{2d,d+1}$.

Putting $d=2,3,4,5$ in (\ref{eq:hyperelliptictorsion}), we verify the numbers $n_7^{4,0}=24$, $n_{13}^{6,0}=0, n_{21}^{8,0}=64,n_{31}^{10,0}=0$ in Table~\ref{tab:nc0} and $n_7^{4,1}=0, n_{13}^{6,1}=40, n_{21}^{8,1}=0,n_{31}^{10,1}=96$ in Table~\ref{tab:nc1}.

We next turn to $n^{g_{\rm max}-1}_{2d}$, as usual starting with $X_{\text{def.}}$ before turning our attention to $\widehat{X}$ and identifying the charge sectors.  We use the method of \cite{Katz:1999xq}, which can be made rigorous using Donaldson-Thomas-type invariants \cite{MT,Zhao} (see also \cite{PT}).   We recall the general formula from \cite{Katz:1999xq}.  Denote by $\mathcal{C}\to\mathcal{M}$ the universal curve of our family $\mathcal{M}$ of curves of genus $g=g_{\rm max}$.  Then we have
\begin{equation}\label{eq:kkv1}
n_{g-1}^\beta=\left(-1\right)^{\dim \mathcal{M}+1}\left(e(\mathcal{C})+\left(2g-2\right)e(\mathcal{M})
\right)\,.
\end{equation}
There are correction terms to (\ref{eq:kkv1}) when the family of curves contains sufficiently many reducible curves.  We simplify the discussion by simply asserting that these correction terms do not occur in the present case, and refer the reader to  (\ref{eq:kkv1}) for more details, or to \cite{Zhao} for a mathematical proof of the validity of the correction terms.

We start with $d=1$.  To apply (\ref{eq:kkv1}), we only have to identify $\mathcal{C}$.  A point of $\mathcal{C}$ consists of the data of a curve $C=\pi^{-1}(L)$ and a point $p\in C$.  We define a mapping $\mathcal{C}\to X_{\text{def.}}$ by sending the data $p\in\pi^{-1}(L)$ to $p$, i.e.\ forgetting $L$.  The fiber of $\mathcal{C}\to X_{\text{def.}}$ is the set of all $L$ such that $p\in\pi^{-1}(L)$, or equivalently, the set of all $L$ containing $\pi(p)$.  This set is isomorphic to $\mathbb{P}^2$, so we have $e(\mathcal{C})=e(\mathbb{P}^2) e(X_{\text{def.}})=-888$.  Applying (\ref{eq:kkv1}) we get $n_2^2=864$.

Turning to $\widehat{X}$, we claim that 864 divides into the two charge sectors as $n_2^{2,0}=504$ and $n_2^{2,1}=360$ in agreement with Tables~\ref{tab:nc0} and \ref{tab:nc1}.  The calculation of $n_2^{2,1}$ requires no new ideas, recalling that we have already shown that the curves $\pi^{-1}(\ell)$ are all in charge sector 1.  We simply recompute $e(\mathcal{C})$ for $\widehat{X}$ as $e(\mathcal{C})=e(\mathbb{P}^2) e(\widehat{X})=-384$, and then (\ref{eq:kkv1}) gives $n_2^{2,1}=360$.

For $n_2^{2,0}$, the lines $L$ containing a conifold point  $p$ play a special role.  To see why, note that for $\pi:\widehat{X}\to \mathbb{P}^3$, the curves $\pi^{-1}(L)$ are reducible as they contain $L_p$, so that $\pi^{-1}(L)=C\cup L_p$ for some curve $C$.  Since $\pi^{-1}(L)$ has degree 2 and charge 1 while $L_p$ has degree 0 and charge 1, we conclude that $C$ has degree 2 and charge 0.  Furthermore, $C$ has genus 2, since it is a double cover of $L$ branched at the 6 points where $L$ meets $B$ outside of $p$.  Thus $C$ contributes to $n_2^{2,0}$, and it is straightforward to check that all curves $C$ of genus 2, degree 2, and charge 0 are of this type.

The moduli space $\mathcal{M}$ of lines containing $p$ is isomorphic to $\mathbb{P}^2$, so each line contributes 3.  However, in keeping with our proposal, we have to consider the flop of $C_p$ as well.  In general, when a curve is flopped, the homology class of the curve changes sign.  However, since the class of each exceptional curve $C_p$ is 2-torsion, we have $-[C_p]=[C_p]$ and flopped curves have the same homology class as the original curve.  So we get a second contribution of 3 by repeating this analysis on a flop, for a total contribution of 6.  Considering the contribution of 6 for each of the 84 conifolds, we finally get $n_2^{2,0}=6\cdot 84=504$.

The same method works for $n_{g_{\rm max}-1}^{2d,0}$ and $n_{g_{\rm max}-1}^{2d,1}$ for any $d>1$.  A point of the universal curve $\mathcal{C}$ corresponds to a complete intersection curve $C\subset \widehat{X}$ of degrees $1,d$ and a point $p\in C$.   Forgetting $C$, we get a map $\mathcal{C}\to \widehat{X}$ whose fiber over $p$ is the space of complete intersection curves containing $p$.  This space in turn fibers over the $\mathbb{P}^2$ of hyperplanes in $\mathbb{P}^3$ containing $\pi(p)$.  The fiber is the codimension~1 subspace of the $\mathbb{P}^{d^2-d+3}$ of those degree $d$ hypersurfaces in $\mathbb{P}_{1114}$ which contain $p$, a $\mathbb{P}^{d^2-d+2}$-bundle.   This gives $e(\mathcal{C})=3e(\widehat{X})(d^2-d+3)=-384(d^2-d+3)$.   Then (\ref{eq:kkv1}) gives, using (\ref{eq:hyperelliptictorsion}) and (\ref{eq:gmax2d}) 
\begin{equation}
n_{g_{\rm max}-1}^{2d,d}=-8d^4+360d^2-416d+1152\,,
\end{equation}
agreeing with Table~\ref{tab:nc0} for $d=2,4$ and Table \ref{tab:nc1} for $d=3,5$. 

In passing, we can do the same calculation on $X_{\text{def.}}$, simply replacing $e(\widehat{X})$ by $e(X_{\text{def.}})$.  The result is 
\begin{equation}
n_{g_{\rm max}-1}^{2d}=n_{g_{\rm max}-1}^{2d,d}+n_{g_{\rm max}-1}^{2d,d+1}=-8 (d^4- 108 d^2 + 115 d-333 )\,,
\end{equation}
agreeing with the sum of the relevant invariants from Tables~\ref{tab:nc0} and \ref{tab:nc1}.

For the other charge sector, repeating the above argument, we have to identify the space of $(1,d)$ complete intersection curves containing one of the 84 singular points.  As just seen above, this space is a $\mathbb{P}^{d^2-d+2}$-bundle over the $\mathbb{P}^2$ of 2-planes containing $\pi(p)$, so has Euler characteristic $3(d^2-d+3)$.  Taking into account the factor of 2 explained above, we get
\begin{equation}
n_{g_{\rm max}-1}^{2d,d+1}=504(d^2-d+3)\,,
\end{equation}
agreeing with  Table \ref{tab:nc1} for $d=2,4$ and Table \ref{tab:nc0} for $d=3,5$.

We next turn to curves of odd degree $\beta=2d+1$ with $d\ge1$, starting with $d=1$, so $\beta=3$.  It is straightforward to see that curves saturating the Castelnuovo bound arise from gluing a $\beta=1$ curve to a $\beta=2$ curve of maximal genus 3 at a point, arriving at a $\beta=3$ curve of genus 3.  Since the $\beta=2$ curves of maximal genus have charge 1, then if we use one of the 14752 $\beta=1$ curves of charge 0, we get a $\beta=3$ curve of charge 1.  Similarly,  if we use one of the 14752 $\beta=1$ curves of charge 1, we get a $\beta=3$ curve of charge 0, and in particular $n_3^{3,0}=n_3^{3,1}$.

To compute either of these numbers, we fix a $\beta=1$ curve $C$, which is isomorphic to $\mathbb{P}^1$ and projects to a line $\ell\subset \mathbb{P}^3$.  We can attach a $\beta=2$ curve $\pi^{-1}(L)$ whenever the line $L$ intersects $\ell$.  The set of all such $L$ is the Schubert cycle $\sigma_1(\ell)\subset G(2,4)$, which is therefore the component of the moduli space of $\beta=3$ curves under consideration which contain the fixed $\beta=1$ curve.  It is well-known that $\sigma_1(\ell)$ is a threefold with a conifold singularity at the point $L=\ell$ and is otherwise smooth.  Since $\sigma_1(\ell)$ is singular, we cannot simply take its signed Euler characteristic to obtained its contribution to the GV invariant.  However, the mathematical definition of the GV invariant in \cite{MT} reduces in this case to taking the signed Euler characteristic of the intersection cohomology $\operatorname{IH}^*(\sigma_1(\ell))$.  It is also well-known that the intersection cohomology of a threefold with conifolds is isomorphic the ordinary cohomology of any small resolution.  Finally, it is also well-known that the set of pairs $\{(p,L)\vert p\in L\cap\ell\}$ is a small resolution $\widetilde{\sigma_1(\ell)}$ of $\sigma_1(\ell)$.  

It only remains to compute the Euler characteristic of $\widetilde{\sigma_1(\ell)}$.  There is a natural projection $\widetilde{\sigma_1(\ell)}\to C$ taking $(p,L)$ to $p$, with fiber the set of all $L$ which contain $p$.  Such $L$ are parametrized by $\mathbb{P}^2$.  The moduli space is therefore a 3-dimensional space of Euler characteristic 6, giving a contribution of $-6$ to $n_3^3$, for each of the 14752 curves.  We conclude that
\begin{equation}
n_3^{3,0}=n_3^{3,1}=-88512\,,
\end{equation}
in agreement with Tables \ref{tab:nc0} and \ref{tab:nc1}.

Turning to $n^{g_{\rm max}}_{2d+1}$ for $d>1$, the Castelnuovo bound is again saturated by gluing a $\beta=1$ curve lying over a line $L$ to a $\beta=2d$ curve $\pi^{-1}(C)$, where $C$ is a plane curve of degree $d$.   It is possible for these two curves to intersect in as many as $d$ points, lying above the $d$ points $L\cap C$.  The maximum genus is realized by curves with the maximum number $d$ of intersection points, which increases the maximum genus (\ref{eq:hyperelliptic}) of $\beta=2d$ curves by $d-1$.  This gives
\begin{equation}
g_{\rm max}=d^2+2d\,.
\end{equation}
Since we have an equal number of $\beta=1$ curves of each charge, we get $n^{2d+1,0}_{d^2+2d}=n^{2d+1,1}_{d^2+2d}$.

Next, $C$ meets $L$ in $d$  points if and only if the 2-plane $P$ containing $C$ also contains $L$.  The moduli space of such $C$ fibers over the space of all 2-planes $P$ containing $L$, which is a $\mathbb{P}^1$.  The fiber is the $\mathbb{P}^{d(d+3)/2}$ of plane curves in $P$.  This moduli space has Euler characteristic $d^2+3d+2$ and dimension $(d^2+3d+2)/2$.  Including this contribution for each of the 14752 $\beta=1$ curves in either charge sector, we get
\begin{equation}
n^{2d+1,0}_{d^2+2d}=n^{2d+1,1}_{d^2+2d}=\left(-1\right)^{(d^2+3d+2)/2}14752\left(d^2+3d+2\right)\,.
\end{equation}
 In particular, for $d=2$ we get $n^{5,0}_8=n^{5,1}_8=177024$, in agreement with Tables \ref{tab:nc0} and \ref{tab:nc1}.

For completeness, we point out that in the cases where we did not consider a separate contribution from flops, this can be rigorously justified since the relevant sheaves on $\widehat{X}$ and its flops can be checked to correspond under the standard derived equivalence $\psi:D^b(\widehat{X})\to D^b(\widehat{X}')$ described at the end of Section~\ref{subsec:derived}. We content ourselves with working out two illustrative examples.

The first example is the 4-tangent lines, $n_0^{1,0}=n_0^{1,1}=14752$.  Recall that $\widehat{X}$ and $\widehat{X}'$ are canonically identified after removing the exceptional curves, and the derived equivalence $\psi$ leaves sheaves whose support is disjoint from the exceptional curves unchanged with this identification.  In particular, $\psi$ takes the structure sheaf of a 4-tangent line in $\widehat{X}$ to the structure sheaf of a 4-tangent line in $\widehat{X}'$.   So for the degree 1 curve classes, it suffices to compute in $\widehat{X}$ as we have done above.  More generally, families of curves which do not intersection any of the exceptional curves can be analyzed on one $\widehat{X}$ without considering any flops.

A more interesting example is degree $n_3^{2,1}$, which we have seen corresponds to the inverse image of a line.  We set up some notation, considering the diagram
\begin{equation}
\begin{array}{ccc}
\widehat{X}\times\widehat{X}' & \stackrel{\pi_2} \to & \widehat{X}'\\
\phantom{\pi_1}\downarrow\pi_1&&\phantom{\sigma_2}\downarrow\sigma_2\\
\widehat{X}& \stackrel{\sigma_1}{\to} & \mathbb{P}^3
\end{array}\,,
\end{equation}
where $\widehat{X}'$ is obtained from $\widehat{X}$ by flopping any set of exceptional curves $C_p$.
Let $\Gamma=\widehat{X}\times_X\widehat{X}'\subset \widehat{X}\times\widehat{X}'$.  Then $\psi:D^b(\widehat{X})\to D^b(\widehat{X}')$ is given by the Fourier-Mukai transform
\begin{align}
\psi(F^\bullet)=R{\pi_2}_*\left(L\pi_1^*F^\bullet\stackrel{\mathrm{L}}{\otimes}\mathcal{O}_\Gamma\,.
\right)
\end{align}
Now consider $\mathcal{O}_{\sigma_1^{-1}}(\ell)$, the inverse image of a line $\ell\subset\mathbb{P}^3$.  As $\ell$ is the complete intersection of two hyperplanes, we can resolve $\mathcal{O}_{\ell}$ by a Koszul complex
\begin{align}
0\to \mathcal{O}_{\mathbb{P}^3}(-2)\to \mathcal{O}_{\mathbb{P}^3}(-1)^2\to \mathcal{O}_{\mathbb{P}^3}\to \mathcal{O}_{\ell}\to 0\,,
\end{align}
which allows us to represent its derived pullback $L\sigma_1^*\mathcal{O}_{\ell}$ by the complex
\begin{align}\label{eq:derpull}
0\to \sigma_1^*\mathcal{O}_{\mathbb{P}^3}(-2)\to \sigma_1^*\mathcal{O}_{\mathbb{P}^3}(-1)^2\to \sigma_1^*\mathcal{O}_{\mathbb{P}^3}\to 0\,.
\end{align}
However, $\sigma_1^{-1}(\ell)$ is the complete intersection of the pullback of the two hyperplanes to $\widehat{X}$, and we recognize (\ref{eq:derpull}) as the corresponding Koszul resolution of $\mathcal{O}_{\sigma_1^{-1}}(\ell)$.  It follows that
\begin{align}
L\sigma_1^*\mathcal{O}_{\ell}=\mathcal{O}_{\sigma_1^{-1}}(\ell)\,.
\end{align}
We can then compute the Fourier-Mukai transform as
\begin{align}
\psi(\mathcal{O}_{\sigma_1^{-1}(\ell)})=R{\pi_2}_*\left(L\pi_1^*L\sigma_1^*\mathcal{O}_\ell
\stackrel{\mathrm{L}}{\otimes}\mathcal{O}_\Gamma
\right)
\,,
\end{align}
which we have to show coincides with $\mathcal{O}_{\sigma_2^{-1}(\ell)}=L\sigma_2^*\mathcal{O}_\ell$.

Pulling back (\ref{eq:derpull}) by $\pi_1$ and restricting to $\Gamma$, we can represent $L\pi_1^*L\sigma_1^*\mathcal{O}_\ell
\stackrel{\mathrm{L}}{\otimes}\mathcal{O}_\Gamma$ by the complex
\begin{align}
0\to \mathcal{O}_{\Gamma}(-2)\to \mathcal{O}_{\Gamma}(-1)^2\to \mathcal{O}_{\Gamma}\to 0\,,
\end{align}
where $\mathcal{O}(-1)$ and $\mathcal{O}(-2)$ are understood as having been pulled back from $\mathbb{P}^3$ to $\Gamma$.

Note that $\Gamma=\widehat{X}\times_X\widehat{X}'$ is just the ``large resolution,'' the blowup of the conifolds with exceptional divisors $\mathbb{P}^1\times\mathbb{P}^1$, and the projections $\widehat{X}\times_X\widehat{X}\to \widehat{X},\widehat{X}'$ are the partial blowdowns to the respective small resolutions.  The fibers of $\pi_2$ are $\mathbb{P}^1$s over points of the exceptional curves of $\widehat{X}'$, and are points elsewhere.  Furthermore, $\mathcal{O}_\Gamma(k)$ restricts to $\mathcal{O}_{\mathbb{P}^1}$ on each of the 1-dimensional fibers, for $k=0,-1,-2$, and therefore has no higher cohomology.  It follows that
\begin{align}
R{\pi_2}_*\mathcal{O}_\Gamma(k)\simeq {\pi_2}_*\mathcal{O}_\Gamma(k) \simeq \mathcal{O}_{\widehat{X}'}(k)\,.
\end{align}

It then follows without difficulty that $R{\pi_2}_*\left(L\pi_1^*L\sigma_1^*\mathcal{O}_\ell
\stackrel{\mathrm{L}}{\otimes}\mathcal{O}_\Gamma
\right)$ can be represented by the complex
\begin{align}
0\to \mathcal{O}_{\widehat{X}'}(-2) \to \mathcal{O}_{\widehat{X}'}(-1)^2 \to \mathcal{O}_{\widehat{X}'} \to 0\,,
\end{align}
which represents $L_{\sigma_2}^*\mathcal{O}_\ell=\mathcal{O}_{\sigma_2^{-1}(\ell)}$ by the same reasoning that led to (\ref{eq:derpull}).  It follows that
\begin{align}
\psi(\mathcal{O}_{\sigma_1^{-1}(\ell)})=\mathcal{O}_{\sigma_2^{-1}(\ell)}\,,
\end{align} 
and the two families $\mathcal{O}_{\sigma_i^{-1}(\ell)}$ for $i=1,2$ are identified under Fourier-Mukai transform as claimed.
We conclude that we only need to study one small resolution $\hat{X}$ for the $\beta=1$ GV invariants, independent of whether or not $\ell$ contains a conifold point.

Our final geometric check is an explanation of the identity 
\begin{align}
	n_g^{\beta,l}=n_g^{\beta,\beta+l}\,.
	\label{eqn:involution}
\end{align}
which is apparent from the tables.

We let $\sigma:X\to X$ be the involution of the double cover $\pi$ and let $\rho:\widehat{X}\to X$ be a small resolution.  We form the fiber product
\begin{equation}
    \begin{array}{ccc}
    \widehat{X}' &\stackrel{\sigma'}{\to}& \widehat{X}\\
    \rho'\downarrow\phantom{\rho'}&&\phantom{\rho}\downarrow\rho\\
    X&\stackrel{\sigma}{\to}&X
    \end{array}
    \label{eq:pullback}
\end{equation}
We claim that $\sigma\circ \rho':\widehat{X}'\to X$ is the small resolution of $X$ obtained by flopping all of the exceptional curves of $\rho$.  We only need to check this locally near any conifold.

Writing the local equation of the octic surface near a node as $x^2-wz$, then the equation of $X$ near the corresponding conifold is
$y^2=x^2-wz$.  Rewriting this as
\begin{equation}
    wz=(x+y)(x-y)\,,
\end{equation}
we see that one small resolution $\widehat{X}$ is obtained by introducing a $\mathbb{P}^1$ with affine coordinate $t=(x+y)/w$ and the flopped small resolution $\widehat{X}'$ is obtained by introducing a $\mathbb{P}^1$ with affine coordinate $t'=(x-y)/w$.

The involution $\sigma$ takes $(w,x,y,z)$ to $(w,x,-y,z)$, and therefore takes $t$ to $t'$.  This proves the claim.  

We conclude that the distinct small resolutions $\widehat{X}'$ and $\widehat{X}$ are isomorphic via $\sigma'$.  Considering all small resolutions simultaneously, the maps $\sigma'$ induce an involution of the disjoint union
\begin{equation}
    \sigma:\bigsqcup_{\widehat{X}\in\mathcal{S}}{\widehat{X}} \to \bigsqcup_{\widehat{X}\in\mathcal{S}}{\widehat{X}}\,,
\end{equation}
hence we get a corresponding involution identity of GV invariants, as our nc GV invariants utilize \emph{all} small resolutions.  To check that the identity takes the form (\ref{eqn:involution}), we only have to check on generators, which we can take to be a $\beta=1$ curve and an exceptional curve $C_p$.
When we apply $\sigma$ to a $\beta=1$ curve, we have seen that we get a $\beta=1$ curve of the opposite $\mathbb{Z}_2$ charge.  Applying $\sigma$ to an exceptional curve $C_p$, we get a flopped curve $C_p'$.  Both $C_p$ and $C_p'$ have $\beta=0$ and charge 1.  It follows immediately that the involution induces the claimed identity (\ref{eqn:involution}).

This argument shows more generally that we have similar identities between the torsion refined GV invariants of any double cover with terminal nodal singularities.

\subsection{An example by Calabrese-Thomas}
We next turn to the Calabrese-Thomas example discussed in Section~\ref{subsec:CalabreseThomas}.
Following~\cite{Schimannek:2021pau}, the GV-invariants have been computed using mirror symmetry and the modular bootstrap and are displayed in Tables~\ref{tab:ctg0c0}--\ref{tab:ctg1c1}. In this section, we verify some of these numbers directly from the geometry of non-K\"ahler small resolutions.  

Recall that this example is elliptically fibered over $\mathbb{P}^2$, with 144 $I_2$ fibers.  The  singular Calabi-Yau has 66 conifolds \cite{Calabrese2016DerivedEC}.  We focus on multiples of the fiber class.   Since the fiber projects to a point in $\mathbb{P}^2$, we have $d_2=0$.  Because of the double cover, the fiber class has $d_1=2$, so each component of an $I_2$ fiber has $d_1=1$.

Let's look at the GV invariants $n_g^{(d_1,d_2)}$ of $X_{\text{def.}}$ first.   Later, we will refine these invariants as
\begin{equation}\label{eq:refinect}
    n_g^{(d_1,d_2)}=n_g^{(d_1,d_2),0}+n_g^{(d_1,d_2),1}\,.
\end{equation}
From the 144 $I_2$ fibers, we get $n^{(1,0)}_0=288$.  From $\chi(X_{\text{def.}})=-252$, we get $n^{(2,0)}_0=252$.  Since adding any multiple of the fiber class does not change the GV invariants, we get
$n^{(2d-1,0)}_0=288$ and $n^{(2d,0)}_0=252$ for any $d\ge1$.

Since the base of the elliptic fibration is $\mathbb{P}^2$, which has Euler characteristic 3, we get $n^{(2d-1,0)}_1=0$ and $n^{(2d,0)}_1=3$ for any $d\ge1$.

From Tables~\ref{tab:ctg0c0} and \ref{tab:ctg0c1}, we infer that the two components of each $I_2$ fiber have opposite $\mathbb{Z}_2$ charges.  We accept this and verify the other $d_2=0$ invariants without any additional assumptions.

Since the fiber class can be represented by the sum of the two curves classes of an $I_2$ fiber, we see that the fiber class is given by $((d_1,d_2),j)=((1,0),0)+((1,0),1)=((2,0),1)$.  This tells us $n^{(2,0)}_1=3$ refines to $n^{(2,0),0}_1=0$ and $n^{(2,0),1}_1=3$.  Adding on multiples of the fiber class, we get $n^{(2d,0),d}_1=3$ for any $d\ge 1$, and $n^{(d_1,0),j}=0$ unless $d_1=2d$ is even and $j\equiv d$ (mod 2).

From (\ref{eq:conifoldtrans}), we easily get $\chi(\widehat{X})=\chi(X_{\text{def.}})+2\cdot 66+0=-120$.  Thus $n^{(2,0),1}_0=120$.  From (\ref{eq:refinect}) we then get $n^{(2,0),0}_0=132$.  The remaining genus 0, $d_2=0$ GV invariants arise from the above by adding multiples of the fiber class.

\subsection{A ${\mathbb Z}_5$ example}

We next turn to the example of an elliptically fibered Calabi-Yau over $\mathbb{P}^2$.  It is the Jacobian fibration of either of two genus 1 fibrations with a 5-section but no section, both of which have been constructed in~\cite{Knapp:2021vkm}, and so has discrete gauge group $\mathbb{Z}_5$.  
In this case, the singular geometry is not a double cover and the non-commutative resolutions are not of Clifford type but fall under what we refer to as the General type nc-resolutions.

The refined invariants were computed in~\cite{Schimannek:2021pau}, where the singular Jacobian fibration was denoted by $X_0^{(5)}$ while the genus 1 fibrations were denoted by $X_1^{(5)}$ and $X_2^{(5)}$.  The invariants are listed in \cite[Tables~34--38]{Schimannek:2021pau}.
We proceed to check some of the invariants against our proposal.

We first collect some facts about the geometry of the $X_i^{(5)}$ from the above references.  For ease of notation, we simply denote $X_0^{(5)}$ by $X$ in this section.  The Euler characteristics of $X_1^{(5)}$ and $X_2^{(5)}$ are both equal to $-90$.

There are 225 conifold singularities of $X$. Each conifold is situated on the fiber corresponding to $I_2$ fibers of the original genus one fibered Calabi-Yau manifolds.  The $I_2$ fibers are partitioned into two classes of cardinalities $n_{\pm1}=100$ and $n_{\pm2}=125$.  For present purposes, we translate the distinction between the geometry of the two different types of $I_2$ fibers into the geometry of the exceptional curves in small and non-K\"ahler resolutions $\widehat{X}$.  

Each exceptional curve is torsion, and we have $H_2(\widehat{X},\mathbb{Z})_{\mathrm{tors}}=\mathbb{Z}_5$.  If an exceptional curve $C_i$ has homology class $[C_i]=j\in \mathbb{Z}_5$, then the flopped curve has homology class $-j\in \mathbb{Z}_5$.  In the previous cases with $\mathbb{Z}_2$-torsion, we had $j=-j\in \mathbb{Z}_2$ for any torsion class $j$, but for $\mathbb{Z}_5$, $j$ and $-j$ are distinct for nontrivial $j$.  Since we are interested in the torsion refined invariants, we consider both small resolutions for each conifold.  Among the 225 conifolds, $n_{\pm1}=100$ of them have exceptional curves with torsion class $\pm1$ and $n_{\pm2}=125$ of them have exceptional curves with torsion class $\pm2$.\footnote{A convenient choice of generator of $\mathbb{Z}_5$ has been made here.}

We have $H_2(\widehat{X},\mathbb{Z})\simeq \mathbb{Z}^2\oplus \mathbb{Z}_5$ and we identify curve class with labels $((d_1,d_2),j)$ with $j\in \mathbb{Z}_5$.  Here $d_1$ is the base degree and $d_2$ is the fiber degree.  

We begin with the fiber classes, which have $(d_1,d_2)=(1,0)$ and $j=0\in\mathbb{Z}_5$.  Since $\widehat{X}$ has Euler characteristic $-90$, we conclude that $n_0^{(1,0),0}=90$,\footnote{In \cite{Schimannek:2021pau}, the GV invariants were written as $n^0_{(d_1,d_2),j}$.} in agreement with \cite[Table~34]{Schimannek:2021pau}.  It follows that $n_0^{(d_1,0),0}=90$ for all $d_1>0$, also in agreement with \cite[Table~34]{Schimannek:2021pau}.  

Since we will need to adapt the usual argument presently, we review it briefly.  Given a point $p$ in a fiber $f$, we have the line bundle $\mathcal{O}_f(p)$ on $f$, which may be viewed as a 1-dimensional sheaf on $\widehat{X}$ of Euler characteristic 1.\footnote{The construction is readily adapted if $p$ is a singular point of a singular fiber.}  It readily follows that the moduli space of these sheaves is isomorphic to $\widehat{X}$, and the genus~0 Gopakumar-Vafa invariant is the Donaldson-Thomas invariant of this moduli space, which is the negative of the Euler characteric of $\widehat{X}$.  For general $d_1$, it is shown inductively following a construction of of Atiyah that all of the moduli spaces are isomorphic to $\widehat{X}$, hence have the same genus 0 GV invariant. Letting $p\in\widehat{X}$, we let $f_p$ be the fiber containing $p$ and we can define the rank 2 bundle\footnote{For $p$ a singular point of a singular fiber we get a rank 2 sheaf rather than a vector bundle.} $E_2(p)$ by the extension
\begin{equation}\label{eq:atiyah}
    0\to \mathcal{O}_{f_p}\to E_2(p)\to \mathcal{O}_{f_p}(p)\to 0.
\end{equation}
Again, the moduli space of sheaves with fiber degree 2 and Euler characteristic 1 is isomorphic to $\widehat{X}$, with $p\in\widehat{X}$ parametrizing the sheaf $E_2(p)$.  So we again get $-\chi_{\rm top}(\widehat{X})=90$ for the GV invariant $n_0^{(2,0),0}$.  

Inductively, we can define $E_{d_1+1}(p)$ as an extension of $E_{d_1}(p)$ by $\mathcal{O}_{f_p}$ just as was done in \ref{eq:atiyah} for $E_1(p)=\mathcal{O}_{f_p}(p)$, and the moduli space is again $\widehat{X}$.

We now compute the GV invariants $n_0^{(0,0),j}$ for $j\ne0$.
Consider an exceptional curve $C_i$ with class $[C_i]=j\in\mathbb{Z}_5$ corresponding to one of our $100+125$ conifolds $p_i$.\footnote{In our description, we do not associate a GV invariant to $C_i$ itself but view it as a contribution to what we have called the constant map contribution in Section~\ref{sec:constantmap}.}  The original singular containing $p_i$ becomes an $I_2$ fiber after performing the small resolution, equal to the union of $C_i$ with another curve $D_i$, the proper transform of the original singular fiber. Since all fibers have class $((1,0),0)$, the curve $D_i$ must have class $((1,0),-j)$.

To compute our GV invariants, we have to consider flops. If we flop $C_i$, the new exceptional curve $C_i'$ has class $((0,0),-j)$, hence the new curve $D_i'$ has class $((1,0),j)$.  Thus each conifold of this type contributes 1 to each of $n_0^{(1,0),\pm j}$.  We conclude that $n_0^{(1,0),\pm1}=n_{\pm1}=100$ and  $n_0^{(1,0),\pm2}=n_{\pm2}=125$, in agreement with \cite[Tables~35--36]{Schimannek:2021pau}.

We adapt the construction (\ref{eq:atiyah}) to the curves $D_i$.  Letting $f_i=C_i\cup D_i$ be the $I_2$  fiber containing $C_i$, we look for extensions \footnote{Note that $\mathcal{O}_{D_i}$ already has Euler characteristic 1 as $D_i\simeq\mathbb{P}^1$, so we don't need a point $p$ as in situation of (\ref{eq:atiyah}).}
\begin{equation}\label{eq:diext}
    0\to \mathcal{O}_{f_i}\to E_2\to \mathcal{O}_{D_i}\to 0\,.
\end{equation}
These are parametrized by $\operatorname{Ext}^1(\mathcal{O}_{D_i}(p),\mathcal{O}_{f_i})$, where the ext group is computed on $f_i$, and this group is dual to $\operatorname{Hom}(\mathcal{O}_{f_i},\mathcal{O}_{D_i})$, which is one dimensional.  Thus there is a unique nontrivial extension (\ref{eq:diext}) up to isomorphism, and this process of taking extensions by $\mathcal{O}_{f_i}$ can be iterated.  We conclude that $n_0^{(d_1,0),\pm1}=100$ and  $n_0^{(d_1,0),\pm2}=125$ for all $d_1>0$, in agreement with \cite[Tables~35--36]{Schimannek:2021pau}.

If we have a class $\beta=((0,d_2),j)$, then $\mathbb{P}^2\cdot C =-3d_2<0$.  It follows readily that $C\subset\mathbb{P}^2$.  The numbers $n_0^{(0,d_2),0}=3,-6,27,\ldots$ are well-known to have been verified geometrically from the geometry of local $\mathbb{P}^2$, agreeing with \cite[Table~34]{Schimannek:2021pau}.  Since these plane curves represent all of the curves independent of torsion, it follows that $n_0^{(0,d_2),j}=0$ for $j\ne 0$, agreeing with \cite[Tables~35--36]{Schimannek:2021pau}.

We next enumerate curves with $d_1=1$.  For $j=0$, these curves are plane curves $C$ of degree $d_2$ glued to a fiber.  For $d_2=1,2$, the moduli space of stable sheaves of Euler characteristic 1 consists of sheaves $\mathcal{O}_f(p)$ glued to $\mathcal{O}_C$ at a point.  To describe the moduli space, we first choice $p$ (parametrized by $\widehat{X}$) and then we choose a curve $C$ containing the point $q$ where the fiber meets $\mathbb{P}^2$.  Lines through a point are parametrized by $\mathbb{P}^1$ and degree 2 curves through a point are parametrized by $\mathbb{P}^4$.  So we get a $\mathbb{P}^1$-bundle over $X$ for $d_2=1$ and a $\mathbb{P}^4$-bundle over $X$ for $d_2=2$.  So we multiply $-\chi_{\rm top}(X)=90$ by the signed Euler characteristic $-2$ (resp. $5$) to get the GV invariants $-180$ and $450$ respectively.  It is obvious that the factors of $-2$ and $5$ are universal, depending only on the base $\mathbb{P}^2$ but not on the nature of the elliptic fibration.  It is also clear that the factors for any $d_2$ are similarly universal.  The factors $-2, 5, -32, 286\ldots$ for $d_2=1,2,3,4,\ldots$ have appeared in the literature before, for instance for the generic elliptic fibration over $\mathbb{P}^2$, so we do not repeat the verification.  The resulting GV invariants $n_0^{(1,d_2),0}=-180,450,-2880,25740$ for $1\le d_2\le 4$ agree with \cite[Table~34]{Schimannek:2021pau}.

Instead of gluing a fiber to a plane curve, we can glue the curves $D_i$ of class $((1,0),-j)$ (and their counterparts on a flop of class $((1,0),j)$). Since $D_i\simeq\mathbb{P}^1$, $\mathcal{O}_{D_i}$ already has Euler characteristic 1, so we don't need to choose a point $p$.  Instead of a factor of 90, we get one contribution for each curve $D_i$, so either $n_{\pm1}=100$ or $n_{\pm2}=125$ counting all curves for each discrete charge $\pm j$.  Multiplying the sequence $-2, 5, -32, 286$ by 100, we reproduce the $d_1=1$ row of \cite[Table~35]{Schimannek:2021pau}, while multiplying by 125 instead reproduces the $d_1=1$ row of \cite[Table~36]{Schimannek:2021pau}.

Next, we enumerate the unions of a line in $\mathbb{P}^2$ with $g$ fibers.  These curves have class $((g,1),0)$ and arithmetic genus $0+g\cdot 1=g$.  So $n_g^{(g,1),0}$ is just the signed Euler characteristic of the moduli space of these curves.  The moduli space has a natural fiber structure.  The base of the fibration is the dual projective plane parametrizing lines in $\mathbb{P}^2$.  The fiber is the $g$-fold symmetric product of the line, isomorphic to $\mathbb{P}^g$.  We get $n_g^{(g,1),0}=(-1)^{g}3(g+1)$, agreeing with \cite[Table~37]{Schimannek:2021pau}.

A simple variant of the above construction comes from gluing $g$ fibers as well as a curve $D_i$ (or its counterpart on a flop) to a line in $\mathbb{P}^2$.  Now the line is constrained to contain the point $q$ where $D_i$ meets $\mathbb{P}^2$.  So the base of the fibration is now $\mathbb{P}^1$ rather than $\mathbb{P}^2$.  The arithmetic genus is still $g$.  For each $D_i$, the fiber of the moduli space is still $\mathbb{P}^g$.  Multiplying as well by the appropriate factor $n_{\pm j}$ which counts the $D_i$, we get $n_g^{(g+1,1),\pm j}=(-1)^{g+1}2n_{\pm j}(g+1)$, which agrees with \cite[Table~38]{Schimannek:2021pau} for $j=1$ and with \cite[Table~39]{Schimannek:2021pau} for $j=2$.

\subsection{A weighted complete intersection}

Our final example will be a weighted complete intersection $X$ of hypersurfaces of degrees 2 and 6 in $\mathbb{P}_{111113}$ with 70 nodes, which is labeled as AESZ 211 in Appendix~\ref{app:otherexamples}.  The GV invariants which we verify geometrically below have been put in boldface type in the corresponding table.

The degree 2 hypersurface necessarily only involves the first 5 variables so can be identified with a quadric hypersurface $Q\subset\mathbb{P}^4$.  Changing coordinates if necessary, the degree 6 hypersurface $Y$ can be put into the form
\begin{equation}
    x_6^2=f_6(x_1,\ldots,x_5)\,,
\end{equation}
where $f_6$ is a degree 6 polynomial in $x_1,\ldots,x_5$.  Thus $X$ is a double cover $\pi:X\to Q$ branched over a $(2,6)$ complete intersection in $\mathbb{P}^4$.  

As in the case of the octic double solid, a degree~1 curve can only be one of the $L_i$ which project isomorphically onto a line $L\subset Q$ with the property that $\pi^{-1}(L)=L_1\cup L_2$ is a union of two curves.  The reducibility of $\pi^{-1}(L)$ can only happen when $L$ intersects $Y$ in three points of tangency.  A straightfoward dimension count shows that this is expected to happen for only finitely many $L$. By standard techniques of classical algebraic geometry, the number of such $L$ is 2496.  Examining the GV invariants of $X$ we see that for each such $L$, one of the curves $L_1,L_2$ has charge 0 and the other has charge 1.

The analysis of the degree 2 GV invariants is entirely analogous to the case of the octic double solid, supplemented with information about the variety of lines on $Q$. The curves $\pi^{-1}(L)$ have degree 2.  Since $\pi^{-1}(L)=L_1\cup L_2$ has charge 1 for these special $L$, it follows that $\pi^{-1}(L)$ has charge 1 for any line $L\subset Q$.  In general, $L$ intersects $Y$ in 6 point, so $\pi^{-1}(L)$ is a double cover of $L$ branched over 6 points and so has genus 2 by Riemann-Hurwitz. So $n_2^{2,1}$ is determined by the space of lines $L\subset Q$.  It is known that this space is parametrized by $\mathbb{P}^3$ (see e.g.\ \cite{Harris}).  It follows that $n_2^{2,1}=-4$ and $n_2^{2,0}=0$.

Similarly, the methods used for the octic double solid can be applied to compute $n_1^{2,0}=-280$.  Whenever $L$ contains one of the 70 conifolds $p$, then $\pi^{-1}(L)=C\cup L_p$ for $L_p$ the exceptional $\mathbb{P}^1$ of the small resolution $\widehat{X}$ and the curve $C$ having class $(2,0)$.  Any such line $L$ is necessarily contained in the tangent hyperplane $H_p$ to $Q$ at $p$, which intersects $Q$ in a singular quadric surface, generically a quadric cone, $p$ being identified with the vertex of the cone and the base of the cone being a plane conic curve.  The set of lines in this cone is clearly parametrized by the conic curve, which is isomorphic to $\mathbb{P}^1$, giving a contribution of $-2$ for each of the 70 conifolds.  Our now familiar argument tells us that we have to consider both small resolutions, giving $n_1^{2,0}=2(-2)(70)=-280$.  Note that the fact that $p$ is a conifold does not affect the description of the lines in $Q$ containing $p$. 

For $n_2^{2,1}$, we again use (\ref{eq:kkv1}).  The universal curve $\mathcal{C}$ fibers over $\widehat{X}$ with fiber equal to the set of lines through a fixed point $p\in Q$, which we have just seen is a $\mathbb{P}^1$.  We have $e(X_{\text{def.}})=-256$, so that $e(\widehat{X})=256+2\cdot 70=-116$ because there are 70 conifolds. We conclude from (\ref{eq:kkv1}) that $n_2^{2,1}=-224$. 

In degree 4, the Castelnuovo bound is achieved by double covers of degree 2 plane curves $D$.  Since $D$ meets $Y$ in 12 points, the branched double cover $\pi^{-1}(D)$ has genus 5.  The curves $D$ which are contained in $Q$ are all complete intersections of $Q$ with a 2-plane $P\subset\mathbb{P}^4$.  These $P$ are parametrized by $G(3,5)$, which has euler characteristic 10.  Thus $n_5^{4,0}=10$.

We can then compute $n_4^{4,0}$ by (\ref{eq:kkv1}).  The universal curve $\mathcal{C}$ fibers over $\widehat{X}$ with fiber equal to the set of plane conics through a fixed point $p\in Q$.  It remains to describe the 2-planes $P$ which contain $\pi(p)$.  By projection from $p$ to a $\mathbb{P}^3$, this moduli space is identified with the moduli space $G(2,4)$ of lines in $\mathbb{P}^3$. So $\mathcal{C}$ is a $G(2,4)$-bundle over $\widehat{X}$, which has Euler characteristic $6(-116)=-696$.  Then (\ref{eq:kkv1}) gives $n_4^{4,0}=616$.

For $n_4^{4,1}$, we look for plane conics $D$ containing a conifold $p$, since $\pi^{-1}(D)=C\cup L_p$ for a curve $C$ of class $(4,1)$.  As we have just seen, the 2-planes $P$ containing $\pi(p)$ are parametrized by $G(2,4)$, so we associate to $\widehat{X}$ the GV invariant $6\cdot 70=420$ by applying this construction to each of the 70 conifolds.  To get the GV invariants of $X$, we have to consider both small resolutions, so we count 420 twice to obtain $n_4^{4,1}=840$.

Just as in the case of the octic double solid, the Castelnuovo bound for odd degrees is obtained by gluing a degree 1 curve to Castelnuovo curves of even degree.  In degree 5, we glue a degree 1 curve ($L_1$ or $L_2$) to any curve $\pi^{-1}(D)$ as above which intersects it.  We describe the moduli space of these curves $\pi^{-1}(D)\cup L_i$ beginning with the point $p=L_i\cap \pi^{-1}(D)$, which is parametrized by $L_i\simeq\mathbb{P}^1$.   Using the above description of the moduli space of 2-planes containing a fixed point, we see that the moduli space of 2-planes which meet $L_i$ is a $G(2,4)$-bundle over $\mathbb{P}^1$, which has signed Euler characteristic $-12$ owing to its odd dimension.  We have such a moduli space for each degree 1 curve, and so we get $n_3^{2,0}=n_3^{2,1}=-12(2496)=-29952$.

In degree 3, we glue a degree 1 curve ($L_1$ or $L_2$) to any curve $\pi^{-1}(L)$ with $L$ a line which intersects $L_i$.  The moduli space is the set of lines $L\subset Q$ which meet the chosen $L_i$.  The moduli space is a subset of the $\mathbb{P}^3$ of all lines in $Q$, and it follows readily from the discussion in \cite{Harris} that the moduli space is $\mathbb{P}^2$. We conclude that $n_2^{2,0}=n_2^{2,1}=(2496)=7488$.
\section{Outlook}
\label{sec:outlook}

In this paper we have started to investigate the general properties of the topological string A-model on non-commutative resolutions of compact Calabi-Yau threefolds with terminal nodal singularities, building on earlier GLSM constructions in~\cite{Caldararu:2010ljp} and torus fibered examples studied in~\cite{Schimannek:2021pau}.
There are many possible future directions that remain to be addressed.  We list a few possibilities below:
\begin{itemize}
    \item \textbf{Strings on non-K\"ahler Calabi-Yau manifolds:}
    From the physical perspective, we have proposed that a discrete gauge symmetry arises in M-theory on Calabi-Yau threefolds $X$ with terminal nodal singularities, corresponding to the torsion of the exceptional curves in a small non-K\"ahler resolution $\widehat{X}$.
    This generalizes earlier observations in the context of F-theory on certain singular torus fibrations, see e.g.~\cite{Braun:2014oya}.
    On the other hand, in the Type IIA compactification we propose that a fractional B-field that stabilizes the singularities can also be interpreted in terms of a corresponding B-field associated to a non-trivial class $\alpha\in\text{Br}(\widehat{X})$.
    Because the volume of the exceptional curves decouples in the topological A-model, it is not entirely clear whether the sigma model on $(\widehat{X},\alpha)$ should be interpreted as a UV theory that flows in the IR to a sigma model on $(X,\alpha)$ or whether the B-field turns the non-K\"ahler space $\widehat{X}$ itself into an admissible string background, analogous to e.g.~\cite{Rocek:1991ze,Gurrieri:2002wz,Hitchin:2003cxu,STROMINGER1986253}.
    It would be very interesting to understand the effect of the renormalization group flow.
    
    \item \textbf{Bridgeland stability and PT/DT:}
    We have proposed a definition of the torsion-refined Gopakumar-Vafa invariants associated to a Calabi-Yau threefold with terminal nodal singularities in terms of a derived category of a small non-K\"ahler resolution $\widehat{X}$.
    For each fractional B-field background $\alpha\in\text{Br}(\widehat{X})$ we expect there to be an analogue of the MNOP relation~\cite{mnop1}, using some suitably modified definition of the corresponding Pandharipande or Donaldson-Thomas invariants.
    The latter should be defined in terms of a Bridgeland stability condition on the twisted derived category $D^b(\widehat{X},\alpha)$
    and should be seen as a generalization to compact Calabi-Yau manifolds of Szendrői's non-commutative DT-invariants of the conifold~\cite{szendrHoi2008non}.
    However, it is unclear how to incorporate the twisting in the usual definition of these invariants in terms of stable pairs or D6-D2-D0 bound states.
    We leave further investigation of this problem for future work.
    
    \item \textbf{Clifford nc-resolutions and FJRW theory:}
    In case of the Clifford type nc-resolutions, there often (conjecturally, always) exists a  construction of the corresponding closed string worldsheet theories in terms of fixed points of hybrid gauged linear sigma models~\cite{Caldararu:2010ljp}.
    The corresponding A-model topological string amplitudes can then be interpreted in terms of FJRW theory~\cite{fjrw1,zbMATH05782775,cir14,Clader:2013tn,Erkinger:2022sqs}, which generalizes the usual Gromov-Witten invariants.
    The expansion of the topological string partition function in terms of torsion-refined Gopakumar-Vafa invariants, that has been proposed in~\cite{Schimannek:2021pau} and that we further corroborate, therefore provides a GV-FJRW correspondence that generalizes the well-known GV-GW correspondence.
    This opens up many interesting questions, for example about the relation between the charge under the discrete symmetry on the GV side and the Hilbert space of states that is given in terms of the Chen-Ruan cohomology of an inertia stack on the FJRW side.
    
    \item \textbf{GLSMs for general type nc-resolutions:}
    For the general type nc-resolutions that appear in the context of torus fibrations, the corresponding large volume limits do not seem to correspond to phases in the GLSM associated to the smooth dual $Y$ but are instead located at phase boundaries where the GLSM description becomes singular~\cite{Schimannek:2021pau}.
    Nevertheless, there could be a dual description in which the corresponding closed string worldsheet theories can again be obtained as fixed points of some other GLSM.
    Perhaps recent ideas from~\cite{Faraggi:2022gkt} about the GLSM realization of discrete torsion can help to address this problem. 
    
    \item \textbf{Fractional B-fields and categorical resolutions:} The mathematical notions of non-commutative crepant resolutions~\cite{vdB:nc-rev} and crepant categorical resolutions~\cite{kuzso} are very general. However, determining whether a given sheaf of algebras falls under either of the two definitions is quite non-trivial in practice.  
    In the context of Calabi-Yau manifolds $X$ with terminal nodal singularities, our analysis leads to small analytic resolutions $\widehat{X}$ that are equipped with a fractional B-field $\alpha\in\text{Br}(\widehat{X})$, giving rise to a sheaf of non-commutative algebras on $\widehat{X}$.
    We have proposed that the pair $(\widehat{X},\alpha)$ can be thought of as representing a non-commutative resolution of $X$ itself, as also suggested in~\cite{caldararu2001derived}.
    Locally, via the results from~\cite{szendrHoi2008non} and also~\cite{Hori:2011pd}, this naturally reduces to the non-commutative conifold around the nodes of $X$.
    It would be interesting to understand whether the associated twisted derived category $D^b(\widehat{X},\alpha)$ satisfies the conditions on a strongly crepant categorical resolution as defined e.g.~in~\cite{kuzso}.
    We further speculate that it may be possible to use a pair $(\widehat{X},\alpha)$ to construct a sheaf of non-commutative algebras on $X$ and therefore an actual non-commutative resolution of $X$ itself.
    Locally near each conifold this sheaf of nc algebras is expected to be Morita-equivalent to the NCCR resolution of the conifold.
      We leave these considerations to future work.

    \item \textbf{General classification:}
    In this paper we have mostly focused on Clifford type nc-resolutions associated to singular double covers, for which GLSMs had been constructed in~\cite{Caldararu:2010ljp}, while topological strings on general type nc-resolutions that arise in the context of torus fibered Calabi-Yau manifolds were investigated in~\cite{Schimannek:2021pau}.
    It would be very interesting to find other types of general type nc-resolutions and eventually to arrive at a general classification.
    Already in the context of Clifford type nc-resolutions, additional GLSM examples have been suggested in e.g.~\cite{Halverson:2013eua,Chen:2020iyo,Guo:2021aqj} and would also be interesting to investigate.
    Let us finally remark that the number of  non commutative MUM points in the list~\cite{aesz} described in~\cite{MR3822913} is considerable, three of them are discussed in detail in Appendix~\ref{app:otherexamples}. Moreover,  MUM-points with algebraic BPS expansions, so called irrational MUMs, can be systematically constructed.\footnote{D. van Straten and J. Walcher in private communication.}
    These could potentially be associated to one parameter slices that intersect non-commutative MUM-points in higher-dimensional moduli spaces of torus fibered Calabi-Yau manifolds, with $\mathbb{Z}_n$-torsion for $n\ge 5$.
    While it is not yet clear to us how ubiquitous the phenomenon is, we take these as indications that a string Landscape without the inclusion of non-commutative Calabi-Yau categories might be seriously incomplete.  
\end{itemize}

\section{Acknowledgements}

We would like to thank N.~Addington, A.~Chiodo, R.~Donagi, J.~Knapp, R.~Pandharipande, T.~Pantev, B.~Pioline, E.~Scheidegger, E.~Segal, and V.~Shende
for useful conversations. The research of S.K. was supported by NSF grants DMS-1802242 and DMS-2201203.
The research of T.S.~is supported by the Agence Nationale de la Recherche (ANR) under contract number ANR-21-CE31-0021. E.S.~is partially supported by 
NSF grant PHY-2014086. A.K.~likes to thank Dr. Max R\"ossler, the Walter Haefner Foundation and the ETH Z\"urich Foundation for support.

\appendix

\section{The Brauer group of the singular octic double solid}\label{app:brauer}

In this appendix, we follow a suggestion of N.~Addington and show that for any small resolution $\widehat{X}$ of the determinantal octic double solid we have
\begin{align}
\operatorname{Br}(\widehat{X})=H_2(\widehat{X},\mathbb{Z})_{\mathrm{tors}}=H^3(\widehat{X},\mathbb{Z})_{\mathrm{tors}}=\mathbb{Z}_2\,.
\end{align}

We first construct a nontrivial Brauer class $\alpha$.  Recall  the complete intersection of quadrics $Y\subset \mathbb{P}^7$ associated to $X$ discussed in Section~\ref{subsec:spinor}.  It is well-known (from mirror symmetry, or directly from enumerative geometry, see e.g.\ \cite{Katz:1986}) that $Y$ contains 512 lines $L$.  Pick any such $L$.  We represent $\alpha$ by constructing a $\mathbb{P}^1$-bundle on $\widehat{X}$.  Once we construct this $\mathbb{P}^1$ bundle, it will be clear that it is a reduction of the $\mathrm{OGr}^+(4,8)$-bundle described in Section~\ref{subsec:spinor}, so the Brauer class $\alpha$ that we describe here coincides with the Brauer class associated to the universal twisted spinor sheaf.  Since $\alpha$ is represented by a $\mathbb{P}^1$-bundle, it is obviously 2-torsion, i.e. $\alpha^2=0$.

Let us first construct the $\mathbb{P}^1$-bundle over the complement of the exceptional curves $C_p$.  Such points of $\widehat{X}$ correspond to a point $p\in \mathbb{P}^3$ and a family of $\mathbb{P}^3$s contained in $Q_p$ (recall that there are two such families if $Q_p$ is smooth and one family if $Q_p$ has corank 1).

Write $L=\mathbb{P}(V)$, where  $V\subset \mathbb{C}^8$ is a  two-dimensional subspace. We investigate linear subspaces $P\simeq\mathbb{P}^3$ with $L\subset P\subset Q_p$, or equivalently 4-dimensional isotropic subspaces of $\mathbb{C}^8$ which contain $V$.   Let $V_p^\perp$ be the six-dimensional subspace orthogonal to $V$ with respect to the quadratic form associated to $p$.  We have $V\subset V_p^\perp$ since $V$ is isotropic.  We get an induced quadratic form on the four-dimensional vector space $V_p^\perp/V$.  Then 4-dimensional isotropic subspaces of the original quadratic form which contain $V$ correspond to two-dimensional isotropic subspaces of $V_p^\perp/V$.

It is easy to see that this quadratic form has maximal rank if $Q_p$ is smooth and has corank~1 if $Q_p$ has corank 1.  In the first case, there are two families of two-dimensional isotropic subspaces of $V_p^\perp/V$, each parametrized by $\mathbb{P}^1$.  In the second case, there is one family of two-dimensional isotropic subspaces of $V_p^\perp/V$, again parametrized by $\mathbb{P}^1$.   There is a natural map from the union of all of these $\mathbb{P}^1$s to $\widehat{X}$, forming a $\mathbb{P}^1$-bundle.

We next extend this $\mathbb{P}^1$-bundle over the exceptional curves $C_p$.  Now $Q_p$ has corank~2.  Let $q\in C_p\subset\widehat{X}$.  Recalling the identification of $C_p$ with the singular $\mathbb{P}^1$ of $Q_p$, we write $q=\mathbb{P}(U)$ with $U\subset \mathbb{C}^8$ the appropriate 1-dimensional subspace.  We project $Q_p$ from $q$ to a corank 1 quadric $Q_q'\subset \mathbb{P}(\mathbb{C}^8/U)\simeq\mathbb{P}^6$.  Since $Q_p$ has corank 2, it follows that $Q_q'$ has corank 1.  Under this projection, the line $L$ projects to a line in $L'\subset Q_q'$.  The 5-dimensional quadric $Q_q'$ contains $\mathbb{P}^2$s.  Since $Q_q'$ is a cone over a smooth 4-dimensional quadric, there are two connected families of $\mathbb{P}^2$s.  However, one of them has already been distinguished by the choice of small resolution $\widehat{X}$.

 The cone from $q$ over any of these $\mathbb{P}^2$s is a $\mathbb{P}^3$ contained in $Q_p'$.  So we have to construct the family of distinguished $\mathbb{P}^2$s in $Q_q$ containing $L'$ and show that it is parametrized by $\mathbb{P}^1$ in order to complete the $\mathbb{P}^1$-bundle over $C_p$.

For this, we repeat an argument already given.  Writing $L'=\mathbb{P}(V')$ with a 2-dimensional $V'\subset\mathbb{C}^7$, we get a corank~1 quadric in $(V')^\perp/V'\simeq \mathbb{C}^3$.  Projectivizing, we get a corank~1 quadric in $\mathbb{P}^2$, which is a union of two $\mathbb{P}^1$s, one of which has been distinguished by the choice of small resolution $\widehat{X}$.  If we pick a point of the distinguished $\mathbb{P}^1$, it together with $L'$ span the required $\mathbb{P}^2$ in $Q_q'$.  So the familiy of such $\mathbb{P}^2$s is indeed parametrized by $\mathbb{P}^1$ and we are done.

Since $\alpha$ is nontrivial and $\alpha^2=0$, $\mathrm{Br}(X)$ contains a $\mathbb{Z}_2$ subgroup.

\medskip

We now show that $\mathrm{Br}(X)=\mathbb{Z}_2$.  It follows that $H_2(\widehat{X},\mathbb{Z})_{\mathrm{tors}}=H^3(\widehat{X},\mathbb{Z})_{\mathrm{tors}}=\mathbb{Z}_2$ as well.  The calculation is similar to that in \cite[Section 1]{AW}, to which we refer the reader for additional details and references.

The derived equivalence $D^b(\widehat{X},\alpha^{-1})=D^b(Y)$ can be shown to imply the identications of topological K-theory
\begin{align}\label{eq:keq}
K^i(\widehat{X},\bar{\alpha}^{-1}) = K^i(Y)\,.
\end{align}
Here $\bar{\alpha}\in H^3(\widehat{X},\mathbb{Z})_{\mathrm{tors}}$ is the class associated to $\alpha$, and $K^i(\widehat{X},\bar{\alpha}^{-1})$ denotes twisted K-theory.  For ease of notation, we put $T=H^3(\widehat{X},\mathbb{Z})_{\mathrm{tors}}$.

The idea is to use the Atiyah-Hirzebruch spectral sequence to show that
\begin{align}\label{eq:tork1}
K^1(\widehat{X},\bar{\alpha}^{-1})_{\mathrm{tors}} \simeq T/\langle \bar{\alpha}^{-1}\rangle\,,
\end{align}
while $K^1(Y)$ is torsion-free.  It follows that $T = \langle\bar{\alpha}^{-1}\rangle\simeq\mathbb{Z}_2$.

The Atiyah-Hirzebruch spectral sequence for twisted K-theory has
\begin{align}
E_2^{p,q}=\left\{
\begin{array}{cl}
H^p(\widehat{X},\mathbb{Z})&q{\rm \ even}\\
0 & q{\rm \ odd}
\end{array}
\right.\,.
\end{align}
The $d_2$ differentials vanish, and $d_3:E_3^{0,2p}\to E_3^{3,2p-1}$ is identified with multiplication by $\bar\alpha^{-1}$ in
\begin{align}
H^0(\widehat{X},\mathbb{Z}) \to H^3(\widehat{X},\mathbb{Z})\,.
\end{align}
Since the only odd cohomology of $\widehat{X}$ is in $H^3$, we see that
\begin{align}
K^1(\widehat{X},\bar{\alpha}^{-1})\simeq H^3(\widehat{X},\mathbb{Z})/\langle \bar{\alpha}^{-1}\rangle\,,
\end{align}
and (\ref{eq:tork1}) follows immediately.

A similar but simpler calculation shows that $K^1(Y)\simeq H^3(Y,\mathbb{Z})$, which is torsion free since $Y$ is a complete intersection.  We conclude that $T=H^3(\widehat{X},\mathbb{Z})_{\mathrm{tors}}\simeq \langle \bar{\alpha}^{-1}\rangle\simeq\mathbb{Z}_2$, as claimed.

The same argument shows more generally that if $D^b(\widehat{X},\alpha)\simeq D^b(Y)$, where $\widehat{X}$ is a simply connected smooth compact threefold and $Y$ is a simply connected smooth compact threefold without torsion in its cohomology, then $\alpha$ generates $H^3(\widehat{X},\mathbb{Z})_{\mathrm{tors}}$.
\section{Direct integration and algebraic K\"ahler gauge choices}
\label{app:algebraicKahlergauge}

The non-holomophic generators $\{G\}$ can be derived from 
special geometry.
Such generators were first found in~\cite{Bershadsky:1993cx} and subsequently different choices have been discussed in~\cite{Yamaguchi:2004bt,Alim:2007qj,Hosono:2008np}, see also~\cite{Grimm:2007tm}.
The most convenient
choice for our purpose is a modification of the propagators $S^{ij},S^i,S$ of~\cite{Bershadsky:1993cx} that has been found in~\cite{Alim:2007qj}. These 
propagators were introduced in~\cite{Bershadsky:1993cx} 
to obtain $Z=\exp(F(\lambda, \uz))$ from an action principle.  
They are defined by successive integration
\be 
\partial_{\bar \jmath} S^{ij}=C^{ij}_{\bar \jmath}, \quad 
\partial_{\bar \jmath} S^j= G_{i \bar \jmath  }S^{ij}\quad 
\partial_{\bar \jmath} S= G_{i \bar \jmath  }S^{i}\,,
\ee
of $C^{ij}_{\bar \jmath}=e^{2K} C_{\bar \imath \bar k \bar \jmath} G^{\bar \imath i} G^{\bar k k}$. As such, they transform as sections of ${\cal L}^2 \otimes {\rm Sym}^\bullet(T{\cal M}(\mM))$ and 
are determined by genus zero data, up to a choice of holomorphic propagator ambiguities that will be further discussed below.

The integration follows from the special geometry relations. 
Firstly, since $C_{ijk}$ is holomorphic, $S^{ij}$ is determined from integrating 
\eqref{eq:specialKaehler} w.r.t. $\bar \jmath$, giving
\begin{align}
        \Gamma^l_{ij}=\delta^l_iK_j+\delta^l_jK_i-C_{ijk}S^{kl}+s^l_{ij}\, ,
\label{eq:Sij} 
\end{align}
up to holomorphic ambiguities $s^l_{ij}$. One solves for $S^{kl}$
by inverting $C_{(i)jk}$. This matrix is invertible at least for one index $i$
due to the local Torelli theorem.
Calculating $\partial_{\bar \jmath} D_i S^{kl}$ 
using \eqref{eq:specialKaehler} and \eqref{eq:Sij} implies that $D_i S^{kl}$ 
closes in $S^{ij},S^i$ up to a further holomorphic ambiguity. The corresponding 
relation can be solved for $S^i$ and calculating $\partial_{\bar \jmath} D_i S^{j}$ 
one sees that $D_i S^j$ closes on $S^{ij},S^i,S,K$ up to yet another holomorphic ambiguity. 
Repeating this procedure for $D_i S$ and $D_i K_i$, where $K_i=\partial_i K$, one finally gets the so-called BCOV ring
\begin{align}
        \begin{split}
                D_iS^{jk}=&\delta^j_iS^k+\delta^k_iS^j-C_{imn}S^{mj}S^{nk}+h^{jk}_i\,,\\
                D_iS^j=&2\delta^j_iS-C_{imn}S^mS^{nj}+h^{jk}_iK_k+h^j_i\,,\\
                D_iS=&-\frac12C_{imn}S^mS^n+\frac12h^{mn}_iK_mK_n+h^j_iK_j+h_i\,,\\
                D_iK_j=&-K_iK_j-C_{ijk}S^k+C_{ijk}S^{kl}K_l+h_{ij}\,.
        \end{split}
        \label{eqn:bcovder}
\end{align}

Since these equations are covariant, one can deduce from them the tensor  
transformation of the ambiguities $h$ under coordinate and under 
K\"ahler transformations.
The latter take the form
\begin{align}
        \mathfrak{X}^I\rightarrow f(z)\mathfrak{X}^I \,,\quad e^{-K}\rightarrow f(z)e^{-K}\,,\quad K_i\rightarrow K_i-\partial_i\log f(z)\,,
\end{align}
for non-vanishing holomorphic functions $f(\uz)$, and are important to combine global information from different local 
expansions in the moduli space.
Using~\eqref{eq:Sij} and~\eqref{eqn:bcovder}, we find that the propagator ambiguities change under K\"ahler transformations as
\begin{align}
        \begin{split}
               s^l_{ij}\rightarrow& s^l_{ij}+\left(\delta^l_i\partial_j+\delta^l_j\partial_i\right)\log f\,,\\
                h^{jk}_i\rightarrow& f^{-2}h^{jk}_i\,,\\
                h^j_i\rightarrow& f^{-2}\left(h^j_i+h^{jk}_i\partial_k\log f\right)\,,\\
                h_i\rightarrow&f^{-2}\left(h_i+\frac12h^{mn}_i\left(\partial_m\log f\right)\left(\partial_n\log f\right)+h^j_i\partial_j\log f\right)\,,\\
                h_{ij}\rightarrow&h_{ij}+s^l_{ij}\partial_l\log f\,.
        \end{split}
\end{align}
On the other hand, the transformations of the holomorphic ambiguities of the free energies take the form
\begin{align}
f_1\rightarrow& f_1+\left(1-\frac{\chi}{24}\right)\log f\,,\quad f_{(g\ge 2)}\rightarrow f^{2-2g}f_{g\ge 2}\,,
\end{align}
and follow from~\eqref{eq:holF1} and the fact that the K\"ahler weight of $F_{g\ge2}$ is $2g-2$.

Coming back to the  direct integration step, it is convenient to introduce the shifted propagators~\cite{Alim:2007qj}
\begin{equation} 
\ts^{ij}=S^{ij}\,,\qquad   \ts^{i}=S^{i}-S^{ia} K_a\,,\qquad \ts=S-S^a K_a+
\frac{1}{2} S^{ab} K_a K_b \,.
\label{tildeS}
\end{equation}

The $\ts^{ij},\ts^i,\ts$ are not sections of ${\cal L}^2$, due to their explicit dependence on $K$, and they transforms as  
\begin{equation}
\begin{gathered}
    \tilde{S}^{ij}\rightarrow f(z)^{-2}\tilde{S}^{ij}\,,\quad \tilde{S}^i\rightarrow  f(z)^{-2}\left(\tilde{S}^i-\tilde{S}^{ij}\partial_j\log f(z)\right)\,,\\
    \tilde{S}\rightarrow f(z)^{-2}\left(\tilde{S}-\tilde{S}^i\partial_i\log f(z)+\frac12(\partial_i\log f(z))(\partial_j\log f(z))\tilde{S}^{ij}\right)\,.
\end{gathered}
\end{equation}
However, using the relations $\bar {\partial}_{\bar \imath } \ts^{ij}= C_{\bar \imath }^{ij}$, 
$\bar {\partial}_{\bar \imath } \ts^i= -C_{\bar \imath}^{ia}K_a$, 
$\bar {\partial}_{\bar \imath} \ts = \frac{1}{2} C_{\bar \imath}^{ab} K_a K_b$, one can show inductively that the non-holomorphic dependence of $F_{g\ge2}$ 
is entirely absorbed in their dependence  on the shifted generators 
$\tilde S^{ij},\tilde S^j,\tilde S$.
In particular, there is no explicit dependence on $K$.

The $\partial_{\bar \jmath}$ derivative in \eqref{eq:holan} becomes a derivative in the generators and the factor $C_{\bar \imath }^{ij}$ 
cancels. Assuming that the propagators and $K_a$ are functionally independent,
one gets by comparing coefficients in $K_a$ the form \cite{Huang:2015sta}
\begin{align}
\begin{split}
\label{partial2.14}
\frac{\partial {F}_g }{\partial \tilde{S}^{ij}}  =& \frac{1}{2} \partial_i(\partial _j^{\prime} 
{F}_{g-1} ) 
+\frac{1}{2} (C_{ijl }\tilde{S}^{lk} - s^k_{ij} )\partial_k^{\prime}  {F}_{(g-1)}   
 \\
&+\frac{1}{2}  (C_{ijk}\tilde{S}^k - h_{ij} ) c_{g-1} + \frac{1}{2} \sum_{h=1}^{g-1} \partial_i^{\prime}  {F}_{h}   \partial_j^{\prime}  
{F}_{g- h}\,, \\
\frac{\partial {F}_{g} }{\partial \tilde{S}^{i}}  =& (2g-3) \partial_i^{\prime}  {F}_{g-1} 
+ \sum_{h=1}^{g-1} c_h   \partial_i ^{\prime} {F}_{g- h}\,, \\
\frac{\partial {F}_{g} }{\partial \tilde{S} }  =& (2g-3) c_{g-1} + \sum_{h=1}^{g-1} c_h c_{g-h}\,, 
\end{split}
\end{align}
where the $c_{g}$ and the action of $\partial^{\prime} $on the free energies are defined as
\begin{align}  \label{cg3.36}
c_g = \left\{
\begin{array}{cl}
 \frac{\chi}{24} -1    &   g=1   \\
 (2g-2) F_{g}             &   g>1   
\end{array}    
\right.\,,\quad \partial_i^{\prime}  {F}_g  = \left\{
\begin{array}{cl}
 \partial_i {F}_g  +  (\frac{\chi}{24}-1) K_i    &   g=1   \\
 \partial_i {F}_g             &   g>1
\end{array}    
\right.\,.
\end{align}

Using~\eqref{tildeS} and~\eqref{eq:Sij}, one can rewrite the BCOV ring~\eqref{eqn:bcovder} in terms of the shifted propagators
\begin{align}
\begin{split} 
\partial_i \ts^{jk}=&C_{imn}\ts^{mj} \ts^{nk}+\delta_i^j \ts^k+\delta_i^k\ts^j- s_{im}^j \ts^{mk}-s_{im}^k \ts^{mj}+ h_i^{jk}\, , \\
\partial_i \ts^{j}=&C_{imn}\ts^{mj} \ts^{n}+2\delta_i^j \ts- s_{im}^j \ts^{m}-h_{im} \ts^{mj}+ h_i^{j}\, , \\
\partial_i \ts =&\frac{1}{2}C_{imn}\tilde{S}^m\tilde{S}^n-h_{ij}\tilde{S}^j+h_i\,,\\
                \partial_iK_j=&K_iK_j-C_{ijn}\tilde{S}^{mn}K_m+s^m_{ij}K_m-C_{ijk}\tilde{S}^k+h_{ij}\,.
        \end{split}
\end{align}
The last equation is not necessary for the direct integration but, together with  
the other relations, can be used to construct a compatible choice for the propagator ambiguities $s^i_{jk}$, $h^{ij}_k,h_i^j,h_{ij}$ and $h_i$, which turn out to be rational functions in $\uz$. 

One can absorb their pole and zero structure 
in $z_i=0$ divisors by rescaling them as $s^i_{jk}=\tilde s^{i}_{jk}\frac{z_i}{z_jz_k}$, 
$h^{ij}_k=\tilde h^{ij}_k\frac{z_i z_j}{z_k},h_i^j=\tilde h_i^j \frac{z_j}{z_j}, h_{ij}=\tilde h_{ij}\frac{1}{z_i z_j}$ and $h_{i}=\tilde h_i \frac{1}{z_i}$. Note that here repeated indices are not summed over. At least for the models discussed in this paper, the  $\tilde s^i_{jk}$, $\tilde h^{ij}_k,\tilde h_i^j,\tilde h_{ij}$ and $\tilde h_i$ can then be chosen to be polynomial in $\uz$.
\section{Non-commutative duals of Pfaffian Calabi-Yau threefolds}
\label{app:otherexamples}
Additional examples of Clifford type nc-resolutions appear in the context of Pfaffian Calabi-Yau threefolds that are not complete intersections in toric ambient spaces.
As a result, they do not fall under the toric double mirror construction from~\cite{borisovzhan}.
In this appendix we briefly summarize the results that we obtain for three examples.

The Pfaffian Calabi-Yaus, each with $h^{1,1}=1$, have been constructed in~\cite{Kanazawa:2012xya} and corresponding non-Abelian gauged linear sigma models were studied in~\cite{Hori:2013gga}.
In three cases the strongly coupled phase has been identified as a hybrid model that consists of a $\mathbb{Z}_2$ Landau-Ginzburg orbifold fibered over a Fano threefold $B$.
We interpret this as corresponding to a non-commutative resolution of a singular double cover over the same base.

In each of those three cases one therefore again obtains a triple of geometries, $Y$, $X_{\text{n.c.}}$ and $X_{\text{def.}}$, that are in turn associated to a singular double cover $X$ of some Fano threefold $B$.
The Pfaffian Calabi-Yau threefold $Y$ is smooth and twisted derived equivalent to the nc-resolution $X_{\text{n.c.}}$ of $X$ that, for the purpose of the topological string A-model, can in turn be replaced by a small analytic resolution $\widehat{X}$ of $X$ with a fractional B-field background $\alpha$.
In particular, we expect that $H^3(\widehat{X},\mathbb{Z})\simeq\mathbb{Z}\oplus\mathbb{Z}_2$ and that there is a twisted derived equivalence $D^b(\widehat{X},\alpha)\simeq D^b(Y)$.
As usual we denote a generic smooth deformation of $X$ by $X_{\text{def.}}$.

Since the geometric analysis of the ramification loci of the singular double covers from the non-Abelian GLSMs is quite non-trivial, we will in each case only deduce the number of nodes from the obstructed complex structure deformations.
A highly non-trivial confirmation is then provided by the constant map contributions to the topological string free energies.

It turns out that for one of the examples, $X$ is again a singular double cover of $\mathbb{P}^3$ but with a different ramification locus and $n_s=88$ nodes instead of $84$.
The geometric calculations from Section~\ref{sec:geometric} directly carry over and confirm many of the invariants.
In the second example, $X$ is a singular double cover of a quadric in $\mathbb{P}^4$ with $n_s=70$ nodes and the smooth deformation $X_{\text{def.}}$ is a degree $(2,6)$ complete intersection $X_{2,6}$ in $\mathbb{P}(111113)$, which belongs to the 13 well-studied hypergeometric families.
For the third example, the base $B=V_5$ is a degree $5$ Fano threefold that can in turn be realized as a complete intersection in the Grassmanian $G(2,5)$ and we predict that the singular double cover $X$ has $50$ nodal singularities.

In the following we just summarize basic data for each of the relevant large volume limits and list some of the associated Gopakumar-Vafa invariants up to genus $10$.
If available, we also provide the corresponding identifier of the Picard-Fuchs operator in the table of Calabi-Yau operators from~\cite{aesz}.
For more information on the Pfaffian Calabi-Yaus we refer to~\cite{Kanazawa:2012xya} and for the associated gauged linear sigma models to~\cite{Hori:2013gga}.

\newpage
\begin{tabular}{|p{\linewidth}|}
                \hline AESZ 99 - Pfaffian in $\mathbb{P}^6$ ($A^2_{(-2)^7,3,1^4},\,r\ll 0$ in~\cite{Hori:2013gga}) \cite{Kanazawa:2012xya,Tonoli:2004}\\\hline
                {\begin{align*}
                        \mathcal{D}=&169 \Theta ^4-13 z \left(260+2158 \Theta +6679 \Theta ^2+9042 \Theta ^3+4569 \Theta ^4\right)\\
                        &+16 z^2 \left(-2119-11596 \Theta -17898 \Theta ^2-1774 \Theta ^3+6386 \Theta ^4\right)\\
                        &+256 z^3 \left(26+312 \Theta +1091 \Theta ^2+1248 \Theta ^3+67 \Theta ^4\right)\\
                        &-4096 z^4 (1+2
                        \Theta )^4\\[.8em]
                        \Delta=&1-349 z-256 z^2
                \end{align*}}
                \begin{center}
                $\chi=-120\,,\quad h^{1,1}=1\,,\quad h^{2,1}=61\,,\quad J^3=13\,,\quad  c_2\cdot J=58$
                \end{center}\\\hline
                \begin{equation*}
                \begin{gathered}
                        s^z_{zz}=-\frac{30}{17}\frac{1}{z}\,,\quad h^{zz}_z=\frac{z(31+328z)}{1445}\,,\quad h_{zz}=\frac{13}{85}\frac{1}{z^2}\,,\\ h_z^z=0\,,\quad h_{z}=\frac{13-361 z}{14450 z}
                \end{gathered}
                \end{equation*}\\\hline
	{\begin{gather*}
\begin{array}{c|cccccc}
n^{\beta}_g&\beta=1&2&3&4&5&6\\\hline
g=0&647&16166&942613&80218296&8418215008&1010635571652\\
1&0&0&176&164696&78309518&26889884396\\
2&0&0&0&0&15399&65099510\\
3&0&0&0&0&0&432\\
4&0&0&0&0&0&-4\\
5&0&0&0&0&0&0\\
6&0&0&0&0&0&0\\
7&0&0&0&0&0&0
\end{array}\\
\begin{array}{c|ccccc}
n^{\beta}_g&\beta=7&8&9&10&11\\\hline
4&1736&79462734&1255525330364&4040105994946376&6399989953841040582\\
5&-4&46197&73497804&8375280645340&55360448310349962\\
6&0&0&494088&-1044672290&78614584131689\\
7&0&0&0&2876415&-44378756124\\
8&0&0&0&1056&4158795\\
9&0&0&0&0&96996\\
10&0&0&0&0&0
\end{array}
	\end{gather*}}\\
                \hline
\end{tabular}

\begin{tabular}{|p{\linewidth}|}
        \hline AESZ 207 - nc-res. of $X_8$ w/ $88$ nodes  ($A^2_{(-2)^7,3,1^4},\,r\gg 0$ in~\cite{Hori:2013gga}) \cite{Kanazawa:2012xya,Tonoli:2004} \\
        The nodal CY 3-fold $X_8$ is a degree $8$ hypersurface in $\mathbb{P}_{11114}$.\\\hline
                                \begin{center}
                Hybrid phase of AESZ 99: $v=i 2^{-8} z^{-\frac12}\,,\quad f(v)=2^{-7}v^{-1}$
        \end{center}
                {\begin{align*}
                        \mathcal{D}=&\Theta ^4-256 v^2 \left(145+988 \Theta +2722 \Theta ^2+2228 \Theta ^3-67 \Theta ^4\right)\\
                        &+2097152 v^4 \left(1397+7114 \Theta +11316 \Theta ^2-14546 \Theta ^3-3193 \Theta ^4\right)\\
                        &-55834574848 v^6 \left(97+192 \Theta -122 \Theta ^2+192 \Theta ^3+4569 \Theta ^4\right)\\
                        &-47569271064100864 v^8 (1+\Theta )^4\\[.8em]
                        \Delta=&1-89344 v^2-16777216 v^4
                \end{align*}}
                \begin{center}
                $\chi=-120\,,\quad h^{1,1}=1\,,\quad h^{2,1}=61\,,\quad J^3=2\,,\quad  c_2\cdot J=44$
                \end{center}\\\hline
                \begin{equation*}
                \begin{gathered}
                        s^v_{vv}=-\frac{25}{17}\frac{1}{v}\,,\quad h^{vv}_v=\frac{v \left(41-253952 v^2\right)}{1445}\,,\quad h_{vv}=\frac{7}{85}\frac{1}{v^2}\,,\\[.4em]
                        h_v^v=\frac{-41+253952 v^2}{1445}\,,\quad h_v=\frac{49-3391488 v^2}{28900 v}
                \end{gathered}
                \end{equation*}\\\hline
	{\small\begin{gather*}
\begin{array}{c|ccccc}
n^{\beta,0}_g&\beta=1&2&3&4&5\\\hline
g=0&\textbf{14752}&64424480&711860273440&11596528017757856&233938237312624658400\\
1&\textbf{0}&19712&10732175296&902646051735936&50712027457008177856\\
2&\textbf{0}&\textbf{528}&-8275872&6249830789008&2700746768622436448\\
3&\textbf{0}&\textbf{0}&\textbf{-88512}&-87428815312&10292236849965248\\
4&\textbf{0}&\textbf{0}&\textbf{0}&197980392&-337281112359424\\
5&\textbf{0}&\textbf{0}&\textbf{0}&161706&6031964134528\\
6&\textbf{0}&\textbf{0}&\textbf{0}&\textbf{1512}&-43153905216\\
7&\textbf{0}&\textbf{0}&\textbf{0}&\textbf{24}&18764544\\
8&\textbf{0}&\textbf{0}&\textbf{0}&\textbf{0}&\textbf{177024}\\
9&\textbf{0}&\textbf{0}&\textbf{0}&\textbf{0}&\textbf{0}
\end{array}\\
\begin{array}{c|ccccc}
n^{\beta,1}_g&\beta=1&2&3&4&5\\\hline
g=0&\textbf{14752}&64410432&711860273440&11596528007035456&233938237312624658400\\
1&\textbf{0}&21600&10732175296&902646040969920&50712027457008177856\\
2&\textbf{0}&\textbf{336}&-8275872&6249836488736&2700746768622436448\\
3&\textbf{0}&\textbf{6}&\textbf{-88512}&-87430688512&10292236849965248\\
4&\textbf{0}&\textbf{0}&\textbf{0}&198235408&-337281112359424\\
5&\textbf{0}&\textbf{0}&\textbf{0}&139744&6031964134528\\
6&\textbf{0}&\textbf{0}&\textbf{0}&\textbf{2640}&-43153905216\\
7&\textbf{0}&\textbf{0}&\textbf{0}&\textbf{0}&18764544\\
8&\textbf{0}&\textbf{0}&\textbf{0}&\textbf{0}&\textbf{177024}\\
9&\textbf{0}&\textbf{0}&\textbf{0}&\textbf{0}&\textbf{0}
\end{array}
	\end{gather*}}\\
                \hline
\end{tabular}

\begin{tabular}{|p{\linewidth}|}
        \hline AESZ 210 - Pfaffian in $\mathbb{P}^6_{1111112}$ ($A^2_{(-1)^6,(-2),1^4,0},\,r\ll 0$ in~\cite{Hori:2013gga}), \cite{Kanazawa:2012xya}\\\hline
                {\begin{align*}
                        \mathcal{D}=&25 \Theta ^4+20 z \left(35+305 \Theta +981 \Theta ^2+1352 \Theta ^3+688 \Theta ^4\right)\\
                        &-16 z^2 \left(265+1260 \Theta -96 \Theta ^2-7008 \Theta ^3-5856 \Theta ^4\right)\\
                        &+1024 z^3 \left(5+30 \Theta +69 \Theta ^2+120 \Theta ^3+176 \Theta ^4\right)+4096 z^4 (1+2 \Theta )^4\\[.8em]
                        \Delta=&1-544 z+256 z^2
                \end{align*}}
                \begin{center}
                $\chi=-116\,,\quad h^{1,1}=1\,,\quad h^{2,1}=59\,,\quad J^3=10\,,\quad  c_2\cdot J=52$
                \end{center}\\\hline
                \begin{equation*}
                \begin{gathered}
                        s^z_{zz}=-\frac{13}{8}\frac{1}{z}\,,\quad h^{zz}_z=\frac{1}{384} z (7-176 z)\,,\quad h_{zz}=\frac{5}{48}\frac{1}{z^2}\,,\\ h_z^z=0\,,\quad h_{z}=\frac{5+464 z}{9216 z}
                \end{gathered}
                \end{equation*}\\\hline
                {\tiny\begin{gather*}\begin{array}{c|cccccc}
n^{\beta}_g&\beta=1&2&3&4&5&6\\\hline
g=0&888&33084&3003816&399931068&65736977760&12365885835028\\
1&0&1&2496&2089393&1210006912&571433267123\\
2&0&0&0&140&1738912&4201109240\\
3&0&0&0&0&0&2103394\\
4&0&0&0&0&0&420\\
5&0&0&0&0&0&0
\end{array}\\
\begin{array}{c|ccccc}
n^{\beta}_g&\beta=7&8&9&10&11\\\hline
4&2980704&169405136180&1152960131742408&3316792484212678360&5990889032340795572144\\
5&7488&-18725646&2029562641360&30712200099903339&143418263870959095504\\
6&0&99528&-1351743312&36741280919032&1224321329388729640\\
7&0&6&231264&-75191142894&1121354031039824\\
8&0&0&5328&-743012&-4046013868992\\
9&0&0&0&20306&2016598080\\
10&0&0&0&860&-867792
\end{array}\end{gather*}}\\
                \hline
\end{tabular}

\begin{tabular}{|p{\linewidth}|}
        \hline AESZ 211 - nc-res. of $X_{26}$ in $\mathbb{P}^5_{111113}$ w/ $70$ nodes ($A^2_{(-1)^6,(-2),1^4,0},\,r\gg 0$ in~\cite{Hori:2013gga}),
\cite{Kanazawa:2012xya}\\
        The nodal CY 3-fold $X_{26}$ is a degree $(2,6)$ complete intersection in $\mathbb{P}_{111113}$.\\\hline
                                \begin{center}
                Hybrid phase of AESZ 210: $v=2^{-6} z^{-\frac12}\,,\quad f(v)=2^{-5}v^{-1}$
        \end{center}
                {\begin{align*}
                        \mathcal{D}=&\Theta ^4-256 v^2 \left(145+988 \Theta +2722 \Theta ^2+2228 \Theta ^3-67 \Theta ^4\right)\\
                        &+2097152 v^4 \left(1397+7114 \Theta +11316 \Theta ^2-14546 \Theta ^3-3193 \Theta ^4\right)\\
                        &-55834574848 v^6 \left(97+192 \Theta -122 \Theta ^2+192 \Theta ^3+4569 \Theta ^4\right)\\
                        &-47569271064100864 v^8 (1+\Theta )^4\\[.8em]
                        \Delta=&(1 - 96 v + 256 v^2) (1 + 96 v + 256 v^2)
                \end{align*}}
                \begin{center}
                $\chi=-116\,,\quad h^{1,1}=1\,,\quad h^{2,1}=59\,,\quad J^3=4\,,\quad  c_2\cdot J=52$
                \end{center}\\\hline
                \begin{center}
                        $s^v_{vv}=-\frac{7}{4}\frac{1}{v}\,,\quad h^{vv}_v=\frac{1}{192} v \left(11-1792 v^2\right)\,,\quad h_{vv}=\frac{1}{6}\frac{1}{v^2}\,,$\\[.4em]
                        $h_v^v=\frac{1}{192} \left(-11+1792 v^2\right)\,,\quad h_v=\frac{1-1664 v^2}{288 v}$
                \end{center}\\\hline
	{\small\begin{gather*}
\begin{array}{c|ccccc}
n^{\beta,0}_g&\beta=1&2&3&4&5\\\hline
g=0&\textbf{2496}&1194672&1366030016&2299808590320&4789856923533120\\
1&\textbf{0}&\textbf{-280}&614016&39637899984&316537005217920\\
2&\textbf{0}&\textbf{0}&\textbf{7488}&-6555216&1960917715008\\
3&\textbf{0}&\textbf{0}&\textbf{0}&44536&-2865875584\\
4&\textbf{0}&\textbf{0}&\textbf{0}&\textbf{616}&-3549312\\
5&\textbf{0}&\textbf{0}&\textbf{0}&\textbf{10}&\textbf{-29952}\\
6&\textbf{0}&\textbf{0}&\textbf{0}&\textbf{0}&\textbf{0}\\
\end{array}\\
\begin{array}{c|ccccc}
n^{\beta,1}_g&\beta=1&2&3&4&5\\\hline
g=0&\textbf{2496}&1194096&1366030016&2299807973904&4789856923533120\\
1&\textbf{0}&\textbf{-224}&614016&39637764816&316537005217920\\
2&\textbf{0}&\textbf{-4}&\textbf{7488}&-6543472&1960917715008\\
3&\textbf{0}&\textbf{0}&\textbf{0}&42840&-2865875584\\
4&\textbf{0}&\textbf{0}&\textbf{0}&\textbf{840}&-3549312\\
5&\textbf{0}&\textbf{0}&\textbf{0}&\textbf{0}&\textbf{-29952}\\
6&\textbf{0}&\textbf{0}&\textbf{0}&\textbf{0}&\textbf{0}\\
\end{array}
	\end{gather*}}\\
                \hline
\end{tabular}

\begin{tabular}{|p{\linewidth}|}
        \hline AESZ 238 - Pfaffian in $\mathbb{P}^6_{1111222}$  ($A^2_{(-1)^4,(-2)^3,1^5},\,r\ll 0$ in~\cite{Hori:2013gga}), \cite{Kanazawa:2012xya}\\\hline
                {\begin{align*}
                        \mathcal{D}=&\Theta ^4-4 z \left(19+189 \Theta +677 \Theta ^2+976 \Theta ^3+500 \Theta ^4\right)\\
                        &-16 z^2 \left(177+1164 \Theta +1336 \Theta ^2-3968 \Theta ^3-3968 \Theta ^4\right)\\
                        &-1024 z^3 \left(3+6 \Theta -37 \Theta ^2+24 \Theta ^3+500 \Theta ^4\right)+4096 z^4 (1+2 \Theta )^4\\[.8em]
                        \Delta=&1-1968 z+256 z^2
                \end{align*}}
                \begin{center}
                $\chi=-100\,,\quad h^{1,1}=1\,,\quad h^{2,1}=51\,,\quad J^3=5\,,\quad  c_2\cdot J=38$
                \end{center}\\\hline
                \begin{equation*}
                \begin{gathered}
                        s^z_{zz}=-\frac{3}{2}\frac{1}{z}\,,\quad h^{zz}_z=\frac{1}{44} z (1-16 z)\,,\quad h_{zz}=\frac{3}{44}\frac{1}{z^2}\,,\\ h_z^z=0\,,\quad h_{z}=\frac{9+736 z}{19360 z}
                \end{gathered}
                \end{equation*}\\\hline
                {\tiny\begin{gather*}
\begin{array}{c|cccccc}
n^{\beta}_g&\beta=1&2&3&4&5&6\\\hline
g=0&2220&285520&95254820&47164553340&28906372957040&20291945542090480\\
1&0&460&873240&1498922677&2306959237408&3311397402449740\\
2&0&0&460&7244660&44839436460&168980152301480\\
3&0&0&0&2020&146398880&2684226893050\\
4&0&0&0&0&-47520&5659459680\\
5&0&0&0&0&240&-14812160\\
6&0&0&0&0&6&-13020\\
7&0&0&0&0&0&0
\end{array}\\
\begin{array}{c|cccc}
n^{\beta}_g&\beta=7&8&9&10\\\hline
6&-3483897700&145720463906800&25318518791299955638&696922725378245090263272\\
7&3565520&-666685721430&109075901600511700&15690823158071372378100\\
8&3680&5248416060&-94782885774180&122115907922680670540\\
9&0&-4739915&2656979965100&169414363367059780\\
10&0&-4600&-22060519240&1167202925388120
\end{array}
                \end{gather*}}\\
                \hline
\end{tabular}

\begin{tabular}{|p{\linewidth}|}
        \hline nc-res. of sing. double cover of $V_5$ w/ $50$ nodes ($A^2_{(-1)^4,(-2)^3,1^5},\,r\gg 0$ in~\cite{Hori:2013gga}), \cite{Kanazawa:2012xya}\\The Fano 3-fold $V_5$ is an intersection of three hyperplanes in $G(2,5)\subset\mathbb{P}^6$.\\\hline
                                {\small\begin{center}
                Hybrid phase of AESZ 238: $v=-2^{-4} z^{-\frac12}\,,\quad f(v)=2^{-3}v^{-1}$
        \end{center}
                {\begin{align*}
                        \mathcal{D}=&\Theta ^4-256 v^2 \left(145+988 \Theta +2722 \Theta ^2+2228 \Theta ^3-67 \Theta ^4\right)\\
                        &+2097152 v^4 \left(1397+7114 \Theta +11316 \Theta ^2-14546 \Theta ^3-3193 \Theta ^4\right)\\
                        &-55834574848 v^6 \left(97+192 \Theta -122 \Theta ^2+192 \Theta ^3+4569 \Theta ^4\right)\\
                        &-47569271064100864 v^8 (1+\Theta )^4\\[.8em]
                        \Delta=&\left(1-44 v-16 v^2\right) \left(1+44 v-16 v^2\right)
                \end{align*}}
                \begin{center}
                $\chi=-100\,,\quad h^{1,1}=1\,,\quad h^{2,1}=51\,,\quad J^3=10\,,\quad  c_2\cdot J=64$
                \end{center}}\\\hline
                {\small\begin{center}
                        $s^v_{vv}=-\frac{2}{v}\,,\quad h^{vv}_v=\frac{1}{22} v (1-4 v) (1+4 v)\,,\quad h_{vv}=\frac{3}{11}\frac{1}{v^2}\,,$\\[.4em]
                        $h_v^v=\frac{1}{22} (-1+4 v) (1+4 v)\,,\quad h_v=\frac{(3-32 v) (3+32 v)}{2420 v}$
                \end{center}}\\\hline
                {\small\begin{gather*}
                    \begin{array}{c|cccccc}
n^{\beta,0}_g&\beta=1&2&3&4&5&6\\\hline
g=0&460&25480&2926980&499942620&105270795760&25363208008280\\
1&0&0&-920&170629&973633792&823834160680\\
2&0&0&0&280&-173420&529051120\\
3&0&0&0&5&-1840&515340\\
4&0&0&0&0&0&800
\end{array}\\
\begin{array}{c|cccccc}
n^{\beta,1}_g&\beta=1&2&3&4&5&6\\\hline
g=0&460&25040&2926980&499877500&105270795760&25363184926320\\
1&0&3&-920&169380&973633792&823834715633\\
2&0&0&0&300&-173420&529098160\\
3&0&0&0&0&-1840&519360\\
4&0&0&0&0&0&720
\end{array}\\
\begin{array}{c|ccccc}
n^{\beta,0}_g&\beta=7&8&9&10&11\\\hline
5&-5520&16447785&-216664462920&9421399885445840&205509186464370358520\\
6&0&151680&-94370380&-19206528701880&250629169915167300\\
7&0&30&465520&-1424809220&-990189801672040\\
8&0&0&4600&-2125580&323552730460\\
9&0&0&0&71000&77098760\\
10&0&0&0&1500&-1529500
\end{array}\\
\begin{array}{c|ccccc}
n^{\beta,1}_g&\beta=7&8&9&10&11\\\hline
5&-5520&16408620&-216664462920&9421399622679790&205509186464370358520\\
6&0&154680&-94370380&-19206500030352&250629169915167300\\
7&0&0&465520&-1424838360&-990189801672040\\
8&0&0&4600&-2137980&323552730460\\
9&0&0&0&73635&77098760\\
10&0&0&0&1080&-1529500
\end{array}
                \end{gather*}}\\
                \hline
\end{tabular}

\begin{tabular}{|p{\linewidth}|}
        \hline AESZ 51 - Smooth double cover of $V_5$\\\hline
                {\begin{align*}
                        \mathcal{D}=&\Theta ^4-4 z (1+4 \Theta ) (3+4 \Theta ) \left(3+11 \Theta +11 \Theta ^2\right)\\
                        &-16 z^2 (1+4 \Theta ) (3+4 \Theta ) (5+4 \Theta ) (7+4 \Theta )\\[.8em]
                        \Delta=&1-704 z-4096 z^2
                \end{align*}}
                \begin{center}
                $\chi=-200\,,\quad h^{1,1}=1\,,\quad h^{2,1}=101\,,\quad J^3=10\,,\quad  c_2\cdot J=64$
                \end{center}\\\hline
                \begin{center}
                        $s^z_{zz}=-\frac{2}{z}\,,\quad h^{zz}_z=\frac{1}{16} z (1-96 z)\,,\quad h_{zz}=\frac{3}{16}\frac{1}{z^2}\,,\quad h_z^z=0\,,\quad h_{z}=\frac{9 (1+320 z)}{5120 z}$
                \end{center}\\\hline
                {\small\begin{gather*}
\begin{array}{c|cccccc}
n^{\beta}_g&\beta=1&2&3&4&5&6\\\hline
g=0&920&50520&5853960&999820120&210541591520&50726392934600\\
1&0&3&-1840&340009&1947267584&1647668876313\\
2&0&0&0&580&-346840&1058149280\\
3&0&0&0&5&-3680&1034700\\
4&0&0&0&0&0&1520\\
5&0&0&0&0&0&10\\
6&0&0&0&0&0&0
\end{array}\\
\begin{array}{c|ccccc}
n^{\beta}_g&\beta=7&8&9&10&11\\\hline
5&-11040&32856405&-433328925840&18842799508125630&411018372928740717040\\
6&0&306360&-188740760&-38413028732232&501258339830334600\\
7&0&30&931040&-2849647580&-1980379603344080\\
8&0&0&9200&-4263560&647105460920\\
9&0&0&0&144635&154197520\\
10&0&0&0&2580&-3059000
\end{array}
\end{gather*}}\\
                \hline
\end{tabular}

\section{GV-invariants for an example by Calabrese-Thomas}
\label{app:ctgv}

\begin{table}[h!]
        {\tiny\begin{align*}
                \begin{array}{c|ccccc}
                n^{(0)}_{(d_1,d_2),0}&d_2=0&1&2&3&4\\\hline
                        d_1=0 &0& 1248 & 112176 & 19318752 & 4550120880 \\
                        1&144 & 108000 & 76015872 & 50010357888 & 31665114116352 \\
                        2&132 & 3195648 & 9819373632 & 17416078783488 & 23521041703147584 \\
                        3&144 & 52492416 & 590225421312 & 2481268636257888 & 6512923728628208640 \\
                        4&120 & 604668672 & 21599507584896 & 200528969095590912 & 968692939329591904128 \\
                        5&144 & 5458960704 & 553633466992128 & 10754603799591217920 & 92103549441254145750528 \\
                \end{array}
        \end{align*}}
        \caption{Genus 0 GV-invariants for singular double cover of $\mathbb{P}^1\times\mathbb{P}^2$ with $\mathbb{Z}_2$ charge 0.} \label{tab:ctg0c0}
\end{table}
\begin{table}[h!]
        {\tiny\begin{align*}
                \begin{array}{c|ccccc}
                n^{(0)}_{(d_1,d_2),1}&d_2=0&1&2&3&4\\\hline
                        d_1=0 &0& 1248 & 111576 & 19318752 & 4550104104 \\
                        1&144 & 108576 & 76015872 & 50010687360 & 31665114116352 \\
                        2&120 & 3195648 & 9819273216 & 17416078783488 & 23521041440903040 \\
                        3&144 & 52502400 & 590225421312 & 2481268714751904 & 6512923728628208640 \\
                        4&132 & 604668672 & 21599502154176 & 200528969095590912 & 968692939259940720192 \\
                        5&144 & 5459022528 & 553633466992128 & 10754603805582771456 & 92103549441254145750528 \\
                \end{array}
        \end{align*}}
        \caption{Genus 0 GV-invariants for singular double cover of $\mathbb{P}^1\times\mathbb{P}^2$ with $\mathbb{Z}_2$ charge 1.} \label{tab:ctg0c1}
\end{table}
\begin{table}[h!]
        {\tiny\begin{align*}
                \begin{array}{c|ccccc}
                n^{(1)}_{(d_1,d_2),0}&d_2=0&1&2&3&4\\\hline
 d_1=0 & 0 & 0 & -264 & -732992 & -521459592 \\
 1 & 0 & 0 & 34560 & 132625152 & 180488387520 \\
 2 & 0 & -2496 & 52493136 & 628502038272 & 2145399128655120 \\
 3 & 0 & -216000 & 5765767680 & 175671831427776 & 1143574332713684736 \\
 4 & 3 & -6398784 & 291281518752 & 21326329807527168 & 253191167583151600992 \\
 5 & 0 & -105636288 & 9098772169728 & 1531300785023621376 & 32102687260579360247424 \\
                \end{array}
        \end{align*}}
        \caption{Genus 1 GV-invariants for singular double cover of $\mathbb{P}^1\times\mathbb{P}^2$ with $\mathbb{Z}_2$ charge 0.}
\end{table}
\begin{table}[h!]
        {\tiny\begin{align*}
                \begin{array}{c|ccccc}
                n^{(1)}_{(d_1,d_2),1}&d_2=0&1&2&3&4\\\hline
 d_1=0 & 0 & 0 & -228 & -732992 & -521483928 \\
 1 & 0 & 0 & 34560 & 132611328 & 180488387520 \\
 2 & 3 & -2496 & 52489152 & 628502038272 & 2145399116501472 \\
 3 & 0 & -217152 & 5765767680 & 175671825357120 & 1143574332713684736 \\
 4 & 0 & -6398784 & 291281413488 & 21326329807527168 & 253191167570157714768 \\
 5 & 0 & -105652800 & 9098772169728 & 1531300784441988864 & 32102687260579360247424 \\
                \end{array}
        \end{align*}}
        \caption{Genus 1 GV-invariants for singular double cover of $\mathbb{P}^1\times\mathbb{P}^2$ with $\mathbb{Z}_2$ charge 1.}
\end{table}
\begin{table}[h!]
        \begin{align*}
                \begin{array}{c|ccccc}
                n^{(g)}_{(d_1,1),0}&g=0&1&2&3&4\\\hline
 d_1=0 & 1248 & 0 & 0 & 0 & 0 \\
 1 & 108576 & 0 & 0 & 0 & 0 \\
 2 & 3195648 & -2496 & 0 & 0 & 0 \\
 3 & 52502400 & -216000 & 0 & 0 & 0 \\
 4 & 604668672 & -6398784 & 3744 & 0 & 0 \\
 5 & 5459022528 & -105636288 & 325728 & 0 & 0 \\
 6 & 41139649536 & -1228521216 & 9606912 & -4992 & 0 \\
 7 & 269250770688 & -11233799808 & 159235200 & -432000 & 0 \\
 8 & 1571999613696 & -85932893760 & 1865192544 & -12820032 & 6240 \\
                \end{array}
        \end{align*}
        \caption{Genus $g$ GV-invariants at base degree $1$ for singular double cover of $\mathbb{P}^1\times\mathbb{P}^2$ with $\mathbb{Z}_2$ charge 0.}
\end{table}
\begin{table}[h!]
        \begin{align*}
                \begin{array}{c|ccccc}
                n^{(g)}_{(d_1,1),1}&g=0&1&2&3&4\\\hline
 d_1=0 & 1248 & 0 & 0 & 0 & 0 \\
 1 & 108000 & 0 & 0 & 0 & 0 \\
 2 & 3195648 & -2496 & 0 & 0 & 0 \\
 3 & 52492416 & -217152 & 0 & 0 & 0 \\
 4 & 604668672 & -6398784 & 3744 & 0 & 0 \\
 5 & 5458960704 & -105652800 & 324000 & 0 & 0 \\
 6 & 41139649536 & -1228521216 & 9606912 & -4992 & 0 \\
 7 & 269250394368 & -11233868160 & 159214464 & -434304 & 0 \\
 8 & 1571999613696 & -85932893760 & 1865192544 & -12820032 & 6240 \\
                \end{array}
        \end{align*}
        \caption{Genus $g$ GV-invariants at base degree $1$ for singular double cover of $\mathbb{P}^1\times\mathbb{P}^2$ with $\mathbb{Z}_2$ charge 1.} \label{tab:ctg1c1}
\end{table}

\addcontentsline{toc}{section}{References}
\bibliography{names}


\begin{thebibliography}{160}
\ifx \bisbn   \undefined \def \bisbn  #1{ISBN #1}\fi
\ifx \binits  \undefined \def \binits#1{#1}\fi
\ifx \bauthor  \undefined \def \bauthor#1{#1}\fi
\ifx \batitle  \undefined \def \batitle#1{#1}\fi
\ifx \bjtitle  \undefined \def \bjtitle#1{#1}\fi
\ifx \bvolume  \undefined \def \bvolume#1{\textbf{#1}}\fi
\ifx \byear  \undefined \def \byear#1{#1}\fi
\ifx \bissue  \undefined \def \bissue#1{#1}\fi
\ifx \bfpage  \undefined \def \bfpage#1{#1}\fi
\ifx \blpage  \undefined \def \blpage #1{#1}\fi
\ifx \burl  \undefined \def \burl#1{\textsf{#1}}\fi
\ifx \doiurl  \undefined \def \doiurl#1{\url{https://doi.org/#1}}\fi
\ifx \betal  \undefined \def \betal{\textit{et al.}}\fi
\ifx \binstitute  \undefined \def \binstitute#1{#1}\fi
\ifx \binstitutionaled  \undefined \def \binstitutionaled#1{#1}\fi
\ifx \bctitle  \undefined \def \bctitle#1{#1}\fi
\ifx \beditor  \undefined \def \beditor#1{#1}\fi
\ifx \bpublisher  \undefined \def \bpublisher#1{#1}\fi
\ifx \bbtitle  \undefined \def \bbtitle#1{#1}\fi
\ifx \bedition  \undefined \def \bedition#1{#1}\fi
\ifx \bseriesno  \undefined \def \bseriesno#1{#1}\fi
\ifx \blocation  \undefined \def \blocation#1{#1}\fi
\ifx \bsertitle  \undefined \def \bsertitle#1{#1}\fi
\ifx \bsnm \undefined \def \bsnm#1{#1}\fi
\ifx \bsuffix \undefined \def \bsuffix#1{#1}\fi
\ifx \bparticle \undefined \def \bparticle#1{#1}\fi
\ifx \barticle \undefined \def \barticle#1{#1}\fi
\bibcommenthead
\ifx \bconfdate \undefined \def \bconfdate #1{#1}\fi
\ifx \botherref \undefined \def \botherref #1{#1}\fi
\ifx \url \undefined \def \url#1{\textsf{#1}}\fi
\ifx \bchapter \undefined \def \bchapter#1{#1}\fi
\ifx \bbook \undefined \def \bbook#1{#1}\fi
\ifx \bcomment \undefined \def \bcomment#1{#1}\fi
\ifx \oauthor \undefined \def \oauthor#1{#1}\fi
\ifx \citeauthoryear \undefined \def \citeauthoryear#1{#1}\fi
\ifx \endbibitem  \undefined \def \endbibitem {}\fi
\ifx \bconflocation  \undefined \def \bconflocation#1{#1}\fi
\ifx \arxivurl  \undefined \def \arxivurl#1{\textsf{#1}}\fi
\csname PreBibitemsHook\endcsname

\bibitem{hori2003mirror}
\begin{bbook}
\bauthor{\bsnm{Hori}, \binits{K.}},
\bauthor{\bsnm{Katz}, \binits{S.}},
\bauthor{\bsnm{Vafa}, \binits{C.}},
\bauthor{\bsnm{Thomas}, \binits{R.}},
\bauthor{\bsnm{Pandharipande}, \binits{R.}},
\bauthor{\bsnm{Klemm}, \binits{A.}}:
\bbtitle{Mirror Symmetry}.
\bsertitle{Clay mathematics monographs}.
\bpublisher{American Mathematical Society},
\blocation{Providence, RI}
(\byear{2003})
\end{bbook}
\endbibitem

\bibitem{Neitzke:2004ni}
\begin{botherref}
\oauthor{\bsnm{Neitzke}, \binits{A.}},
\oauthor{\bsnm{Vafa}, \binits{C.}}:
{Topological strings and their physical applications}
(2004)
{\href{https://arxiv.org/abs/hep-th/0410178}{{arXiv:hep-th/0410178}}}
\end{botherref}
\endbibitem

\bibitem{Aspinwall:1993nu}
\begin{barticle}
\bauthor{\bsnm{Aspinwall}, \binits{P.S.}},
\bauthor{\bsnm{Greene}, \binits{B.R.}},
\bauthor{\bsnm{Morrison}, \binits{D.R.}}:
\batitle{{Calabi-Yau moduli space, mirror manifolds and space-time topology
  change in string theory}}.
\bjtitle{Nucl. Phys. B}
\bvolume{416},
\bfpage{414}--\blpage{480}
(\byear{1994})
{\href{https://arxiv.org/abs/hep-th/9309097}{{arXiv:hep-th/9309097}}}.
\doiurl{10.1016/0550-3213(94)90321-2}
\end{barticle}
\endbibitem

\bibitem{Witten:1993yc}
\begin{barticle}
\bauthor{\bsnm{Witten}, \binits{E.}}:
\batitle{{Phases of N=2 theories in two-dimensions}}.
\bjtitle{Nucl. Phys. B}
\bvolume{403},
\bfpage{159}--\blpage{222}
(\byear{1993})
{\href{https://arxiv.org/abs/hep-th/9301042}{{arXiv:hep-th/9301042}}}.
\doiurl{10.1016/0550-3213(93)90033-L}
\end{barticle}
\endbibitem

\bibitem{Schimannek:2021pau}
\begin{barticle}
\bauthor{\bsnm{Schimannek}, \binits{T.}}:
\batitle{{Modular curves, the Tate-Shafarevich group and Gopakumar-Vafa
  invariants with discrete charges}}.
\bjtitle{JHEP}
\bvolume{02},
\bfpage{007}
(\byear{2022})
{\href{https://arxiv.org/abs/2108.09311}{{arXiv:2108.09311}}}
{[hep-th]}.
\doiurl{10.1007/JHEP02(2022)007}
\end{barticle}
\endbibitem

\bibitem{kuznetsov2008lefschetz}
\begin{barticle}
\bauthor{\bsnm{Kuznetsov}, \binits{A.}}:
\batitle{Lefschetz decompositions and categorical resolutions of
  singularities}.
\bjtitle{Selecta Math. (N.S.)}
\bvolume{13}(\bissue{4}),
\bfpage{661}--\blpage{696}
(\byear{2008})
{\href{https://arxiv.org/abs/math/0609240}{{arXiv:math/0609240}}}.
\doiurl{10.1007/s00029-008-0052-1}
\end{barticle}
\endbibitem

\bibitem{thomas-rev}
\begin{bchapter}
\bauthor{\bsnm{Thomas}, \binits{R.P.}}:
\bctitle{Notes on homological projective duality}.
In: \bbtitle{Algebraic Geometry: {S}alt {L}ake {C}ity 2015}.
\bsertitle{Proc. Sympos. Pure Math.},
vol. \bseriesno{97},
pp. \bfpage{585}--\blpage{609}.
\bpublisher{Amer. Math. Soc.},
\blocation{Providence, RI}
(\byear{2018})
\end{bchapter}
\endbibitem

\bibitem{bondalorlov}
\begin{botherref}
\oauthor{\bsnm{Bondal}, \binits{A.}},
\oauthor{\bsnm{Orlov}, \binits{D.}}:
Derived categories of coherent sheaves
(2002).
\doiurl{10.48550/ARXIV.MATH/0206295}
\end{botherref}
\endbibitem

\bibitem{vdb02}
\begin{bchapter}
\bauthor{\bparticle{van~den} \bsnm{Bergh}, \binits{M.}}:
\bctitle{Non-commutative crepant resolutions}.
In: \bbtitle{The Legacy of {Niels} {Henrik} {Abel}},
pp. \bfpage{749}--\blpage{770}.
\bpublisher{Springer},
\blocation{Berlin}
(\byear{2004})
\end{bchapter}
\endbibitem

\bibitem{vdB:nc-rev}
\begin{botherref}
\oauthor{\bparticle{Van~den} \bsnm{Bergh}, \binits{M.}}:
Non-commutative crepant resolutions, an overview
{\href{https://arxiv.org/abs/2207.09703}{{arXiv:2207.09703}}}
{[math]}
\end{botherref}
\endbibitem

\bibitem{Roggenkamp:2003qp}
\begin{barticle}
\bauthor{\bsnm{Roggenkamp}, \binits{D.}},
\bauthor{\bsnm{Wendland}, \binits{K.}}:
\batitle{{Limits and degenerations of unitary conformal field theories}}.
\bjtitle{Commun. Math. Phys.}
\bvolume{251},
\bfpage{589}--\blpage{643}
(\byear{2004})
{\href{https://arxiv.org/abs/hep-th/0308143}{{arXiv:hep-th/0308143}}}.
\doiurl{10.1007/s00220-004-1131-6}
\end{barticle}
\endbibitem

\bibitem{Roggenkamp:2008jm}
\begin{barticle}
\bauthor{\bsnm{Roggenkamp}, \binits{D.}},
\bauthor{\bsnm{Wendland}, \binits{K.}}:
\batitle{{Decoding the geometry of conformal field theories}}.
\bjtitle{Bulg. J. Phys.}
\bvolume{35},
\bfpage{139}--\blpage{150}
(\byear{2008})
{\href{https://arxiv.org/abs/0803.0657}{{arXiv:0803.0657}}}
{[hep-th]}
\end{barticle}
\endbibitem

\bibitem{Connes:1994yd}
\begin{bbook}
\bauthor{\bsnm{Connes}, \binits{A.}}:
\bbtitle{Noncommutative Geometry}.
\bpublisher{Academic Press},
\blocation{San Diego}
(\byear{1994})
\end{bbook}
\endbibitem

\bibitem{Seiberg:1999vs}
\begin{barticle}
\bauthor{\bsnm{Seiberg}, \binits{N.}},
\bauthor{\bsnm{Witten}, \binits{E.}}:
\batitle{{String theory and noncommutative geometry}}.
\bjtitle{JHEP}
\bvolume{09},
\bfpage{032}
(\byear{1999})
{\href{https://arxiv.org/abs/hep-th/9908142}{{arXiv:hep-th/9908142}}}.
\doiurl{10.1088/1126-6708/1999/09/032}
\end{barticle}
\endbibitem

\bibitem{Connes:1987ue}
\begin{barticle}
\bauthor{\bsnm{Connes}, \binits{A.}},
\bauthor{\bsnm{Rieffel}, \binits{M.A.}}:
\batitle{{Yang-Mills for noncommutative two-tori}}.
\bjtitle{Contemp. Math.}
\bvolume{62},
\bfpage{237}--\blpage{266}
(\byear{1987})
\end{barticle}
\endbibitem

\bibitem{Connes:1997cr}
\begin{barticle}
\bauthor{\bsnm{Connes}, \binits{A.}},
\bauthor{\bsnm{Douglas}, \binits{M.R.}},
\bauthor{\bsnm{Schwarz}, \binits{A.S.}}:
\batitle{{Noncommutative geometry and matrix theory: Compactification on
  tori}}.
\bjtitle{JHEP}
\bvolume{02},
\bfpage{003}
(\byear{1998})
{\href{https://arxiv.org/abs/hep-th/9711162}{{arXiv:hep-th/9711162}}}.
\doiurl{10.1088/1126-6708/1998/02/003}
\end{barticle}
\endbibitem

\bibitem{Douglas:1997fm}
\begin{barticle}
\bauthor{\bsnm{Douglas}, \binits{M.R.}},
\bauthor{\bsnm{Hull}, \binits{C.M.}}:
\batitle{{D-branes and the noncommutative torus}}.
\bjtitle{JHEP}
\bvolume{02},
\bfpage{008}
(\byear{1998})
{\href{https://arxiv.org/abs/hep-th/9711165}{{arXiv:hep-th/9711165}}}.
\doiurl{10.1088/1126-6708/1998/02/008}
\end{barticle}
\endbibitem

\bibitem{Schomerus:1999ug}
\begin{barticle}
\bauthor{\bsnm{Schomerus}, \binits{V.}}:
\batitle{{D-branes and deformation quantization}}.
\bjtitle{JHEP}
\bvolume{06},
\bfpage{030}
(\byear{1999})
{\href{https://arxiv.org/abs/hep-th/9903205}{{arXiv:hep-th/9903205}}}.
\doiurl{10.1088/1126-6708/1999/06/030}
\end{barticle}
\endbibitem

\bibitem{Kapustin:1999di}
\begin{barticle}
\bauthor{\bsnm{Kapustin}, \binits{A.}}:
\batitle{{D-branes in a topologically nontrivial B field}}.
\bjtitle{Adv. Theor. Math. Phys.}
\bvolume{4},
\bfpage{127}--\blpage{154}
(\byear{2000})
{\href{https://arxiv.org/abs/hep-th/9909089}{{arXiv:hep-th/9909089}}}.
\doiurl{10.4310/ATMP.2000.v4.n1.a3}
\end{barticle}
\endbibitem

\bibitem{Berenstein:2000ux}
\begin{barticle}
\bauthor{\bsnm{Berenstein}, \binits{D.}},
\bauthor{\bsnm{Jejjala}, \binits{V.}},
\bauthor{\bsnm{Leigh}, \binits{R.G.}}:
\batitle{{Marginal and relevant deformations of N=4 field theories and
  noncommutative moduli spaces of vacua}}.
\bjtitle{Nucl. Phys. B}
\bvolume{589},
\bfpage{196}--\blpage{248}
(\byear{2000})
{\href{https://arxiv.org/abs/hep-th/0005087}{{arXiv:hep-th/0005087}}}.
\doiurl{10.1016/S0550-3213(00)00394-1}
\end{barticle}
\endbibitem

\bibitem{Berenstein:2000jh}
\begin{barticle}
\bauthor{\bsnm{Berenstein}, \binits{D.}},
\bauthor{\bsnm{Leigh}, \binits{R.G.}}:
\batitle{{Noncommutative Calabi-Yau manifolds}}.
\bjtitle{Phys. Lett. B}
\bvolume{499},
\bfpage{207}--\blpage{214}
(\byear{2001})
{\href{https://arxiv.org/abs/hep-th/0009209}{{arXiv:hep-th/0009209}}}.
\doiurl{10.1016/S0370-2693(01)00005-3}
\end{barticle}
\endbibitem

\bibitem{Berenstein:2001jr}
\begin{barticle}
\bauthor{\bsnm{Berenstein}, \binits{D.}},
\bauthor{\bsnm{Leigh}, \binits{R.G.}}:
\batitle{{Resolution of stringy singularities by noncommutative algebras}}.
\bjtitle{JHEP}
\bvolume{06},
\bfpage{030}
(\byear{2001})
{\href{https://arxiv.org/abs/hep-th/0105229}{{arXiv:hep-th/0105229}}}.
\doiurl{10.1088/1126-6708/2001/06/030}
\end{barticle}
\endbibitem

\bibitem{Vafa:1994rv}
\begin{barticle}
\bauthor{\bsnm{Vafa}, \binits{C.}},
\bauthor{\bsnm{Witten}, \binits{E.}}:
\batitle{{On orbifolds with discrete torsion}}.
\bjtitle{J. Geom. Phys.}
\bvolume{15},
\bfpage{189}--\blpage{214}
(\byear{1995})
{\href{https://arxiv.org/abs/hep-th/9409188}{{arXiv:hep-th/9409188}}}.
\doiurl{10.1016/0393-0440(94)00048-9}
\end{barticle}
\endbibitem

\bibitem{Douglas:1998xa}
\begin{botherref}
\oauthor{\bsnm{Douglas}, \binits{M.R.}}:
{D-branes and discrete torsion}
(1998)
{\href{https://arxiv.org/abs/hep-th/9807235}{{arXiv:hep-th/9807235}}}
\end{botherref}
\endbibitem

\bibitem{Sharpe:2000ki}
\begin{barticle}
\bauthor{\bsnm{Sharpe}, \binits{E.R.}}:
\batitle{{Discrete torsion}}.
\bjtitle{Phys. Rev. D}
\bvolume{68},
\bfpage{126003}
(\byear{2003})
{\href{https://arxiv.org/abs/hep-th/0008154}{{arXiv:hep-th/0008154}}}.
\doiurl{10.1103/PhysRevD.68.126003}
\end{barticle}
\endbibitem

\bibitem{Aspinwall:1995rb}
\begin{barticle}
\bauthor{\bsnm{Aspinwall}, \binits{P.S.}},
\bauthor{\bsnm{Morrison}, \binits{D.R.}},
\bauthor{\bsnm{Gross}, \binits{M.}}:
\batitle{{Stable singularities in string theory}}.
\bjtitle{Commun. Math. Phys.}
\bvolume{178},
\bfpage{115}--\blpage{134}
(\byear{1996})
{\href{https://arxiv.org/abs/hep-th/9503208}{{arXiv:hep-th/9503208}}}.
\doiurl{10.1007/BF02104911}
\end{barticle}
\endbibitem

\bibitem{caldararuThesis}
\begin{bbook}
\bauthor{\bsnm{C\u{a}ld\u{a}raru}, \binits{A.H.}}:
\bbtitle{Derived Categories of Twisted Sheaves on {C}alabi-{Y}au Manifolds},
p. \bfpage{196}.
\bpublisher{ProQuest LLC},
\blocation{Ann Arbor, MI}
(\byear{2000}).
\bcomment{\url{http://www.math.wisc.edu/~andreic/publications/ThesisSingleSpaced.pdf}}
\end{bbook}
\endbibitem

\bibitem{caldararu2001derived}
\begin{barticle}
\bauthor{\bsnm{C\u{a}ld\u{a}raru}, \binits{A.}}:
\batitle{Derived categories of twisted sheaves on elliptic threefolds}.
\bjtitle{J. Reine Angew. Math.}
\bvolume{544},
\bfpage{161}--\blpage{179}
(\byear{2002})
{\href{https://arxiv.org/abs/math/0012083}{{arXiv:math/0012083}}}.
\doiurl{10.1515/crll.2002.022}
\end{barticle}
\endbibitem

\bibitem{addington2009derived}
\begin{botherref}
\oauthor{\bsnm{Addington}, \binits{N.}}:
The derived category of the intersection of four quadrics
(2009)
{\href{https://arxiv.org/abs/0904.1764}{{arXiv:0904.1764}}}
{[math.AG]}
\end{botherref}
\endbibitem

\bibitem{szendrHoi2008non}
\begin{barticle}
\bauthor{\bsnm{Szendr{\H{o}}i}, \binits{B.}}:
\batitle{Non-commutative {Donaldson}--{Thomas} invariants and the conifold}.
\bjtitle{Geom. Topol.}
\bvolume{12}(\bissue{2}),
\bfpage{1171}--\blpage{1202}
(\byear{2008})
{\href{https://arxiv.org/abs/0705.3419}{{arXiv:0705.3419}}}.
\doiurl{10.2140/gt.2008.12.1171}
\end{barticle}
\endbibitem

\bibitem{Caldararu:2010ljp}
\begin{barticle}
\bauthor{\bsnm{C\u{a}ld\u{a}raru}, \binits{A.}},
\bauthor{\bsnm{Distler}, \binits{J.}},
\bauthor{\bsnm{Hellerman}, \binits{S.}},
\bauthor{\bsnm{Pantev}, \binits{T.}},
\bauthor{\bsnm{Sharpe}, \binits{E.}}:
\batitle{{Non-birational twisted derived equivalences in abelian GLSMs}}.
\bjtitle{Commun. Math. Phys.}
\bvolume{294},
\bfpage{605}--\blpage{645}
(\byear{2010})
{\href{https://arxiv.org/abs/0709.3855}{{arXiv:0709.3855}}}
{[hep-th]}.
\doiurl{10.1007/s00220-009-0974-2}
\end{barticle}
\endbibitem

\bibitem{buchweitz-eisenbud-herzog}
\begin{bchapter}
\bauthor{\bsnm{Buchweitz}, \binits{R.-O.}},
\bauthor{\bsnm{Eisenbud}, \binits{D.}},
\bauthor{\bsnm{Herzog}, \binits{J.}}:
\bctitle{Cohen-{M}acaulay modules on quadrics}.
In: \bbtitle{Singularities, Representation of Algebras, and Vector Bundles
  ({L}ambrecht, 1985)}.
\bsertitle{Lecture Notes in Math.},
vol. \bseriesno{1273},
pp. \bfpage{58}--\blpage{116}.
\bpublisher{Springer},
\blocation{Berlin}
(\byear{1987}).
\doiurl{10.1007/BFb0078838}.
\burl{https://doi-org.ezproxy.lib.vt.edu/10.1007/BFb0078838}
\end{bchapter}
\endbibitem

\bibitem{Kapustin:2002bi}
\begin{barticle}
\bauthor{\bsnm{Kapustin}, \binits{A.}},
\bauthor{\bsnm{Li}, \binits{Y.}}:
\batitle{{D branes in Landau-Ginzburg models and algebraic geometry}}.
\bjtitle{JHEP}
\bvolume{12},
\bfpage{005}
(\byear{2003})
{\href{https://arxiv.org/abs/hep-th/0210296}{{arXiv:hep-th/0210296}}}.
\doiurl{10.1088/1126-6708/2003/12/005}
\end{barticle}
\endbibitem

\bibitem{dyckerhoff-cg}
\begin{barticle}
\bauthor{\bsnm{Dyckerhoff}, \binits{T.}}:
\batitle{Compact generators in categories of matrix factorizations}.
\bjtitle{Duke Math. J.}
\bvolume{159}(\bissue{2}),
\bfpage{223}--\blpage{274}
(\byear{2011}).
\doiurl{10.1215/00127094-1415869}
\end{barticle}
\endbibitem

\bibitem{teleman-mf}
\begin{barticle}
\bauthor{\bsnm{Teleman}, \binits{C.}}:
\batitle{Matrix factorisation of morse-bott functions}.
\bjtitle{Duke Math. J.}
\bvolume{169}(\bissue{3}),
\bfpage{533}--\blpage{549}
(\byear{2020}).
\doiurl{10.1215/00127094-2019-0048}
\end{barticle}
\endbibitem

\bibitem{yoshino}
\begin{bbook}
\bauthor{\bsnm{Yoshino}, \binits{Y.}}:
\bbtitle{Cohen-{M}acaulay Modules over {C}ohen-{M}acaulay rings}.
\bsertitle{London Mathematical Society Lecture Note Series},
vol. \bseriesno{146},
p. \bfpage{177}.
\bpublisher{Cambridge University Press},
\blocation{Cambridge}
(\byear{1990}).
\doiurl{10.1017/CBO9780511600685}.
\burl{https://doi-org.ezproxy.lib.vt.edu/10.1017/CBO9780511600685}
\end{bbook}
\endbibitem

\bibitem{kuz0}
\begin{botherref}
\oauthor{\bsnm{Kuznetsov}, \binits{A.}}:
Homological projective duality.
Publ. Math. Inst. Hautes \'{E}tudes Sci.
(105),
157--220
(2007)
{\href{https://arxiv.org/abs/math/0507292}{{arXiv:math/0507292}}}.
\doiurl{10.1007/s10240-007-0006-8}
\end{botherref}
\endbibitem

\bibitem{kuznetsov2008derived}
\begin{barticle}
\bauthor{\bsnm{Kuznetsov}, \binits{A.}}:
\batitle{Derived categories of quadric fibrations and intersections of
  quadrics}.
\bjtitle{Adv. Math.}
\bvolume{218}(\bissue{5}),
\bfpage{1340}--\blpage{1369}
(\byear{2008})
{\href{https://arxiv.org/abs/math/0510670}{{arXiv:math/0510670}}}.
\doiurl{10.1016/j.aim.2008.03.007}
\end{barticle}
\endbibitem

\bibitem{kuzso}
\begin{bchapter}
\bauthor{\bsnm{Kuznetsov}, \binits{A.}}:
\bctitle{Semiorthogonal decompositions in algebraic geometry}.
In: \bbtitle{Proceedings of the {I}nternational {C}ongress of
  {M}athematicians---{S}eoul 2014. {V}ol. {II}},
pp. \bfpage{635}--\blpage{660}.
\bpublisher{Kyung Moon Sa},
\blocation{Seoul}
(\byear{2014})
\end{bchapter}
\endbibitem

\bibitem{borisovzhan}
\begin{barticle}
\bauthor{\bsnm{Borisov}, \binits{L.A.}},
\bauthor{\bsnm{Li}, \binits{Z.}}:
\batitle{On {C}lifford double mirrors of toric complete intersections}.
\bjtitle{Adv. Math.}
\bvolume{328},
\bfpage{300}--\blpage{355}
(\byear{2018})
{\href{https://arxiv.org/abs/1601.00809}{{arXiv:1601.00809}}}.
\doiurl{10.1016/j.aim.2018.01.017}
\end{barticle}
\endbibitem

\bibitem{Hori:2013gga}
\begin{barticle}
\bauthor{\bsnm{Hori}, \binits{K.}},
\bauthor{\bsnm{Knapp}, \binits{J.}}:
\batitle{{Linear sigma models with strongly coupled phases - one parameter
  models}}.
\bjtitle{JHEP}
\bvolume{11},
\bfpage{070}
(\byear{2013})
{\href{https://arxiv.org/abs/1308.6265}{{arXiv:1308.6265}}}
{[hep-th]}.
\doiurl{10.1007/JHEP11(2013)070}
\end{barticle}
\endbibitem

\bibitem{Sharpe:2012ji}
\begin{barticle}
\bauthor{\bsnm{Sharpe}, \binits{E.}}:
\batitle{{Predictions for Gromov-Witten invariants of noncommutative
  resolutions}}.
\bjtitle{J. Geom. Phys.}
\bvolume{74},
\bfpage{256}--\blpage{265}
(\byear{2013})
{\href{https://arxiv.org/abs/1212.5322}{{arXiv:1212.5322}}}
{[hep-th]}.
\doiurl{10.1016/j.geomphys.2013.08.012}
\end{barticle}
\endbibitem

\bibitem{CANDELAS199121}
\begin{barticle}
\bauthor{\bsnm{Candelas}, \binits{P.}},
\bauthor{\bsnm{{De La Ossa}}, \binits{X.C.}},
\bauthor{\bsnm{Green}, \binits{P.S.}},
\bauthor{\bsnm{Parkes}, \binits{L.}}:
\batitle{A pair of calabi-yau manifolds as an exactly soluble superconformal
  theory}.
\bjtitle{Nuclear Physics B}
\bvolume{359}(\bissue{1}),
\bfpage{21}--\blpage{74}
(\byear{1991}).
\doiurl{10.1016/0550-3213(91)90292-6}
\end{barticle}
\endbibitem

\bibitem{Hosono:1993qy}
\begin{barticle}
\bauthor{\bsnm{Hosono}, \binits{S.}},
\bauthor{\bsnm{Klemm}, \binits{A.}},
\bauthor{\bsnm{Theisen}, \binits{S.}},
\bauthor{\bsnm{Yau}, \binits{S.-T.}}:
\batitle{{Mirror symmetry, mirror map and applications to Calabi-Yau
  hypersurfaces}}.
\bjtitle{Commun. Math. Phys.}
\bvolume{167},
\bfpage{301}--\blpage{350}
(\byear{1995})
{\href{https://arxiv.org/abs/hep-th/9308122}{{arXiv:hep-th/9308122}}}.
\doiurl{10.1007/BF02100589}
\end{barticle}
\endbibitem

\bibitem{Hosono:1994ax}
\begin{barticle}
\bauthor{\bsnm{Hosono}, \binits{S.}},
\bauthor{\bsnm{Klemm}, \binits{A.}},
\bauthor{\bsnm{Theisen}, \binits{S.}},
\bauthor{\bsnm{Yau}, \binits{S.-T.}}:
\batitle{{Mirror symmetry, mirror map and applications to complete intersection
  Calabi-Yau spaces}}.
\bjtitle{Nucl. Phys. B}
\bvolume{433},
\bfpage{501}--\blpage{554}
(\byear{1995})
{\href{https://arxiv.org/abs/hep-th/9406055}{{arXiv:hep-th/9406055}}}.
\doiurl{10.1016/0550-3213(94)00440-P}
\end{barticle}
\endbibitem

\bibitem{Huang:2006hq}
\begin{barticle}
\bauthor{\bsnm{Huang}, \binits{M.-x.}},
\bauthor{\bsnm{Klemm}, \binits{A.}},
\bauthor{\bsnm{Quackenbush}, \binits{S.}}:
\batitle{{Topological string theory on compact Calabi-Yau: Modularity and
  boundary conditions}}.
\bjtitle{Lect. Notes Phys.}
\bvolume{757},
\bfpage{45}--\blpage{102}
(\byear{2009})
{\href{https://arxiv.org/abs/hep-th/0612125}{{arXiv:hep-th/0612125}}}.
\doiurl{10.1007/978-3-540-68030-7_3}
\end{barticle}
\endbibitem

\bibitem{Gopakumar:1998ii}
\begin{botherref}
\oauthor{\bsnm{Gopakumar}, \binits{R.}},
\oauthor{\bsnm{Vafa}, \binits{C.}}:
{M theory and topological strings. 1.}
(1998)
{\href{https://arxiv.org/abs/hep-th/9809187}{{arXiv:hep-th/9809187}}}
\end{botherref}
\endbibitem

\bibitem{Gopakumar:1998jq}
\begin{botherref}
\oauthor{\bsnm{Gopakumar}, \binits{R.}},
\oauthor{\bsnm{Vafa}, \binits{C.}}:
{M theory and topological strings. 2.}
(1998)
{\href{https://arxiv.org/abs/hep-th/9812127}{{arXiv:hep-th/9812127}}}
\end{botherref}
\endbibitem

\bibitem{gvintegral}
\begin{barticle}
\bauthor{\bsnm{Ionel}, \binits{E.-N.}},
\bauthor{\bsnm{Parker}, \binits{T.}}:
\batitle{{The Gopakumar-Vafa formula for symplectic manifolds}}.
\bjtitle{Annals of Mathematics}
\bvolume{187}(\bissue{1}),
\bfpage{1}--\blpage{64}
(\byear{2018}).
\doiurl{10.4007/annals.2018.187.1.1}
\end{barticle}
\endbibitem

\bibitem{gvfinite}
\begin{botherref}
\oauthor{\bsnm{Doan}, \binits{A.}},
\oauthor{\bsnm{Ionel}, \binits{E.-N.}},
\oauthor{\bsnm{Walpuski}, \binits{T.}}:
The Gopakumar-Vafa finiteness conjecture.
arXiv
(2021).
\doiurl{10.48550/ARXIV.2103.08221}.
\url{https://arxiv.org/abs/2103.08221}
\end{botherref}
\endbibitem

\bibitem{Katz:1999xq}
\begin{barticle}
\bauthor{\bsnm{Katz}, \binits{S.H.}},
\bauthor{\bsnm{Klemm}, \binits{A.}},
\bauthor{\bsnm{Vafa}, \binits{C.}}:
\batitle{{M theory, topological strings and spinning black holes}}.
\bjtitle{Adv. Theor. Math. Phys.}
\bvolume{3},
\bfpage{1445}--\blpage{1537}
(\byear{1999})
{\href{https://arxiv.org/abs/hep-th/9910181}{{arXiv:hep-th/9910181}}}.
\doiurl{10.4310/ATMP.1999.v3.n5.a6}
\end{barticle}
\endbibitem

\bibitem{MT}
\begin{barticle}
\bauthor{\bsnm{Maulik}, \binits{D.}},
\bauthor{\bsnm{Toda}, \binits{Y.}}:
\batitle{Gopakumar-{V}afa invariants via vanishing cycles}.
\bjtitle{Invent. Math.}
\bvolume{213}(\bissue{3}),
\bfpage{1017}--\blpage{1097}
(\byear{2018})
{\href{https://arxiv.org/abs/1610.07303}{{arXiv:1610.07303}}}.
\doiurl{10.1007/s00222-018-0800-6}
\end{barticle}
\endbibitem

\bibitem{Zhao}
\begin{botherref}
\oauthor{\bsnm{Zhao}, \binits{L.}}:
Gopakumar-vafa invariant and macdonald formula.
PhD thesis,
University of Illinois at Urbana-Champaign
(2021)
\end{botherref}
\endbibitem

\bibitem{Dedushenko:2014nya}
\begin{barticle}
\bauthor{\bsnm{Dedushenko}, \binits{M.}},
\bauthor{\bsnm{Witten}, \binits{E.}}:
\batitle{{Some Details On The Gopakumar-Vafa and Ooguri-Vafa Formulas}}.
\bjtitle{Adv. Theor. Math. Phys.}
\bvolume{20},
\bfpage{1}--\blpage{133}
(\byear{2016})
{\href{https://arxiv.org/abs/1411.7108}{{arXiv:1411.7108}}}
{[hep-th]}.
\doiurl{10.4310/ATMP.2016.v20.n1.a1}
\end{barticle}
\endbibitem

\bibitem{Braun:2007tp}
\begin{barticle}
\bauthor{\bsnm{Braun}, \binits{V.}},
\bauthor{\bsnm{Kreuzer}, \binits{M.}},
\bauthor{\bsnm{Ovrut}, \binits{B.A.}},
\bauthor{\bsnm{Scheidegger}, \binits{E.}}:
\batitle{{Worldsheet instantons, torsion curves, and non-perturbative
  superpotentials}}.
\bjtitle{Phys. Lett. B}
\bvolume{649},
\bfpage{334}--\blpage{341}
(\byear{2007})
{\href{https://arxiv.org/abs/hep-th/0703134}{{arXiv:hep-th/0703134}}}.
\doiurl{10.1016/j.physletb.2007.03.066}
\end{barticle}
\endbibitem

\bibitem{Braun:2007xh}
\begin{barticle}
\bauthor{\bsnm{Braun}, \binits{V.}},
\bauthor{\bsnm{Kreuzer}, \binits{M.}},
\bauthor{\bsnm{Ovrut}, \binits{B.A.}},
\bauthor{\bsnm{Scheidegger}, \binits{E.}}:
\batitle{{Worldsheet instantons and torsion curves, part A: Direct
  computation}}.
\bjtitle{JHEP}
\bvolume{10},
\bfpage{022}
(\byear{2007})
{\href{https://arxiv.org/abs/hep-th/0703182}{{arXiv:hep-th/0703182}}}.
\doiurl{10.1088/1126-6708/2007/10/022}
\end{barticle}
\endbibitem

\bibitem{Braun:2007vy}
\begin{barticle}
\bauthor{\bsnm{Braun}, \binits{V.}},
\bauthor{\bsnm{Kreuzer}, \binits{M.}},
\bauthor{\bsnm{Ovrut}, \binits{B.A.}},
\bauthor{\bsnm{Scheidegger}, \binits{E.}}:
\batitle{{Worldsheet instantons and torsion curves, Part B: Mirror symmetry}}.
\bjtitle{JHEP}
\bvolume{10},
\bfpage{023}
(\byear{2007})
{\href{https://arxiv.org/abs/0704.0449}{{arXiv:0704.0449}}}
{[hep-th]}.
\doiurl{10.1088/1126-6708/2007/10/023}
\end{barticle}
\endbibitem

\bibitem{Dierigl:2022zll}
\begin{botherref}
\oauthor{\bsnm{Dierigl}, \binits{M.}},
\oauthor{\bsnm{Oehlmann}, \binits{P.-K.}},
\oauthor{\bsnm{Schimannek}, \binits{T.}}:
The discrete {G}reen-{S}chwarz mechanism in 6d {F}-theory and elliptic genera
  of non-critical strings
(2022)
{\href{https://arxiv.org/abs/2212.04503}{{arXiv:2212.04503}}}
{[hep-th]}
\end{botherref}
\endbibitem

\bibitem{Addington:2012zv}
\begin{barticle}
\bauthor{\bsnm{Addington}, \binits{N.M.}},
\bauthor{\bsnm{Segal}, \binits{E.P.}},
\bauthor{\bsnm{Sharpe}, \binits{E.}}:
\batitle{{D-brane probes, branched double covers, and noncommutative
  resolutions}}.
\bjtitle{Adv. Theor. Math. Phys.}
\bvolume{18}(\bissue{6}),
\bfpage{1369}--\blpage{1436}
(\byear{2014})
{\href{https://arxiv.org/abs/1211.2446}{{arXiv:1211.2446}}}
{[hep-th]}.
\doiurl{10.4310/ATMP.2014.v18.n6.a5}
\end{barticle}
\endbibitem

\bibitem{Bershadsky:1993cx}
\begin{barticle}
\bauthor{\bsnm{Bershadsky}, \binits{M.}},
\bauthor{\bsnm{Cecotti}, \binits{S.}},
\bauthor{\bsnm{Ooguri}, \binits{H.}},
\bauthor{\bsnm{Vafa}, \binits{C.}}:
\batitle{{Kodaira-Spencer theory of gravity and exact results for quantum
  string amplitudes}}.
\bjtitle{Commun. Math. Phys.}
\bvolume{165},
\bfpage{311}--\blpage{428}
(\byear{1994})
{\href{https://arxiv.org/abs/hep-th/9309140}{{arXiv:hep-th/9309140}}}.
\doiurl{10.1007/BF02099774}
\end{barticle}
\endbibitem

\bibitem{datalink}
\begin{botherref}
{One Parameter Calabi-Yau higher genus data}
(\url{http://www.th.physik.uni-bonn.de/Groups/Klemm/data.php})
\end{botherref}
\endbibitem

\bibitem{vdb02-3d}
\begin{barticle}
\bauthor{\bparticle{Van~den} \bsnm{Bergh}, \binits{M.}}:
\batitle{Three-dimensional flops and noncommutative rings}.
\bjtitle{Duke Math. J.}
\bvolume{122}(\bissue{3}),
\bfpage{423}--\blpage{455}
(\byear{2004})
{\href{https://arxiv.org/abs/math/0207170}{{arXiv:math/0207170}}}.
\doiurl{10.1215/S0012-7094-04-12231-6}
\end{barticle}
\endbibitem

\bibitem{kuzgrass}
\begin{botherref}
\oauthor{\bsnm{Kuznetsov}, \binits{A.}}:
Homological projective duality for {Grassmannians} of lines
{\href{https://arxiv.org/abs/math.AG/0610957}{{arXiv:math.AG/0610957}}}
\end{botherref}
\endbibitem

\bibitem{perry1}
\begin{barticle}
\bauthor{\bsnm{Perry}, \binits{A.}}:
\batitle{Noncommutative homological projective duality}.
\bjtitle{Adv. Math.}
\bvolume{350},
\bfpage{877}--\blpage{972}
(\byear{2019})
{\href{https://arxiv.org/abs/1804.00132}{{arXiv:1804.00132}}}.
\doiurl{10.1016/j.aim.2019.04.052}
\end{barticle}
\endbibitem

\bibitem{kuzp1}
\begin{barticle}
\bauthor{\bsnm{Kuznetsov}, \binits{A.}},
\bauthor{\bsnm{Perry}, \binits{A.}}:
\batitle{Categorical joins}.
\bjtitle{J. Amer. Math. Soc.}
\bvolume{34}(\bissue{2}),
\bfpage{505}--\blpage{564}
(\byear{2021})
{\href{https://arxiv.org/abs/1804.00144}{{arXiv:1804.00144}}}.
\doiurl{10.1090/jams/963}
\end{barticle}
\endbibitem

\bibitem{kuzp2}
\begin{botherref}
\oauthor{\bsnm{Kuznetsov}, \binits{A.}},
\oauthor{\bsnm{Perry}, \binits{A.}}:
Categorical cones and quadratic homological projective duality
{\href{https://arxiv.org/abs/1902.09824}{{arXiv:1902.09824}}}
\end{botherref}
\endbibitem

\bibitem{kuzp3}
\begin{barticle}
\bauthor{\bsnm{Kuznetsov}, \binits{A.}},
\bauthor{\bsnm{Perry}, \binits{A.}}:
\batitle{Homological projective duality for quadrics}.
\bjtitle{J. Algebraic Geom.}
\bvolume{30}(\bissue{3}),
\bfpage{457}--\blpage{476}
(\byear{2021})
{\href{https://arxiv.org/abs/1902.09832}{{arXiv:1902.09832}}}.
\doiurl{10.1090/jag/767}
\end{barticle}
\endbibitem

\bibitem{kuz4}
\begin{botherref}
\oauthor{\bsnm{Kuznetsov}, \binits{A.}}:
Semiorthogonal decompositions in families
{\href{https://arxiv.org/abs/2111.00527}{{2111.00527:2111.00527}}}
\end{botherref}
\endbibitem

\bibitem{Douglas:1996sw}
\begin{botherref}
\oauthor{\bsnm{Douglas}, \binits{M.R.}},
\oauthor{\bsnm{Moore}, \binits{G.W.}}:
{D-branes, quivers, and ALE instantons}
(1996)
{\href{https://arxiv.org/abs/hep-th/9603167}{{arXiv:hep-th/9603167}}}
\end{botherref}
\endbibitem

\bibitem{Leuschke+2012+293+364}
\begin{bbook}
\bauthor{\bsnm{Leuschke}, \binits{G.J.}}:
In: \beditor{\bsnm{Francisco}, \binits{C.}},
\beditor{\bsnm{Klingler}, \binits{L.C.}},
\beditor{\bsnm{Sather-Wagstaff}, \binits{S.}},
\beditor{\bsnm{Vassilev}, \binits{J.C.}} (eds.)
\bbtitle{Non-commutative Crepant Resolutions: Scenes From Categorical
  Geometry},
pp. \bfpage{293}--\blpage{364}.
\bpublisher{De Gruyter},
\blocation{Berlin, Boston}
(\byear{2012}).
\doiurl{10.1515/9783110250404.293}.
\burl{https://doi.org/10.1515/9783110250404.293}
\end{bbook}
\endbibitem

\bibitem{wemyss}
\begin{botherref}
\oauthor{\bsnm{{Wemyss}}, \binits{M.}}:
{Lectures on Noncommutative Resolutions}.
arXiv e-prints,
1210--2564
(2012)
{\href{https://arxiv.org/abs/1210.2564}{{arXiv:1210.2564}}}
{[math.RT]}
\end{botherref}
\endbibitem

\bibitem{Klebanov:1998hh}
\begin{barticle}
\bauthor{\bsnm{Klebanov}, \binits{I.R.}},
\bauthor{\bsnm{Witten}, \binits{E.}}:
\batitle{{Superconformal field theory on three-branes at a Calabi-Yau
  singularity}}.
\bjtitle{Nucl. Phys. B}
\bvolume{536},
\bfpage{199}--\blpage{218}
(\byear{1998})
{\href{https://arxiv.org/abs/hep-th/9807080}{{arXiv:hep-th/9807080}}}.
\doiurl{10.1016/S0550-3213(98)00654-3}
\end{barticle}
\endbibitem

\bibitem{Gopakumar:1998ki}
\begin{barticle}
\bauthor{\bsnm{Gopakumar}, \binits{R.}},
\bauthor{\bsnm{Vafa}, \binits{C.}}:
\batitle{{On the gauge theory / geometry correspondence}}.
\bjtitle{Adv. Theor. Math. Phys.}
\bvolume{3},
\bfpage{1415}--\blpage{1443}
(\byear{1999})
{\href{https://arxiv.org/abs/hep-th/9811131}{{arXiv:hep-th/9811131}}}.
\doiurl{10.4310/ATMP.1999.v3.n5.a5}
\end{barticle}
\endbibitem

\bibitem{MORRISON20122065}
\begin{barticle}
\bauthor{\bsnm{Morrison}, \binits{A.}},
\bauthor{\bsnm{Mozgovoy}, \binits{S.}},
\bauthor{\bsnm{Nagao}, \binits{K.}},
\bauthor{\bsnm{Szendr{\H o}i}, \binits{B.}}:
\batitle{Motivic {D}onaldson--{T}homas invariants of the conifold and the
  refined topological vertex}.
\bjtitle{Adv. Math.}
\bvolume{230}(\bissue{4}),
\bfpage{2065}--\blpage{2093}
(\byear{2012})
{\href{https://arxiv.org/abs/1107.5017}{{arXiv:1107.5017}}}.
\doiurl{10.1016/j.aim.2012.03.030}
\end{barticle}
\endbibitem

\bibitem{Aspinwall:2010mw}
\begin{barticle}
\bauthor{\bsnm{Aspinwall}, \binits{P.S.}},
\bauthor{\bsnm{Morrison}, \binits{D.R.}}:
\batitle{Quivers from matrix factorizations}.
\bjtitle{Commun. Math. Phys.}
\bvolume{313},
\bfpage{607}--\blpage{633}
(\byear{2012})
{\href{https://arxiv.org/abs/1005.1042}{{arXiv:1005.1042}}}
{[hep-th]}.
\doiurl{10.1007/s00220-012-1520-1}
\end{barticle}
\endbibitem

\bibitem{Mozgovoy:2020has}
\begin{botherref}
\oauthor{\bsnm{Mozgovoy}, \binits{S.}},
\oauthor{\bsnm{Pioline}, \binits{B.}}:
{Attractor invariants, brane tilings and crystals}
(2020)
{\href{https://arxiv.org/abs/2012.14358}{{arXiv:2012.14358}}}
{[hep-th]}
\end{botherref}
\endbibitem

\bibitem{Hori:2011pd}
\begin{barticle}
\bauthor{\bsnm{Hori}, \binits{K.}}:
\batitle{Duality in two-dimensional (2,2) supersymmetric non-abelian gauge
  theories}.
\bjtitle{JHEP}
\bvolume{10},
\bfpage{121}
(\byear{2013})
{\href{https://arxiv.org/abs/1104.2853}{{arXiv:1104.2853}}}
{[hep-th]}.
\doiurl{10.1007/JHEP10(2013)121}
\end{barticle}
\endbibitem

\bibitem{Halverson:2013eua}
\begin{barticle}
\bauthor{\bsnm{Halverson}, \binits{J.}},
\bauthor{\bsnm{Kumar}, \binits{V.}},
\bauthor{\bsnm{Morrison}, \binits{D.R.}}:
\batitle{New methods for characterizing phases of 2d supersymmetric gauge
  theories}.
\bjtitle{JHEP}
\bvolume{09},
\bfpage{143}
(\byear{2013})
{\href{https://arxiv.org/abs/1305.3278}{{arXiv:1305.3278}}}
{[hep-th]}.
\doiurl{10.1007/JHEP09(2013)143}
\end{barticle}
\endbibitem

\bibitem{Sharpe:2013bwa}
\begin{barticle}
\bauthor{\bsnm{Sharpe}, \binits{E.}}:
\batitle{{A few Ricci-flat stacks as phases of exotic GLSM's}}.
\bjtitle{Phys. Lett. B}
\bvolume{726},
\bfpage{390}--\blpage{395}
(\byear{2013})
{\href{https://arxiv.org/abs/1306.5440}{{arXiv:1306.5440}}}
{[hep-th]}.
\doiurl{10.1016/j.physletb.2013.08.013}
\end{barticle}
\endbibitem

\bibitem{Sharpe:2010iv}
\begin{barticle}
\bauthor{\bsnm{Sharpe}, \binits{E.}}:
\batitle{{GLSM}'s, gerbes, and {Kuznetsov's} homological projective duality}.
\bjtitle{J. Phys. Conf. Ser.}
\bvolume{462}(\bissue{1}),
\bfpage{012047}
(\byear{2013})
{\href{https://arxiv.org/abs/1004.5388}{{arXiv:1004.5388}}}
{[hep-th]}.
\doiurl{10.1088/1742-6596/462/1/012047}
\end{barticle}
\endbibitem

\bibitem{Sharpe:2010zz}
\begin{barticle}
\bauthor{\bsnm{Sharpe}, \binits{E.}}:
\batitle{{Landau-Ginzburg models, gerbes, and Kuznetsov's homological
  projective duality}}.
\bjtitle{Proc. Symp. Pure Math.}
\bvolume{81},
\bfpage{237}--\blpage{249}
(\byear{2010}).
\doiurl{10.1090/pspum/081/2681766}
\end{barticle}
\endbibitem

\bibitem{Ballard:2013fxa}
\begin{barticle}
\bauthor{\bsnm{Ballard}, \binits{M.}},
\bauthor{\bsnm{Deliu}, \binits{D.}},
\bauthor{\bsnm{Favero}, \binits{D.}},
\bauthor{\bsnm{Isik}, \binits{M.U.}},
\bauthor{\bsnm{Katzarkov}, \binits{L.}}:
\batitle{{Homological projective duality via variation of geometric invariant
  theory quotients}}.
\bjtitle{J. Eur. Math. Soc.}
\bvolume{19}(\bissue{4}),
\bfpage{1127}--\blpage{1158}
(\byear{2017})
{\href{https://arxiv.org/abs/1306.3957}{{arXiv:1306.3957}}}
{[math.AG]}.
\doiurl{10.4171/jems/689}
\end{barticle}
\endbibitem

\bibitem{Hori:2016txh}
\begin{botherref}
\oauthor{\bsnm{Hori}, \binits{K.}},
\oauthor{\bsnm{Knapp}, \binits{J.}}:
{A pair of Calabi-Yau manifolds from a two parameter non-Abelian gauged linear
  sigma model}
(2016)
{\href{https://arxiv.org/abs/1612.06214}{{arXiv:1612.06214}}}
{[hep-th]}
\end{botherref}
\endbibitem

\bibitem{Wong:2017cqs}
\begin{barticle}
\bauthor{\bsnm{Wong}, \binits{K.}}:
\batitle{{Two-dimensional gauge dynamics and the topology of singular
  determinantal varieties}}.
\bjtitle{JHEP}
\bvolume{03},
\bfpage{132}
(\byear{2017})
{\href{https://arxiv.org/abs/1702.00730}{{arXiv:1702.00730}}}
{[hep-th]}.
\doiurl{10.1007/JHEP03(2017)132}
\end{barticle}
\endbibitem

\bibitem{Kapustka:2017jyt}
\begin{barticle}
\bauthor{\bsnm{Kapustka}, \binits{M.}},
\bauthor{\bsnm{Rampazzo}, \binits{M.}}:
\batitle{{Torelli problem for Calabi\textendash{}Yau threefolds with GLSM
  description}}.
\bjtitle{Commun. Num. Theor. Phys.}
\bvolume{13}(\bissue{4}),
\bfpage{725}--\blpage{761}
(\byear{2019})
{\href{https://arxiv.org/abs/1711.10231}{{arXiv:1711.10231}}}
{[math.AG]}.
\doiurl{10.4310/CNTP.2019.v13.n4.a2}
\end{barticle}
\endbibitem

\bibitem{Parsian:2018fhm}
\begin{barticle}
\bauthor{\bsnm{Parsian}, \binits{H.}},
\bauthor{\bsnm{Sharpe}, \binits{E.}},
\bauthor{\bsnm{Zou}, \binits{H.}}:
\batitle{{(0,2) versions of exotic (2,2) GLSMs}}.
\bjtitle{Int. J. Mod. Phys. A}
\bvolume{33}(\bissue{18n19}),
\bfpage{1850113}
(\byear{2018})
{\href{https://arxiv.org/abs/1803.00286}{{arXiv:1803.00286}}}
{[hep-th]}.
\doiurl{10.1142/S0217751X18501130}
\end{barticle}
\endbibitem

\bibitem{Chen:2018qww}
\begin{barticle}
\bauthor{\bsnm{Chen}, \binits{Z.}},
\bauthor{\bsnm{Pantev}, \binits{T.}},
\bauthor{\bsnm{Sharpe}, \binits{E.}}:
\batitle{{Landau\textendash{}Ginzburg models for certain fiber products with
  curves}}.
\bjtitle{J. Geom. Phys.}
\bvolume{137},
\bfpage{204}--\blpage{211}
(\byear{2019})
{\href{https://arxiv.org/abs/1806.01283}{{arXiv:1806.01283}}}
{[hep-th]}.
\doiurl{10.1016/j.geomphys.2018.11.012}
\end{barticle}
\endbibitem

\bibitem{Chen:2020iyo}
\begin{barticle}
\bauthor{\bsnm{Chen}, \binits{Z.}},
\bauthor{\bsnm{Guo}, \binits{J.}},
\bauthor{\bsnm{Romo}, \binits{M.}}:
\batitle{A {GLSM} view on homological projective duality}.
\bjtitle{Commun. Math. Phys.}
\bvolume{394}(\bissue{1}),
\bfpage{355}--\blpage{407}
(\byear{2022})
{\href{https://arxiv.org/abs/2012.14109}{{arXiv:2012.14109}}}
{[hep-th]}.
\doiurl{10.1007/s00220-022-04401-1}
\end{barticle}
\endbibitem

\bibitem{Guo:2021aqj}
\begin{botherref}
\oauthor{\bsnm{Guo}, \binits{J.}},
\oauthor{\bsnm{Romo}, \binits{M.}}:
{Hybrid models for homological projective duals and noncommutative resolutions}
(2021)
{\href{https://arxiv.org/abs/2111.00025}{{arXiv:2111.00025}}}
{[hep-th]}
\end{botherref}
\endbibitem

\bibitem{Hellerman:2006zs}
\begin{barticle}
\bauthor{\bsnm{Hellerman}, \binits{S.}},
\bauthor{\bsnm{Henriques}, \binits{A.}},
\bauthor{\bsnm{Pantev}, \binits{T.}},
\bauthor{\bsnm{Sharpe}, \binits{E.}},
\bauthor{\bsnm{Ando}, \binits{M.}}:
\batitle{{Cluster decomposition, T-duality, and gerby CFT's}}.
\bjtitle{Adv. Theor. Math. Phys.}
\bvolume{11}(\bissue{5}),
\bfpage{751}--\blpage{818}
(\byear{2007})
{\href{https://arxiv.org/abs/hep-th/0606034}{{arXiv:hep-th/0606034}}}.
\doiurl{10.4310/ATMP.2007.v11.n5.a2}
\end{barticle}
\endbibitem

\bibitem{Clemens}
\begin{barticle}
\bauthor{\bsnm{Clemens}, \binits{C.H.}}:
\batitle{Double solids}.
\bjtitle{Adv. in Math.}
\bvolume{47}(\bissue{2}),
\bfpage{107}--\blpage{230}
(\byear{1983}).
\doiurl{10.1016/0001-8708(83)90025-7}
\end{barticle}
\endbibitem

\bibitem{Calabrese2016DerivedEC}
\begin{barticle}
\bauthor{\bsnm{Calabrese}, \binits{J.R.}},
\bauthor{\bsnm{Thomas}, \binits{R.P.}}:
\batitle{Derived equivalent {C}alabi-{Y}au threefolds from cubic fourfolds}.
\bjtitle{Math. Ann.}
\bvolume{365}(\bissue{1-2}),
\bfpage{155}--\blpage{172}
(\byear{2016})
{\href{https://arxiv.org/abs/1408.4063}{{arXiv:1408.4063}}}.
\doiurl{10.1007/s00208-015-1260-6}
\end{barticle}
\endbibitem

\bibitem{borisov2018clifford}
\begin{barticle}
\bauthor{\bsnm{Borisov}, \binits{L.A.}},
\bauthor{\bsnm{Li}, \binits{Z.}}:
\batitle{On clifford double mirrors of toric complete intersections}.
\bjtitle{Advances in Mathematics}
\bvolume{328},
\bfpage{300}--\blpage{355}
(\byear{2018})
\end{barticle}
\endbibitem

\bibitem{Kawamata}
\begin{barticle}
\bauthor{\bsnm{Kawamata}, \binits{Y.}}:
\batitle{Crepant blowing-up of 3-dimensional canonical singularities and its
  application to degenerations of surfaces}.
\bjtitle{Annals of Mathematics}
\bvolume{127}(\bissue{1}),
\bfpage{93}--\blpage{163}
(\byear{1988})
\end{barticle}
\endbibitem

\bibitem{Namikawa1995}
\begin{barticle}
\bauthor{\bsnm{Namikawa}, \binits{Y.}},
\bauthor{\bsnm{Steenbrink}, \binits{J.H.}}:
\batitle{Global smoothing of {C}alabi-{Y}au threefolds.}
\bjtitle{Inventiones mathematicae}
\bvolume{122}(\bissue{2}),
\bfpage{403}--\blpage{420}
(\byear{1995})
\end{barticle}
\endbibitem

\bibitem{Arras:2016evy}
\begin{barticle}
\bauthor{\bsnm{Arras}, \binits{P.}},
\bauthor{\bsnm{Grassi}, \binits{A.}},
\bauthor{\bsnm{Weigand}, \binits{T.}}:
\batitle{Terminal singularities, {Milnor} numbers, and matter in {F}-theory}.
\bjtitle{J. Geom. Phys.}
\bvolume{123},
\bfpage{71}--\blpage{97}
(\byear{2018})
{\href{https://arxiv.org/abs/1612.05646}{{arXiv:1612.05646}}}
{[hep-th]}.
\doiurl{10.1016/j.geomphys.2017.09.001}
\end{barticle}
\endbibitem

\bibitem{ARRAS201871}
\begin{barticle}
\bauthor{\bsnm{Arras}, \binits{P.}},
\bauthor{\bsnm{Grassi}, \binits{A.}},
\bauthor{\bsnm{Weigand}, \binits{T.}}:
\batitle{Terminal singularities, milnor numbers, and matter in f-theory}.
\bjtitle{Journal of Geometry and Physics}
\bvolume{123},
\bfpage{71}--\blpage{97}
(\byear{2018}).
\doiurl{10.1016/j.geomphys.2017.09.001}
\end{barticle}
\endbibitem

\bibitem{Werner}
\begin{bbook}
\bauthor{\bsnm{Werner}, \binits{J.}}:
\bbtitle{Kleine {A}ufl\"{o}sungen Spezieller Dreidimensionaler
  {V}ariet\"{a}ten}.
\bsertitle{Bonner Mathematische Schriften [Bonn Mathematical Publications]},
vol. \bseriesno{186},
p. \bfpage{119}.
\bpublisher{Universit\"{a}t Bonn, Mathematisches Institut},
\blocation{Bonn}
(\byear{1987}).
\bcomment{Dissertation, Rheinische Friedrich-Wilhelms-Universit\"{a}t, Bonn,
  1987}
\end{bbook}
\endbibitem

\bibitem{DG}
\begin{barticle}
\bauthor{\bsnm{Dolgachev}, \binits{I.}},
\bauthor{\bsnm{Gross}, \binits{M.}}:
\batitle{Elliptic threefolds. {I}. {O}gg-{S}hafarevich theory}.
\bjtitle{J. Algebraic Geom.}
\bvolume{3}(\bissue{1}),
\bfpage{39}--\blpage{80}
(\byear{1994})
\end{barticle}
\endbibitem

\bibitem{Braun:2014oya}
\begin{barticle}
\bauthor{\bsnm{Braun}, \binits{V.}},
\bauthor{\bsnm{Morrison}, \binits{D.R.}}:
\batitle{F-theory on genus-one fibrations}.
\bjtitle{JHEP}
\bvolume{08},
\bfpage{132}
(\byear{2014})
{\href{https://arxiv.org/abs/1401.7844}{{arXiv:1401.7844}}}
{[hep-th]}.
\doiurl{10.1007/JHEP08(2014)132}
\end{barticle}
\endbibitem

\bibitem{Weigand}
\begin{barticle}
\bauthor{\bsnm{Mayrhofer}, \binits{C.}},
\bauthor{\bsnm{Palti}, \binits{E.}},
\bauthor{\bsnm{Till}, \binits{O.}},
\bauthor{\bsnm{Weigand}, \binits{T.}}:
\batitle{On discrete symmetries and torsion homology in {F}-theory}.
\bjtitle{Journal of High Energy Physics}
\bvolume{2015}(\bissue{6}),
\bfpage{29}
(\byear{2015}).
\doiurl{10.1007/JHEP06(2015)029}
\end{barticle}
\endbibitem

\bibitem{Freed:2006yc}
\begin{barticle}
\bauthor{\bsnm{Freed}, \binits{D.S.}},
\bauthor{\bsnm{Moore}, \binits{G.W.}},
\bauthor{\bsnm{Segal}, \binits{G.}}:
\batitle{{Heisenberg Groups and Noncommutative Fluxes}}.
\bjtitle{Annals Phys.}
\bvolume{322},
\bfpage{236}--\blpage{285}
(\byear{2007})
{\href{https://arxiv.org/abs/hep-th/0605200}{{arXiv:hep-th/0605200}}}.
\doiurl{10.1016/j.aop.2006.07.014}
\end{barticle}
\endbibitem

\bibitem{Candelas:1987kf}
\begin{barticle}
\bauthor{\bsnm{Candelas}, \binits{P.}},
\bauthor{\bsnm{Dale}, \binits{A.M.}},
\bauthor{\bsnm{Lutken}, \binits{C.A.}},
\bauthor{\bsnm{Schimmrigk}, \binits{R.}}:
\batitle{Complete intersection {Calabi}-{Yau} manifolds}.
\bjtitle{Nucl. Phys. B}
\bvolume{298},
\bfpage{493}
(\byear{1988}).
\doiurl{10.1016/0550-3213(88)90352-5}
\end{barticle}
\endbibitem

\bibitem{Strominger:1995cz}
\begin{barticle}
\bauthor{\bsnm{Strominger}, \binits{A.}}:
\batitle{{Massless black holes and conifolds in string theory}}.
\bjtitle{Nucl. Phys. B}
\bvolume{451},
\bfpage{96}--\blpage{108}
(\byear{1995})
{\href{https://arxiv.org/abs/hep-th/9504090}{{arXiv:hep-th/9504090}}}.
\doiurl{10.1016/0550-3213(95)00287-3}
\end{barticle}
\endbibitem

\bibitem{C-TS}
\begin{bbook}
\bauthor{\bsnm{Colliot-Th\'{e}l\`ene}, \binits{J.-L.}},
\bauthor{\bsnm{Skorobogatov}, \binits{A.N.}}:
\bbtitle{The {B}rauer-{G}rothendieck Group}.
\bsertitle{Ergebnisse der Mathematik und ihrer Grenzgebiete. 3. Folge. A Series
  of Modern Surveys in Mathematics [Results in Mathematics and Related Areas.
  3rd Series. A Series of Modern Surveys in Mathematics]},
vol. \bseriesno{71},
p. \bfpage{453}.
\bpublisher{Springer},
\blocation{Berlin}
(\byear{2021}).
\doiurl{10.1007/978-3-030-74248-5}.
\burl{https://doi.org/10.1007/978-3-030-74248-5}
\end{bbook}
\endbibitem

\bibitem{Huang:2020dbh}
\begin{barticle}
\bauthor{\bsnm{Huang}, \binits{M.-X.}},
\bauthor{\bsnm{Katz}, \binits{S.}},
\bauthor{\bsnm{Klemm}, \binits{A.}}:
\batitle{{Towards refining the topological strings on compact Calabi-Yau
  3-folds}}.
\bjtitle{JHEP}
\bvolume{03},
\bfpage{266}
(\byear{2021})
{\href{https://arxiv.org/abs/2010.02910}{{arXiv:2010.02910}}}
{[hep-th]}.
\doiurl{10.1007/JHEP03(2021)266}
\end{barticle}
\endbibitem

\bibitem{Li:1998hba}
\begin{barticle}
\bauthor{\bsnm{Li}, \binits{A.-M.}},
\bauthor{\bsnm{Ruan}, \binits{Y.}}:
\batitle{Symplectic surgery and {G}romov-{W}itten invariants of {C}alabi-{Y}au
  3-folds}.
\bjtitle{Invent. Math.}
\bvolume{145}(\bissue{1}),
\bfpage{151}--\blpage{218}
(\byear{2001})
{\href{https://arxiv.org/abs/math/9803036}{{arXiv:math/9803036}}}.
\doiurl{10.1007/s002220100146}
\end{barticle}
\endbibitem

\bibitem{maulik2006gromov}
\begin{barticle}
\bauthor{\bsnm{Maulik}, \binits{D.}},
\bauthor{\bsnm{Nekrasov}, \binits{N.}},
\bauthor{\bsnm{Okounkov}, \binits{A.}},
\bauthor{\bsnm{Pandharipande}, \binits{R.}}:
\batitle{Gromov--{W}itten theory and {D}onaldson--{T}homas theory, {I}}.
\bjtitle{Compositio Mathematica}
\bvolume{142}(\bissue{5}),
\bfpage{1263}--\blpage{1285}
(\byear{2006})
{\href{https://arxiv.org/abs/math/0312059}{{arXiv:math/0312059}}}
\end{barticle}
\endbibitem

\bibitem{Batyrev:1993oya}
\begin{barticle}
\bauthor{\bsnm{Batyrev}, \binits{V.V.}}:
\batitle{{Dual polyhedra and mirror symmetry for Calabi-Yau hypersurfaces in
  toric varieties}}.
\bjtitle{J. Alg. Geom.}
\bvolume{3},
\bfpage{493}--\blpage{545}
(\byear{1994})
{\href{https://arxiv.org/abs/alg-geom/9310003}{{arXiv:alg-geom/9310003}}}
\end{barticle}
\endbibitem

\bibitem{HARRIS198471}
\begin{barticle}
\bauthor{\bsnm{Harris}, \binits{J.}},
\bauthor{\bsnm{Tu}, \binits{L.W.}}:
\batitle{On symmetric and skew-symmetric determinantal varieties}.
\bjtitle{Topology}
\bvolume{23}(\bissue{1}),
\bfpage{71}--\blpage{84}
(\byear{1984}).
\doiurl{10.1016/0040-9383(84)90026-0}
\end{barticle}
\endbibitem

\bibitem{Addington:2009thesis}
\begin{botherref}
\oauthor{\bsnm{Addington}, \binits{N.M.}}:
Spinor sheaves and complete intersections of quadrics.
PhD thesis,
University of Wisconsin-Madison
(2009).
Ph.D. thesis
\end{botherref}
\endbibitem

\bibitem{ACGH}
\begin{bbook}
\bauthor{\bsnm{Arbarello}, \binits{E.}},
\bauthor{\bsnm{Cornalba}, \binits{M.}},
\bauthor{\bsnm{Griffiths}, \binits{P.A.}},
\bauthor{\bsnm{Harris}, \binits{J.}}:
\bbtitle{Geometry of Algebraic Curves. {V}ol. {I}}.
\bsertitle{Grundlehren der mathematischen Wissenschaften [Fundamental
  Principles of Mathematical Sciences]},
vol. \bseriesno{267},
p. \bfpage{386}.
\bpublisher{Springer}, \blocation{???}
(\byear{1985}).
\doiurl{10.1007/978-1-4757-5323-3}.
\burl{https://doi.org/10.1007/978-1-4757-5323-3}
\end{bbook}
\endbibitem

\bibitem{GdJ}
\begin{botherref}
\oauthor{\bparticle{de} \bsnm{Jong}, \binits{A.J.}}:
A result of {G}abber.
preprint at \url{https://www.math.columbia.edu/~dejong/papers/2-gabber.pdf}
\end{botherref}
\endbibitem

\bibitem{schroer}
\begin{barticle}
\bauthor{\bsnm{Schr\"{o}er}, \binits{S.}}:
\batitle{Topological methods for complex-analytic {B}rauer groups}.
\bjtitle{Topology}
\bvolume{44}(\bissue{5}),
\bfpage{875}--\blpage{894}
(\byear{2005}).
\doiurl{10.1016/j.top.2005.02.005}
\end{barticle}
\endbibitem

\bibitem{Katz:2002gh}
\begin{barticle}
\bauthor{\bsnm{Katz}, \binits{S.H.}},
\bauthor{\bsnm{Sharpe}, \binits{E.}}:
\batitle{{D-branes, open string vertex operators, and Ext groups}}.
\bjtitle{Adv. Theor. Math. Phys.}
\bvolume{6},
\bfpage{979}--\blpage{1030}
(\byear{2003})
{\href{https://arxiv.org/abs/hep-th/0208104}{{arXiv:hep-th/0208104}}}.
\doiurl{10.4310/ATMP.2002.v6.n6.a1}
\end{barticle}
\endbibitem

\bibitem{lawson-michelsohn}
\begin{bbook}
\bauthor{\bsnm{Lawson}, \binits{H.B.} \bsuffix{Jr.}},
\bauthor{\bsnm{Michelsohn}, \binits{M.-L.}}:
\bbtitle{Spin Geometry}.
\bsertitle{Princeton Mathematical Series},
vol. \bseriesno{38},
p. \bfpage{427}.
\bpublisher{Princeton University Press, Princeton, NJ},
\blocation{Princeton, NJ}
(\byear{1989})
\end{bbook}
\endbibitem

\bibitem{MR717607}
\begin{bchapter}
\bauthor{\bsnm{Bryant}, \binits{R.L.}},
\bauthor{\bsnm{Griffiths}, \binits{P.A.}}:
\bctitle{Some observations on the infinitesimal period relations for regular
  threefolds with trivial canonical bundle}.
In: \bbtitle{Arithmetic and Geometry, {V}ol. {II}}.
\bsertitle{Progr. Math.},
vol. \bseriesno{36},
pp. \bfpage{77}--\blpage{102}.
\bpublisher{Birkh\"{a}user},
\blocation{Boston, MA}
(\byear{1983})
\end{bchapter}
\endbibitem

\bibitem{Bershadsky:1993ta}
\begin{barticle}
\bauthor{\bsnm{Bershadsky}, \binits{M.}},
\bauthor{\bsnm{Cecotti}, \binits{S.}},
\bauthor{\bsnm{Ooguri}, \binits{H.}},
\bauthor{\bsnm{Vafa}, \binits{C.}}:
\batitle{{Holomorphic anomalies in topological field theories}}.
\bjtitle{Nucl. Phys. B}
\bvolume{405},
\bfpage{279}--\blpage{304}
(\byear{1993})
{\href{https://arxiv.org/abs/hep-th/9302103}{{arXiv:hep-th/9302103}}}.
\doiurl{10.1016/0550-3213(93)90548-4}
\end{barticle}
\endbibitem

\bibitem{Yamaguchi:2004bt}
\begin{barticle}
\bauthor{\bsnm{Yamaguchi}, \binits{S.}},
\bauthor{\bsnm{Yau}, \binits{S.-T.}}:
\batitle{{Topological string partition functions as polynomials}}.
\bjtitle{JHEP}
\bvolume{07},
\bfpage{047}
(\byear{2004})
{\href{https://arxiv.org/abs/hep-th/0406078}{{arXiv:hep-th/0406078}}}.
\doiurl{10.1088/1126-6708/2004/07/047}
\end{barticle}
\endbibitem

\bibitem{Grimm:2007tm}
\begin{barticle}
\bauthor{\bsnm{Grimm}, \binits{T.W.}},
\bauthor{\bsnm{Klemm}, \binits{A.}},
\bauthor{\bsnm{Marino}, \binits{M.}},
\bauthor{\bsnm{Weiss}, \binits{M.}}:
\batitle{Direct integration of the topological string}.
\bjtitle{JHEP}
\bvolume{08},
\bfpage{058}
(\byear{2007})
{\href{https://arxiv.org/abs/hep-th/0702187}{{arXiv:hep-th/0702187}}}.
\doiurl{10.1088/1126-6708/2007/08/058}
\end{barticle}
\endbibitem

\bibitem{Alim:2007qj}
\begin{barticle}
\bauthor{\bsnm{Alim}, \binits{M.}},
\bauthor{\bsnm{Lange}, \binits{J.D.}}:
\batitle{Polynomial structure of the (open) topological string partition
  function}.
\bjtitle{JHEP}
\bvolume{10},
\bfpage{045}
(\byear{2007})
{\href{https://arxiv.org/abs/0708.2886}{{arXiv:0708.2886}}}
{[hep-th]}.
\doiurl{10.1088/1126-6708/2007/10/045}
\end{barticle}
\endbibitem

\bibitem{Ghoshal:1995wm}
\begin{barticle}
\bauthor{\bsnm{Ghoshal}, \binits{D.}},
\bauthor{\bsnm{Vafa}, \binits{C.}}:
\batitle{{C = 1 string as the topological theory of the conifold}}.
\bjtitle{Nucl. Phys. B}
\bvolume{453},
\bfpage{121}--\blpage{128}
(\byear{1995})
{\href{https://arxiv.org/abs/hep-th/9506122}{{arXiv:hep-th/9506122}}}.
\doiurl{10.1016/0550-3213(95)00408-K}
\end{barticle}
\endbibitem

\bibitem{Vafa:1995ta}
\begin{barticle}
\bauthor{\bsnm{Vafa}, \binits{C.}}:
\batitle{A stringy test of the fate of the conifold}.
\bjtitle{Nucl. Phys. B}
\bvolume{447},
\bfpage{252}--\blpage{260}
(\byear{1995})
{\href{https://arxiv.org/abs/hep-th/9505023}{{arXiv:hep-th/9505023}}}.
\doiurl{10.1016/0550-3213(95)00279-2}
\end{barticle}
\endbibitem

\bibitem{MR3965409}
\begin{bchapter}
\bauthor{\bsnm{Klemm}, \binits{A.}}:
\bctitle{The {B}-model approach to topological string theory on {C}alabi-{Y}au
  n-folds}.
In: \bbtitle{B-model {G}romov-{W}itten Theory}.
\bsertitle{Trends Math.},
pp. \bfpage{79}--\blpage{397}.
\bpublisher{Birkh\"{a}user/Springer, Cham},
\blocation{Boston}
(\byear{2018})
\end{bchapter}
\endbibitem

\bibitem{Antoniadis:1993ze}
\begin{barticle}
\bauthor{\bsnm{Antoniadis}, \binits{I.}},
\bauthor{\bsnm{Gava}, \binits{E.}},
\bauthor{\bsnm{Narain}, \binits{K.S.}},
\bauthor{\bsnm{Taylor}, \binits{T.R.}}:
\batitle{{Topological amplitudes in string theory}}.
\bjtitle{Nucl. Phys. B}
\bvolume{413},
\bfpage{162}--\blpage{184}
(\byear{1994})
{\href{https://arxiv.org/abs/hep-th/9307158}{{arXiv:hep-th/9307158}}}.
\doiurl{10.1016/0550-3213(94)90617-3}
\end{barticle}
\endbibitem

\bibitem{MR1728879}
\begin{barticle}
\bauthor{\bsnm{Faber}, \binits{C.}},
\bauthor{\bsnm{Pandharipande}, \binits{R.}}:
\batitle{Hodge integrals and {G}romov-{W}itten theory}.
\bjtitle{Invent. Math.}
\bvolume{139}(\bissue{1}),
\bfpage{173}--\blpage{199}
(\byear{2000}).
\doiurl{10.1007/s002229900028}
\end{barticle}
\endbibitem

\bibitem{BHTS}
\begin{botherref}
\oauthor{\bsnm{Alexandrov}, \binits{S.}},
\oauthor{\bsnm{Feyzbakhsh}, \binits{S.}},
\oauthor{\bsnm{Klemm}, \binits{A.}},
\oauthor{\bsnm{Pioline}, \binits{B.}},
\oauthor{\bsnm{Schimannek}, \binits{T.}}:
{Quantum geometry from stability and modularity (work in progress)}
\end{botherref}
\endbibitem

\bibitem{MR4012553}
\begin{barticle}
\bauthor{\bsnm{Kerr}, \binits{M.}},
\bauthor{\bsnm{Pearlstein}, \binits{G.J.}},
\bauthor{\bsnm{Robles}, \binits{C.}}:
\batitle{Polarized relations on horizontal {$\rm SL(2)$}'s}.
\bjtitle{Doc. Math.}
\bvolume{24},
\bfpage{1295}--\blpage{1360}
(\byear{2019})
\end{barticle}
\endbibitem

\bibitem{Grimm:2018ohb}
\begin{barticle}
\bauthor{\bsnm{Grimm}, \binits{T.W.}},
\bauthor{\bsnm{Palti}, \binits{E.}},
\bauthor{\bsnm{Valenzuela}, \binits{I.}}:
\batitle{Infinite distances in field space and massless towers of states}.
\bjtitle{JHEP}
\bvolume{08},
\bfpage{143}
(\byear{2018})
{\href{https://arxiv.org/abs/1802.08264}{{arXiv:1802.08264}}}
{[hep-th]}.
\doiurl{10.1007/JHEP08(2018)143}
\end{barticle}
\endbibitem

\bibitem{Joshi:2019nzi}
\begin{barticle}
\bauthor{\bsnm{Joshi}, \binits{A.}},
\bauthor{\bsnm{Klemm}, \binits{A.}}:
\batitle{Swampland distance conjecture for one-parameter {C}alabi-{Y}au
  threefolds}.
\bjtitle{JHEP}
\bvolume{08},
\bfpage{086}
(\byear{2019})
{\href{https://arxiv.org/abs/1903.00596}{{arXiv:1903.00596}}}
{[hep-th]}.
\doiurl{10.1007/JHEP08(2019)086}
\end{barticle}
\endbibitem

\bibitem{MR3822913}
\begin{bchapter}
\bauthor{\bparticle{van} \bsnm{Straten}, \binits{D.}}:
\bctitle{Calabi-{Y}au operators}.
In: \bbtitle{Uniformization, {R}iemann-{H}ilbert Correspondence, {C}alabi-{Y}au
  Manifolds \& {P}icard-{F}uchs equations}.
\bsertitle{Adv. Lect. Math. (ALM)},
vol. \bseriesno{42},
pp. \bfpage{401}--\blpage{451}.
\bpublisher{Int. Press},
\blocation{Somerville, MA}
(\byear{2018})
\end{bchapter}
\endbibitem

\bibitem{Bonisch:2022mgw}
\begin{botherref}
\oauthor{\bsnm{B\"onisch}, \binits{K.}},
\oauthor{\bsnm{Klemm}, \binits{A.}},
\oauthor{\bsnm{Scheidegger}, \binits{E.}},
\oauthor{\bsnm{Zagier}, \binits{D.}}:
D-brane masses at special fibres of hypergeometric families of {C}alabi-{Y}au
  threefolds, modular forms, and periods
(2022)
{\href{https://arxiv.org/abs/2203.09426}{{arXiv:2203.09426}}}
{[hep-th]}
\end{botherref}
\endbibitem

\bibitem{Aspinwall:2004jr}
\begin{bchapter}
\bauthor{\bsnm{Aspinwall}, \binits{P.S.}}:
\bctitle{{D-branes on Calabi-Yau manifolds}}.
In: \bbtitle{Theoretical {A}dvanced {S}tudy {I}nstitute in {E}lementary
  {P}article {P}hysics ({TASI} 2003): {R}ecent Trends in String Theory},
pp. \bfpage{1}--\blpage{152}
(\byear{2004}).
\doiurl{10.1142/9789812775108_0001}
\end{bchapter}
\endbibitem

\bibitem{Cota:2019cjx}
\begin{barticle}
\bauthor{\bsnm{Cota}, \binits{C.F.}},
\bauthor{\bsnm{Klemm}, \binits{A.}},
\bauthor{\bsnm{Schimannek}, \binits{T.}}:
\batitle{{Topological strings on genus one fibered Calabi-Yau 3-folds and
  string dualities}}.
\bjtitle{JHEP}
\bvolume{11},
\bfpage{170}
(\byear{2019})
{\href{https://arxiv.org/abs/1910.01988}{{arXiv:1910.01988}}}
{[hep-th]}.
\doiurl{10.1007/JHEP11(2019)170}
\end{barticle}
\endbibitem

\bibitem{Gerhardus:2016iot}
\begin{barticle}
\bauthor{\bsnm{Gerhardus}, \binits{A.}},
\bauthor{\bsnm{Jockers}, \binits{H.}}:
\batitle{{Quantum periods of Calabi\textendash{}Yau fourfolds}}.
\bjtitle{Nucl. Phys. B}
\bvolume{913},
\bfpage{425}--\blpage{474}
(\byear{2016})
{\href{https://arxiv.org/abs/1604.05325}{{arXiv:1604.05325}}}
{[hep-th]}.
\doiurl{10.1016/j.nuclphysb.2016.09.021}
\end{barticle}
\endbibitem

\bibitem{IRITANI20091016}
\begin{barticle}
\bauthor{\bsnm{Iritani}, \binits{H.}}:
\batitle{An integral structure in quantum cohomology and mirror symmetry for
  toric orbifolds}.
\bjtitle{Advances in Mathematics}
\bvolume{222}(\bissue{3}),
\bfpage{1016}--\blpage{1079}
(\byear{2009}).
\doiurl{10.1016/j.aim.2009.05.016}
\end{barticle}
\endbibitem

\bibitem{Halverson:2013qca}
\begin{barticle}
\bauthor{\bsnm{Halverson}, \binits{J.}},
\bauthor{\bsnm{Jockers}, \binits{H.}},
\bauthor{\bsnm{Lapan}, \binits{J.M.}},
\bauthor{\bsnm{Morrison}, \binits{D.R.}}:
\batitle{Perturbative corrections to {K}aehler moduli spaces}.
\bjtitle{Commun. Math. Phys.}
\bvolume{333}(\bissue{3}),
\bfpage{1563}--\blpage{1584}
(\byear{2015})
{\href{https://arxiv.org/abs/1308.2157}{{arXiv:1308.2157}}}
{[hep-th]}.
\doiurl{10.1007/s00220-014-2157-z}
\end{barticle}
\endbibitem

\bibitem{Knapp:2020oba}
\begin{barticle}
\bauthor{\bsnm{Knapp}, \binits{J.}},
\bauthor{\bsnm{Romo}, \binits{M.}},
\bauthor{\bsnm{Scheidegger}, \binits{E.}}:
\batitle{{D-Brane Central Charge and Landau\textendash{}Ginzburg Orbifolds}}.
\bjtitle{Commun. Math. Phys.}
\bvolume{384}(\bissue{1}),
\bfpage{609}--\blpage{697}
(\byear{2021})
{\href{https://arxiv.org/abs/2003.00182}{{arXiv:2003.00182}}}
{[hep-th]}.
\doiurl{10.1007/s00220-021-04042-w}
\end{barticle}
\endbibitem

\bibitem{schubert:1879}
\begin{bbook}
\bauthor{\bsnm{Schubert}, \binits{H.}}:
\bbtitle{Kalk\"{u}l der Abz\"{a}hlenden {G}eometrie},
p. \bfpage{349}.
\bpublisher{Springer},
\blocation{Berlin-New York}
(\byear{1979}).
\bcomment{Reprint of the 1879 original, With an introduction by Steven L.
  Kleiman}
\end{bbook}
\endbibitem

\bibitem{PT}
\begin{barticle}
\bauthor{\bsnm{Pandharipande}, \binits{R.}},
\bauthor{\bsnm{Thomas}, \binits{R.P.}}:
\batitle{Stable pairs and {BPS} invariants}.
\bjtitle{J. Amer. Math. Soc.}
\bvolume{23}(\bissue{1}),
\bfpage{267}--\blpage{297}
(\byear{2010})
{\href{https://arxiv.org/abs/0711.3899}{{arXiv:0711.3899}}}.
\doiurl{10.1090/S0894-0347-09-00646-8}
\end{barticle}
\endbibitem

\bibitem{Knapp:2021vkm}
\begin{botherref}
\oauthor{\bsnm{Knapp}, \binits{J.}},
\oauthor{\bsnm{Scheidegger}, \binits{E.}},
\oauthor{\bsnm{Schimannek}, \binits{T.}}:
{On genus one fibered Calabi-Yau threefolds with 5-sections}
(2021)
{\href{https://arxiv.org/abs/2107.05647}{{arXiv:2107.05647}}}
{[hep-th]}
\end{botherref}
\endbibitem

\bibitem{Harris}
\begin{bbook}
\bauthor{\bsnm{Harris}, \binits{J.}}:
\bbtitle{Algebraic Geometry}.
\bsertitle{Graduate Texts in Mathematics},
vol. \bseriesno{133},
p. \bfpage{328}.
\bpublisher{Springer},
\blocation{New York}
(\byear{1995}).
\bcomment{A first course, Corrected reprint of the 1992 original}
\end{bbook}
\endbibitem

\bibitem{Rocek:1991ze}
\begin{barticle}
\bauthor{\bsnm{Rocek}, \binits{M.}}:
\batitle{{Modified Calabi-Yau manifolds with torsion}}.
\bjtitle{AMS/IP Stud. Adv. Math.}
\bvolume{9},
\bfpage{421}--\blpage{429}
(\byear{1998})
\end{barticle}
\endbibitem

\bibitem{Gurrieri:2002wz}
\begin{barticle}
\bauthor{\bsnm{Gurrieri}, \binits{S.}},
\bauthor{\bsnm{Louis}, \binits{J.}},
\bauthor{\bsnm{Micu}, \binits{A.}},
\bauthor{\bsnm{Waldram}, \binits{D.}}:
\batitle{{Mirror symmetry in generalized Calabi-Yau compactifications}}.
\bjtitle{Nucl. Phys. B}
\bvolume{654},
\bfpage{61}--\blpage{113}
(\byear{2003})
{\href{https://arxiv.org/abs/hep-th/0211102}{{arXiv:hep-th/0211102}}}.
\doiurl{10.1016/S0550-3213(03)00045-2}
\end{barticle}
\endbibitem

\bibitem{Hitchin:2003cxu}
\begin{barticle}
\bauthor{\bsnm{Hitchin}, \binits{N.}}:
\batitle{{Generalized Calabi-Yau manifolds}}.
\bjtitle{Quart. J. Math.}
\bvolume{54},
\bfpage{281}--\blpage{308}
(\byear{2003})
{\href{https://arxiv.org/abs/math/0209099}{{arXiv:math/0209099}}}.
\doiurl{10.1093/qjmath/54.3.281}
\end{barticle}
\endbibitem

\bibitem{STROMINGER1986253}
\begin{barticle}
\bauthor{\bsnm{Strominger}, \binits{A.}}:
\batitle{Superstrings with torsion}.
\bjtitle{Nuclear Physics B}
\bvolume{274}(\bissue{2}),
\bfpage{253}--\blpage{284}
(\byear{1986}).
\doiurl{10.1016/0550-3213(86)90286-5}
\end{barticle}
\endbibitem

\bibitem{mnop1}
\begin{barticle}
\bauthor{\bsnm{Maulik}, \binits{D.}},
\bauthor{\bsnm{Nekrasov}, \binits{N.}},
\bauthor{\bsnm{Okounkov}, \binits{A.}},
\bauthor{\bsnm{Pandharipande}, \binits{R.}}:
\batitle{Gromov–{W}itten theory and {D}onaldson–{T}homas theory, {I}}.
\bjtitle{Compositio Mathematica}
\bvolume{142}(\bissue{5}),
\bfpage{1263}--\blpage{1285}
(\byear{2006})
{\href{https://arxiv.org/abs/math/0312059}{{arXiv:math/0312059}}}.
\doiurl{10.1112/S0010437X06002302}
\end{barticle}
\endbibitem

\bibitem{fjrw1}
\begin{barticle}
\bauthor{\bsnm{Fan}, \binits{H.}},
\bauthor{\bsnm{Jarvis}, \binits{T.}},
\bauthor{\bsnm{Ruan}, \binits{Y.}}:
\batitle{The {W}itten equation, mirror symmetry, and quantum singularity
  theory}.
\bjtitle{Ann. of Math. (2)}
\bvolume{178}(\bissue{1}),
\bfpage{1}--\blpage{106}
(\byear{2013}).
\doiurl{10.4007/annals.2013.178.1.1}
\end{barticle}
\endbibitem

\bibitem{zbMATH05782775}
\begin{barticle}
\bauthor{\bsnm{Chiodo}, \binits{A.}},
\bauthor{\bsnm{Ruan}, \binits{Y.}}:
\batitle{Landau-{Ginzburg}/{Calabi}-{Yau} correspondence for quintic
  three-folds via symplectic transformations}.
\bjtitle{Invent. Math.}
\bvolume{182}(\bissue{1}),
\bfpage{117}--\blpage{165}
(\byear{2010}).
\doiurl{10.1007/s00222-010-0260-0}
\end{barticle}
\endbibitem

\bibitem{cir14}
\begin{barticle}
\bauthor{\bsnm{Chiodo}, \binits{A.}},
\bauthor{\bsnm{Iritani}, \binits{H.}},
\bauthor{\bsnm{Ruan}, \binits{Y.}}:
\batitle{Landau-{G}inzburg/{C}alabi-{Y}au correspondence, global mirror
  symmetry and {O}rlov equivalence}.
\bjtitle{Publications math{\'e}matiques de l'IH{\'E}S}
\bvolume{119}(\bissue{1}),
\bfpage{127}--\blpage{216}
(\byear{2014}).
\doiurl{10.1007/s10240-013-0056-z}
\end{barticle}
\endbibitem

\bibitem{Clader:2013tn}
\begin{barticle}
\bauthor{\bsnm{Clader}, \binits{E.}}:
\batitle{{Landau\textendash{}Ginzburg/Calabi\textendash{}Yau correspondence for
  the complete intersections $X_{3,3}$ and $X_{2,2,2,2}$}}.
\bjtitle{Adv. Math.}
\bvolume{307},
\bfpage{1}--\blpage{52}
(\byear{2017})
{\href{https://arxiv.org/abs/1301.5530}{{arXiv:1301.5530}}}
{[math.AG]}.
\doiurl{10.1016/j.aim.2016.11.010}
\end{barticle}
\endbibitem

\bibitem{Erkinger:2022sqs}
\begin{botherref}
\oauthor{\bsnm{Erkinger}, \binits{D.}},
\oauthor{\bsnm{Knapp}, \binits{J.}}:
{On genus-0 invariants of Calabi-Yau hybrid models}
(2022)
{\href{https://arxiv.org/abs/2210.01226}{{arXiv:2210.01226}}}
{[hep-th]}
\end{botherref}
\endbibitem

\bibitem{Faraggi:2022gkt}
\begin{botherref}
\oauthor{\bsnm{Faraggi}, \binits{A.E.}},
\oauthor{\bsnm{Nibbelink}, \binits{S.G.}},
\oauthor{\bsnm{Heredia}, \binits{M.H.}}:
{The fate of discrete torsion on resolved heterotic Z2xZ2 orbifolds using (0,2)
  GLSMs}
(2022)
{\href{https://arxiv.org/abs/2211.01397}{{arXiv:2211.01397}}}
{[hep-th]}
\end{botherref}
\endbibitem

\bibitem{aesz}
\begin{botherref}
\oauthor{\bsnm{Almkvist}, \binits{G.}},
\oauthor{\bparticle{van} \bsnm{Enckevort}, \binits{C.}},
\oauthor{\bparticle{van} \bsnm{Straten}, \binits{D.}},
\oauthor{\bsnm{Zudilin}, \binits{W.}}:
Tables of {C}alabi--{Y}au equations
(2005)
{\href{https://arxiv.org/abs/math/0507430}{{arXiv:math/0507430}}}
{[math.AG]}
\end{botherref}
\endbibitem

\bibitem{Katz:1986}
\begin{barticle}
\bauthor{\bsnm{Katz}, \binits{S.}}:
\batitle{Lines on complete intersection threefolds with {$K=0$}}.
\bjtitle{Math. Z.}
\bvolume{191}(\bissue{2}),
\bfpage{293}--\blpage{296}
(\byear{1986}).
\doiurl{10.1007/BF01164033}
\end{barticle}
\endbibitem

\bibitem{AW}
\begin{barticle}
\bauthor{\bsnm{Addington}, \binits{N.}},
\bauthor{\bsnm{Wray}, \binits{A.}}:
\batitle{Twisted {Fourier}-{Mukai} partners of {Enriques} surfaces}.
\bjtitle{Math. Z.}
\bvolume{297}(\bissue{3-4}),
\bfpage{1239}--\blpage{1247}
(\byear{2021})
{\href{https://arxiv.org/abs/1803.03250}{{arXiv:1803.03250}}}.
\doiurl{10.1007/s00209-020-02555-z}
\end{barticle}
\endbibitem

\bibitem{Hosono:2008np}
\begin{barticle}
\bauthor{\bsnm{Hosono}, \binits{S.}}:
\batitle{{BCOV ring and holomorphic anomaly equation}}.
\bjtitle{Adv. Stud. Pure Math.}
\bvolume{59},
\bfpage{79}
(\byear{2008})
{\href{https://arxiv.org/abs/0810.4795}{{arXiv:0810.4795}}}
{[math.AG]}
\end{barticle}
\endbibitem

\bibitem{Huang:2015sta}
\begin{barticle}
\bauthor{\bsnm{Huang}, \binits{M.-x.}},
\bauthor{\bsnm{Katz}, \binits{S.}},
\bauthor{\bsnm{Klemm}, \binits{A.}}:
\batitle{Topological string on elliptic {CY} 3-folds and the ring of {J}acobi
  forms}.
\bjtitle{JHEP}
\bvolume{10},
\bfpage{125}
(\byear{2015})
{\href{https://arxiv.org/abs/1501.04891}{{arXiv:1501.04891}}}
{[hep-th]}.
\doiurl{10.1007/JHEP10(2015)125}
\end{barticle}
\endbibitem

\bibitem{Kanazawa:2012xya}
\begin{barticle}
\bauthor{\bsnm{Kanazawa}, \binits{A.}}:
\batitle{{Pfaffian Calabi-Yau threefolds and mirror symmetry}}.
\bjtitle{Commun. Num. Theor. Phys.}
\bvolume{6},
\bfpage{661}--\blpage{696}
(\byear{2012})
{\href{https://arxiv.org/abs/1006.0223}{{arXiv:1006.0223}}}.
\doiurl{10.4310/CNTP.2012.v6.n3.a3}
\end{barticle}
\endbibitem

\bibitem{Tonoli:2004}
\begin{barticle}
\bauthor{\bsnm{Tonoli}, \binits{F.}}:
\batitle{Construction of {C}alabi-{Y}au 3-folds in {${\mathbb P}^6$}}.
\bjtitle{J. Algebraic Geom.}
\bvolume{13}(\bissue{2}),
\bfpage{209}--\blpage{232}
(\byear{2004}).
\doiurl{10.1090/S1056-3911-03-00371-0}
\end{barticle}
\endbibitem

\end{thebibliography}
\end{document}